\documentclass{AIMS}
\usepackage{amsmath}
  \usepackage{paralist}
  \usepackage{graphics} 
  \usepackage{epsfig} 
\usepackage{graphicx}  \usepackage{epstopdf}
 \usepackage[colorlinks=true]{hyperref}
\hypersetup{urlcolor=blue, citecolor=red}
\def\currentvolume{X}

    \def\ppages{X--XX}

\theoremstyle{definition}

\usepackage{mathtools,amssymb,bm}
\usepackage{subfigmat}
\usepackage{wrapfig}
\usepackage{float}
\usepackage{mathtools}
\usepackage{multirow}
\usepackage{xcolor}
\usepackage{mwe}
\usepackage{mathrsfs} 
\usepackage{paralist}
\usepackage{blindtext}
\usepackage[ruled,vlined]{algorithm2e}
\usepackage[noend]{algpseudocode}
\title[Bayesian Multiscale Deep Learning Framework] 
      {A Bayesian multiscale deep learning framework for flows in random media}
\author[Govinda Anantha Padmanabha and Nicholas Zabaras]{}
 \email{ganantha@nd.edu}
 \email{nzabaras@gmail.com}
\thanks{$^*$ Corresponding author: Nicholas Zabaras}
\begin{document}
\maketitle
\centerline{\scshape Govinda Anantha Padmanabha}
\medskip
{\footnotesize
 \centerline{Scientific Computing and Artificial Intelligence (SCAI) Laboratory}
   \centerline{311I Cushing Hall, University of Notre Dame, Notre Dame, IN 46556, USA}
} 
\medskip
\centerline{\scshape Nicholas Zabaras$^*$}
\medskip
{\footnotesize
 \centerline{Scientific Computing and Artificial Intelligence (SCAI) Laboratory}
   \centerline{311I Cushing Hall, University of Notre Dame, Notre Dame, IN 46556, USA}
}
\bigskip
\begin{abstract}
Fine-scale simulation of complex systems governed by multiscale partial differential equations (PDEs) is computationally expensive and various multiscale methods  have been developed for addressing such problems. In addition, it is challenging to develop accurate surrogate and uncertainty quantification models for high-dimensional problems  governed by stochastic multiscale PDEs using limited training data.
In this work to address these challenges, we introduce a novel hybrid deep-learning and multiscale approach for stochastic multiscale PDEs with limited training data.  For demonstration purposes, we focus on a porous media flow problem. We use an image-to-image supervised deep learning model to learn the mapping between the input permeability field and the multiscale basis functions. We introduce a Bayesian approach to this hybrid framework to allow us to perform uncertainty quantification and propagation tasks. The performance of this hybrid approach is evaluated with varying intrinsic dimensionality of the permeability field. Numerical results indicate that the hybrid network can efficiently predict well for high-dimensional inputs.
\end{abstract}
%
%
%
%
\section{Introduction} 
Uncertainty quantification is an increasingly vital aspect of various research areas of engineering and the sciences, e.g., for robust complex engineering designs or weather forecasting. Predicting fluid flows through porous media is another area where quantifying the uncertainty in permeability and propagating it to the system response is important. For example, predicting flow through an oil reservoir is a rather challenging problem. 
\par
One of the simplest ways to solve the forward uncertainty quantification (UQ) problem is using the Monte Carlo (MC) method. However, this method is not suitable for problems with expensive computer codes since it requires a large number of samples to obtain convergent statistics. An approach to address this problem is by developing a surrogate model for uncertainty quantification and uncertainty propagation~\cite{zhu2018bayesian,tripathy2018deep,mo2019deep}. Zhu and Zabaras~\cite{zhu2018bayesian} developed surrogate models using a dense encoder-decoder network (DenseED)~\cite{huang2017densely} for uncertainty quantification and propagation in problems governed by stochastic PDEs with high-dimensional stochastic input. They have adopted an image-to-image regression approach to solve the problem with a moderate size dataset. Another recent approach is to combine aspects of multiscale methods with deep-learning techniques~\cite{chan2018machine, wang2020deep}.
\par
In a multiscale method, the fine-scale problem is decomposed into a coarse-scale problem and then numerically solved as a local problem to obtain the basis functions. Then the local solution is mapped to the fine-scale to obtain the approximate fine-scale solution. These mappings between the fine-scale to coarse-scale or coarse-scale to fine-scale are obtained by using the restriction and prolongation operators, respectively. The prolongation operator is constructed by combining the basis functions, and the restriction operator is constructed either by control volume summation or a Galerkin operator formulation~\cite{moyner2016multiscale}. The most commonly used multiscale methods for solving flow through porous media include the multi-scale finite volume method (MsFVM)~\cite{jenny2003multi}, the multi-scale mixed finite element method (MsMFEM)~\cite{aarnes2008mixed} and the multiscale restricted-smoothed basis (MsRSB)~\cite{moyner2016multiscale, shah2016multiscale}. M{\o}yner and Lie~\cite{moyner2016multiscale} introduced a multiscale restricted-smoothed basis method for porous media flow where an iterative process is adopted to solve the local problem. In this work, we implement the MsRSB method in our hybrid framework. The second component of our hybrid framework is deep neural networks. 
\par
 Deep neural networks (DNN) are widely used for surrogate modeling and have been applied to various fields in science and engineering~\cite{goodfellow2016deep, lecun2015deep}. For example, flow through porous media~\cite{zhu2018bayesian, tripathy2018deep, mo2019deep, zhu2019physics} or Reynolds-Averaged Navier-Stokes (RANS) simulations~\cite{geneva2019quantifying, thuerey2020deep}. In the deep learning community, there are various convolutional neural network architectures (CNN)~\cite{zeiler2014visualizing} that have been successfully implemented to solve these problems~\cite{radford2015unsupervised, mo2020integration}. For example, in the computer vision field, various networks such as AlexNET~\cite{krizhevsky2017imagenet}, DenseNet~\cite{huang2017densely}, ResNet~\cite{he2016deep}, and many more have been successfully implemented for supervised and unsupervised tasks~\cite{radford2015unsupervised, alom2018history}. Recently, DenseNet~\cite{huang2017densely} has been successfully implemented to capture the complex nonlinear mapping between a high-dimensional input and outputs~\cite{mo2019deep, geneva2020modeling} and was also extended to a Bayesian formulation for uncertainty quantification and propagation tasks~\cite{zhu2018bayesian}. In a Bayesian neural network, the network parameters, namely weights and biases, are considered as random variables, and the task is to carry out Bayesian inference on those parameters conditioned on the data~\cite{blundell2015weight, kingma2015variational, gal2016dropout, hernandez2015probabilistic}. 
 \par
In this work, we combine the multiscale and deep learning approaches. Several works related to this hybrid framework have been explored~\cite{chan2018machine, wang2019prediction, zhang2019deep}. For example, Chan and Elsheikh~\cite{chan2018machine} developed a framework where the basis functions are trained using a fully-connected network (FCN) for flow through porous media. Wang et al.~\cite{wang2019prediction} developed a hybrid approach for flow problems that involve heterogeneous and high-contrast porous media by combining the generalized multiscale finite element method (MsFEM) and a fully-connected network. The works mentioned above emphasize fast computation using deep learning techniques, however, they are not focused on quantifying model uncertainty.
\par 
The novel contributions of this work are as follows. First, we make use of a data-driven model with a hybrid framework by combining multiscale methods and deep learning to perform uncertainty quantification and propagation in problems governed by stochastic PDEs. In the MsRSB method, the fine-scale problem is decomposed into a coarse-scale problem, and then the local problem is solved in an iterative process to obtain the basis functions. Instead, we use the dense convolutional encoder-decoder network (Dense-ED) to learn the basis functions. We define the loss function using the predicted pressure from the hybrid framework and the fine-scale pressure. We refer to this model as Hybrid Multiscale DenseED (HM-DenseED). A Bayesian approach is then introduced to the hybrid framework for uncertainty quantification and propagation for a two-dimensional, single-phase, steady-state flow through a random permeability field. In this work, we evaluated our model using datasets produced with permeability of varying intrinsic dimensionality: $100$, $1000$, and $16384$ (for Gaussian random fields, GRFs) and channelized fields. The intrinsic dimensionality is obtained using the Karhunen-Lo\`{e}ve expansion (KLE)~\cite{van2009dimensionality} of the input permeability field. Note that this model reduction for GRFs is only introduced to produce training data and for exploring the performance of the model for data of given dimensionality. Finally, we compare the HM-DenseED hybrid method with the DenseED surrogate model in~\cite{zhu2018bayesian}. For a limited size dataset, we will show that the hybrid approach leads to  more accurate predictions.
\par
The paper is structured as follows. In Section~\ref{sec:PD}, we provide the problem definition. In Section~\ref{sec:SL2}, we review the details of the fine-scale system, the multiscale method, the DenseED network, the hybrid multiscale-neural network model and its Bayesian formulation.  In Section~\ref{sec:SL3}, we present the results obtained using the hybrid approach for both non-Bayesian and Bayesian approaches and different sizes of training data: $32$, $64$, $96$ (for KLE$-100$); $64$, $96$, $128$ (for KLE$-1000$); $96$, $128$, $160$ (for KLE$-16384$) and $160$ (for channelized flow). Finally, we conclude with a summary of the work in Section~\ref{sec:Conclusions}. 
\section{Problem definition}
\label{sec:PD}
Consider a two-dimensional, single-phase, steady-state porous media flow~\cite{aarnes2007modelling, jenny2003multi, wan2013probabilistic, zhu2019physics}. The relationship between the pressure field $p$, velocity field $\bm{u}$ and permeability $K$ is given by Darcy's law. For a unit square spatial domain $\Omega = [0,1]^2$, the governing equations considered are given as~\cite{aarnes2007modelling}: 
\begin{eqnarray}
    \bm{u}(s) &=& -K(s)\triangledown p(s),~s \in {\Omega}, \label{Eq:DarcyA}\\
    \triangledown \cdot \bm{u}(s) &=& q(s),~s \in {\Omega}, \label{Eq:DarcyB}
    \end{eqnarray}
where $q$ is a source term. In addition,  appropriate boundary conditions are applied to ensure the well-posedness of the boundary value problem.  
\par
Combining Eqs.~(\ref{Eq:DarcyA}) and~(\ref{Eq:DarcyB})  allows us to write the pressure field equation over the domain $\Omega$ as:
\begin{equation}
    -\triangledown \cdot(K(s)\triangledown p(s)) = q(s), \hspace{1cm} s \in {\Omega}. \label{Eq:Darcy}
\end{equation}
The domain $\Omega$ is discretized with non-overlapping elements $f_i$ to obtain the fine-grid $\{\Omega_i\}_{i=1}^{n_f}$, where $n_f$ is the number of fine-cells and $f_i$ is the fine cell. The corresponding discretized version of the permeability is given by: $\bm{x} = [K(s_1), K(s_2),\ldots,K(s_{n_f})]$. Since no dimensionality reduction is applied beyond the representation on the fine-grid, we consider $n_f$ to define the dimensionality of the permeability field (input space). We denote $p(s)$ as the solution of Eq.~(\ref{Eq:Darcy}) with the corresponding discretized version  given by $\bm{p_{f}} = [p(s_1),p(s_2),\ldots,p(s_{n_f})]$.  To solve the above defined problem, one can implement finite-volume discretization of Eq.~(\ref{Eq:Darcy}) to obtain  $\bm{p_{f}}$ as  the solution of a linear system of equations: 
\begin{equation}
\label{Eq:FS_eq}
    \bm{A \thinspace p_{f}} = \bm{q},
\end{equation} 
where $\bm{A}$ is a coefficient matrix  (also called matrix$-\bm{A} \in \mathbb{R}^{n_f \times n_f}$),  $\bm{q} \in \mathbb{R}^{n_f\times 1}$  is the source term and $\bm{p_{f}} \in \mathbb{R}^{n_f \times 1}$ is the discrete pressure solution (taken as the primary output variable for our system). Once the pressure is computed, the flux is constructed as follows:  
\begin{equation}
    v_{l m}=-T_{l m}\left(\left(\bm{p}_{f}\right)_{l}-\left(\bm{p}_{f}\right)_{m}\right),
    \label{Eq:Flux}
\end{equation}
where $v_{l m}$ is the flux, and $T_{l m}$ is the transmissibility matrix that is related to the interface between each pair of cells $l$ and $m$. In order to solve Eq.~(\ref{Eq:FS_eq}) for $\bm{p}_{f}$, we need to invert the matrix $\bm{A}$. If the dimension of the input field $n_f$ is high,  inverting $\bm{A}$ and evaluating the pressure is computationally expensive. One  possible way to address this is by using a multiscale approach~\cite{moyner2016multiscale,shah2016multiscale,jenny2003multi}.   
\par
\begin{figure}[H]
\begin{center}
	\centering
	\includegraphics[width=0.65\linewidth]{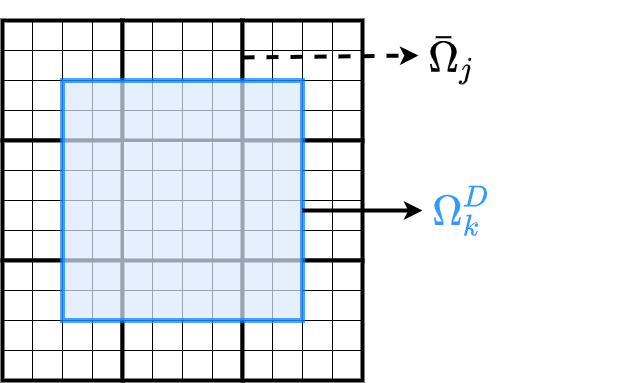}
	\caption{Schematic of the fine-scale and coarse-scale grids. Thick lines represent the primal coarse-grid $\bar{\Omega}_j$, and the blue line indicates a block of the dual coarse-grid $\Omega_k^D$. Thin lines define the fine-scale elements $f_i$ that constitute the fine-grid   $\{\Omega_i\}_{i=1}^{n_f}$.}
	\label{fig:PD_image}
  \end{center}
\end{figure}

 
In a multiscale method, the fine-scale problem as described in Eq.~(\ref{Eq:FS_eq}) is decomposed into local problems using a coarse-grid discretization. Two distinct  coarse-grids are defined: the primal coarse-grid $\{\bar{\Omega}_{j}\}_{j=1}^{n_c}$, where $n_c$ is the number of primal coarse-blocks and the dual coarse-grid $\{\Omega_k^D\}_{k=1}^{m}$, where $m$ is the number of dual coarse-blocks. Fig.~\ref{fig:PD_image} shows a schematic of the primal coarse-blocks and the dual coarse-grid. The primal coarse-grid is constructed as the coarse-scale control volume, whereas the dual coarse-grid is defined by overlapping coarse-blocks on which the local problem will be solved to obtain the multiscale basis functions. Details regarding the steps involved in solving the local problem are given in Section~\ref{sec:formulation}.

\par
Our focus in this work is to address the solution of stochastic multiscale PDEs~\cite{wan2013probabilistic, aarnes2007modelling, jenny2003multi}. In our particular flow in porous media problem this refers to solving Eq.~(\ref{Eq:Darcy}) for a random permeability e.g. a log-permeability represented by a Gaussian Random Field (GRF)~\cite{zhu2018bayesian}. We will focus on the development of surrogate models that are capable of producing estimates of the solution $\bm{p}_{f}$ for any realization of the input field $\bm{x}$ from the underlying input probability model. Such supervised learning models will need to be trained with a collection of input/output data of the form $\mathcal{D}=\{\bm{x}^{(i)},\bm{p}_{f}^{(i)}\}$ obtained from repeated solution of the multiscale PDE in Eq.~(\ref{Eq:Darcy}). Rather than developing a surrogate model that directly maps $\bm{x}$ to $\bm{p}_{f}$ (as for example our work in~\cite{zhu2018bayesian}), we will develop a supervised deep learning based surrogate that learns the multiscale basis functions (which in many ways is one of the most computational expensive components of a multiscale method) and then integrate this model with the remaining components of a multiscale method. This hybrid methodology combines the efficient training and generalization properties of deep learning with the accuracy guarantees of multiscale methods. 
 The developed hybrid surrogate method  maps  the input permeability field $\bm{K}$ to the pressure field $\bm{p}_{f}$ and thus via Eq.~(\ref{Eq:Flux}) to the velocity field $\bm{u}$. 
We will next present the details of non-Bayesian and Bayesian versions of the developed surrogate model.
%
%
%
\section{Methodology}
\label{sec:SL2}
In this section, the details of the multiscale method, DenseED network, stochastic variational gradient descent (SVGD),  and of the hybrid multiscale-neural network model and its Bayesian formulation are presented.  
\subsection{Multiscale model} 
\label{sec:formulation}
\begin{figure}[htp]
\begin{center}
    \includegraphics[scale=0.15]{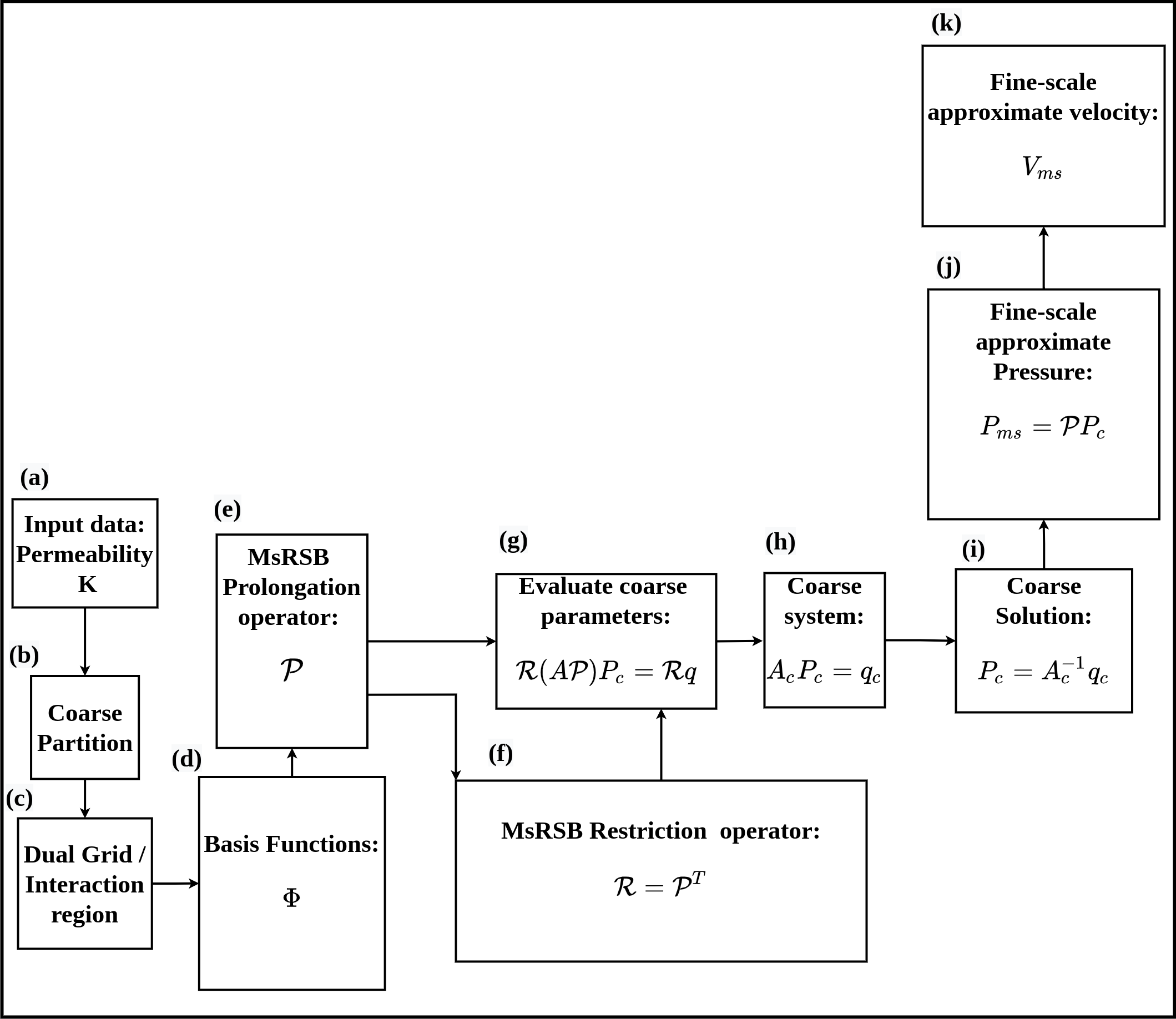}
    \caption{A schematic of the multiscale framework.}
    \label{fig:domain}
  \end{center}
\end{figure}

Simulating subsurface flow with high-precision depends on the amount of the details provided regarding the formation of the reservoir, which is described e.g. using the porosity and permeability. These properties exhibit heterogeneity with multiple length scales. Small length scale heterogeneity is essential to simulate since it has an influence on the flow and transport at a larger scale, thereby impacting the flow dynamics of the reservoir. Therefore, a high-dimensional fine-scale simulation is required to obtain reliable results.  If one is solving  Eq.~(\ref{Eq:FS_eq}) to obtain the fine-scale pressure, there are two basic methods: direct methods and iterative linear solvers. A direct method like LU decomposition can be implemented for solving reservoir simulation~\cite{aziz2002petroleum}. In an iterative method, the simulation begins with an initial estimate of the solution and then continues to iterate until the stop criterion is reached. Freund et al.~\cite{freund1992iterative} provide details regarding these iterative methods. However,  solving Eq.~(\ref{Eq:FS_eq}) using the above methods   is computationally expensive   for problems when $n_f$ is high, and therefore there are various multiscale models~\cite{moyner2016multiscale, jenny2003multi, aarnes2008mixed} that have been developed in the past to obtain an approximation solution to the fine-scale with a reduced computational cost. In this work, we implement the multiscale restriction-smoothed basis (MsRSB) method to  approximate the fine-scale pressure with a low computational cost. We closely follow the work of M\o{}yner and Lie~\cite{moyner2016multiscale}. Fig.~\ref{fig:domain} illustrates the multiscale framework.
\par
\begin{figure}[htp] 
  \begin{minipage}[b]{0.45\linewidth}
    \centering
    \includegraphics[width=.85\linewidth]{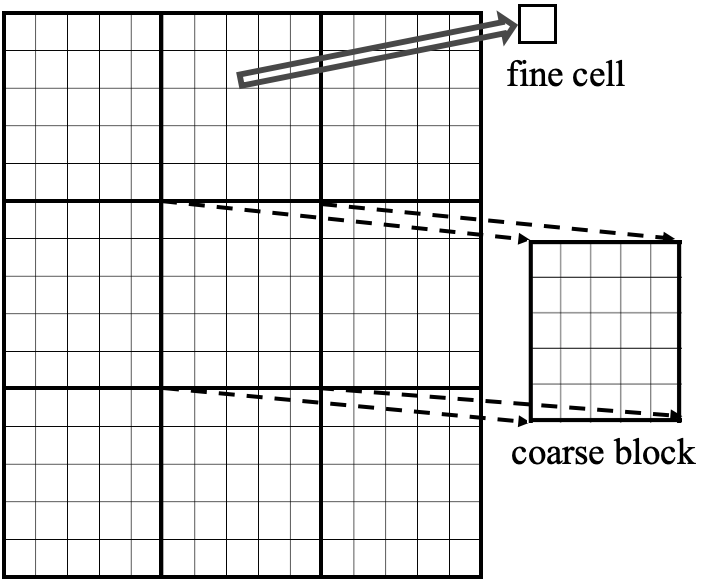} 
    \\{(a)} 
  \end{minipage}
  \begin{minipage}[b]{0.45\linewidth}
    \centering
    \includegraphics[height=4.0cm, width=4.0cm]{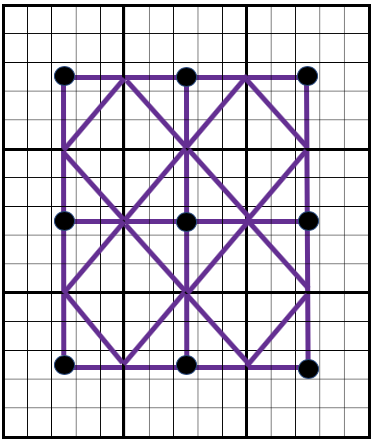} 
    \\{{(b)}} 
  \end{minipage} 
  \begin{minipage}[b]{0.45\linewidth}
    \centering
    \includegraphics[height=4.4cm, width=4.4cm]{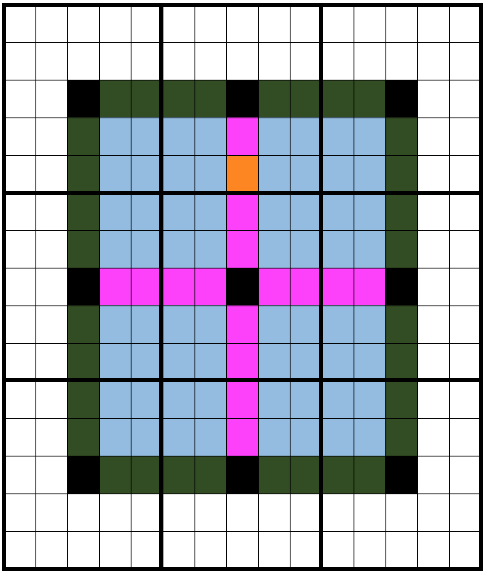} 
    \\{(c)} 
  \end{minipage}
  \begin{minipage}[b]{0.45\linewidth}
    \centering
    \includegraphics[width=.7\linewidth]{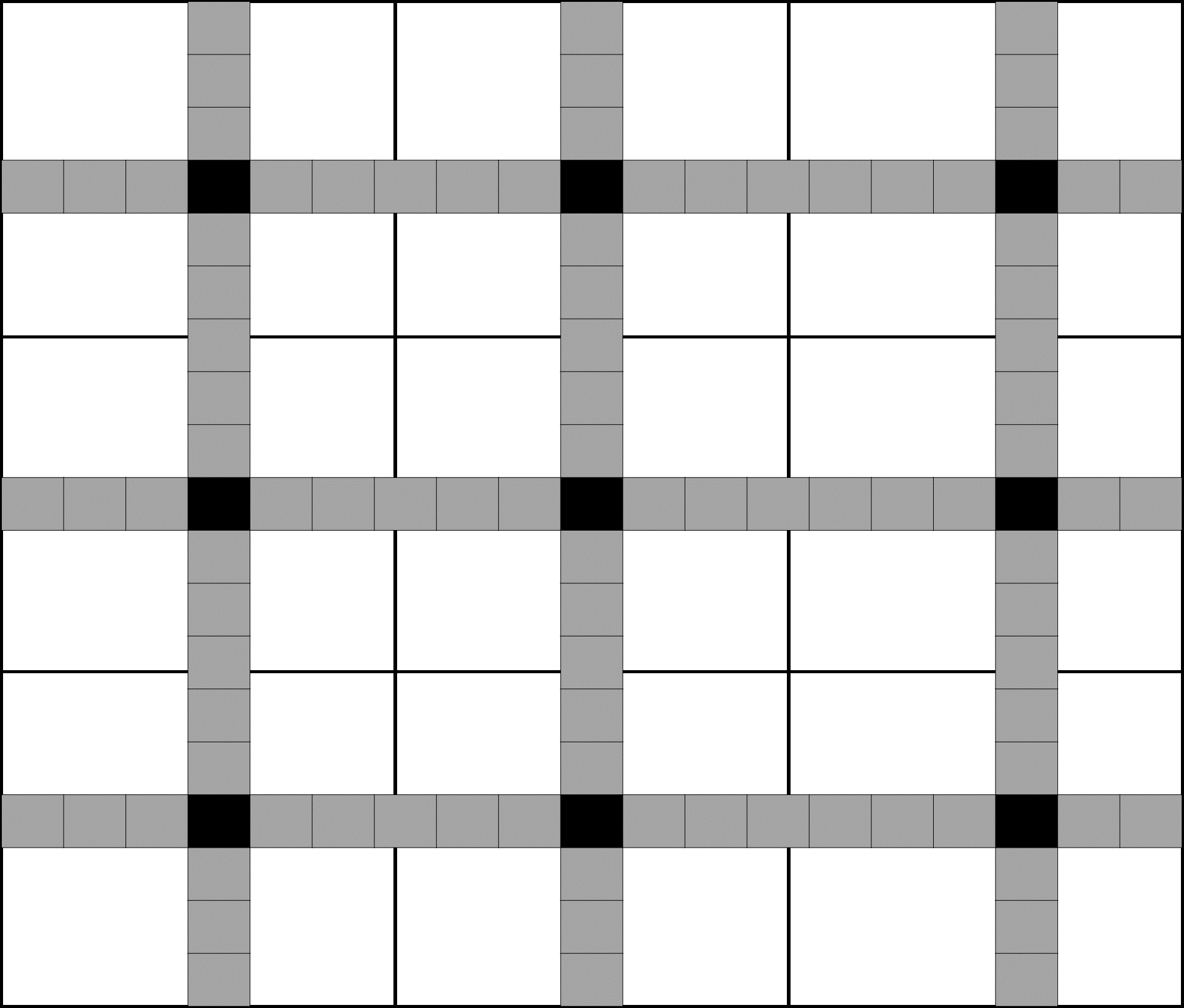} 
    \\{(d)} 
    \vspace{4ex}
  \end{minipage} 
  \caption{(a) Discretization of the domain: fine-scale domain (black bold lines correspond to the coarse-grid ($\bar{\Omega}_j$) and thin lines correspond to the fine-grid  $\Omega_i$), coarse-blocks and fine-cells (b)  Local triangulation (indicated in purple) and coarse-block centers (indicated in black) (c) Cells inside the support region are indicated in blue patch, the support boundary is indicated in green patch and the coarse center node is indicated in black for the corresponding coarse-block and (d) Global boundary (indicated in gray) and coarse-block center (indicated in black).}
  \label{fig:test}
\end{figure}

In a multiscale method, given a fine-grid $\{\Omega_i\}_{i=1}^{n_f}$,  a primal coarse-grid   $\{\bar{\Omega}_j\}_{j=1}^{n_c}$ is defined such that each fine cell in $\Omega_i$ belongs to only one coarse-block in $\bar{\Omega}_j$. This is illustrated in Fig.~\ref{fig:domain} (a)-(b).
 Fig.~\ref{fig:test} (a) illustrates an example of the coarse partition where a $15 \times 15$ fine-grid ($225$ fine-cells) is divided into a $3 \times 3$ coarse-grid.  Let $\Psi$ be the index set of fine-cells and $\xi_{j}$ be the set of fine-scale indices corresponding to the coarse-block $j$ such that:
 \begin{equation}
\xi_{j} \subseteq \Psi, \quad \xi_{j} \cap \xi_{i}=\emptyset \quad \forall \quad i \neq j,
\end{equation}
where, $i, j \in[1, n_c]$ and $|\Psi|=n_f$. We proceed to define for each coarse-block $\bar{\Omega}_j$ the corresponding support region that determines the support of the basis function for that block. Let $I_j$ be the set of all points contained in this support region (see Fig.~\ref{fig:test} (b)) and let $\Phi_j$ be the corresponding basis function. This implies that
\begin{equation}
    \Phi_j(\bm{x})>0  \quad \forall ~\bm{x} \in I_j, \quad \Phi_j(\bm{x})=0~\text{otherwise}.
\end{equation}
To define the support region for the coarse-block $\bar{\Omega}_j$,  we consider all the coarse-blocks that share a coarse-node with $\bar{\Omega}_j$ (i.e. its immediate neighbors) and create a local triangulation by considering the coarse-block centers and the centers of all coarse faces that are shared by any two of these blocks~\cite{moyner2016multiscale, shah2016multiscale}.   The support region $I_j$ of coarse-block $j$ is defined as all the fine-cells within the coarse-block $j$ and the neighboring coarse-blocks whose centroid lies within the triangulation as shown in Fig.~\ref{fig:test} (b). 
For the coarse-block $j$, the support boundary $S_j$ is defined as the set of the fine-cells that are topological neighbors to $I_j$ but not contained in it~\cite{moyner2016multiscale, shah2016multiscale}. The global support boundary $S$ can now be defined as the union of all the support boundary $S_j$ for all the primal coarse-blocks:
\begin{equation}
    B = S_{1} \cup S_{2} \cup \ldots \cup S_{n_c-1} \cup S_{n_c}.
\end{equation}
The support boundary $S_j$ is illustrated in Fig.~\ref{fig:test} (c) (indicated in green patch) and the global boundary in Fig.~\ref{fig:test} (d) (indicated in gray).
Finally, for each fine cell $i$ that belongs to the global support boundary $B$, we define $L_i$ to be the set of  indices of the support regions the cell belongs to (marked orange in Fig.~\ref{fig:test} (c)):
\begin{equation}
L_{i}=\left\{j | i \in I_{j}, i \in B\right\}.
\end{equation}
To introduce the details regarding the multiscale method, we divide the given support region into three regions. As illustrated in Fig.~\ref{fig:test}, consider the support region for the center coarse block number as three parts. Region~$1$ is defined as the set of indices that are members of both the sets $B$ and $I_j$ (shown in a blue patch). Region~$2$ is defined as the support boundary (shown in a green patch), and finally,  region~$3$ is defined as a set of indices contained in the support region but not included in the above mentioned two regions. This process is illustrated in Fig.~\ref{fig:domain} (b)-(c). 
\par
Once the support region is defined, we construct the prolongation operator $\bm{\mathcal{P}}:\{\bar{\Omega}_j\}\rightarrow \{\Omega_i\}$ that maps quantities associated with the coarse-blocks to quantities associated with the fine-cells. The prolongation operator is here represented as the matrix  $\bm{\mathcal{P}} \in \mathbb{R}^{n_f \times n_c}$   that maps quantities from the reduced system of flow Eqs.~(\ref{Eq:MS_eq}) defined on the coarse-grid  $\{\bar{\Omega}_j\}$  to the quantities from the system of flow Eqs.~(\ref{Eq:FS_eq}) defined on the fine-grid $\{\Omega_i\}$, where $i = \{1,2,\ldots,n_f\}$ and $j = \{1,2,\ldots,n_c\}$.
\par
\noindent \textbf{Basis function:} Given a fine-scale grid $\{\Omega_i\}_{i=1}^{n_f}$, where $n_f$ is the number of fine-cells, a coarse-scale grid $\{\bar{\Omega}_j\}_{j=1}^{n_c}$, where $n_c$ is the number of coarse-blocks defined on a Cartesian grid, the prolongation operator $\bm{\mathcal{P}} \in \mathbb{R}^{n_f \times n_c}$  is constructed by combining together a set of basis function $\Phi_j \in \mathbb{R}^{n_f \times 1},~j=1,\ldots,n_c$.  The $j-$th column of the prolongation operator is  $\Phi_j$ associated with each coarse-block $j$, i.e. 
\begin{equation}\label{combine}
    \mathcal{P} _{i,j} = \Phi_{j}(\bm{x}_i),  
\end{equation} 
where $\Phi_{j}(\bm{x}_i)$ is the value of the basis function at the $i-$th fine-cell.
These basis functions are initialized to unity inside the coarse-block and to zero outside that coarse-block. With a local smoothing iteration (Jacobi's method) the basis functions are updated iteratively to obtain the prolongation operator $\mathcal{P}_{i,j}$, where each column represents each learned basis functions that basically maps quantities from the reduced system of flow Eq.~(\ref{Eq:MS_eq}) defined on the coarse-grid $\{\bar{\Omega}_j\}$ to  quantities from the system of flow Eq.~(\ref{Eq:FS_eq}) defined on the fine-grid $\{\Omega_i\}$, i.e. $\bm{p}_f = \bm{\mathcal{P}} \bm{p}_c$. 

\noindent \textit{Solution of  the local problem:} 
We   initialize the basis function  for each coarse-block as follows:
\begin{equation}
\Phi_{j}^{0}({x}_i)=\left\{\begin{array}{ll}{1} & {\text{ if } i \in \xi_{j}} \\ {0} & {\text { otherwise. }}\end{array}\right.
\end{equation}
Similarly to multigrid methods,   we define a local smoothing iteration:
\begin{equation}\label{Eq:local}
\Phi_{j}^{n+1}(\bm{x}_:)=\left\{\begin{array}{ll}{\Phi_{j}^{n}(\bm{x}_:)+\bm{d}_{:j}} & {\text { if }  \text{region 1 or region 3}} \\ {0} & {\text { otherwise, }}\end{array}\right.
\end{equation}
where $d_{i j}$ and the different regions are defined by:
\begin{equation}
{d}_{i j}=\left\{\begin{array}{ll}{\frac{\widehat{d}_{i j}-\Phi_{j}^{n}({x}_i) \sum_{k \in L_{i}} \widehat{d}_{i k}}{1+\sum_{k \in L_{i}} \widehat{d}_{i k}},} & {i \in I_{j}, i \in B \text{   (region 1)}} \\ {\widehat{d}_{i j},} & {i \in I_{j}, i \notin B \text{ (region 3).}} \end{array}\right.
\end{equation}
From the above equation, we restrict the increments $\hat{d}_j$  to have non-zero values only inside the support region. This can be achieved by setting the prolongation operator $\Phi_{j}^{n}(\bm{x}_:)$ of the coarse-block $j$   to zero for region$-2$ ($i \notin I_{j}$) 
and normalize all the other basis functions in region $1$ that have non-zero values at the support boundary of coarse-block $j$:
\begin{equation}
    \widehat{\bm{d}}_{:j}=-\omega D^{-1} A \Phi_{j}^{n}(\bm{x}_:),
\end{equation}
where $A$ is the fine-scale system, $D$ is the diagonal entries of $A$ and $\omega$ is known as the damping parameter. Finally, we define a local error $e_j$ outside the global boundary as:
\begin{equation}
    e_{j}=\max _{i}\left(\left|\widehat{d}_{i j}\right|\right), \quad i \notin G.
\end{equation}
The above iterative process is converged when $\|\mathbf{e}\|_{\infty}<\mathrm{tolerance}$.  This process is illustrated in Fig.~\ref{fig:domain} (c)-(d)-(e) and Algorithm~\ref{Algo_basis}. 
\par
Next, we define a restriction operator $\bm{\mathcal{R}} \in \mathbb{R}^{n_c \times n_f}$  that maps quantities from the system of flow Eq.~(\ref{Eq:FS_eq}) defined on the fine-grid $\{\Omega_i\}$ to the quantities from the reduced system of flow Eq.~(\ref{Eq:MS_eq}) defined over the coarse-grid $\{{\Omega}_j\}$, where $i = \{1,2,\ldots,n_f\}$ and $j = \{1,2,\ldots,n_c\}$.  This process is illustrated in Fig.~\ref{fig:domain} (e)-(f). The restriction operator is constructed here using a Galerkin  approach i.e. as:
    \begin{equation}\label{Eq:restriction_opt}
        \bm{\mathcal{R}} = \bm{\mathcal{P}}^T.
    \end{equation}
Once the prolongation and the restriction operators are constructed, we then compute the pressure on the coarse grid $\bm{p}_c$:
\begin{equation}\label{Eq:coarse_matrix1}
    \bm{\mathcal{R}}(\bm{A}(\bm{\mathcal{P}}\bm{p_c})) = (\bm{\mathcal{R}A\mathcal{P}})\bm{p_c} = \bm{A_c} \bm{p_c} = \bm{\mathcal{R}q} = \bm{q_c},
\end{equation}
where $\bm{A_c}$ is the coarse-scale system matrix and $\bm{q_c}$ is the coarse source term. This process is illustrated in Fig.~\ref{fig:domain} (e)-(g) and Fig.~\ref{fig:domain} (f)-(g). Therefore, we obtain the pressure computed on the coarse grid:
\begin{equation}\label{Eq:MS_eq}
    \bm{A_c} \bm{p_c} = \bm{q_c},
\end{equation} 
\begin{equation}\label{Eq:coarse_matrix}
      \bm{p_c} = \bm{A_c}^{-1} \bm{q_c}. 
\end{equation}
This process is illustrated in Fig.~\ref{fig:domain} (g)-(h) and Fig.~\ref{fig:domain} (h)-(i). Finally combining Eq.~(\ref{Eq:coarse_matrix}) and the prolongation operator, we obtain the approximate fine-scale pressure as:
\begin{align}\label{approx1}
    \bm{p_f} &= \mathcal{P} \bm{A_c}^{-1} \bm{q_c}, \\
    \bm{p_f} &= \mathcal{P} \bm{p_c}.
\end{align}
  This process is illustrated in Fig.~\ref{fig:domain} (i)-(j). Once the pressure is obtained using the coarse pressure and the prolongation operator, the flux is computed using the transmissibility matrix, prolongation operator  and coarse pressure as follows: 
\begin{equation}
    v_{i j}^{m s}=-T_{i j}\left(\left(\bm{\mathcal{P}} \bm{p}_{c}\right)_{i}-\left(\bm{\mathcal{P}} \bm{p}_{c}\right)_{j}\right),
\end{equation}
where $v_{i j}^{m s}$ is the flux evaluated using the  multiscale method, and $T_{i j}$ is the transmissibility matrix that relates with the interface between each pair of cells $i$ and $j$, respectively.  This process is illustrated in Fig.~\ref{fig:domain} (j)-(k).\\
\begin{algorithm}[H]
\KwIn{:$A-$matrix ($\bm{A}$), coarse grid $\bar{\Omega}_j$, total number of coarse blocks $J$,  the diagonal entries $D$ of $A$ and damping parameter $\omega$, set of fine-scale indices $\xi_{j}$  corresponding to coarse block $j$, support boundary $S_j$, global support boundary $B$, number of fine-cells with in a coarse-block $I$ and tolerance $\epsilon$.}
\SetAlgoLined 
\For{j = 1 to J}{
\For{i=1 to I}{
Set $n=0$. \\
$\Phi_{j}^{n}({x}_i)=\left\{\begin{array}{ll}{1} & {\text { if } i \in \xi_{j}} \\ {0} & {\text { otherwise }}\end{array}\right.$ }
\While{$e_{j}>\epsilon$}{
Evaluate the Jacobi increment: $\widehat{\bm{d}}_{:j}=-\omega D^{-1} A \Phi_{j}^{n}(\bm{x}_:),$ \\
\For{i=1 to I}{
Apply the local update: $  {d}_{i j}=\left\{\begin{array}{ll}{\frac{\widehat{d}_{i j}-\Phi_{j}^{n}({x}_i) \sum_{k \in L_{i}} \widehat{d}_{i k}}{1+\sum_{k \in L_{i}} \widehat{d}_{i k}},} & {i \in I_{j}, i \in B \text{   (region 1)}} \\ {\widehat{d}_{i j},} & {i \in I_{j}, i \notin B \text{ (region 3)}} \end{array}\right.$}}
$n \leftarrow n+1$ \\
$\Phi_{j}^{n+1}(\bm{x}_:)=\left\{\begin{array}{ll}{\Phi_{j}^{n}(\bm{x}_:)+\bm{d}_{:j}} & {\text { if }  \text{region 1 or region 3}} \\ {0} & {\text { otherwise, }}\end{array}\right.$ \\
Evaluate the local error: \\
$e_{j}=\max _{i}\left(\left|\widehat{d}_{i j}\right|\right), \quad i \notin G$}
\KwOut {$\Phi_{j}({x}_i)$ (For the coarse-block $j$, $\Phi_{j}({x}_i)$ is the value of the basis function at the $i$-th fine-cell.)}
 \caption{Solving the local problem to obtain the multiscale basis functions~\cite{moyner2016multiscale}.}
 \label{Algo_basis}
\end{algorithm}
\subsection{Interior and non-interior support region}
\begin{figure}[htp] 
\centering
  \begin{minipage}[b]{0.3\linewidth}
    \centering
    \includegraphics[width=.75\linewidth]{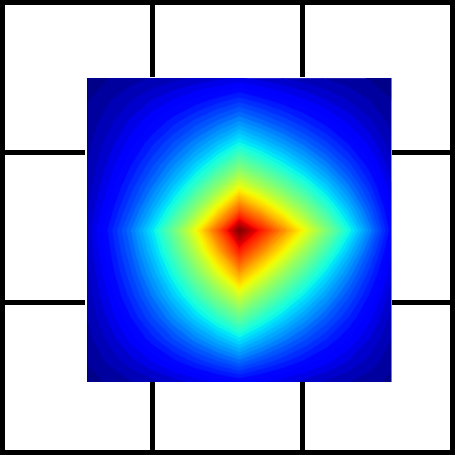} 
    \\{(a)} 
    \vspace{4ex}
  \end{minipage}
  \begin{minipage}[b]{0.3\linewidth}
    \centering
    \includegraphics[width=.75\linewidth]{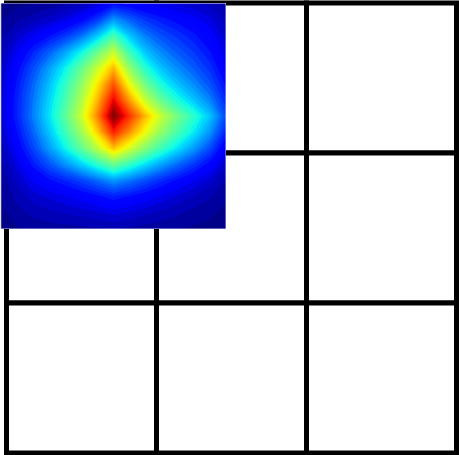} 
    \\{{(b)}} 
    \vspace{4ex}
  \end{minipage} 
    \begin{minipage}[b]{0.3\linewidth}
    \centering
    \includegraphics[width=.75\linewidth]{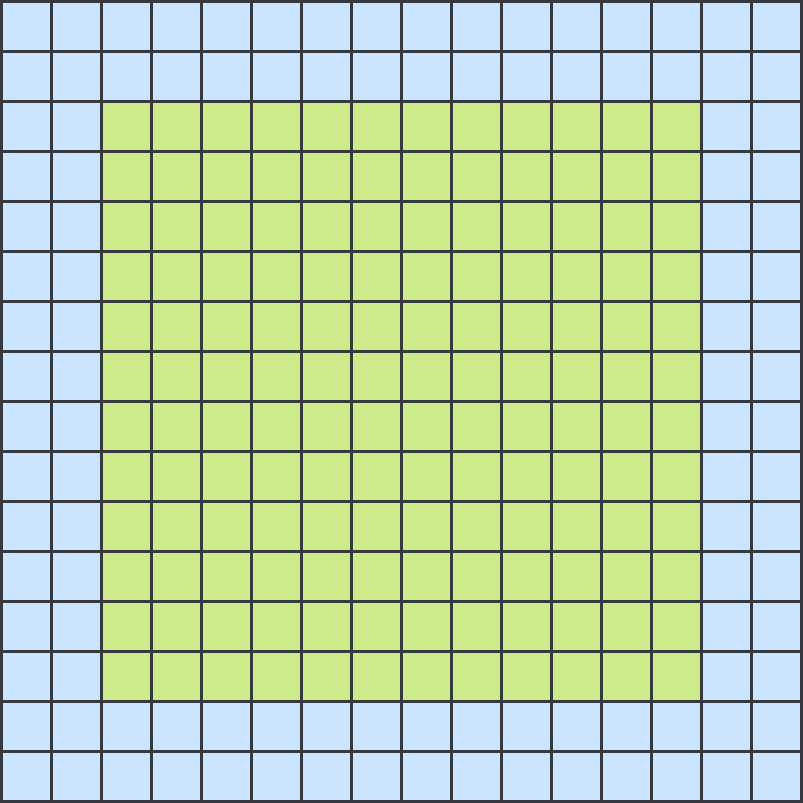} 
    \\{{(c)}} 
    \vspace{4ex}
  \end{minipage} 
  \caption{The basis functions for the interior and non-interior support regions: (a) Coarse-blocks ($3\times3$) and basis function for coarse-block $5$ (basis function for the interior support region), (b) Coarse-blocks ($3\times3$) and basis function for coarse-block $1$ (basis function for the non-interior support region) and (c)  Illustration of the interior support regions (shown in the green patch) where the basis functions are computed using the Deep Learning surrogate, and non-interior support regions (shown in the blue patch) where the basis functions are computed using the multiscale solver.}
  \label{fig:fig12}
\end{figure}

In this section, we enumerate the details of the interior and the non-interior support regions. Once the support region is constructed for a given input data, we divide the constructed support regions for a given domain into the interior and non-interior support regions. Given a set of support regions $\bm{\Upsilon} = \{\bm{\upsilon}_{n}\}_{n=1}^{n_c}$, where $n_c$ is number of support regions, we loop over  each support region ($\bm{\upsilon}_{n}$) and determine if it corresponds to the interior or non-interior support regions as illustrated in Algorithm ~\ref{Algo_interior}.
Given an input field (fine-scale permeability filed), we consider $k^{'}$ as the fine-scale permeability corresponding to the interior support regions ($\Delta_{\text{interior}}$) and $k^{'}_{non-interior}$ as the fine-scale permeability corresponding to the non-interior support regions. Similarly, we consider $\Phi_{interior}$ as the basis functions corresponding to the interior support regions ($\Delta_{\text{interior}}$) and $\Phi_{non-interior}$ as the basis functions corresponding to the non-interior support regions. In this work, we consider a fine-scale of domain size $128 \times 128$ with $8 \times 8$ as the coarse-blocks. Therefore, we have a total of $N_c=256$ coarse-blocks. Fig.~\ref{fig:fig12} (c) shows the interior blocks (green patch) and the non-interior blocks (blue patch). As will be discussed in Section~\ref{sec:SL21}, the calculation of the basis functions on the fixed-size interior blocks will be performed with a deep learning surrogate whereas the basis functions in the non-interior blocks will be computed using the multiscale approach. In this work, the number of support regions is the same as the number of coarse-blocks. In Fig.~\ref{fig:fig12} (c), the permeability $k^{'}$ defined in the support
regions of the $144$ interior blocks is $256$-dimensional. 

\begin{algorithm}[H]
\textbf{Input:} Support regions: $\bm{\Upsilon} = \{\bm{\upsilon}_{n}\}_{n=1}^{n_c}$  and $n_c$ = Number of support regions \\
Reshape $\bm{\Upsilon}$ to a square matrix: $\bm{\Delta}$ with $\beta$ rows and columns; $\beta \in \{1,2,3,\ldots,\sqrt{N_c}\}$ \Comment{Given $\Omega_i$ and $\Omega_j$ are square domains} \\
$\bm{\Xi}_{row} = \{\bm{\Delta}[1,:],\bm{\Delta}[2,:],\bm{\Delta}[\beta-1,:],\bm{\Delta}[\beta,:]\}$ \\ 
$\bm{\Xi}_{column} = \{\bm{\Delta}[:,1],\bm{\Delta}[:,2],\bm{\Delta}[:,\beta-1],\bm{\Delta}[:,\beta]\}$ \\ 
\SetAlgoLined
\For{$i = 1$ to $\sqrt{n_c}$}{
\For{$j = 1$ to $\sqrt{n_c}$}{
\eIf{$(\Delta[i,j] = \{\bm{\Xi}_{\text{row}},\bm{\Xi}_{\text{column}}\})$}{$\Delta_{\text{non-interior}} = \Delta[i,j]$}{$\Delta_{\text{interior}} = \Delta[i,j]$}
}
}
 \caption{Interior and non-interior support regions.}
 \label{Algo_interior}
\end{algorithm}


\par

\noindent \textbf{Remark 1.} In practice, the coarse-grid size plays a vital role in the accuracy of the solution. If the coarse-blocks are large, then the approximate solution will be unsatisfactory and if the coarse-blocks are small then the computational time will not justify the use of the multiscale method. 
In this work, we consider $128\times128$ fine-grid and $8\times8$ as the coarse-grid.

\subsection{DenseED architecture}
Training basis functions is vital as it is the building block for the multiscale system. In this work, we use a Dense convolutional Encoder-Decoder network~\cite{jegou2017one}  (DenseED) to learn the basis functions. The DenseED architecture has been previously used on various tasks in particular for an image-to-image regression problem~\cite{zhu2018bayesian, mo2019deep,geneva2019quantifying}. It consists of three main parts: encoding (or down-sampling), dense block (where the feature size remains the same) and decoding (or up-sampling). Encoding the input information to the dense block is processed by down-sampling, and likewise, the decoding from the dense block to the output block is processed by up-sampling the information. In this work, we learn a map between the  fine-scale permeability $k^{'}_{interior}$ (input to the regression model) corresponding to the interior support regions  and  the basis function corresponding to the interior support regions $\phi_{interior}$ (output to the regression model).
\subsubsection{Encoder-Decoder block}
\label{sec:DenseED}
The Encoder-Decoder architecture is similar to U-net~\cite{ronneberger2015u}. However, there is no concatenation of feature maps, and the encoder and decoder are used to change the size of feature maps, as illustrated in Fig.~\ref{fig:Dense_ED}. In a conventional convolutional network, max-pooling is performed for down-sampling. However, in this architecture, we use convolution of small stride value to reduce the feature size. In this DenseED network, the encoder initially performs a convolution of the input image and then subsequently a series of an encoder and dense blocks are constructed such that the input to the decoder block consists of a better coarser representation of the input. Similar to the encoder, the decoder consists of a series of decoder blocks and dense blocks where the up-sampling of the input data is performed.  The encoding or the down-sampling layers consists of initial batch normalization~\cite{ioffe2015batch}, ReLU~\cite{glorot2011deep} (activation function), convolution or max-pooling and similarly up-sampling consists of batch normalization~\cite{ioffe2015batch}, ReLU~\cite{glorot2011deep} (activation function), and transposed convolution. Both the encoder and decoder are constructed with fully-convolutional neural networks (CNN).
\subsubsection{Dense block}
\begin{figure}[htp]
  \centering
\includegraphics[scale=0.28]{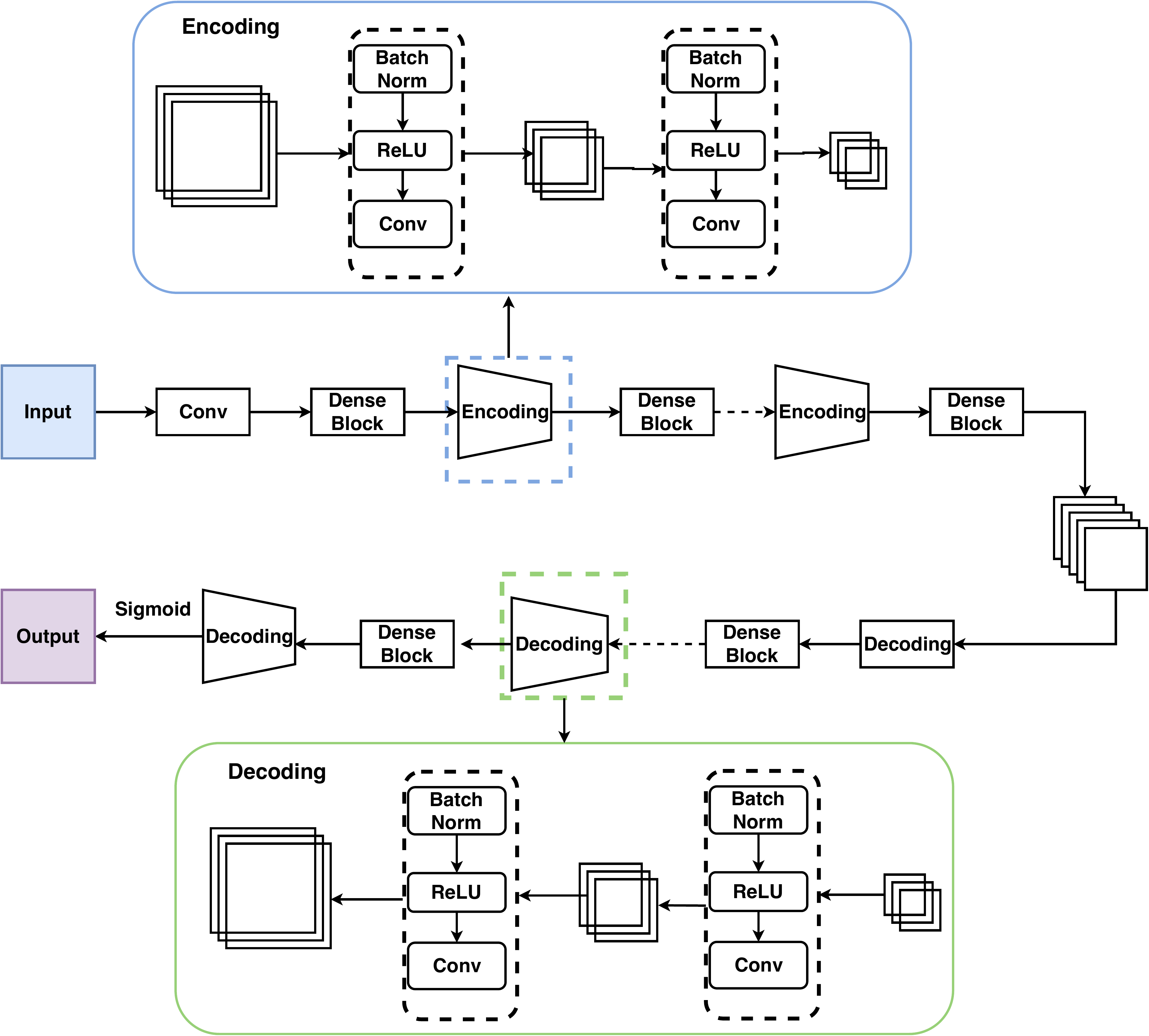}
\\{(a)} \\
\includegraphics[scale=0.18]{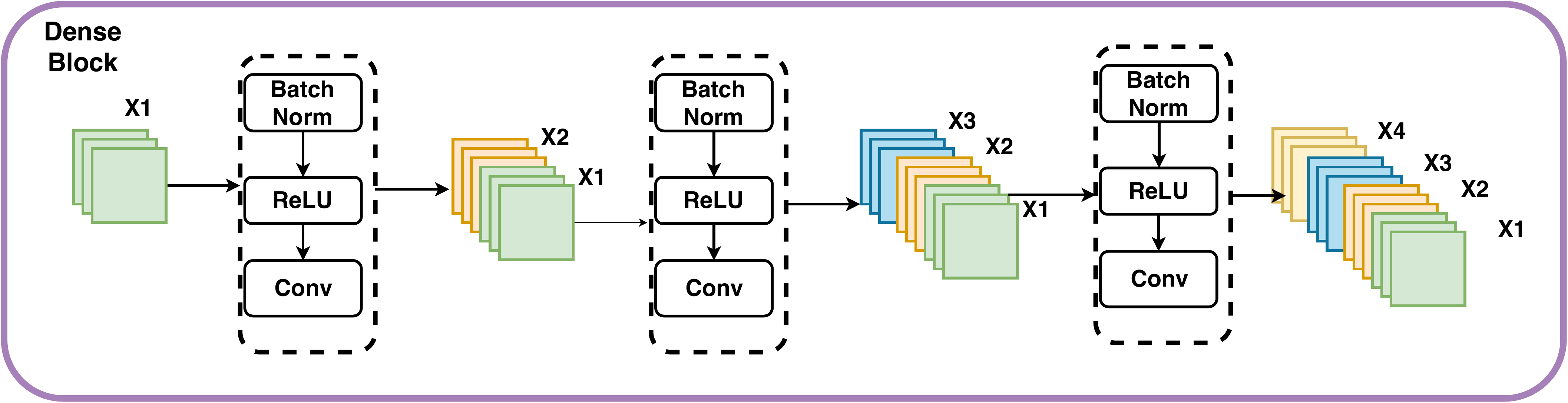}
\\{(b)}
\caption{A schematic of the DenseED network. (a) The top block shows the DenseED architecture with the encoding layers, and the bottom block shows the decoding layers containing convolution, Batch Norm, and ReLU. The convolution in the encoding layer reduces the size of the feature map, and the convolution (ConvT) in the decoding layer performs up-sampling. (b) The dense block also contains convolution, Batch Norm, and ReLU. The main difference between the encoding or decoding layer and the dense block is that the size of the feature maps is the same as the input in the dense block. Lastly, we apply the sigmoid activation function at the end of the last decoding layer.}
\label{fig:Dense_ED}
\end{figure}

A dense block consists of multiple densely connected layers as illustrated in Fig.~\ref{fig:Dense_ED}. In this block, the feature size remains the same, that is the output feature size of the dense block is the same as the input feature size to that block. Also, the input to this block is the concatenation of the previous block output and its input (previous block input). As shown in Fig.~\ref{fig:Dense_ED}, the input to the dense block is $\bm{X}_1$ and the input to the subsequent block is $\bm{X}_2$ (output of the previous block) and $\bm{X}_1$. That is, the input for the $i-$th layer is given by:
\begin{equation}
    \bm{X}_{i} = f([\bm{X}_{i-1},\bm{X}_{i-2},\ldots,\bm{X}_{1}]),
\end{equation}
where $\bm{X}_{i}$ is the output of the $i-$th layer. Inside each block the  following sequence of operations is performed: batch normalization, ReLU (activation function) and convolution. The network architecture is specified in Table~\ref{tab:NetArchitecture}. 
\begin{table}[htp]
\caption{DenseED architecture.}\label{tab:NetArchitecture}
\centering
\begin{tabular}{cccc}
\hline
\textit{Layers}       & $C_f$ & \textit{Resolution $H_f \times W_f$} & \textit{Number of parameters} \\ \hline
Input                 & $1$              & $15 \times 15$                       & -                             \\ 
Convolution k7s2p3    & $48$             & $7 \times 7$                       & $2352$                          \\ 
Dense Block (1) K16L4 & $112$           & $7 \times 7$                       & $42048$                         \\ 
Encoding Layer        & $56$             & $4 \times 4$                       & $34888$                         \\ 
Dense Block (2) K16L8 & $184$         & $4 \times 4$                       & $130944$                         \\ 
Decoding Layer (1)    & $92$             & $8 \times 8$                       & $14276$                         \\ 
Dense Block (3) K16L4 & $156$            & $8 \times 8$                       & $67808$                         \\ 
Decoding Layer (2)    & $1$              & $15 \times 15$                       & $13728$                          \\ \hline
\end{tabular}
{\\ $k =$ kernel size,  $s =$ stride, $p =$ padding, $L =$ Number of layers  and  $K =$ growth rate.}
\end{table}

\subsection{Bayesian neural networks}
\label{sec:BayesianNN}
In this section, we enumerate the details regarding the extension of the hybrid approach to a Bayesian framework~\cite{zhu2018bayesian}. For a regression problem, consider a neural network  $\bm{y}=\bm{f}(\bm{x},\bm{w})$, where $\bm{x}$ is the input, $\bm{y}$ is the output and the parameters $\bm{w}$ include both the weights and biases. In a Bayesian setting, the weight parameters are treated as random variables  to account for epistemic uncertainty induced by the small amount of training dataset. The probabilistic model with an explicit likelihood is expressed as follows:
\begin{equation}
\bm{y} = \bm{f(x,w)}+\bm{n}.
\end{equation}
In this work, we consider   additive noise $\bm{n}$   in the form $\bm{n} = \sigma \bm{\epsilon}$ (same for all output pixels), where  $\sigma$ is a scalar and $\bm{\epsilon}$ is a Gaussian noise, $\bm{\epsilon} \sim \mathcal{N}(\bm{0},\bm{I})$. Here, we consider the noise precision $\beta = 1/\sigma^{2}$ as a random variable with a conjugate prior as follows: $p(\beta) = \text{Gamma}(\beta\mid a_{1},b_{1})$. We assume the output-size noise to have a small variance and therefore, we consider the shape and rate parameters to be $a_1 = 3$ and $b_1 = 1 \times 10^{-6}$. In deep neural networks, there is a large number of parameters $\bm{w}$ and hence we assume that the weights have a probability density function of a fully-factorized Gaussian prior with mean zero and Gamma distributed precision $\alpha$ on $\bm{w}$. The resulting prior is a Student's$-\mathcal{T}$ distribution and this helps to promote sparsity and also prevents over-fitting~\cite{zhu2018bayesian}: 
\begin{equation}
    p(\bm{w}|\alpha) = \mathcal{N}(\bm{w}|0,\alpha^{-1}\bm{I}) \hspace{1cm} p(\alpha) = \text{Gamma}(\alpha|a_0,b_0),
\end{equation}
where $a_0 = 1$ and $b_0 = 0.03$ are the  values taken for the rate and shape parameters.
\subsubsection{Stein Variational Gradient Descent}
\label{sec:SVGD}
In this work, our goal is to obtain a high-dimensional posterior distribution over a large number of random variables using a limited amount of training data. One of the simplest ways to obtain the approximate posterior is to implement Monte Carlo methods. This approach is slow and difficult to converge.  There has been enormous progress made to approximate high-dimensional posterior distributions using variational inference methods. However, these methods restrict the approximate posterior to a certain parametric family.   Recently, Liu and Wang~\cite{liu2016stein} proposed a non-parametric variational inference called Stein Variational Gradient Descent (SVGD). In the current work, we adopt the SVGD approach to approximate the posterior distribution $p(\bm{\theta}|\mathcal{\bm{D}})$, where $\bm{\theta}$ are the uncertain parameters that consist of $\bm{w}$ and $\beta$ (noise precision) and $\mathcal{\bm{D}}$ is the training data, for a given likelihood $p(\bm{y}|\bm{x},\bm{\theta})$ and prior $p_0(\bm{\theta})$. The variational distribution aims to approximate the target distribution with a variational distribution $q^*(\bm{\theta})$ which lies in the restricted set of distributions $\mathscr{Q}$:
\begin{equation}
q^{*}(\bm{\theta})=\underset{q \in \mathcal{Q}}{\arg \min } {KL}(q(\bm{\theta}) \| p(\bm{\theta} | \mathcal{D}))=\underset{q \in \mathcal{Q}}{\arg \min} \mathbb{E}_{q}[\log q(\bm{\theta})-\log \tilde{p}(\bm{\theta} | \mathcal{D})+\log Z],
\end{equation}
where $\tilde{p}(\bm{\theta} | \mathcal{D})=p(\mathcal{D} | \bm{\theta}) p_{0}(\bm{\theta})=\prod_{i=1}^{N} p\left(\mathbf{y}^{i} | \bm{\theta}, \mathbf{x}^{i}\right) p_{0}(\bm{\theta})$ is the unnormalized posterior and $Z$ is the normalization constant. The normalization constant is not considered when we optimize the KL divergence.  
In SVGD, we consider an initial tractable distribution represented in terms of samples and then apply a transformation to each of these samples:
\begin{equation}
\mathbf{T}(\bm{\theta})=\bm{\theta}+\epsilon \bm{\phi}(\bm{\theta}),
\end{equation}
where $\epsilon$ is the step size and $\bm{\phi}(\bm{\theta}) \in \mathcal{F}$ is the perturbation direction within a function space $\mathcal{F}$. Therefore, $T$ transforms the initial density $q(\bm{\theta})$ to $q_{[T]}(\bm{\theta})$  
\begin{equation}
q_{[\mathbf{T}]}(\bm{\theta})=q\left(\mathbf{T}^{-1}(\bm{\theta})\right)\left|\operatorname{det}\left(\nabla \mathbf{T}^{-1}(\bm{\theta})\right)\right|.
\end{equation}

This method uses a particle approximation for the variational posterior rather than   a parametric form. Therefore, we consider a set of samples $\{\bm{\theta}^i\}_{i=1}^{S}$ with empirical measure: $\mu_{S}(d\bm{\theta}) = \frac{1}{S}\sum_{i=1}^{S}\delta(\bm{\theta}-\bm{\theta}^{i})d\bm{\theta}$.  It is important for $\mu$ to weakly converge to the true posterior $\nu_p(d \bm{\theta}) = p(\bm{\theta} \mid \mathcal{D})d\bm{\theta}$. We now apply the transform $\bm{T}$ on the samples: $\mu$ to $\bm{T}\mu$. The next task is to obtain the minimum KL divergence of the variational approximation and the target distribution:
\begin{equation}
    \max_{\bm{\phi} \in \mathcal{F}}\Big[- \frac{d}{d \epsilon} KL (\bm{T}\mu||\nu_{p})|_{\epsilon=0} \Big].
\end{equation}
The above term can also be expressed as~\cite{liu2016stein}:
\begin{equation}
    - \frac{d}{d \epsilon} KL (\bm{T}\mu||\nu_{p})|_{\epsilon=0} = \mathbb{E}_{\mu}[\mathcal{T}_p\bm{\phi}],
\end{equation}
where $\mathcal{T}_{p}$ is the Stein operator associated with the distribution $p$   given by:
\begin{equation}
    \mathcal{T}_{p} \bm{\phi} = \frac{\triangledown \cdot (p\bm{\phi})}{p} = \frac{(\triangledown p)\cdot \bm{\phi}+p(\triangledown\cdot \bm{\phi})}{p} = (\triangledown \log p)\cdot \bm{\phi} + \triangledown \cdot \bm{\phi}.
\end{equation}

The expectation of the above term, i.e, $\mathbb{E}_{\mu}\left[\mathcal{T}_{p} \bm{\phi}\right]$ evaluates the difference between $p$ and $\mu$ and its maximum is defined as the Stein discrepancy $(\mathcal{S}(\mu, p))$:
\begin{equation}
\mathcal{S}(\mu, p)=\max _{\bm{\phi} \in \mathscr{F}} \mathbb{E}_{\mu}\left[\mathcal{T}_{p} \bm{\phi}\right].
\end{equation}

If the functional space $\mathscr{F}$ is chosen to be the unit ball in a product reproducing kernel Hilbert space $\mathcal{H}$ with the positive kernel $k(\bm{x},\bm{x^{'}})$~\cite{liu2016stein}, then the  Stein discrepancy has a closed-form solution:
\begin{equation}
    \bm{\phi}^{*}(\bm{\theta}) \propto \mathbb{E}_{\bm{\theta}^{'}\sim \mu}[\mathcal{T}^{\bm{\theta}^{'}}k(\bm{\theta},\bm{\theta}^{'})] = \mathbb{E}_{\bm{\theta}^{'}\sim \mu}[\triangledown_{\bm{\theta}^{'}}\log p(\bm{\theta}^{'}, \mathcal{D})k(\bm{\theta},\bm{\theta}^{'})+\triangledown_{\bm{\theta}^{'}}k(\bm{\theta},\bm{\theta}^{'})],
\end{equation}
where the term $\nabla_{\bm{\theta}^{\prime}} \log p\left(\bm{\theta}^{\prime}, \mathcal{D}\right) k\left(\bm{\theta}, \bm{\theta}^{\prime}\right)$ is known as the kernel smooth gradient term and $\nabla_{\bm{\theta}} k\left(\bm{\theta}, \bm{\theta}^{\prime}\right)$ is the repulsive force. In this work, we choose a standard radial basis function kernel in the above update procedure.

\begin{algorithm}[htp]
\textbf{Input}: \text{A set of initial particles} $\{\bm{\theta}_0^i\}_{i=1}^{S}$, \text{score function} $\triangledown \log p(\bm{\theta}\mid \mathcal{D})$, \text{kernel} $k(\bm{\theta},\bm{\theta}^{'})$, \text{step-size} $\{\epsilon_t\}$
\SetAlgoLined \\
\For{iteration t}{
$\bm{\phi}(\bm{\theta}_t^{i}) = \frac{1}{S} \sum_{j=1}^{S}\Big[k(\bm{\theta_t^j},\bm{\theta_t^i})\triangledown_{\bm{\theta_t^j}}\log p(\bm{\theta_t^j}, \mathcal{D})+\triangledown_{\bm{\theta_t^j}}k(\bm{\theta_t^j},\bm{\theta_t^i})\Big]$\\
$\bm{\theta}^i_{t+1}$ $\leftarrow$ $\bm{\theta}^i_t+\epsilon_t\bm{ \phi}(\bm{\theta}^i_t)$
}
\textbf{Result:} A set of particles $\bm{\theta}^i$ that approximates the target posterior
 \caption{Stein Variational Gradient Descent~\cite{liu2016stein}.}
 \label{algo-2}
\end{algorithm}
The SVGD procedure is summarized in Algorithm~\ref{algo-2}, where the gradient term $\bm{\phi}(\bm{\theta})$ drives the samples to the high probability posterior region with the kernel smooth gradient, whilst maintaining a degree of diversity using the repulsive force term.
\subsection{Hybrid multiscale-DenseED}\label{sec:SL21}
\begin{figure}[H]
  \centering
\includegraphics[scale=0.04]{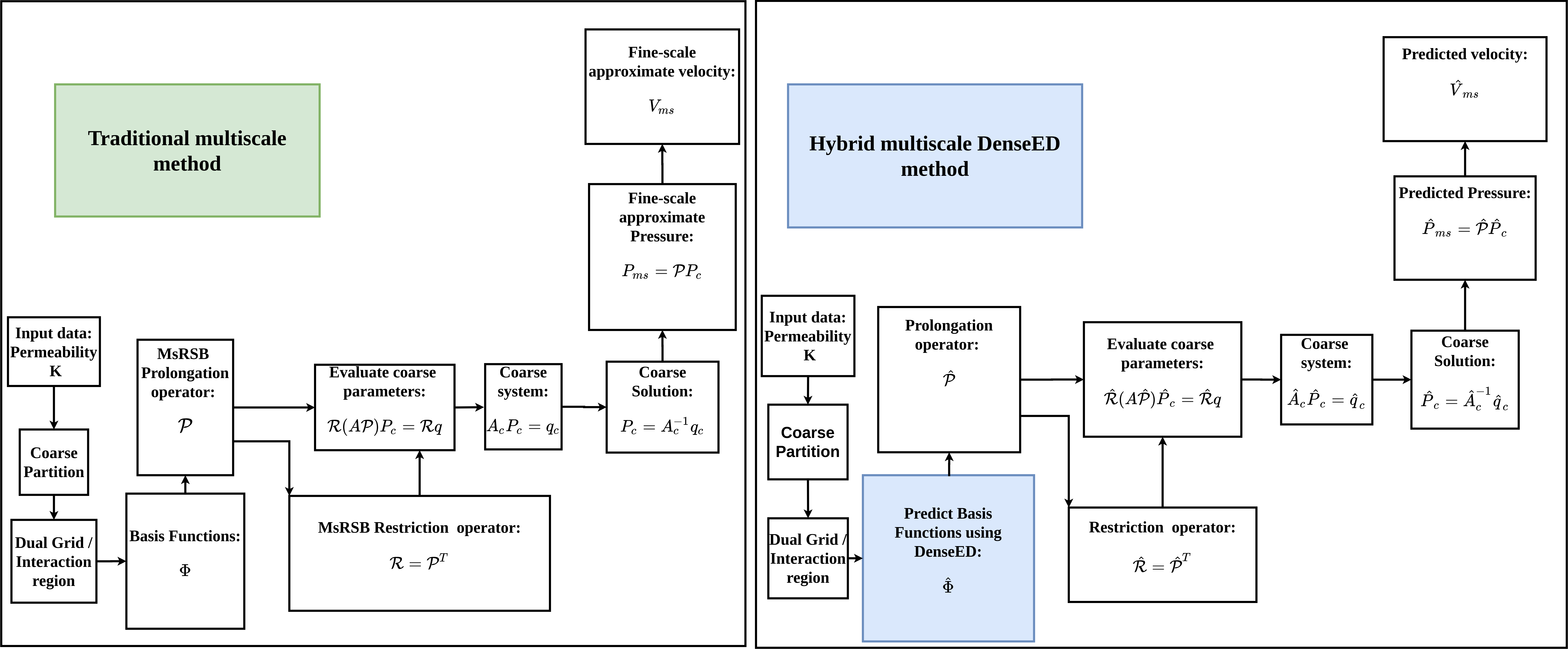}
\caption{Comparison of the standard multiscale framework with the data-driven hybrid multiscale DenseED framework.}
\label{fig:comapre}
\end{figure}

In this work, we aim to solve the stochastic PDE in Eq.~(\ref{Eq:Darcy}). Using a multiscale approach together with  Monte Carlo methods is computationally inefficient as   one has to run thousands of forward simulations to obtain convergent statistics. Thus, to perform the uncertainty quantification and propagation tasks, it is not computationally efficient to solve the local problem and obtain the basis functions for every input permeability field. Therefore, we embed deep learning techniques within the multiscale method for solving the stochastic PDE as given in Eq.~(\ref{Eq:Darcy}). Fig.~\ref{fig:comapre} illustrates a comparison of the standard   multiscale method with the hybrid method in which, in a broad sense, we replace the block that computes the basis functions with the DenseED block and train the model with the loss function between the predicted pressure and fine-scale pressure.
\subsubsection{Deterministic hybrid surrogate model}
To train the system, we first generate the fine-scale pressure by solving the Darcy equation (Eq.~(\ref{Eq:Darcy})) using a  mixed finite element formulation implemented in  FEniCS~\cite{alnaes2015fenics} (python package) with fourth-order discontinuous elements for solving the pressure for a given input field $\bm{K}$. We then obtain the following parameters for the same input field using an open-source reservoir modeling and simulation software MRST~\cite{2015SINTEF}: the $\bm{A}-$ matrix, the $\bm{q}-$ matrix, the permeability corresponding to the interior support region $\{\bm{k}^{\prime}_{j}\}_{j=1}^{n_{I}}$, where $n_{I}$ is the number of permeability field from the interior support regions for a given input field and the basis functions for the non-interior support region $\{\bar{\bm{\Phi}}^{non-int}_{j}\}_{j=1}^{n_{IO}}$, where $n_{IO}$ is the number of basis functions corresponding to the non-interior support region for a given permeability field $\bm{K}$. Note that the total number of support regions (also equal to the total number of coarse blocks) $n_c = n_{I}+n_{NI}$ and $n_{I}>n_{NI}$. MRST is a widely accepted reservoir modeling, and simulation software~\cite{2015SINTEF} and consists of many solvers such as the MsRSB, MsFV, MsFEM, and many more. Before we embark to the training procedure, we first generate the following data using the MRST software for the entire training dataset, i.e., $\{\bm{K}^{(l)}\}_{l=1}^{D}$, where $D$ is the total number of training dataset: the $\{\bm{A}-\text{matrix}^{(l)}\}_{l=1}^{D}$, the $\{\bm{q}-\text{matrix}^{(l)}\}_{l=1}^{D}$, the 
permeability corresponding to the interior support region $\{\bm{k}^{\prime}_{j}{}^{(l)}\}_{l=1,j=1}^{D,n_{I}}$, and the basis functions for the non-interior support region $\{\bar{\bm{\Phi}}^{non-int}_{j}{}^{(l)}\}_{l=1,j=1}^{D,n_{IO}}$.
\subsubsection*{\bf Training:}
During training, we consider the batch size to be one, i.e., we consider one complete permeability field as the input to the HM-DenseED model for a given iteration step (or epoch). Therefore, during training, for a given input field $\bm{K}^{(l)}$, we consider the permeability corresponding to the interior support regions $\bm{k}^{\prime \prime (l)} \in \mathbb{R}^{n_{I} \times C \times H \times W}$ as the input to the Dense-ED network, where $C$, $H$, $W$ are channel, height and width, respectively. In this work, we consider the channel to be one, i.e., $C=1$. The two axes of the spatial domain, i.e., height and width: $H \times W = (n_{FC}-1) \times (n_{FC}-1)$, where $n_{FC}$ is the number of fine cells along each axes for a given coarse block $j$ and we consider $n_{FC}$ to be $16$ in the present study. With $\bm{k}^{\prime \prime (l)}$ as the input, we train a surrogate to predict the basis functions corresponding to the interior support region $\hat{\bm{\Phi}}^{int, (l)} \in \mathbb{R}^{n_{I} \times C \times H \times W}$:
\begin{equation}
\hat{\bm{\Phi}}^{int,(l)} = \text{DenseED}(\bm{k}^{\prime \prime (l)},\bm{\theta}),
\end{equation}
where $\bm{\theta}$ are the model parameters.
\par 
To the predicted basis functions corresponding to the interior support regions $\hat{\bm{\Phi}}^{int,(l)}_{j}$, we append zeros as the outermost fine-cell values for a given support region $n_{I}$ and let us denote it as $\tilde{\Phi}^{int,(l)}$, i.e., $\hat{\bm{\Phi}}^{int,(l)}\in \mathbb{R}^{n_{I} \times C \times H \times W} \rightarrow \tilde{\bm{\Phi}}^{int,(l)}\in \mathbb{R}^{n_{I} \times C \times (H+1) \times (W+1)}$. 
Recall, in Section~\ref{sec:formulation} and Algorithm~\ref{Algo_basis}, the fine-cell values in region$-2$ ($i \notin I_{j}$, see Fig.~\ref{fig:test}(c) indicated in green patch) are set to zero to restrict the increments of $\hat{d}_j$ to have non-zero values only inside the support region.
\par 
Once the model predicts the basis functions corresponding to the interior support regions $\tilde{\bm{\Phi}}^{int,(l)}$ for a given input field $\bm{K}^{(l)}$, we now proceed to obtain the predicted fine-scale pressure as follows. First, we construct a prolongation operator $\widehat{\bm{\mathcal{P}}}^{(l)} \in \mathbb{R}^{n_{f}\times n_{c}}$ by combining the predicted basis functions corresponding to the interior support regions $\tilde{\bm{\Phi}}^{int,(l)}$ and the basis functions for the non-interior support region $\bar{\bm{\Phi}}^{non-int,(l)}$. Note that in the Dense-ED network (see Section~\ref{sec:DenseED}), we apply the sigmoid activation function at the end of the last decoding layer to only ensure that the output values are in the range $[0,1]$ for a given predicted basis functions $\hat{\bm{\Phi}}^{int,(l)}_{j}$ corresponding to the interior support region $j$. However, we need to explicitly satisfy the partition of unity property~\cite{moyner2016multiscale} for the prolongation operator $\widehat{\bm{\mathcal{P}}}^{(l)}$ that was constructed by combining $\tilde{\bm{\Phi}}^{int,(l)}$ and $\bar{\bm{\Phi}}^{non-int,(l)}$. Therefore, to ensure the partition of unity for a given fine-cell $i$, we take the summation over all the coarse blocks $j$ to be equal to $1$. We evaluate $\widetilde{\bm{\mathcal{P}}}^{(l)}$ using the prolongation operator $\widehat{\bm{\mathcal{P}}}^{(l)}$:
\begin{equation}
    \sum_{j}\left(\widetilde{{\mathcal{P}}}^{(l)}_{ij}\right) = 1 \quad \forall \quad i,
\end{equation}
where the $\widetilde{\bm{\mathcal{P}}}^{(l)} \in \mathbb{R}^{n_{f}\times n_{c}}$ is obtained as follows:
\begin{equation}\label{Eq:unity}
    \widetilde{{\mathcal{P}}}^{(l)}_{ij} = \frac{\widehat{{\mathcal{P}}}^{(l)}_{ij}}{\sum_{j}{\widehat{{\mathcal{P}}}^{(l)}_{ij}}}.
\end{equation}
Note that we predict the basis functions corresponding to the interior support regions $\tilde{\bm{\Phi}}^{int,(l)}$ for a given input field $\bm{K}^{(l)}$ all together and combine them with the basis functions for the non-interior support region $\bar{\bm{\Phi}}^{non-int,(l)}$ and satisfy the partition of unity property. 
\par
\begin{figure}[H]
  \centering
    \includegraphics[scale=0.15]{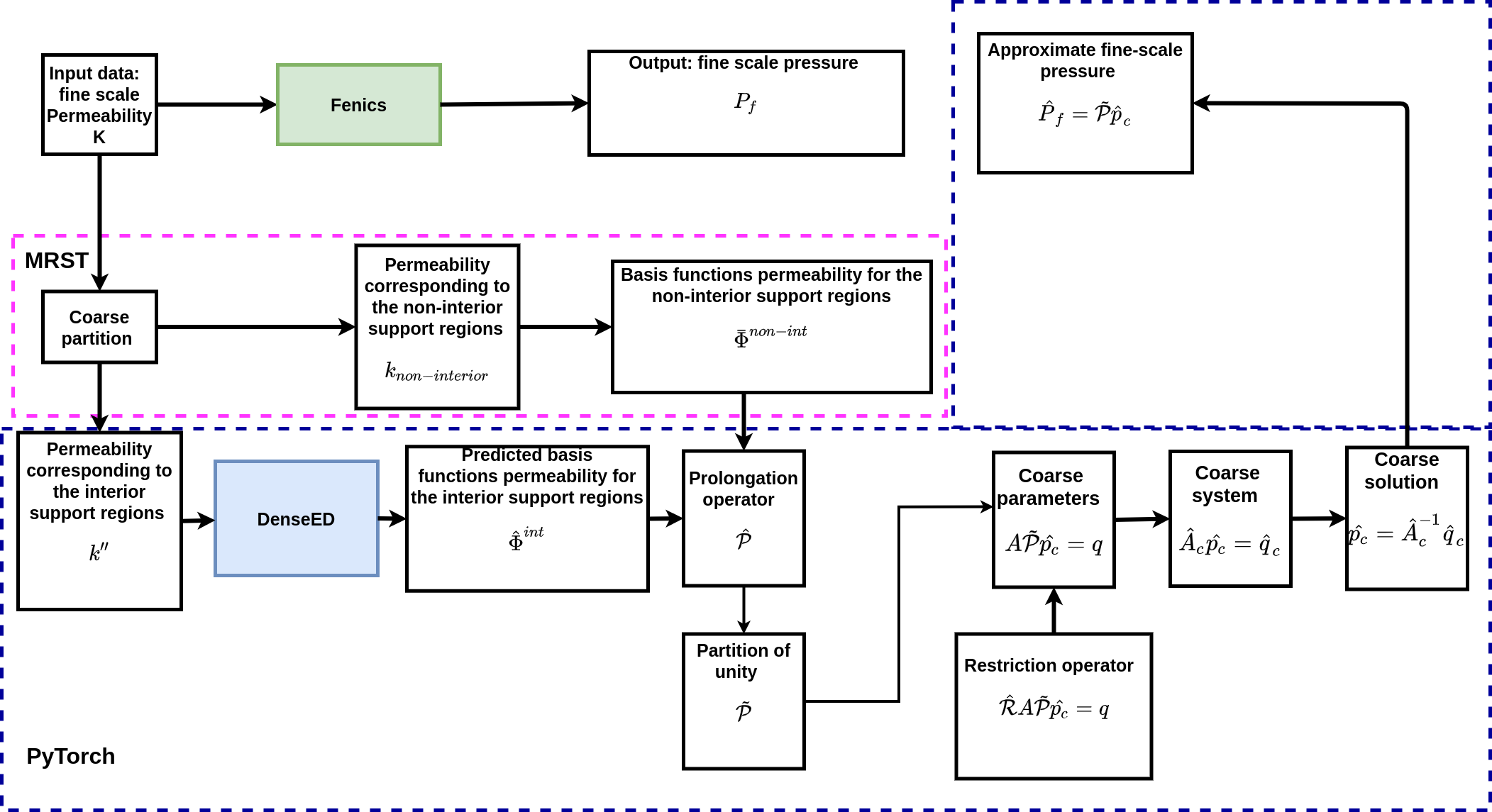}
    \caption{A schematic of the hybrid deep neural network- multiscale framework using DenseED. Parameters $\bm{A}$, $\bm{q}$ and $\bar{\bm{\Phi}}^{non-int}$ are obtained from MRST~\cite{2015SINTEF} (magenta dashed line) and the network is trained using Pytorch (implementation is marked in blue dashed line).}
\label{fig:training}
\end{figure}

 
We now proceed to construct the predicted restriction operator $\widehat{\bm{\mathcal{R}}}^{(l)}$ using the predicted prolongation operator $\widetilde{\bm{\mathcal{P}}}^{(l)}$ as:
\begin{equation}\label{Eq:predict_restriction}
    \widehat{\bm{\mathcal{R}}}^{(l)} = \left(\widetilde{\bm{\mathcal{P}}}^{(l)}\right)^T,
\end{equation}
Given the predicted prolongation operator $\widetilde{\bm{\mathcal{P}}}^{(l)} \in \mathbb{R}^{n_{f}\times n_{c}}$, the predicted restriction operator $\widehat{\bm{\mathcal{R}}}^{(l)}\in \mathbb{R}^{n_{c}\times n_{f}}$ and the A$-matrix$, we compute the $\hat{\bm{A_c}}^{(l)} \in \mathbb{R}^{n_{c}\times n_{c}}$ is the coarse-scale system matrix and $\hat{\bm{q_c}}^{(l)} \in \mathbb{R}^{n_{c}\times 1}$ is the coarse source term as follows:
\begin{equation}
    \widehat{\bm{\mathcal{R}}}^{(l)}(\bm{A}^{(l)}(\widetilde{\bm{\mathcal{P}}}^{(l)}\hat{\bm{p}_c}^{(l)})) = (\widehat{\bm{\mathcal{R}}}^{(l)}\bm{A}^{(l)}\widetilde{\bm{\mathcal{P}}}^{(l)})\hat{\bm{p}_c^{(l)}} = \hat{\bm{A_c}}^{(l)} \hat{\bm{p_c}}^{(l)} = \widehat{\bm{\mathcal{R}}}^{(l)}\bm{q}^{(l)} = \hat{\bm{q_c}}^{(l)}.
\end{equation}
We now obtain the pressure computed on the coarse grid as follows:
\begin{equation}\label{Eq:MS_eq_HM}
    \hat{\bm{A}_c}^{(l)} \hat{\bm{p}_c}^{(l)} = \hat{\bm{q}_c}^{(l)},
\end{equation} 
\begin{equation}\label{Eq:pred_coarse_matrix}
      \hat{\bm{p}_c}^{(l)} = \left(\hat{\bm{A}_c}^{(l)}\right)^{-1} \hat{\bm{q}_c}^{(l)}. 
\end{equation}
Therefore, the predicted fine-scale pressure is given as:
\begin{align}\label{Eq:pred_approx1}
    \widehat{\bm{P}_f}^{(l)} &= \widetilde{\bm{\mathcal{P}}}^{(l)} (\hat{\bm{A}_c}^{(l)})^{-1} \hat{\bm{q}_c}^{(l)}, \\
    \widehat{\bm{P}_f}^{(l)} &= \widetilde{\bm{\mathcal{P}}}^{(l)} \hat{\bm{p}_c}^{(l)}.
\end{align}
The training process is summarized in Algorithm~\ref{algo_hybrid} and the framework is illustrated in Fig.~\ref{fig:training}.
\begin{algorithm}[htp]
\KwIn{Training data \{$\bm{K}^{(l)} ,\bm{P}_f^{(l)}\}_{l=1}^{D}$, number of epochs: $E_{epochs}$,  source term: $\{\bm{q}-\text{matrix}^{(l)}\}_{l=1}^{D}$; $\{\bm{A}-\text{matrix}^{(l)}\}_{l=1}^{D}$, number of interior support regions: $n_I$, number of non-interior support regions:  $n_{IO}$, and learning rate: $\eta$.} 
Given $\bm{K}^{(l)}$, $n_{IO}$, and $n_I$ obtain the permeability corresponding to the interior support regions $\{\bm{k}^{\prime{(l)}}_j\}_{j=1}^{n_I}$ and  permeability corresponding to the non-interior support regions $\{(\bm{k}^{(l)}_{\text{non-interior}})_h\}_{h=1}^{n_{I0}}$ using Algorithm~\ref{Algo_interior}. \\
Evaluate $\{(\bar{\bm{\Phi}}^{non-int,(l)})_j\}_{j=1}^{n_{IO}}$ using MRST \\
  \SetAlgoLined
\For{epoch $=1$ to $E_{epochs}$}{
$\hat{\bm{\Phi}}^{int,(l)} = \text{DenseED}(\bm{k}^{\prime \prime(l)},\bm{\theta})$ \Comment{Input: $\bm{k}^{\prime \prime(l)} \in \mathbb{R}^{n_{I} \times C \times H \times W}$} \\
Append zeros as the outermost fine-cell values for a given support region $n_{I}$: $\hat{\bm{\Phi}}^{int,(l)}\in \mathbb{R}^{n_{I} \times C \times H \times W} \rightarrow \tilde{\bm{\Phi}}^{int,(l)}\in \mathbb{R}^{n_{I} \times C \times (H+1) \times (W+1)}$ \\
Construct the prolongation operator: $\widehat{\bm{\mathcal{P}}}^{(l)}$ by combining the predicted basis functions for the interior support region $\tilde{\bm{\Phi}}^{int,(l)}$ and the basis functions corresponding to the non-interior support region $\bar{\bm{\Phi}}^{non-int,(l)}$. \\
Evaluate $\widetilde{\bm{\mathcal{P}}}^{(l)}$ using the prolongation operator $\widehat{\bm{\mathcal{P}}}^{(l)}$ to satisfy the partition of unity property using Eq.~(\ref{Eq:unity}). \\
Construct the restriction operator: $\widehat{\bm{\mathcal{R}}}^{(l)} = \left(\widetilde{\bm{\mathcal{P}}}^{(l)}\right)^T$ \\
Obtain coarse pressure: $\bm{\hat{p}}_c^{(l)}$ using Eq.~(\ref{Eq:pred_coarse_matrix}) \\
Using the prolongation operator $\widetilde{\bm{\mathcal{P}}}^{(l)}$ map the predicted coarse pressure $\hat{\bm{p}}_c^{(l)}$ to the predicted fine-scale pressure $\widehat{\bm{P}}_f^{(l)}$ using Eq.~(\ref{Eq:pred_approx1})\\
$\mathcal{L}$ = MSE($\widehat{\bm{P}}_f^{(l)}$,$\bm{P}_{f}^{(l)}$) \Comment{MSE: mean square error} \\
$\nabla \bm{\theta} \leftarrow$ \text {Backprop}($\mathcal{L}$)
\\
$\bm{\theta} \leftarrow \bm{\theta}-\eta \nabla \bm{\theta}$
}
\KwOut{Trained hybrid deep learning multiscale model.}
\caption{Training hybrid deep learning multiscale method.}
\label{algo_hybrid}
\end{algorithm}
\subsubsection*{\bf Testing:}
During testing, we generate the following quantities using the MRST for a given set of test data $n=\{1,..,N_{test}\}$: input data $\bm{K}^{\star(n)}$, that was unseen during training, the $\bm{A}^{\star}-\text{matrix}^{(n)}$, the $\bm{q}^{\star}-\text{matrix}^{(n)}$, the permeability corresponding to the interior support region $\{\bm{k}^{\star\prime (n)}_{j}\}_{j=1}^{n_{I}}$,  and the basis functions for the non-interior support region $\{\bar{\bm{\Phi}}^{\star non-int, (n)}_{j}\}_{j=1}^{n_{IO}}$. 
\par
 In testing, we first predict the basis functions for the interior support regions $\{\hat{\bm{\Phi}}^{\star int, (n)}_j\}_{j=1}^{n_{I}}$:
\begin{equation}
\hat{\bm{\Phi}}^{\star int} =  \text{DenseED}(\bm{k}^{\star \prime \prime},\bar{\bm{\theta}}), 
\end{equation}
where $\bm{k}^{\star \prime \prime} \in \mathbb{R}^{(N_{test}\times n_{I}) \times C \times H \times W}$ is the input to the network and $\bar{\bm{\theta}}$ are the trained model parameters. For each $N_{test}$ data, we append zeros as the outermost fine-cell values for a given support region $n_{I}$: $\hat{\bm{\Phi}}^{int,(n)}\in \mathbb{R}^{n_{I} \times C \times H \times W} \rightarrow \tilde{\bm{\Phi}}^{int,(n)}\in \mathbb{R}^{n_{I} \times C \times (H+1) \times (W+1)}$.
\par
We then proceed to construct the prolongation operator $\widehat{\bm{\mathcal{P}}}^{\star} \in \mathbb{R}^{N_{test}\times n_f \times n_c}$ by combining $\tilde{\bm{\Phi}}^{\star int}$ and $\bar{\bm{\Phi}}^{\star non-int}$ and evaluate $\widetilde{\bm{\mathcal{P}}}^{\star} \in \mathbb{R}^{N_{test}\times n_f \times n_c}$ to satisfy the partition of unity property for each $N_{test}$. 
\par
Finally, using the $\bm{A}^{\star}-\text{matrix}\in \mathbb{R}^{N_{test}\times n_f \times n_f}$, the $\bm{q}^{\star}-\text{matrix} \in \mathbb{R}^{N_{test}\times n_f \times 1}$, and the $\widetilde{\bm{\mathcal{P}}}^{\star}$, we  evaluate the restriction operator $\widehat{\bm{\mathcal{R}}}^{\star}\in \mathbb{R}^{N_{test}\times n_c \times n_f}$, coarse pressure: $\bm{\hat{p}}_c^{\star}$ and the predicted fine-scale pressure $\widehat{\bm{P}}_f^{\star}\in \mathbb{R}^{N_{test}\times n_f}$
using Eq.~(\ref{Eq:predict_restriction}), Eq.~(\ref{Eq:pred_coarse_matrix}) and Eq.~(\ref{Eq:pred_approx1}), respectively.
\par
The fluxes using the predicted pressure $\widehat{\bm{P}}_f^{\star}$ are obtained as follows. We apply a Gaussian smoothing on the predicted pressure before evaluating the velocity. Gaussian smoothing is a widely used smoothing method in the field of computer vision and robotics for suppressing the high-frequency components for a given image~\cite{davies2004machine,vernon1991machine}. We mainly implement this in our work to avoid the checkerboard pattern in the predicted velocity that is mainly caused due to the noisy artifacts in the predicted pressure $\widehat{\bm{P}}_f^{\star}$. Recall in Eq.~(\ref{Eq:DarcyA}), the fluxes are calculated by taking the gradient of the pressure and then multiplying with the permeability field. If there are predominant artifacts in the predicted pressure $\widehat{\bm{P}}_f^{\star}$, then taking its gradient and multiplying it with the permeability field causes a checkerboard pattern in the flux prediction. Therefore, we apply Gaussian smoothing to the predicted pressure by convolving over a fixed kernel that is obtained from a truncated Gaussian distribution with mean $0$ and a fixed standard deviation $\sigma$ to obtain a smoothed predicted pressure $\bar{\bm{P}}_f^{\star}$. In this work, we consider the truncation value and the standard deviation to be $4$ and $1.8$, respectively. Using the predicted pressure (smoothed) $\bar{\bm{P}}_f^{\star}$ and the transmissibility matrix $T_{ij}$ we evaluate the fluxes using Eq.(~\ref{Eq:Flux}). Note that we evaluate the fluxes only in the prediction stage.
\par
Zhu and Zabaras~\cite{zhu2018bayesian} developed a surrogate model using a DenseED~\cite{huang2017densely} network for uncertainty quantification and propagation with high-dimensional stochastic input and a limited size dataset. However, it is hard to train and obtain good prediction accuracy using a standard surrogate model with very limited data. Hence, in this work, we introduce a new hybrid deep-learning and multiscale approach to train with a limited dataset and obtain better accuracy. Our hybrid framework is different from the framework of Chan and Elsheikh~\cite{chan2018machine}. The authors developed a framework where the basis functions are trained using a fully-connected network for flow through porous media problems with the loss function defined between the predicted and the actual basis functions generated using the MsFVM approach. Our goal is to develop a hybrid framework to account for the limited data used during training and provide the model with uncertainty estimates. Therefore, we define the loss function between the predicted approximate pressure and the fine-scale pressure and in addition extend this HM-DenseED to a fully Bayesian framework.
\subsubsection*{\textit{Network architecture and hyperparameter tuning}}
In this work, we empirically perform an extensive search for network architecture designs and hyperparameters. As mentioned earlier, in this work, we use a dense encoder-decoder network~\cite{huang2017densely, zhu2018bayesian} for training our HM-DenseED model. We perform an extensive search for the number of encoder blocks, decoder blocks, and dense blocks for the network architecture design. The details regarding the network architecture design are provided in Appendix~\ref{sec:App-B1}. Once we select the DenseED configuration that gives a better performance with fewer parameters to train, we perform an extensive hyperparameter search. Here, we seek to train our HM-DenseED model with a suitable learning-rate, weight decay, and the number of epochs that perform well across a different number of the training dataset, different KLE truncation levels, and channelized field as considered earlier. 
\subsubsection{Bayesian HM-DenseED Surrogate Model}
In this section, we provide the implementation details for the  Bayesian-HM-DenseED which is the extension of the HM-DenseED model (see Section~\ref{sec:SL21}) to the Bayesian framework~\cite{liu2016stein}.
Given a set of training data $l = \{1 \ldots D\}$, where $D$ is the total number of training dataset: i.e., the input data $\mathcal{\bm{D}}=\{\bm{K}^{(l)},\bm{P}^{(l)}\}$, we first generate the following quantities using the MRST software: the $\bm{A}-\text{matrix}^{(l)}$, the $\bm{q}-\text{matrix}^{(l)}$, the permeability corresponding to the interior support region $\{\bm{k}^{\prime}_{j}{}^{(l)}\}_{j=1}^{n_{I}}$ and the basis functions for the non-interior support region $\{\bar{\bm{\Phi}}^{non-int}_{j}{}^{(l)}\}_{j=1}^{n_{IO}}$.
\subsubsection*{\bf Training}
In this work, we consider the batch size to be one, i.e., we consider one complete permeability field as the input to the Bayesian HM-DenseED model for a given iteration step $t$ in Algorithm~\ref{algo-2}. The input for the Bayesian-HM-DenseED model is the following: the permeability corresponding to the interior support regions $\bm{k}^{\prime \prime (l)} \in \mathbb{R}^{n_{I} \times C \times H \times W}$, the $\bm{A}-\text{matrix}^{(l)}$, the $\bm{q}-\text{matrix}^{(l)}$, the permeability corresponding to the interior support region $\bm{k}^{\prime}_{j}{}^{(l)}$, and the basis functions for the non-interior support region $\bar{\bm{\Phi}}^{non-int}_{j}{}^{(l)}$. With the permeability corresponding to the interior support regions $\bm{k}^{\prime \prime (l)}$ as the input to the Bayesian-DenseED model~\cite{zhu2018bayesian}, we train the Bayesian surrogate model to predict the basis functions corresponding to the interior support region $\hat{\bm{\Phi}}^{int, (l)} \in \mathbb{R}^{S \times n_{I} \times C \times H \times W}$: 
\begin{equation}
\bar{\bm{\Phi}}^{int,m,(l)} = \text{Bayesian-DenseED}(\bm{k}^{\prime \prime,m,(l)},\bm{\theta}^{m}),
\end{equation}
where $\bm{\theta}^{m}$ are the particles and $m$ is the model. For each model $m$, we append zeros to the outermost fine-cell values for the predicted basis functions corresponding to the interior support region $\bar{\bm{\Phi}}^{int,m,(l)}$ and let us denote it as $\tilde{\bm{\Phi}}^{int,(l),m}$, i.e., $\hat{\bm{\Phi}}^{int,(l),m}\in \mathbb{R}^{n_{I} \times C \times H \times W} \rightarrow \tilde{\bm{\Phi}}^{int,(l),m}\in \mathbb{R}^{n_{I} \times C \times (H+1) \times (W+1)}$.
\par 
Next, we construct the prolongation operator $\hat{\mathcal{\bm{P}}}^{m,l} \in \mathbb{R}^{n_f \times n_c}$ for each of the models $m$ by combining the predicted basis functions $\tilde{\bm{\Phi}}^{int,m,l}$ for the interior support region and the basis functions corresponding to the non-interior support region $\bm{\Phi}^{non-int,m,l}$. Here, we predict the basis functions corresponding to the interior support regions $\tilde{\bm{\Phi}}^{int,l,m}$ for a given input field $\bm{K}^{(l)}$ all together and combine it with the basis functions for the non-interior support region $\bar{\bm{\Phi}}^{non-int,l,m}$. We explicitly satisfy the partition of unity property~\cite{moyner2016multiscale} for the above constructed prolongation operator $\hat{\bm{\mathcal{{P}}}}^{m,l}$ using Eq.~(\ref{Eq:unity}) to obtain $\widetilde{\bm{\mathcal{{P}}}}^{m,l}$.
The restriction operator is then evaluated using using the predicted prolongation operator $\widetilde{\bm{\mathcal{{P}}}}^{m,l}$ in Eq.~(\ref{Eq:predict_restriction}). Now, we proceed to evaluate the predicted pressure $\widehat{{\bm{P}}_f}^{m,l}$ using Eqs.~(\ref{Eq:pred_coarse_matrix})--(\ref{Eq:pred_approx1}).
\par
Once the predicted pressure $\widehat{{\bm{P}}_f}^{m,l}$ is evaluated, we compute the score function as follows. First, the unnormalized posterior is computed: $p(\bm{\theta}_{t}^{m}, \mathcal{D}) = \prod_{l=1}^{N} p(\widetilde{\bm{\mathcal{P}}}^{(l)}\mid \bm{\theta}_{t}^{m},\bm{K}^{(j)})p(\bm{\theta}_{t}^{m})$ for each $m$ model (HM-DenseED framework) and at each iteration $t$ together with its gradient: $\nabla_{(\bm{\theta}_{t}^{m})} \log p(\bm{\theta}_{t}^{m}, \mathcal{D})$ using the automatic differentiation in PyTorch~\cite{paszke2017automatic}, where $\bm{\theta}$ are the uncertain parameters that consist of $\bm{w}$ and $\beta$ (noise precision).  
\par
Next, we compute the kernel matrix $\Big[k\left({\bm{\theta}_{t}^{m}}, {\bm{\theta}_{t}^{j}}\right)\Big]_{j \in \{1,...,S\}}$, its gradient (i.e., the repulsive force term) $\nabla_{(\bm{\theta}_{t}^{m})}k\left({\bm{\theta}_{t}^{m}}, {\bm{\theta}_{t}^{j}}\right)$ and the kernel smoothed gradient term as follows $k\left({\bm{\theta}_{t}^{m}}, {\bm{\theta}_{t}^{j}}\right) \nabla_{(\bm{\theta}_{t}^{m})} \log p(\bm{\theta}_{t}^{m}, \mathcal{D})$.
\par
Finally, we compute $\{\phi(\bm{\theta}_{t}^{m})\}_{m=1}^{S}$ by adding the repulsive force term and the kernel smoothed gradient term. For each HM-DenseED network, we update locally by sending $\phi(\bm{\theta}_{t}^{m})$ to each HM-DenseED network. We repeat the above steps until convergence. 
The training algorithm from the Bayesian HM-DenseED is summarized in Algorithm~\ref{algo_hybrid_Bayesian}.
\subsubsection*{\bf Testing:}
Once the Bayesian-HM-DenseED is trained, we then perform the testing as follows. First, we generate the following quantities using the MRST for a given set of test data $n=\{1,\ldots,M_{test}\}$ which includes the input data $\bm{K}^{\star(n)}$ that was unseen during training: the $\bm{A}^{\star}-\text{matrix}^{(n)}$, the $\bm{q}^{\star}-\text{matrix}^{(n)}$, the permeability corresponding to the interior support region $\{\bm{k}^{\star\prime (n)}_{j}\}_{j=1}^{n_{I}}$,  and the basis functions for the non-interior support region $\{\bar{\bm{\Phi}}^{\star non-int, (n)}_{j}\}_{j=1}^{n_{IO}}$. 
\par
Next, we predict the basis functions for the interior support regions \\ $\{\hat{\bm{\Phi}}^{\star int, n,m}_j\}_{j=1,n=1,m=1}^{n_{I}, M_{test},S}$: 
\begin{equation}
\hat{\bm{\Phi}}^{\star int} =  \text{Bayesian-DenseED}(\bm{k}^{\star \prime \prime},\tilde{\bm{\theta}}),
\end{equation}
where $\bm{k}^{\star \prime \prime} \in \mathbb{R}^{(M_{test}\times n_{I}) \times C \times H \times W}$ is the input to the network and $\bar{\bm{\theta}} = \{\bm{\theta}^{(m)}\}_{m=1}^{S}$ are the particles.
Now for each model $m$ and for each test data $M_{test}$ data, we first append zeros as the outermost fine-cell values for a given support region $n_{I}$: $\hat{\bm{\Phi}}^{int,(n)}\in \mathbb{R}^{n_{I} \times C \times H \times W} \rightarrow \tilde{\bm{\Phi}}^{int,(n)}\in \mathbb{R}^{n_{I} \times C \times (H+1) \times (W+1)}$. 
\par
Next, for each model $m$, we proceed to construct the prolongation operator $\widehat{\bm{\mathcal{P}}}^{\star,m} \in \mathbb{R}^{M_{test}\times n_f \times n_c}$ by combining $\hat{\bm{\Phi}}^{\star int, m}$ and $\bar{\bm{\Phi}}^{\star non-int, m}$ and evaluate $\widetilde{\bm{\mathcal{P}}}^{\star, m} \in \mathbb{R}^{M_{test}\times n_f \times n_c}$ to satisfy the partition of unity property. 
\par
Finally, using the $\bm{A}^{\star}-\text{matrix}\in \mathbb{R}^{M_{test}\times n_f \times n_f}$, the $\bm{q}^{\star}-\text{matrix} \in \mathbb{R}^{M_{test}\times n_f \times 1}$, and the $\widetilde{\bm{\mathcal{P}}}^{\star,m}$ evaluate the restriction operator $\widehat{\bm{\mathcal{R}}}^{\star,m}\in \mathbb{R}^{M_{test}\times n_c \times n_f}$, coarse pressure: $\bm{\hat{p}}_c^{\star,m}$ and the predicted pressure for each model $m$: $\tilde{\bm{P}}_f^{\star,m}\in \mathbb{R}^{1 \times M_{test}\times n_f}$
using Eq.~(\ref{Eq:predict_restriction}), Eq.~(\ref{Eq:pred_coarse_matrix}) and Eq.~(\ref{Eq:pred_approx1}), respectively. Lastly, we concatenate the pressure $\tilde{\bm{P}}_f^{\star,m}\in \mathbb{R}^{1 \times M_{test}\times n_f}$ from each model $m$   to obtain the predicted fine-scale pressure $\tilde{\bm{P}}_f^{\star}\in \mathbb{R}^{S \times M_{test}\times n_f}$.
\begin{algorithm}[hbt!]
\KwIn{Training data \{$\bm{K}^{(l)} ,\bm{P}_f^{(l)}\}_{l=1}^{D}$, number of epochs: $E_{epochs}$, number of samples $S$, a set of initial particles $\{\theta_{0}^{m}\}_{m=1}^{S}$, source term: $\{\bm{q}-\text{matrix}^{(l)}\}_{l=1}^{D}$; $\{\bm{A}-\text{matrix}^{(l)}\}_{l=1}^{D}$, number of interior support regions: $n_I$, number of non-interior support regions:  $n_{IO}$, and learning rate: $\eta$.} 
Given $\bm{K}^{(l)}$, $n_{IO}$, and $n_I$, obtain the permeability corresponding to the interior support regions $\{\bm{k}^{\prime{(l)}}_j\}_{j=1}^{n_I}$ and  permeability corresponding to the non-interior support regions $\{(\bm{k}^{(l)}_{\text{non-interior}})_h\}_{h=1}^{n_{I0}}$ using Algorithm~\ref{Algo_interior}. \\
Evaluate $\{(\bar{\bm{\Phi}}^{non-int,(l)})_j\}_{j=1}^{n_{IO}}$ using MRST. \\
\SetAlgoLined
\For{t $=1$ to $E_{epochs}$}{
\For{m $=1$ to $S$ }{
$\hat{\bm{\Phi}}^{int,(l),m} = \text{Bayesian-DenseED}(\bm{k}^{\prime \prime,(l),m},\bm{\theta}^{m})$ \\
Append zeros as the outermost fine-cell values for a given support region $n_{I}$: $\hat{\bm{\Phi}}^{int,(l),m}\in \mathbb{R}^{n_{I} \times C \times H \times W} \rightarrow \tilde{\bm{\Phi}}^{int,(l),m}\in \mathbb{R}^{n_{I} \times C \times (H+1) \times (W+1)}$ \\
Construct the prolongation operator: $\widehat{\bm{\mathcal{P}}}^{l,m}$ by combining the predicted basis functions for the interior support region $\tilde{\bm{\Phi}}^{int,(l),m}$ and the basis functions corresponding to the non-interior support region $\bar{\bm{\Phi}}^{non-int,(l),m}$. \\
Evaluate $\widetilde{\bm{\mathcal{P}}}^{l,m}$ using the prolongation operator $\widehat{\bm{\mathcal{P}}}^{l,m}$ to satisfy the partition of unity property using Eq.~(\ref{Eq:unity}). \\
Construct the restriction operator: $\widehat{\bm{\mathcal{R}}}^{l,m} = \left(\widetilde{\bm{\mathcal{P}}}^{l,m}\right)^T$ \\
Obtain coarse pressure: $\bm{\hat{p}}_c^{l,m}$ using Eq.~(\ref{Eq:pred_coarse_matrix}) \\
Using the prolongation operator $\widetilde{\bm{\mathcal{P}}}^{l,m}$ map the predicted coarse pressure $\hat{\bm{p}}_c^{l,m}$ to the predicted fine-scale pressure $\widehat{\bm{P}}_f^{l,m}$ using Eq.~(\ref{Eq:pred_approx1}) \\
Compute the unnormalized posterior: $p(\bm{\theta}_{t}^{m}, \mathcal{D})$ and the gradient $\nabla_{(\bm{\theta}_{t}^{m})} \log p(\bm{\theta}_{t}^{m}, \mathcal{D})$ using  automatic differentiation~\cite{paszke2017automatic} \\
Compute the kernel matrix $k\left({\bm{\theta}_{t}^{m}}, {\bm{\theta}_{t}^{l}}\right)$, its gradient $\nabla_{(\bm{\theta}_{t}^{m})}k\left({\bm{\theta}_{t}^{m}}, {\bm{\theta}_{t}^{l}}\right)$ and the kernel smoothed gradient term $k\left({\bm{\theta}_{t}^{m}}, {\bm{\theta}_{t}^{l}}\right) \nabla_{(\bm{\theta}_{t}^{m})} \log p(\bm{\theta}_{t}^{m}, \mathcal{D})$.\\
$\bm{\phi}(\bm{\theta}_t^{m}) = \frac{1}{S} \sum_{j=1}^{S}\Big[k(\bm{\theta_t^j},\bm{\theta_t^m})\triangledown_{\bm{\theta_t^j}}\log p(\bm{\theta_t^j}, \mathcal{D})+\triangledown_{\bm{\theta_t^j}}k(\bm{\theta_t^j},\bm{\theta_t^m})\Big]$\\
$\bm{\theta}^m_{t+1}$ $\leftarrow$ $\bm{\theta}^m_t+\epsilon_t\bm{ \phi}(\bm{\theta}^m_t)$}
}
\KwOut{Trained Bayesian-hybrid deep learning multiscale model.}
\caption{Training the Bayesian-hybrid deep learning multiscale model.}
\label{algo_hybrid_Bayesian}
\end{algorithm}
\subsubsection*{\bf Uncertainty quantification}
\label{sec:UQ}
Once the Bayesian-HM-DenseED is trained, the predictive distribution for a single input $\bm{K}^{*}$ is 
        \begin{equation}
            p(\widehat{{\bm{P}}_f}^{*}|\bm{K}^{*},\mathcal{\bm{D}}) = \int p(\widehat{{\bm{P}}_f}^{*}|\bm{K}^{*},\bm{\theta})p(\bm{\theta}|\mathcal{\bm{D}})d\bm{\theta}.
        \end{equation}
The predictive mean is 
        \begin{align}
            \mathbb{E} [\widehat{{\bm{P}}_f}^{*}|\bm{K}^{*},\mathcal{\bm{D}}] = \mathbb{E}_{p(\bm{w}, \beta|\mathcal{\bm{D}})}\big[\mathbb{E}[\widehat{{\bm{P}}_f}^{*}|\bm{K}^{*},\bm{\theta}]\big] = \mathbb{E}_{p(\bm{w}|\mathcal{\bm{D}})}\big[f(\bm{K}^{*},\bm{w})\big] \\ \approx \frac{1}{S}\sum_{m=1}^{S}f(\bm{K}^{*},\bm{w}^{m}), \hspace{2cm} \bm{w}^{m} \sim p(\bm{w}|\mathcal{\bm{D}}),
        \end{align}
where $\bm{w}$ corresponds to the weights of the neural networks that are used to train the HM-DenseED network. Here the SVGD algorithm~\cite{liu2016stein} provides samples of the joint posterior of all parameters $p(\bm{w}, \beta|\mathcal{\bm{D}})$. If one is interested to obtain the marginal posterior $p(\bm{w}\mid \mathcal{D})$, then use the samples corresponding to $\bm{w}$~\cite{zhu2018bayesian}. 
The approximation of the predictive covariance for input $\bm{K}^{*}$ is:  
        \begin{multline}
            Cov (\widehat{{\bm{P}}_f}^{*}|\bm{K}^{*},\mathcal{\bm{D}}) = \mathbb{E}_{\bm{w},\beta }[Cov(\widehat{{\bm{P}}_f}^{*}|\bm{K}^{*},\bm{\theta})]+Cov_{\bm{w},\beta}(\mathbb{E}[\widehat{{\bm{P}}_f}^{*}|\bm{K}^{*},\bm{\theta}]) \\
            = \mathbb{E}_{\bm{w},\beta} [\beta^{-1}\bm{I}]+ Cov_{\bm{w},\beta} [\bm{\bm{f}}(\bm{K}^{*},\bm{w})] \\
            = \mathbb{E}_{\beta} [\beta^{-1}\bm{I}] + \mathbb{E}_{\bm{w}}[\bm{\bm{f}}(\bm{K}^{*},\bm{w})\bm{\bm{f}}^{T}(\bm{K}^{*},\bm{w})]-\mathbb{E}_{\bm{w}}[\bm{\bm{f}}(\bm{K}^{*},\bm{w})]\cdot \mathbb{E}_{\bm{w}}^{T}[\bm{\bm{f}}(\bm{K}^{*},\bm{w})] \\
            \approx \frac{1}{S}\sum_{m=1}^{S}\Big((\beta^{m})^{-1}\bm{I}+\bm{\bm{f}}(\bm{K}^{*},\bm{w}^{(m)})\bm{\bm{f}}^{T}(\bm{K}^{*},\bm{w}^{(m)})  \Big) \\ -\Big(\frac{1}{S}\sum_{m=1}^{S}\bm{\bm{f}}(\bm{K}^{*},\bm{w}^{(m)})\Big)\Big(\frac{1}{S}\sum_{m=1}^{S}\bm{\bm{f}}(\bm{K}^{*},\bm{w}^{(m)})\Big)^{T},
        \end{multline}
and therefore the predictive variance is given by:
  \begin{align}
    Var(\widehat{{\bm{P}}_f}^{*}|\bm{K}^{*},\mathcal{\bm{D}}) = \text{diag Cov} (\widehat{{\bm{P}}_f}^{*}|\bm{K}^{*},\mathcal{\bm{D}}) \\
    \approx \frac{1}{S}\sum_{m=1}^{S}\Big((\beta^{m})^{-1}\bm{1}+\bm{f}^2(\bm{K}^{*},\bm{w}^{(m)})\Big)- \Big(\frac{1}{S}\sum_{m=1}^{S}\bm{\bm{f}}(\bm{K}^{*},\bm{w}^{(m)}) \Big)^2.
  \end{align}
  In order to compute the average prediction over the distribution of the uncertain input, we have to compute the output statistics given the data of the uncertain parameters. Let $\theta = \{\bm{w},\beta\}$ be the uncertain parameters where $\bm{\theta}\sim p(\bm{\theta}|\mathcal{D})$:
The conditional predictive mean is given by:
\begin{equation}
    \mathbb{E}[\widehat{{\bm{P}}_f} | \bm{\theta}]=\mathbb{E}_{\bm{K}}[ \mathbb{E}[\widehat{{\bm{P}}_f} | \bm{K}, \bm{\theta}]]=\mathbb{E}_{\bm{K}}[\bm{\bm{f}}(\bm{K}, \bm{w})] \approx \frac{1}{M} \sum_{j=1}^{M} \bm{\bm{f}}\left(\bm{K}^{j}, \bm{w}\right), \quad \bm{K}^{j} \sim p(\bm{K}), 
\end{equation}
  and the   predictive covariance is calculated as:
  \begin{align} 
  \operatorname{Cov}(\widehat{{\bm{P}}_f} | \bm{\theta}) &=\mathbb{E}_{\bm{K}}[\operatorname{Cov}(\widehat{{\bm{P}}_f} | \bm{K}, \bm{\theta})]+\operatorname{Cov}_{\bm{K}}(\mathbb{E}[\widehat{{\bm{P}}_f} | \bm{K}, \bm{\theta}]) \\ &=\mathbb{E}_{\bm{K}}\left[(\beta)^{-1} \bm{I}\right]+\operatorname{Cov}_{\bm{K}}(\bm{\bm{f}}(\bm{K}, \bm{w})) \\ & \approx \beta^{-1} \bm{I}+\frac{1}{M} \sum_{j=1}^{M} \bm{\bm{f}}\left(\bm{K}^{j}, \bm{w}\right) \bm{\bm{f}}^{\top}\left(\bm{K}^{j}, \bm{w}\right) \nonumber \\
  &\qquad -\left(\frac{1}{M} \sum_{j=1}^{M} \bm{\bm{f}}\left(\bm{K}^{j}, \bm{w}\right)\right)\left(\frac{1}{M} \sum_{j=1}^{M} \bm{\bm{f}}\left(\bm{K}^{j}, \bm{w}\right)\right)^{\top}, 
  \end{align}
  and therefore the conditional predictive variance is given by:
  \begin{equation}  \operatorname{Var}(\widehat{{\bm{P}}_f} | \bm{\theta})=\operatorname{diag} \operatorname{Cov}(\widehat{{\bm{P}}_f} | \bm{\theta})=\beta^{-1} \bm{1}+\frac{1}{M} \sum_{j=1}^{M} \bm{\bm{f}}^{2}\left(\bm{K}^{j}, \bm{w}\right)-\left(\frac{1}{M} \sum_{j=1}^{M} \bm{\bm{f}}\left(\bm{K}^{j}, \bm{w}\right)\right)^{2}.
    \end{equation}
 Aforementioned,  we can compute the following statistics: $\mathbb{E}_{\bm{\theta}}[\mathbb{E}[\widehat{{\bm{P}}_f} | \bm{\theta}]]$, $\operatorname{Var}_{\theta}(\mathbb{E}[\widehat{{\bm{P}}_f} | \bm{\theta}])$, $\mathbb{E}_{\theta}(\operatorname{Var}(\widehat{{\bm{P}}_f} | \bm{\theta}))$ and $\operatorname{Var}_{\theta}(\operatorname{Var}(\widehat{{\bm{P}}_f} | \bm{\theta}))$.  For complete details regarding the above method, we direct the readers to Liu and Wang~\cite{liu2016stein} and Zhu and Zabaras~\cite{zhu2018bayesian}.
\subsubsection{Computational time:}
In this section, we compare the prediction computational cost of the HM-DenseED and the Bayesian HM-DenseED models with the fine-scale and the multiscale simulation in Tables~\ref{tab: compare1} and~\ref{tab: compare2}. As mentioned earlier, we perform the fine-scale simulation using Fenics and the multiscale simulation using the MRST software. Here, we observe that the HM-DenseED and Bayesian HM-DenseED perform significantly better than the fine-scale and multiscale simulations with several orders of magnitude in terms of wall-clock time for $100$ test permeability realizations. Note that the computational time for the  HM-DenseED and the Bayesian HM-DenseED includes evaluating the basis functions corresponding to the non-interior support region and predicting the basis functions for the interior support regions from the Dense-ED network. Here, we first obtain the basis functions corresponding to the non-interior support region for all the $100$ test data and then perform prediction for the basis functions corresponding to the interior support using the DenseED network for $100$ test data. For the Bayesian HM-DenseED model, we consider $20$ HM-DenseED models (number of samples $S$). We also notice a substantial improvement in computational time when using a GPU for the HM-DenseED and the Bayesian HM-DenseED. Implementing the Bayesian neural network with GPU is preferred when developing surrogate models for physical systems that map the high-dimensional input fields to the output fields. Since it is computationally expensive to evaluate the output fields using a simulator, one will have a relatively small amount of data to train the surrogate model. Therefore, it is important to quantify the epistemic uncertainty induced by the limited number of data. The Bayesian neural network allows us to quantify the epistemic uncertainty by considering the model parameters, which include the weights and biases, as random variables~\cite{zhu2018bayesian}. 
\begin{table}[H]
\caption{The computational cost for the HM-DenseED, the Bayesian HM-DenseED,  the fine-scale and the multiscale simulation models in-terms of wall-clock time for obtaining the basis functions for $100$ test data. Here,  Matlab* indicates that we only use  Matlab for generating the basis functions for the non-interior support regions and Matlab indicates generating  the basis functions for both the interior and non-interior support regions.}
\begin{tabular}{c|c|c|}
& \textit{Backend}, Hardware& \begin{tabular}[c]{@{}c@{}}Wall-clock (s) for obtaining \\ the basis functions\end{tabular} \\ \hline
Fine-scale& \textit{Fenics}, Intel Xeon $E5-2680$ & - \\ \hline
Multiscale& \textit{Matlab}, Intel Xeon $E5-2680$ & 654.930 \\ \hline
\multirow{2}{*}{HM-DenseED}& \textit{PyTorch}, Intel Xeon $E5-2680$ & \multirow{2}{*}{$\underset{(\text{Matlab*})}{113.23}$+$\underset{\text{(PyTorch)}}{19.5}$=132.73}\\ & \textit{Matlab}$^{*}$, Intel Xeon $E5-2680$ &\\ \hline
\multirow{2}{*}{HM-DenseED} & \textit{PyTorch}, NVIDIA Tesla V100 & \multirow{2}{*}{$\underset{(\text{Matlab*})}{113.23}$+$\underset{\text{(PyTorch)}}{1.9}$=115.13} \\
& \textit{Matlab}$^{*}$, Intel Xeon $E5-2680$ &\\ \hline
\multirow{2}{*}{\begin{tabular}[c]{@{}c@{}}Bayesian \\ HM-DenseED\end{tabular}} & \textit{PyTorch}, Intel Xeon $E5-2680$ & \multirow{2}{*}{$\underset{(\text{Matlab*})}{113.23}$+$\underset{\text{(PyTorch)}}{259.5}$ = 372.73}\\
 & \textit{Matlab}$^{*}$, Intel Xeon $E5-2680$ &\\ \hline
\multirow{2}{*}{\begin{tabular}[c]{@{}c@{}}Bayesian \\ HM-DenseED\end{tabular}} & \textit{PyTorch}, NVIDIA Tesla $V100$          & \multirow{2}{*}{$\underset{(\text{Matlab*})}{113.23}$+$\underset{\text{(PyTorch)}}{15.7}$ = 128.93} \\& \textit{Matlab}$^{*}$, Intel Xeon $E5-2680$ &                                 
\end{tabular}
\label{tab: compare1}
\end{table}
\begin{table}[H]
\caption{The computational cost for the HM-DenseED, the Bayesian HM-DenseED, the fine-scale and the multiscale simulation models in-terms of wall-clock time for obtaining the pressure  for $100$ test data. Here, Matlab* indicates that we only use the Matlab for generating the basis functions for the non-interior support regions and Matlab indicates generating  the basis functions for both the interior and non-interior support regions.}
\begin{tabular}{c|c|c|}
 & \textit{Backend}, Hardware & \begin{tabular}[c]{@{}c@{}}Wall-clock (s) for obtaining \\ the pressure\end{tabular} \\ \hline
Fine-scale & \textit{Fenics}, Intel Xeon $E5-2680$ & 2300.822 \\ \hline
Multiscale & \textit{Matlab}, Intel Xeon $E5-2680$ & 1500.611 \\ \hline
\multirow{2}{*}{HM-DenseED} & \textit{PyTorch}, Intel Xeon $E5-2680$ & \multirow{2}{*}{$\underset{(\text{Matlab*})}{113.23}$+$\underset{\text{(PyTorch)}}{32.5}$=145.73} \\
 & \textit{Matlab}$^{*}$, Intel Xeon E5-2680 &  \\ \hline
\multirow{2}{*}{HM-DenseED} & \textit{PyTorch}, NVIDIA Tesla V100 & \multirow{2}{*}{$\underset{(\text{Matlab*})}{113.23}$+$\underset{\text{(PyTorch)}}{6.04}$=119.27} \\
 & \textit{Matlab}$^{*}$, Intel Xeon $E5-2680$ &  \\ \hline
\multirow{2}{*}{\begin{tabular}[c]{@{}c@{}}Bayesian \\ HM-DenseED\end{tabular}} & \textit{PyTorch}, Intel Xeon $E5-2680$ & \multirow{2}{*}{$\underset{(\text{Matlab*})}{113.23}$+$\underset{\text{(PyTorch)}}{627.20}$= 740.43} \\
 & \textit{Matlab}$^{*}$, Intel Xeon $E5-2680$ &  \\ \hline
\multirow{2}{*}{\begin{tabular}[c]{@{}c@{}}Bayesian \\ HM-DenseED\end{tabular}} & \textit{PyTorch}, NVIDIA Tesla V100 & \multirow{2}{*}{$\underset{(\text{Matlab*})}{113.23}$+$\underset{\text{(PyTorch)}}{61.27}$ = 174.5} \\
 & \textit{Matlab}$^{*}$, Intel Xeon $E5-2680$ & 
\end{tabular}
\label{tab: compare2}
\end{table}
\section{Results:}\label{sec:SL3}
In this section, we enumerate the results obtained from the deterministic and Bayesian hybrid surrogates.  The computer code and data can be found on \href{https://github.com/zabaras/bayesmultiscale/}{https://github.com/zabaras/bayesmultiscale/}.
\subsection*{Dataset}
\begin{figure}[H]
  \centering
\includegraphics[scale=0.3]{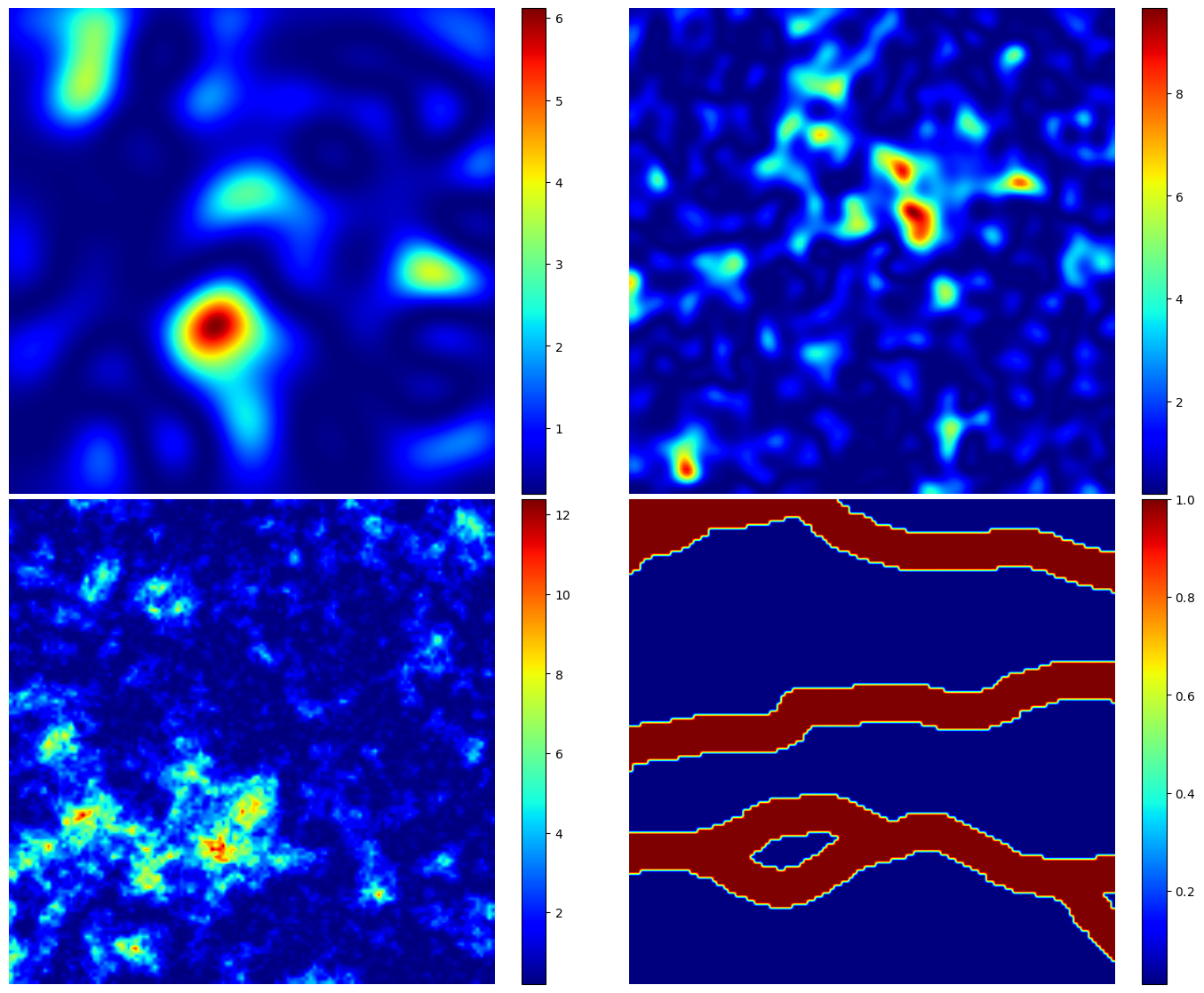}
\caption{Permeability field KLE$-100$ (top left), KLE$-1000$ (top right), KLE$-16384$ (bottom left) and channelized (bottom right).}
\label{fig:Input}
\end{figure}

\begin{figure}[H]
  \centering
\includegraphics[scale=0.35]{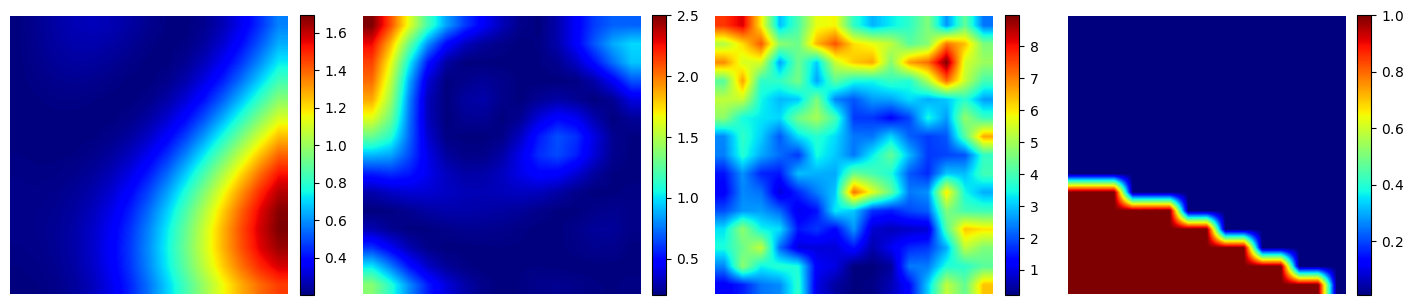}
\includegraphics[scale=0.35]{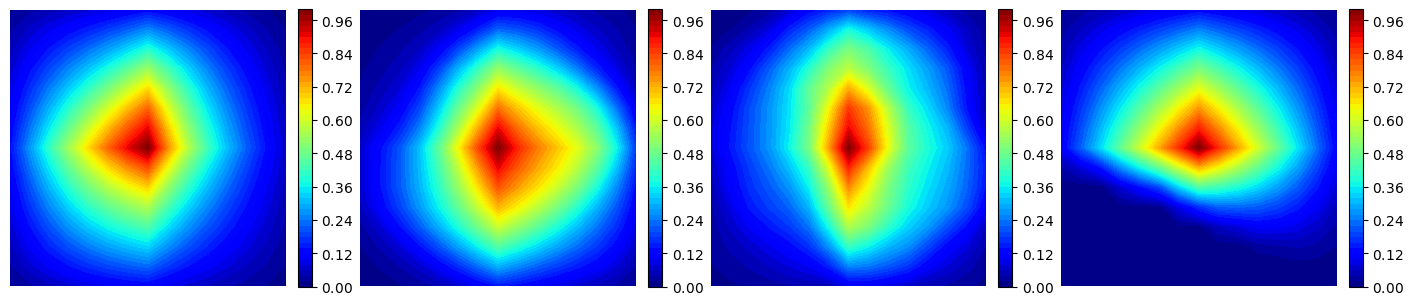}
\caption{Permeability coarse block (top) for KLE$-100$, KLE$-1000$, KLE$-16384$ and channelized field and the corresponding basis functions (bottom).}
\label{fig:Patch}
\end{figure}

 We consider two types of  input permeability dataset, namely, Gaussian random field and channelized field, as shown in Fig.~\ref{fig:Input}. For the first type of the input permeability dataset, the input log-permeability field is considered to be a Gaussian random field with constant mean $m$ and covariance function $k$  specified in the following form using the $L_2$ norm in the exponent 
i.e,
$k(s,s') = exp(-\|s-s'\|/l)$
\begin{equation}\label{Eq:GRF}
    K(s) = exp(G(s)), \hspace{2cm} G(\cdot) \sim \mathcal{N}(m,k(\cdot,\cdot)).
\end{equation}
Fig.~\ref{fig:Patch} shows the permeability field for a given support region and the corresponding basis function. In this work, we consider  a $128 \times 128$ grid and apply a uniformly distributed inflow and outflow conditions on the left and right sides of the domain $[-1,1]$, respectively. The boundary conditions for the upper and lower walls are Neumann with zero flux. Here, an $8 \times 8$ coarse-grid is constructed. Also the number of fine-cells in the support region is $16 \times 16$. 
 For all the GRF datasets considered, the log-permeability field in Eq.~(\ref{Eq:GRF}) was generated with $m=0$ and $l=0.1$. 
 The intrinsic dimensionality is captured  using the  Karhunen-Lo\`{e}ve expansion (KLE)~\cite{van2009dimensionality}. 
 The KLE for the log-permeability field is given as follows:
\begin{equation}
G(\bm{s}) = m+\sum_{r=1}^{R}\sqrt{\lambda_{r}}\xi_{r}\chi_{r}(\bm{s}),
\end{equation}
where $m$ is the mean of the GRF, and $\lambda_{r}$ and $\chi_{r}(\bm{s})$ are the eigenvalues and eigenfunctions of the   covariance matrix for the $128 \times 128$ spatial discretization. Here, $R$ is the number of KLE coefficients and we consider $R = 100$ and $1000$. For the KLE $-16384$ dataset, we sample from the exponential GRF as mentioned in Eq.~(\ref{Eq:GRF})  without any dimensionality reduction. 
For the training and test datasets, we consider a  Latin hypercube design to obtain the samples for $\xi_r$, i.e., $\xi_r = F^{-1}(\rho_r)$, where $\rho_r$ is obtained from a hypercube of $[0,1]^{R}$ and $F$ is the cumulative distribution function   of the standard normal distribution~\cite{zhu2018bayesian}.
\par
The other input field is considered to be a channelized field where the samples of size $128 \times 128$ are cropped from a large training image~\cite{laloy2018training}. The channelized field consists of two facies: high-permeability channels and low-permeability non-channels, which are assigned  values of $1$ and $0$, respectively.
The boundary conditions and coarse-grid size remain the same as mentioned above.  

\subsection{Deterministic hybrid surrogate model}
The hybrid DenseED-multiscale method is trained for flow in porous media using permeability obtained from a GRF with KLE$-100$, KLE$-1000$, and KLE$-16384$. Channelized field is also considered.  
 After extensive network architecture experimentation for training and testing, we implement $4-8-4$ DenseED configurations. The details regarding testing different configurations are given in Appendix~\ref{sec:App-B1}. We train the DenseED with an $L_2$ loss function defined between the approximated hybrid predicted pressure and fine-scale pressure. Adam~\cite{kingma2014adam} optimizer was used for training $200$ epochs, with a learning rate of $1e-5$ and a plateau scheduler on the RMSE. Weight decay was set to $1e-6$. Since we consider $8 \times 8$ as the coarse-grid, a total of $144$ coarse-blocks that correspond to the interior support regions are considered for training.  In this work, a batch size of $1$ input field is considered for training as it consists of $144$ coarse blocks corresponding to the interior support regions. 
 \par
 This is the minimum number of basis that has to be considered since the loss function is between the predicted pressure and fine-scale pressure of one complete image. Here, we consider $256$ test data that was unseen during training to test our network. In Appendix~\ref{sec:App-C}, we compare the DenseED network and a fully-connected network for training the basis functions and we observe that the hybrid DenseED performs better when compared to the hybrid fully-connected model.

\subsubsection{Surrogate (DenseED) model}
Zhu and Zabaras~\cite{zhu2018bayesian}  developed surrogate models using a dense encoder-decoder network (DenseED)~\cite{huang2017densely} for uncertainty quantification and propagation in problems governed by stochastic PDEs  with high-dimensional stochastic input. They have adopted the image-to-image regression approach to solve the problem with a limited dataset. Therefore, in this work, we consider a dense encoder-decoder network (DenseED)~\cite{huang2017densely} as the conventional (baseline) surrogate model and compare its performance with the hybrid multiscale DenseED model. The model is trained for KLE$-100$, KLE$-1000$, KLE$-16384$, and channelized flow with a $2-2-4-2-2$ DenseED configuration. The configuration $2-2-4-2-2$ was found to be the best configuration for the standard surrogate (DenseED) model. Adam~\cite{kingma2014adam} optimizer was used for training $200$ epochs, with a learning rate of $1e-3$ and a plateau scheduler on the RMSE.  Weight decay was set to $5e-4$. $256$ data is considered for testing the network, and it is the same test data that was implemented in the hybrid model. The input dimension for the standard surrogate model is $128 \times 128$ and a single-channel output (pressure) is considered. We adopt a similar method to the multiscale method to obtain the fluxes, i.e., by post-processing the predicted DenseED pressure with the help of the transmissibility matrix and obtaining the predicted velocities.

 In this work, the training data is sampled using Latin hypercube sampling. The data is structured as follows: for training KLE$-100$, we consider $32, 64$, and $96$ training data. KLE$-1000$ is trained with $64, 96$, and $128$ training data, KLE$-16384$ is trained with $96,~128$ and $160$ training data and the  channelized field is trained with $160$ training data. Testing is performed with $500$ data, and for uncertainty propagation, we consider $10000$ data.

\par
\begin{figure}[H]
    \centering
    \includegraphics[scale=0.25]{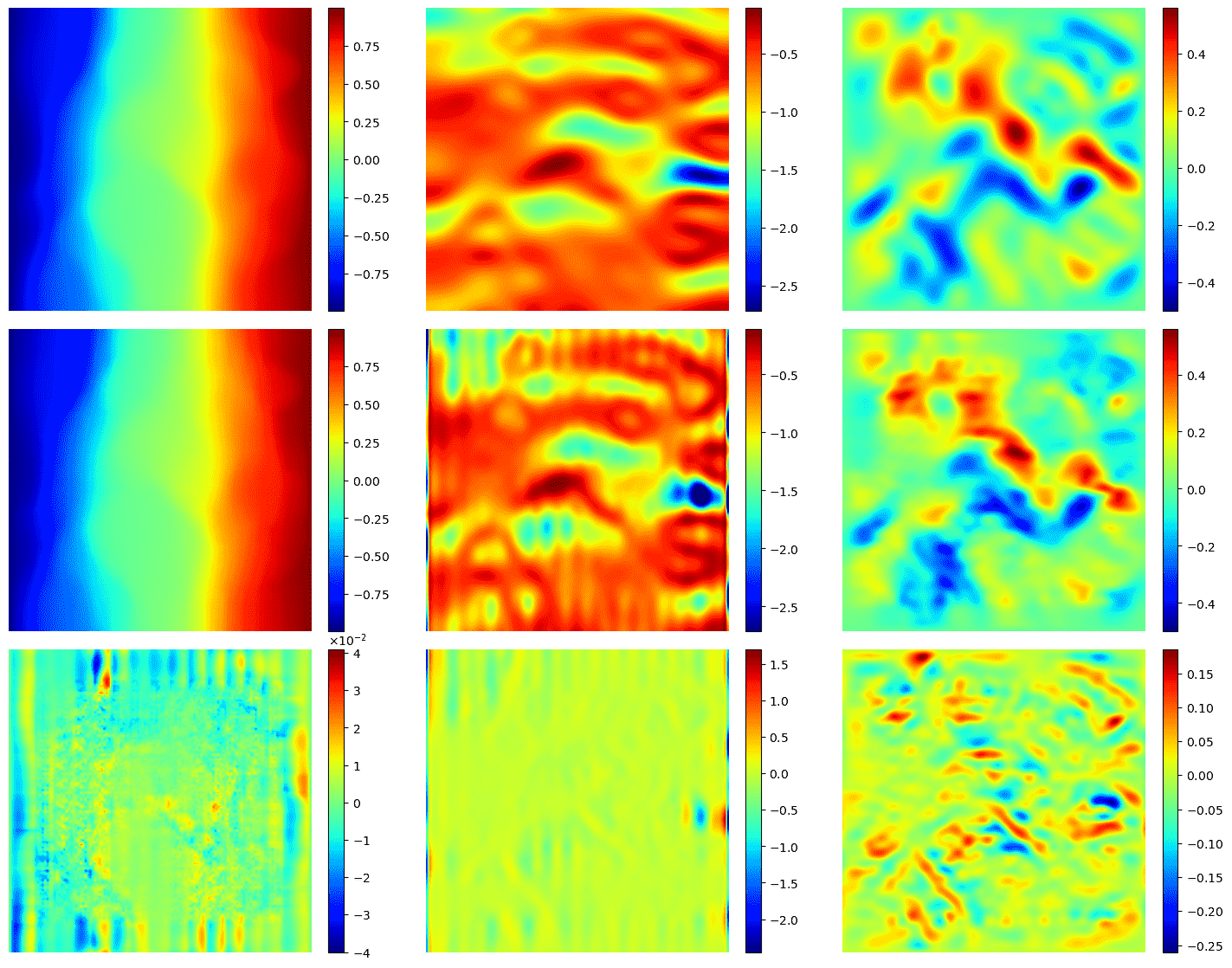}
    \caption{\textbf{HM-DenseED model:} Prediction of KLE$-100$ with $32$ training data: first row (from left to right) shows the target (pressure, $x-$velocity and $y-$velocity components), the second row shows the corresponding predictions, and the last row shows the error between the corresponding targets and predictions. 
    }
    \label{fig:KLE_100_1}
\end{figure}
\begin{figure}[H]
    \centering
    \includegraphics[scale=0.25]{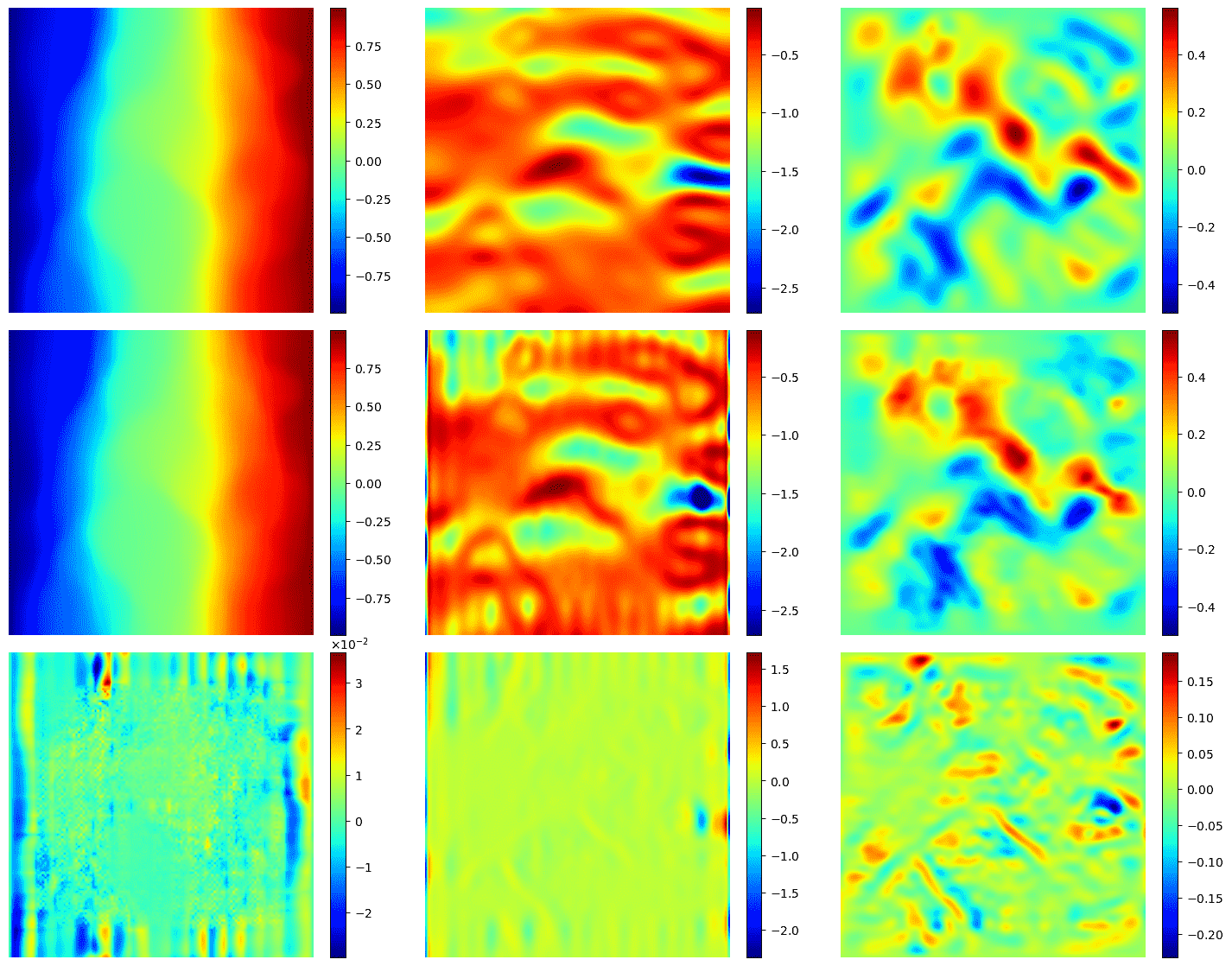}
    \caption{\textbf{HM-DenseED model:} Prediction of KLE$-100$ with $96$ training data: first row (from left to right) shows the target (pressure, $x-$velocity and $y-$velocity components), the second row shows the corresponding predictions, and the last row shows the error between the corresponding targets and predictions. 
    }
    \label{fig:KLE_100_3}
\end{figure}
\begin{figure}[H]
    \centering
    \includegraphics[scale=0.25]{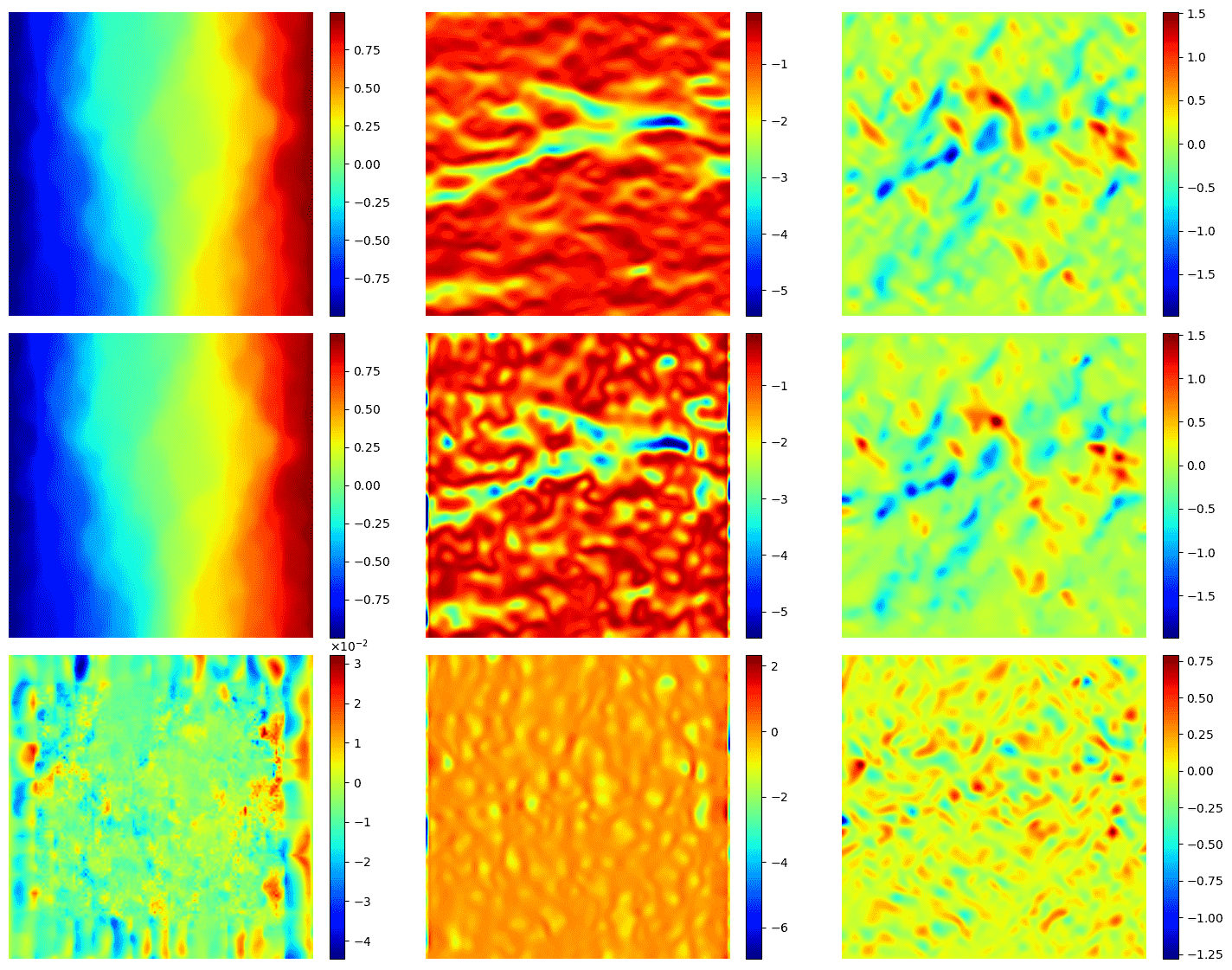}
    \caption{\textbf{HM-DenseED model:} Prediction of KLE$-1000$ with $64$ training data: first row (from left to right) shows the target (pressure, $x-$velocity and $y-$velocity components), the second row shows the corresponding predictions, and the last row shows the error between the corresponding targets and predictions. 
    }
    \label{fig:KLE_1000_1}
\end{figure}
\begin{figure}[H]
    \centering
    \includegraphics[scale=0.25]{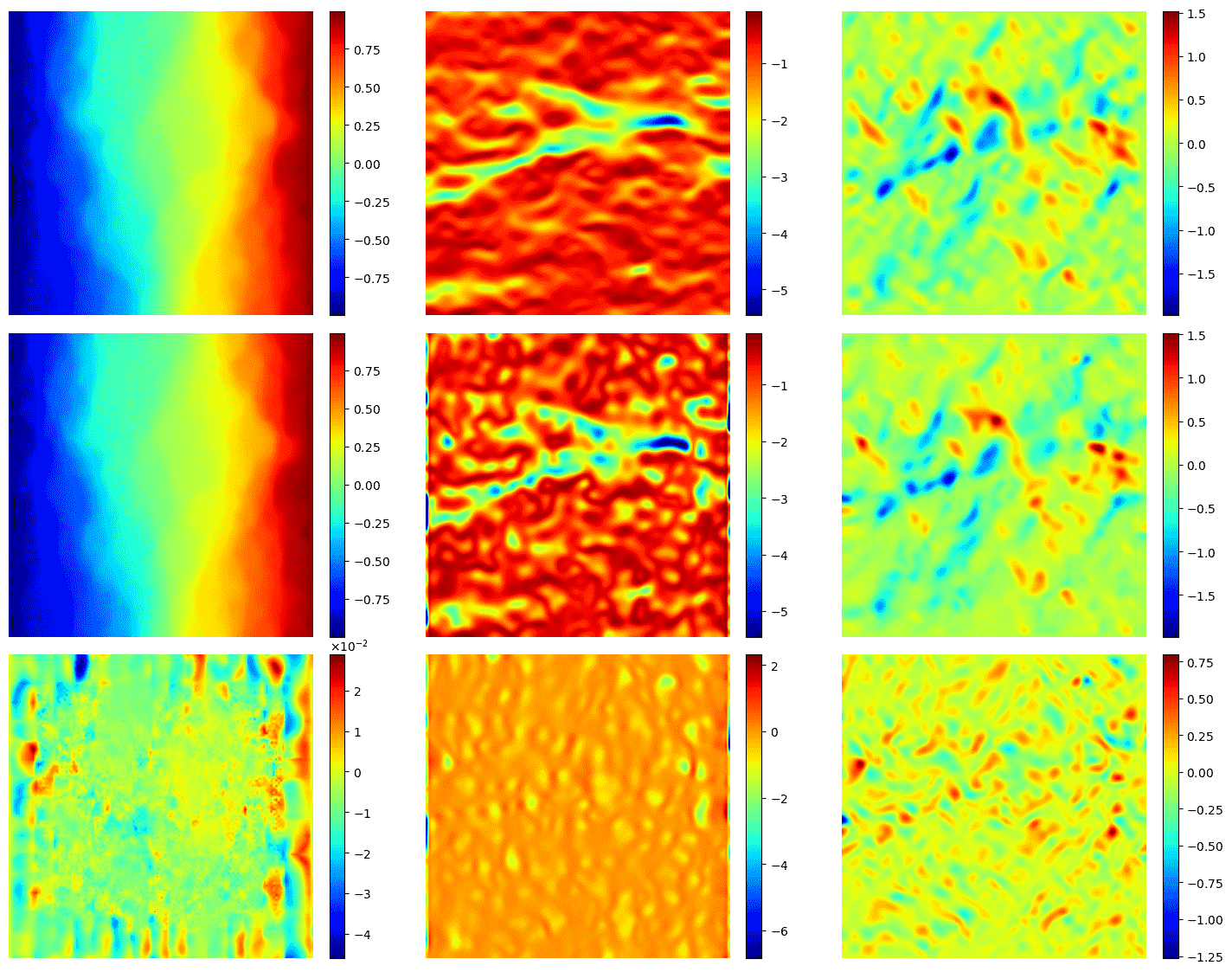}
    \caption{\textbf{HM-DenseED model:} Prediction of KLE$-1000$ with $128$ training data: first row (from left to right) shows the target (pressure, $x-$velocity and $y-$velocity components), the second row shows the corresponding predictions, and the last row shows the error between the corresponding targets and predictions. 
    }
    \label{fig:KLE_1000_3}
\end{figure}
\begin{figure}[H]
    \centering
    \includegraphics[scale=0.25]{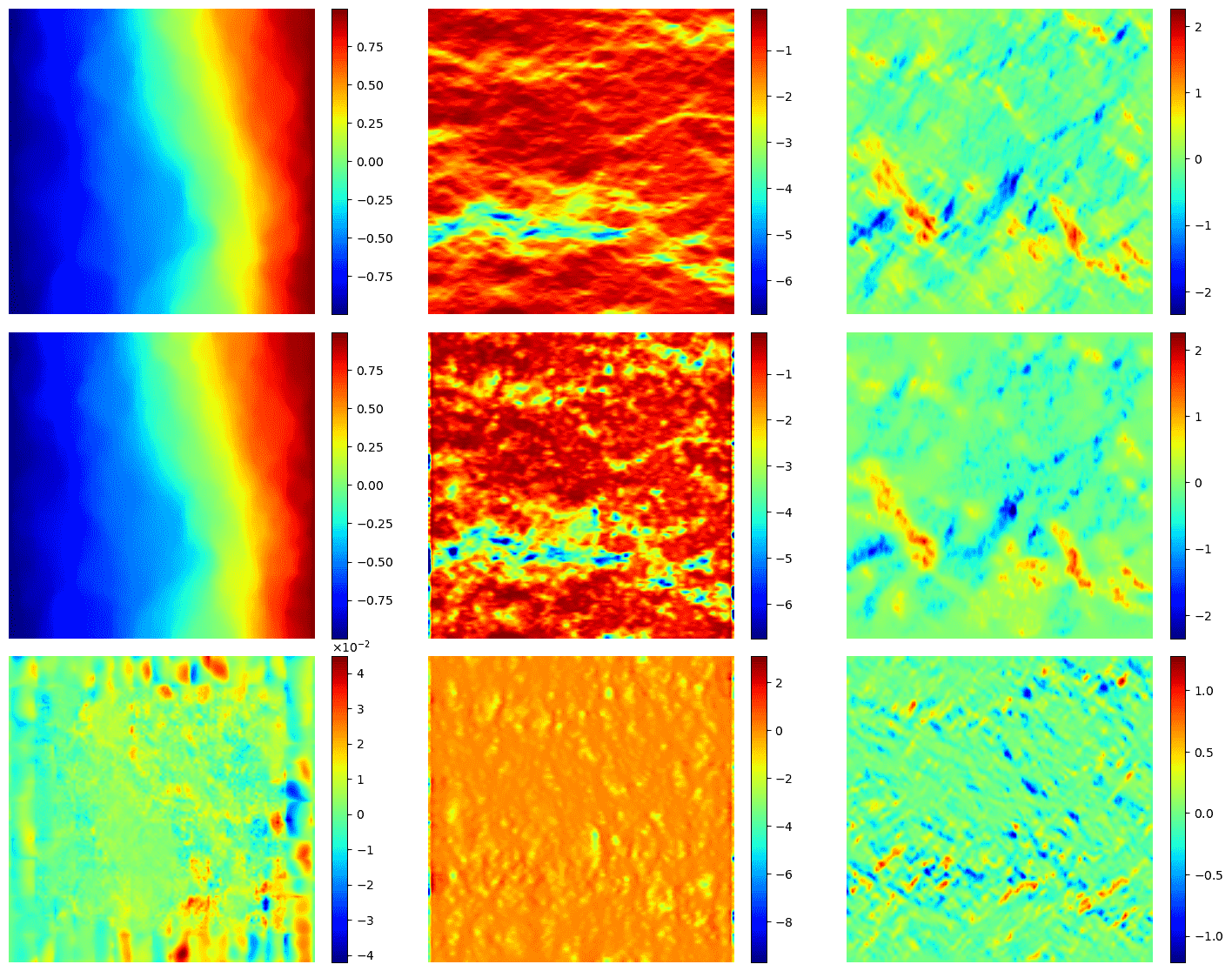}
    \caption{\textbf{HM-DenseED model:} Prediction of KLE$-16384$ with $96$ training data: first row (from left to right) shows the target (pressure, $x-$velocity and $y-$velocity components), the second row shows the corresponding predictions, and the last row shows the error between the corresponding targets and predictions. 
    }
    \label{fig:KLE_16384_1}
\end{figure}
\begin{figure}[H]
    \centering
    \includegraphics[scale=0.25]{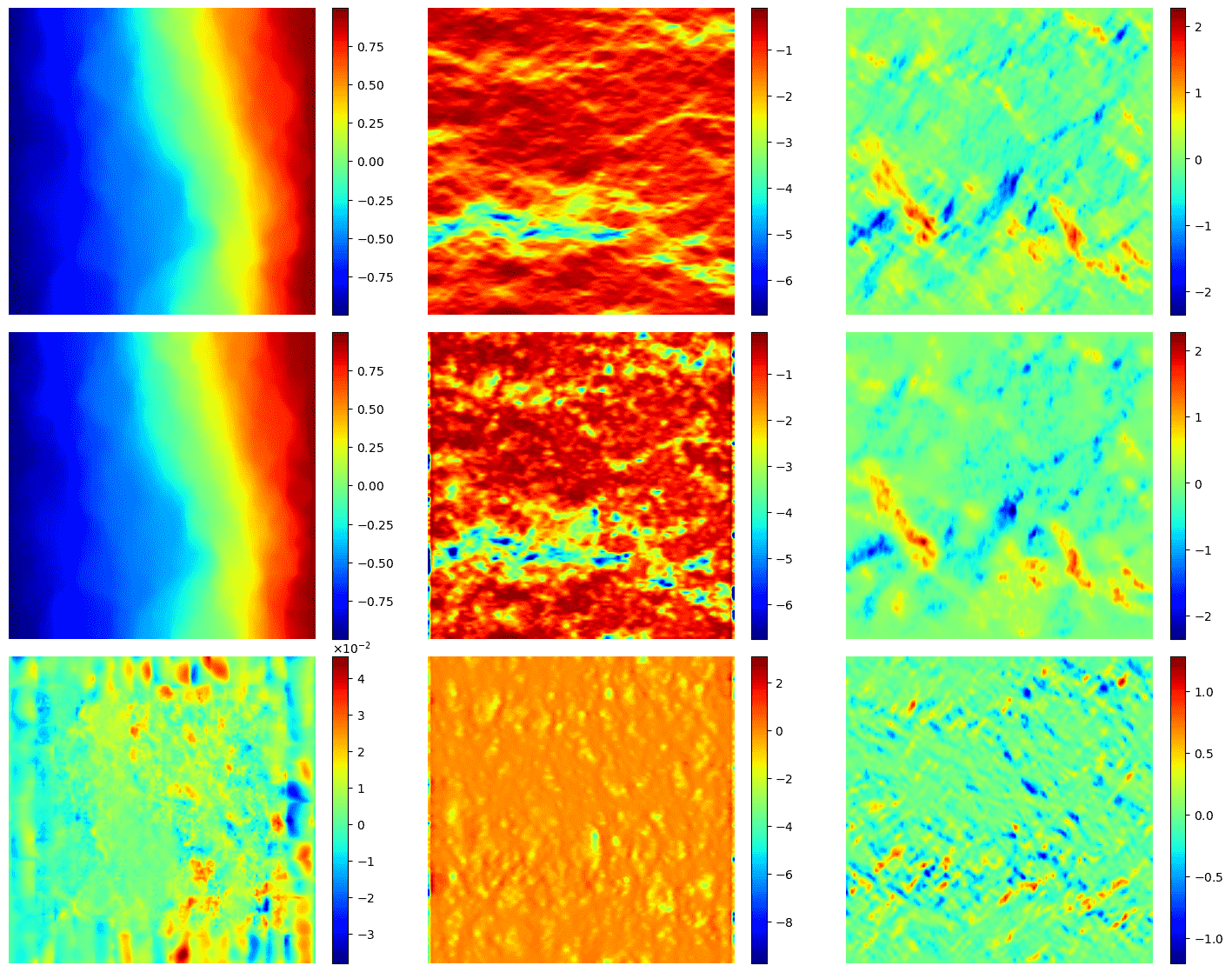}
    \caption{\textbf{HM-DenseED model:} Prediction of KLE$-16384$ with $160$ training data: first row (from left to right) shows the target (pressure, $x-$velocity and $y-$velocity components), the second row shows the corresponding predictions, and the last row shows the error between the corresponding targets and predictions. 
    }
    \label{fig:KLE_16384_3}
\end{figure}
\begin{figure}[H]
    \centering
    \includegraphics[scale=0.25]{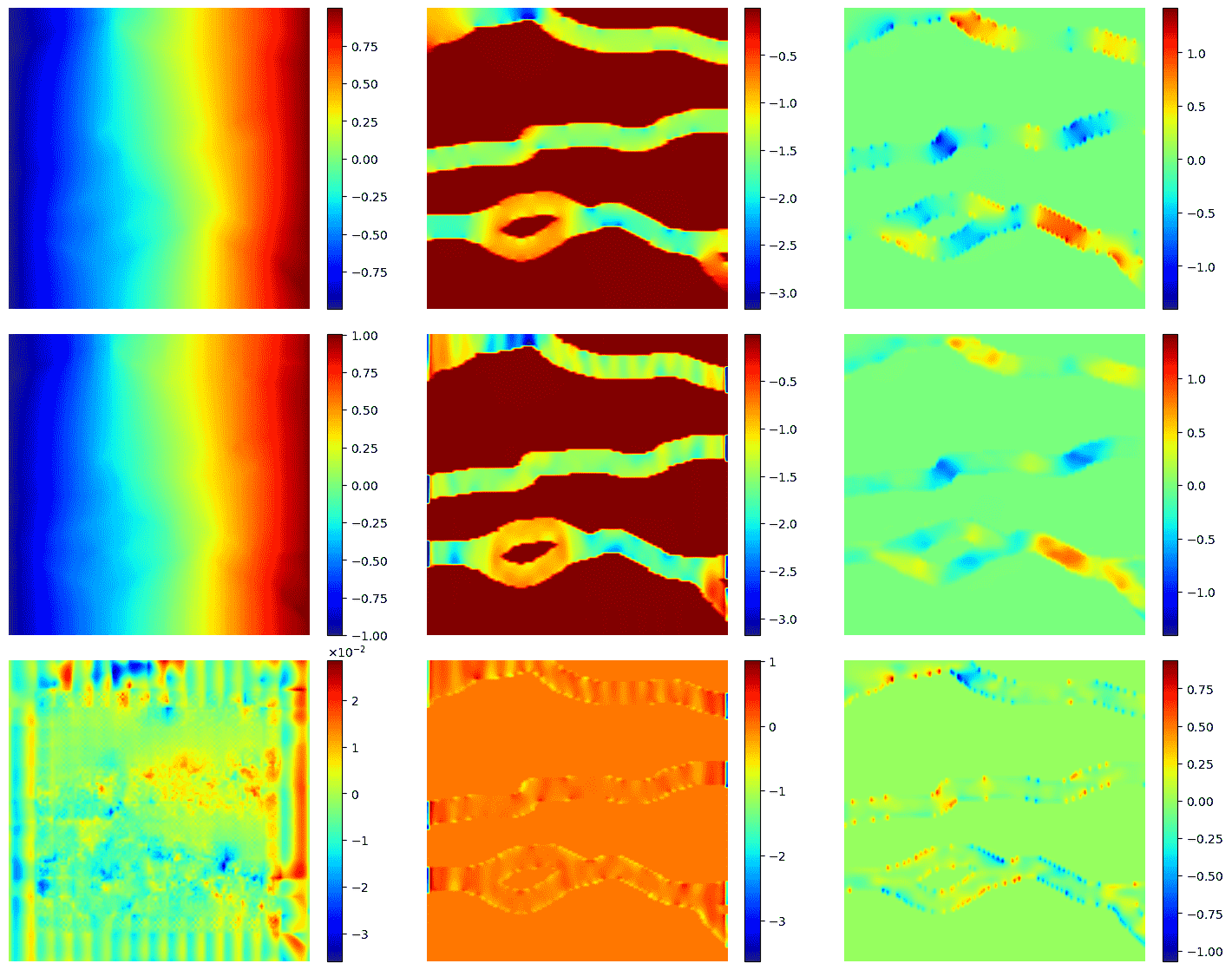}
    \caption{\textbf{HM-DenseED model:} Prediction of channelized field with $160$ training data: first row (from left to right) shows the target (pressure, $x-$velocity and $y-$velocity components), the second row shows the corresponding predictions, and the last row shows the error between the corresponding targets and predictions. 
    }
    \label{fig:channel_result}
\end{figure}

From the aforementioned hybrid DenseED-multiscale model, we obtain $144$ coarse-blocks (interior support regions) per image, and therefore, the total training data is ($144 \times$ number of fine-scale input realizations). That is for KLE$-100$ we have $4608$, $9216$ and $13824$ for $32$, $64$ and $96$ fine-scale data, respectively; for KLE$-1000$, we have $9216$, $13824$ and $18432$ for $64$, $96$ and $128$ fine-scale data; finally, for KLE$-16384$, we have $13824$, $18432$ and $23040$ for $96$, $128$ and $160$ fine-scale data, respectively. The channelized flow is trained with $160$ fine-scale data, that is $18432$ coarse-scale data. This is the advantage of coarsening the given fine-scale image so that we can obtain more data for training. 
As mentioned earlier, the Dense-ED model in the HM-DensED framework maps the fine scale permeability field for a given interior support region to the corresponding basis functions for that interior support regions. Since we train our model with the loss function between the fine scale pressure and the predicted pressure (post-processed using the predicted basis functions for interior support regions and basis functions from non-interior region as mentioned in Sec.~\ref{sec:SL21}), this make it challenging for the hybrid network.  
The input permeabilities for these predictions can be found in the GitHub repository that accompanies this paper. In Figs.~\ref{fig:KLE_100_1}--\ref{fig:channel_result}, we show the prediction of pressure, horizontal flux and vertical flux for both the GRF and channelized permeability field for the HM-DenseED model. We observe that the HM-DenseED prediction for ultra high-dimension (KLE$-16384$) is good even with limited fine-scale data when compared to the standard surrogate model based on the DenseED. We also observe the vertical stripe pattern in the prediction of velocities, and also in the error (predicted velocity versus actual velocity) plot.  The error in this region is high because the hybrid model predicts the approximate pressure, and a small difference in the predicted pressure compared to actual pressure will impact the final predicted velocity in the overlap regions of the support region.
\par
\begin{figure}[H]
\begin{center}
	\centering
	\includegraphics[scale=0.4]{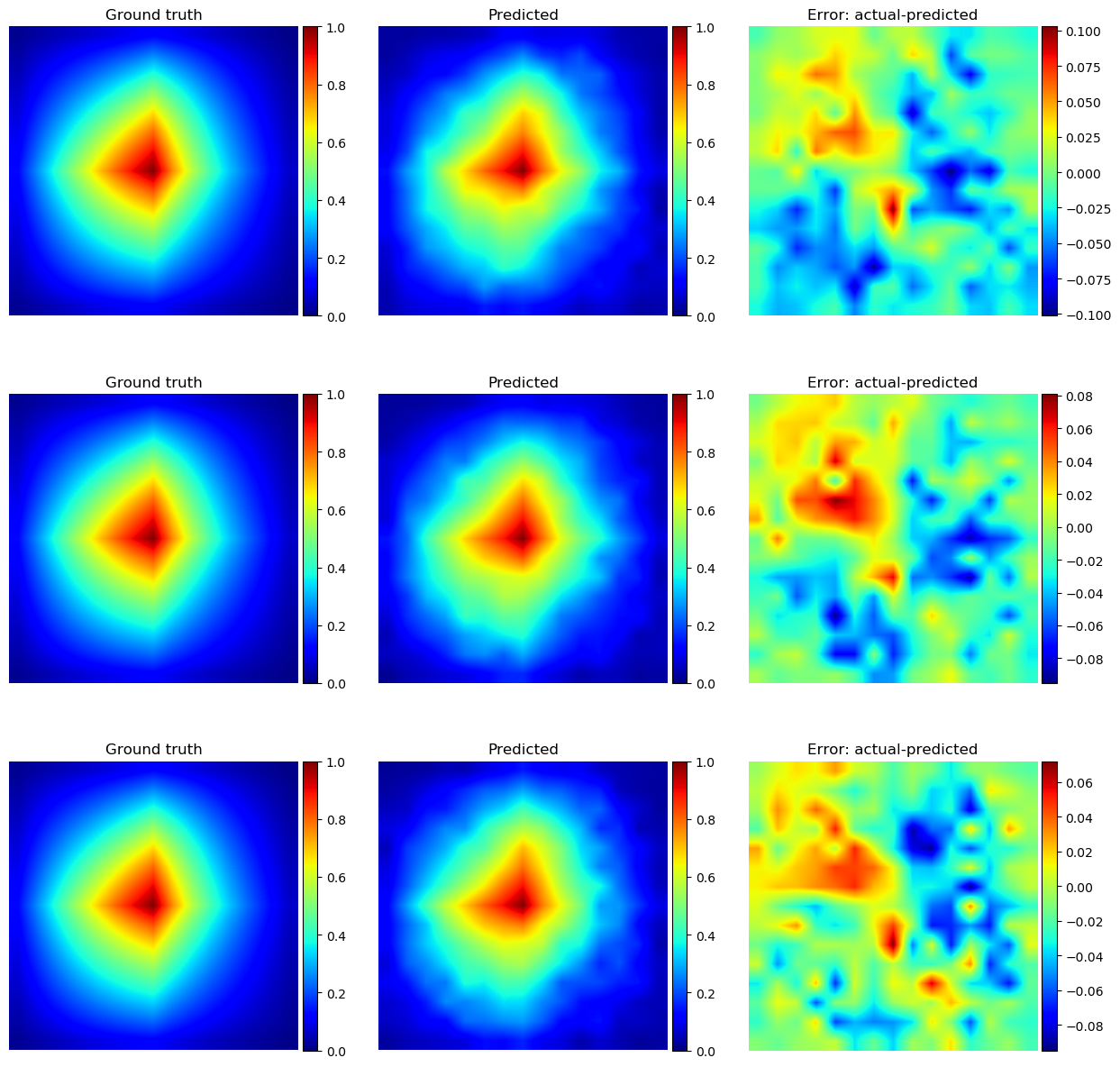}
	\caption{The basis function for KLE$-100$. The first row shows the ground truth basis function, HM-DenseED model predicted basis function, and the error between them for the model trained with $32$ training data. The second row shows the ground truth basis function, HM-DenseED model predicted basis function, and the error between them for the model trained with $64$ training data. The last row shows the ground truth basis function, HM-DenseED model predicted basis function, and the error between them for the model trained with $96$ training data.}
	\label{fig:basis_plot1}
  \end{center}
\end{figure}
\begin{figure}[H]
\begin{center}
	\centering
	\includegraphics[scale=0.4]{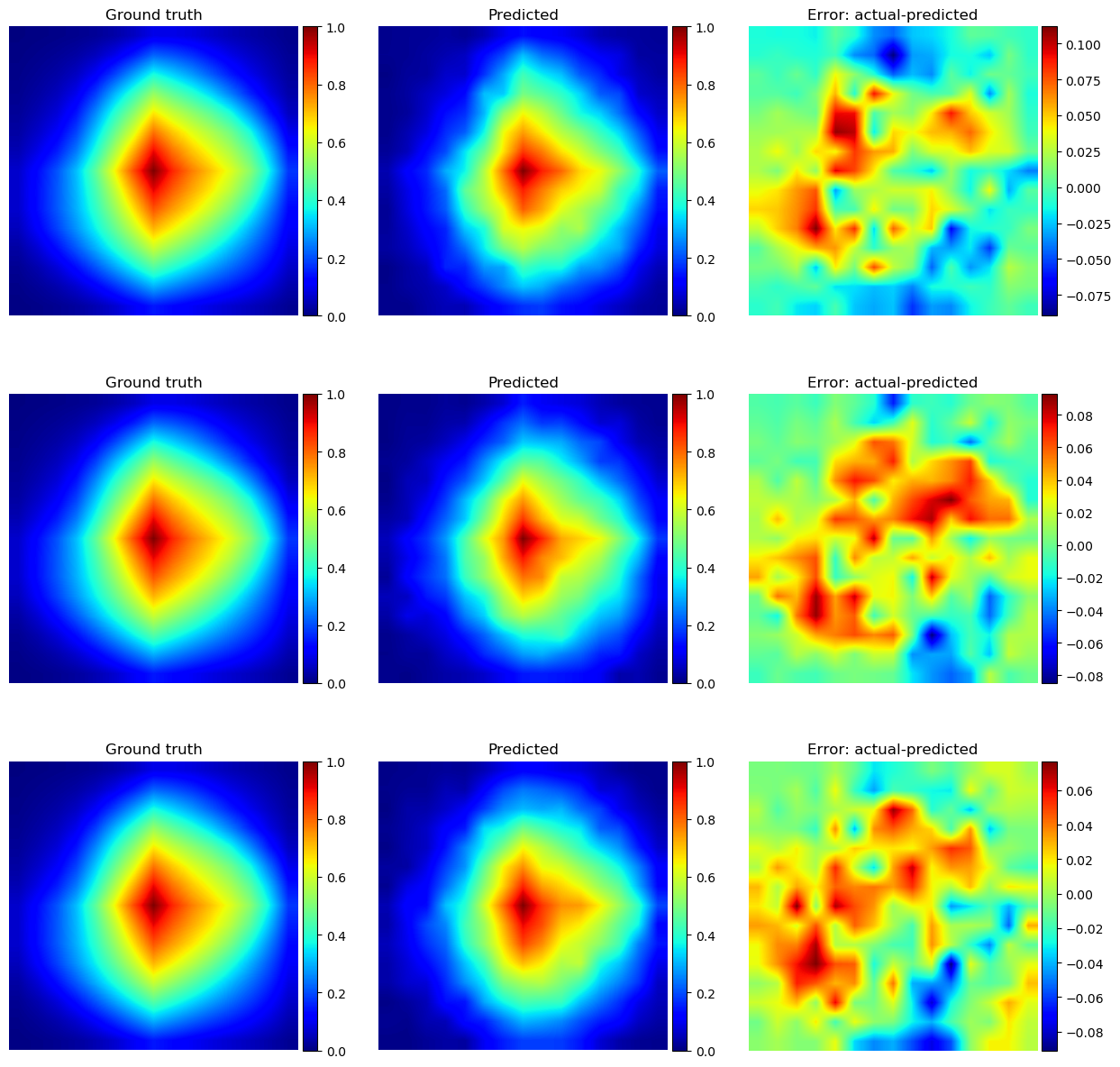}
	\caption{ The basis function for KLE$-1000$. The first row shows the ground truth basis function, HM-DenseED model predicted basis function, and the error between them for the model trained with $64$ training data. The second row shows the ground truth basis function, HM-DenseED model predicted basis function, and the error between them for the model trained with $96$ training data. The last row shows the ground truth basis function, HM-DenseED model predicted basis function, and the error between them for the model trained with $128$ training data.}
	\label{fig:basis_plot2}
  \end{center}
\end{figure}
\begin{figure}[H]
\begin{center}
	\centering
	\includegraphics[scale=0.4]{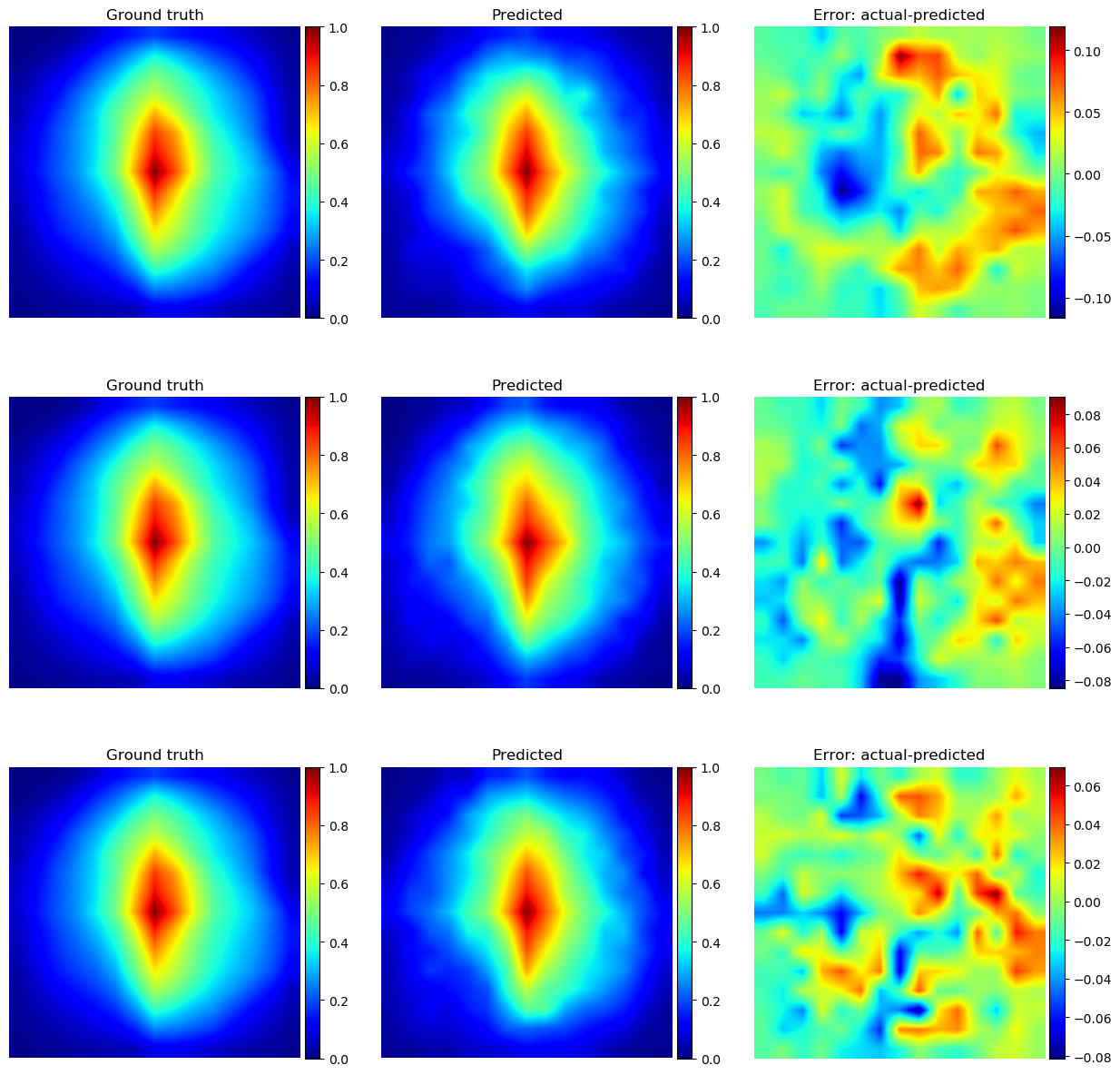}
	\caption{The basis function for KLE$-16384$. The first row shows the ground truth basis function, HM-DenseED model predicted basis function, and the error between them for the model trained with $96$ training data. The second row shows the ground truth basis function, HM-DenseED model predicted basis function, and the error between them for the model trained with $128$ training data. The last row shows the ground truth basis function, HM-DenseED model predicted basis function, and the error between them for the model trained with $160$ training data.}
	\label{fig:basis_plot3}
  \end{center}
\end{figure}
\begin{figure}[H]
\begin{center}
	\centering
	\includegraphics[scale=0.4]{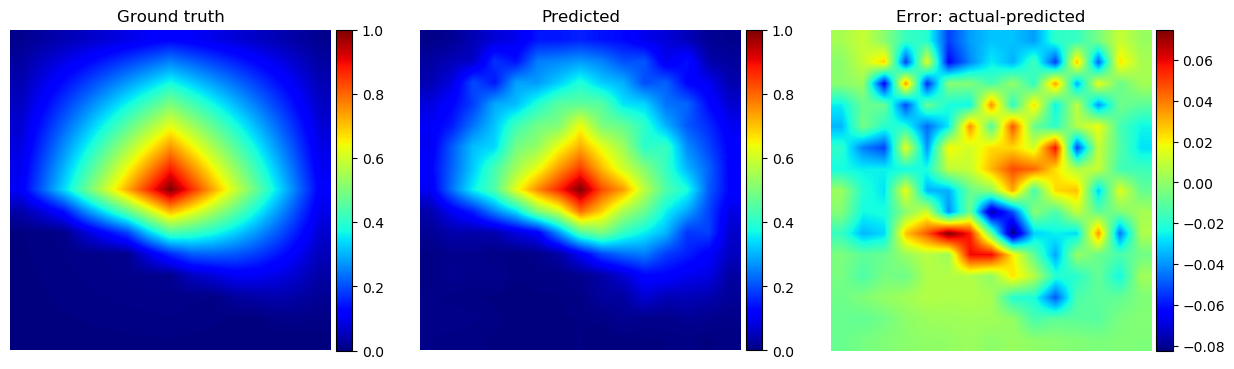}
	\caption{ The basis function for the channelized field. Here, we show the ground truth basis function, HM-DenseED model predicted basis function, and the error between them for the model trained with $160$ training data.}
	\label{fig:basis_plot4}
  \end{center}
\end{figure}

For a test data that was unseen during training, we show the computed basis functions along with the ground truth basis functions and the error between them in Figs~\ref{fig:basis_plot1}--\ref{fig:basis_plot4} for KLE$-100$, KLE$-1000$, KLE$-16384$ and the channelized field. Note that these basis functions are from the interior support region for the test data shown in Figs.~\ref{fig:KLE_100_1}--\ref{fig:channel_result}.

\subsubsection{Comparison of the hybrid DenseED-multiscale and DenseED surrogates}
\begin{figure}[H]
    \centering
    \subfigure[KLE$-100$ ($32-$data)]
    {
        \includegraphics[width=0.3\textwidth]{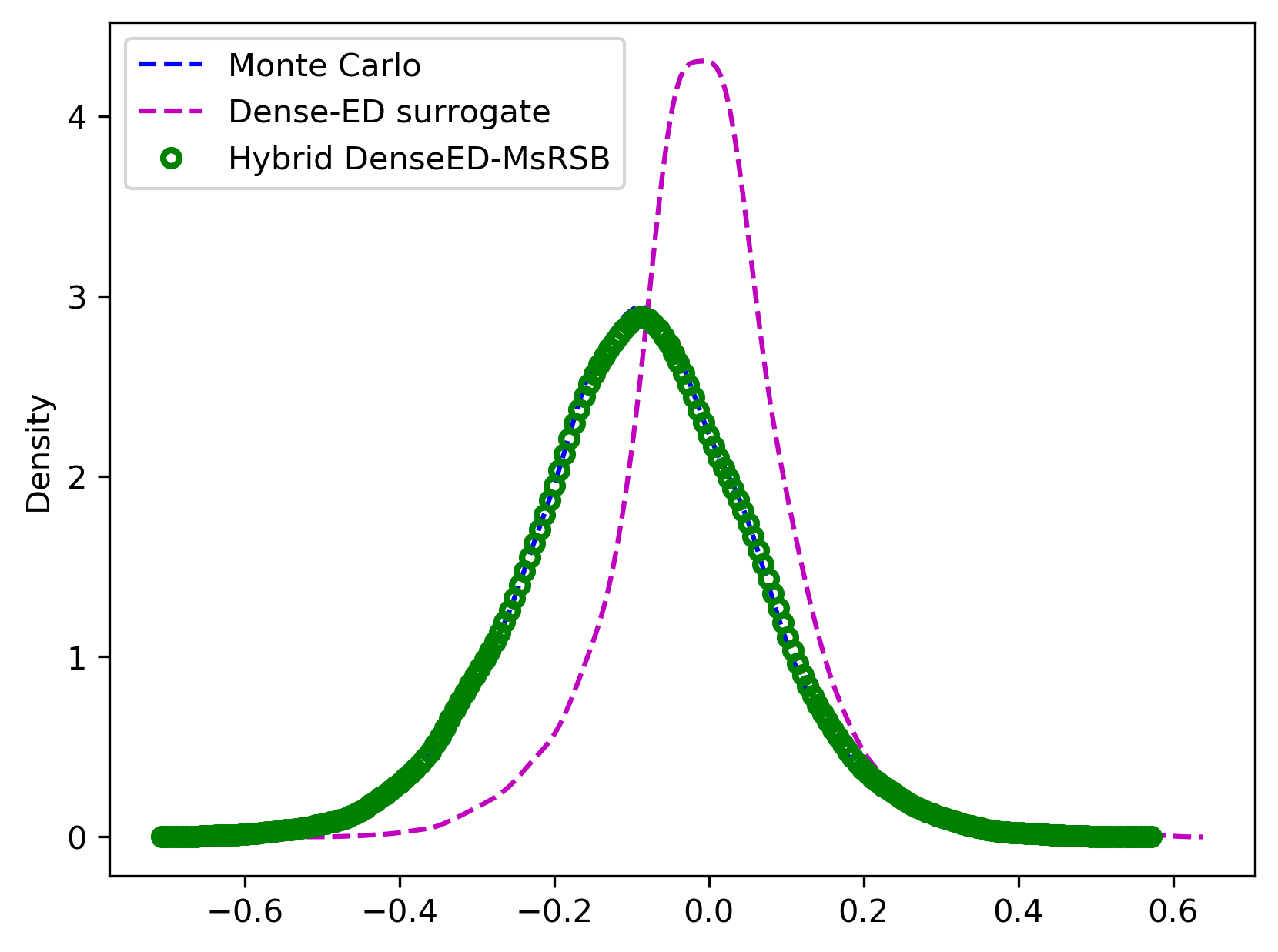}
        \label{fig:first_sub1_1}
    }
    \subfigure[KLE$-100$ ($64-$data)]
    {
        \includegraphics[width=0.3\textwidth]{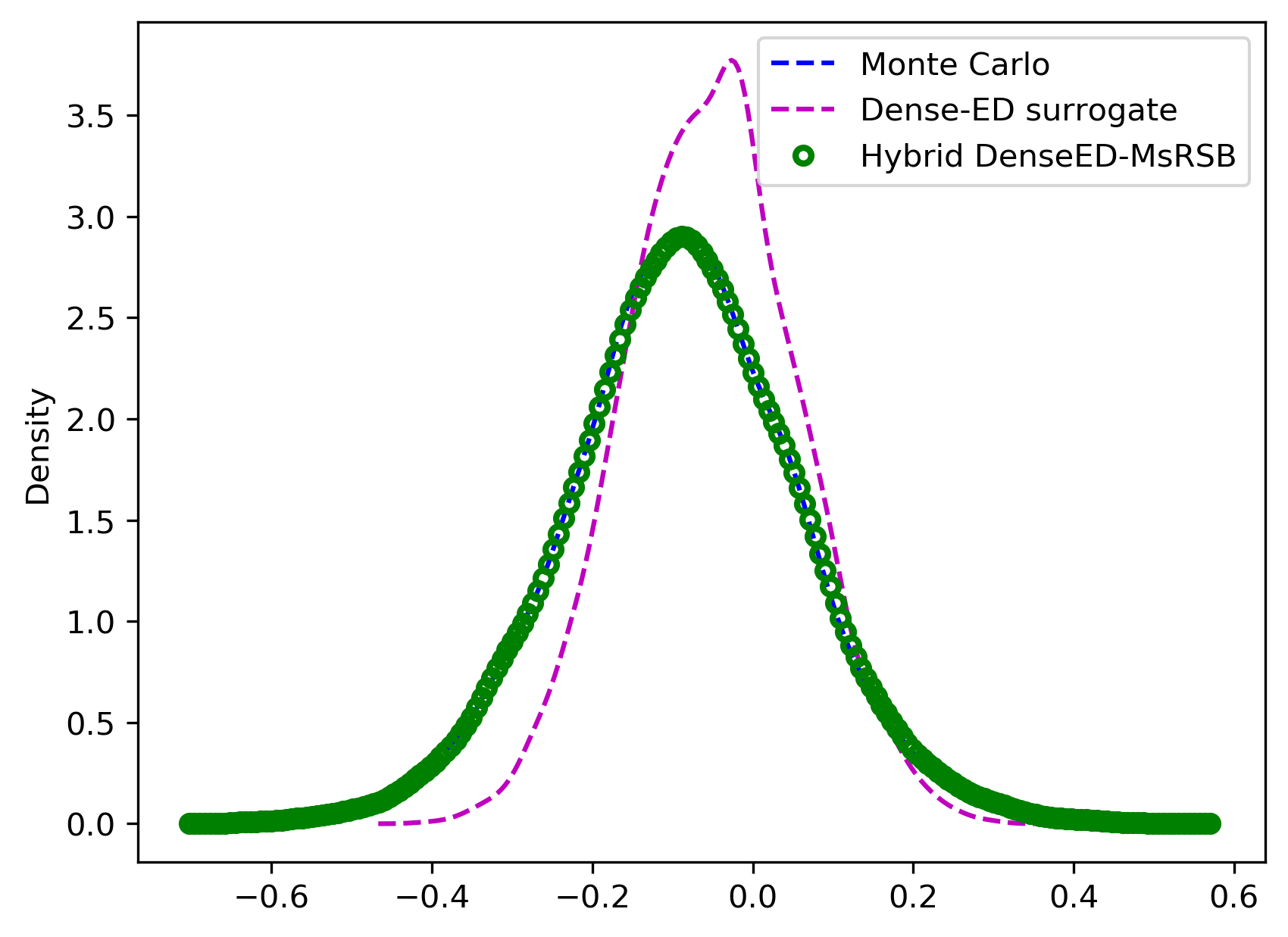}
        \label{fig:first_sub_1}
    }
    \subfigure[KLE$-100$ ($96-$data)]
    {
        \includegraphics[width=0.3\textwidth]{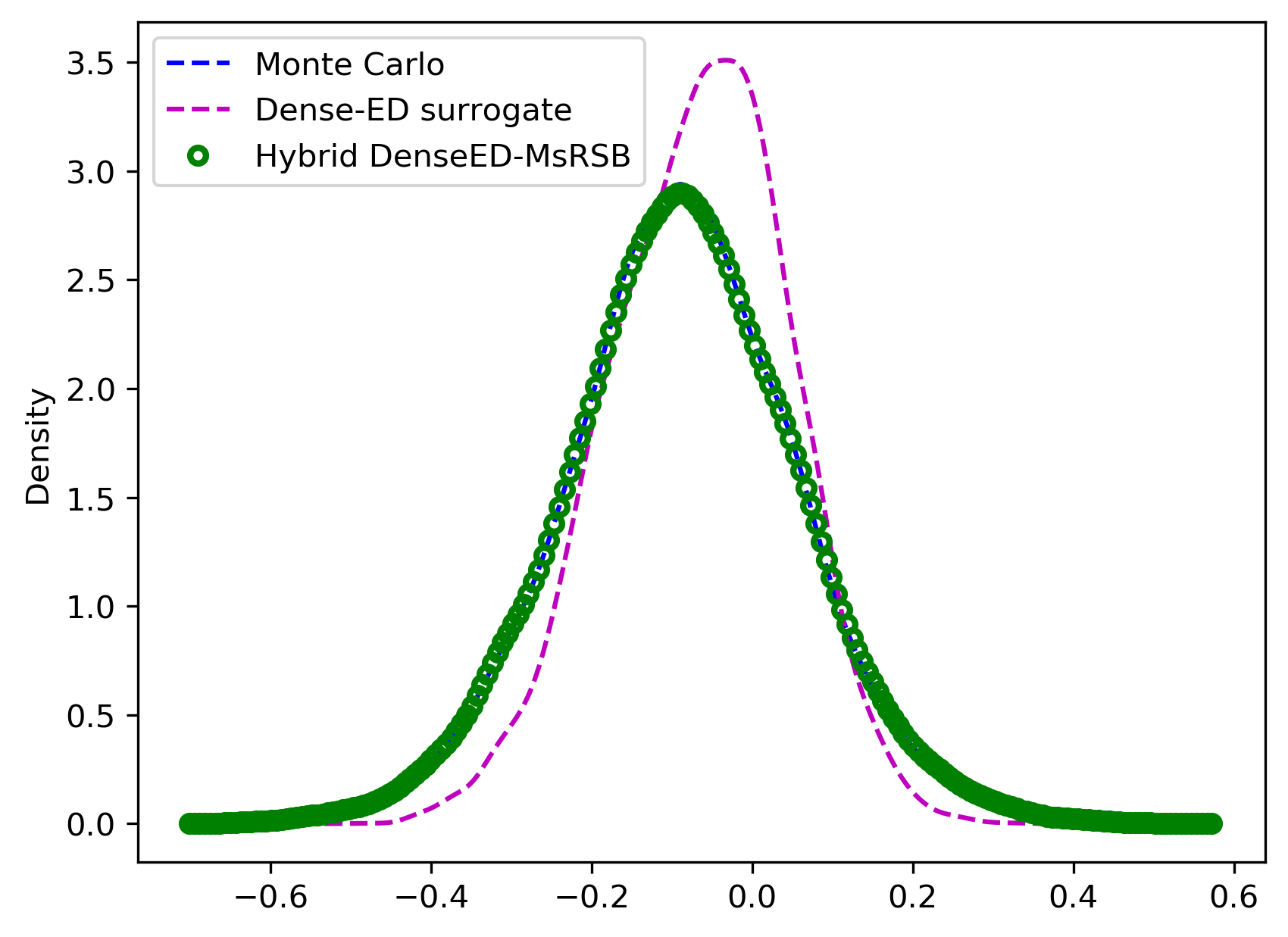}
        \label{fig:first_sub_2}
    }
    \caption{Distribution estimate for the pressure for KLE$-100$. Location: $(0.6,0.4)$. Here, the Monte Carlo result is shown in the blue dashed line; the hybrid DenseED result is shown in green circles, and the DenseED model result is shown in magenta dashed line.}
    \label{fig:pressure_PDF_100}
\end{figure}

\begin{figure}[H]
    \centering
    \subfigure[KLE$-1000$ ($64$~data)]
    {
        \includegraphics[width=0.3\textwidth]{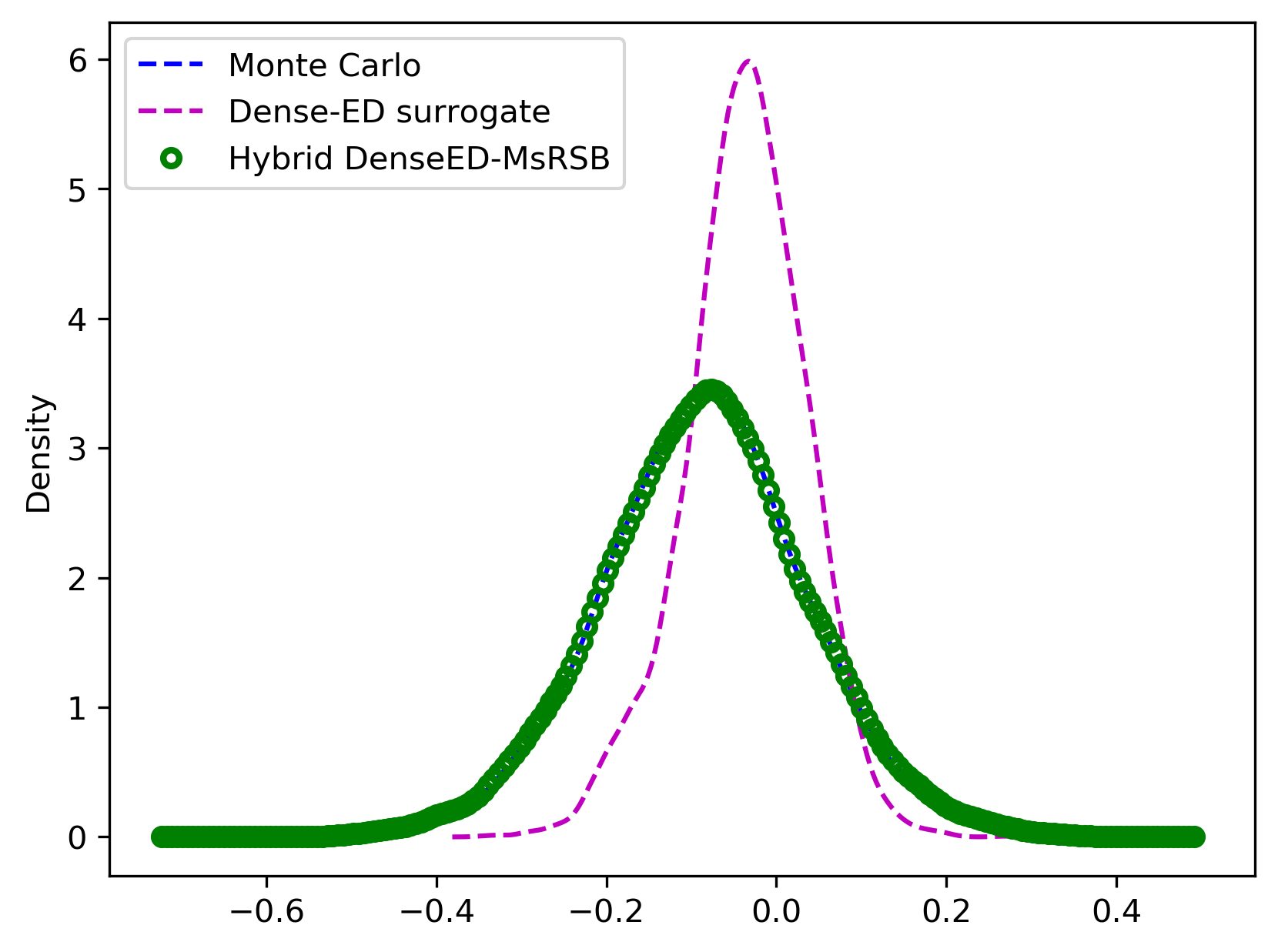}
        \label{fig:first_sub_3}
    }
    \subfigure[KLE$-1000$ ($96$~data)]
    {
        \includegraphics[width=0.3\textwidth]{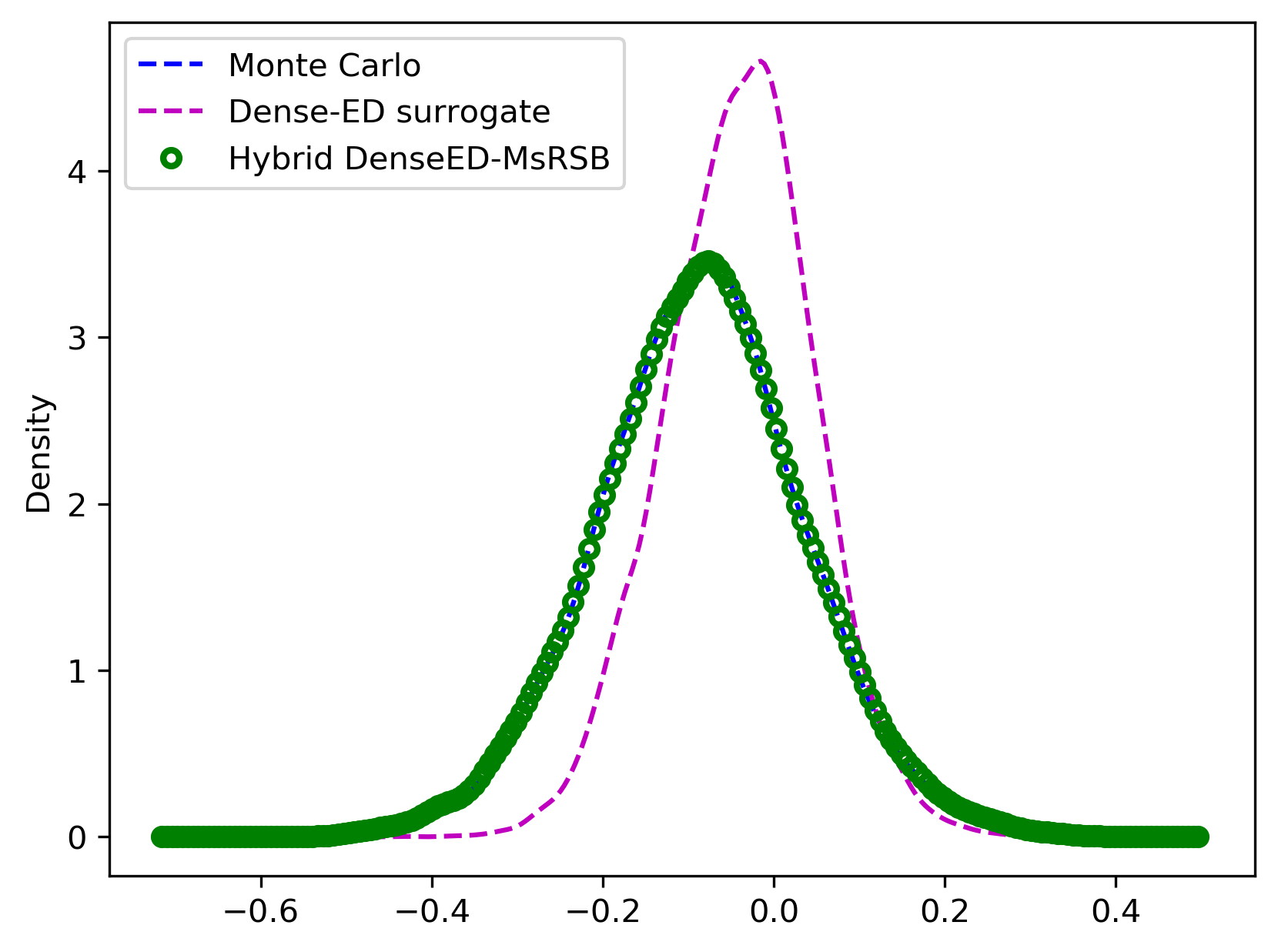}
        \label{fig:first_sub_4}
    }
    \subfigure[KLE$-1000$ ($128$~data)]
    {
        \includegraphics[width=0.3\textwidth]{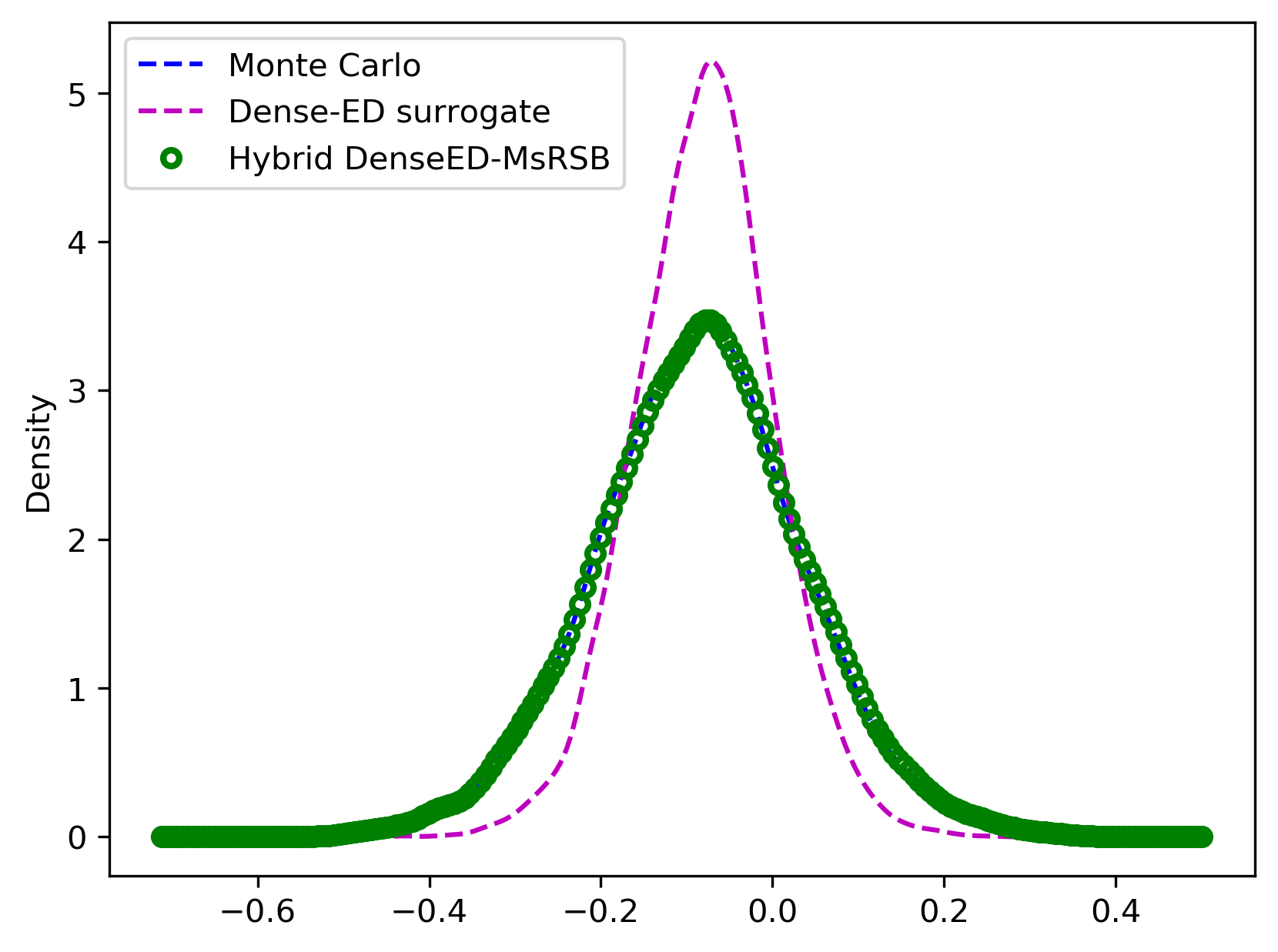}
        \label{fig:first_sub_5}
    }
        \caption{Distribution estimate for the pressure for KLE$-1000$. Location: $(0.6,0.4)$. Here, the Monte Carlo result is shown in the blue dashed line; the hybrid DenseED result is shown in green circles, and the DenseED model result is shown in magenta dashed line.}
    \label{fig:pressure_PDF_1000}
\end{figure}

\begin{figure}[H]
    \centering
  \subfigure[KLE$-16384$ ($96$~data)]
    {
        \includegraphics[width=0.3\textwidth]{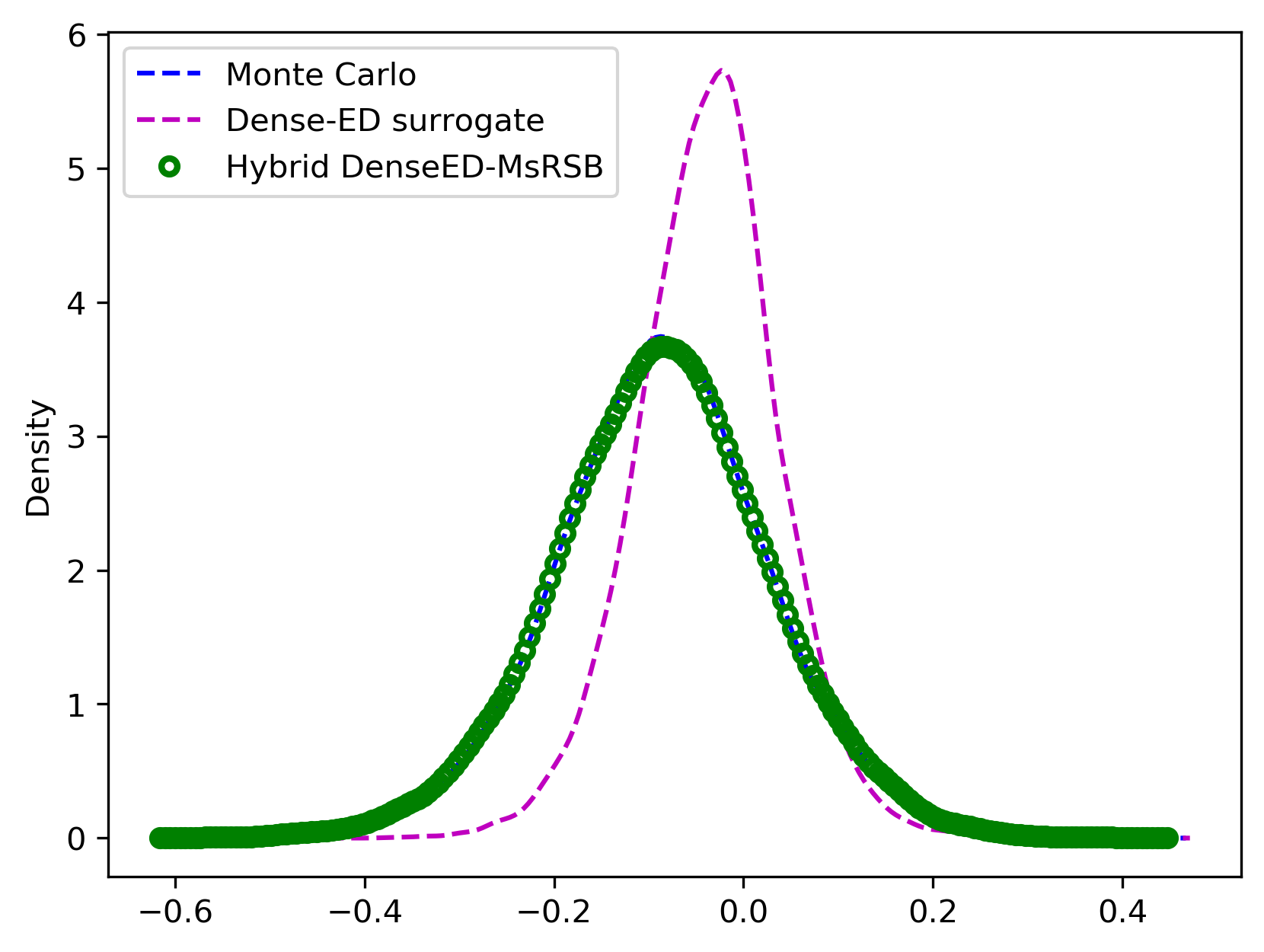}
        \label{fig:first_sub_6}
    }
    \subfigure[KLE$-16384$ ($128$~data)]
    {
        \includegraphics[width=0.3\textwidth]{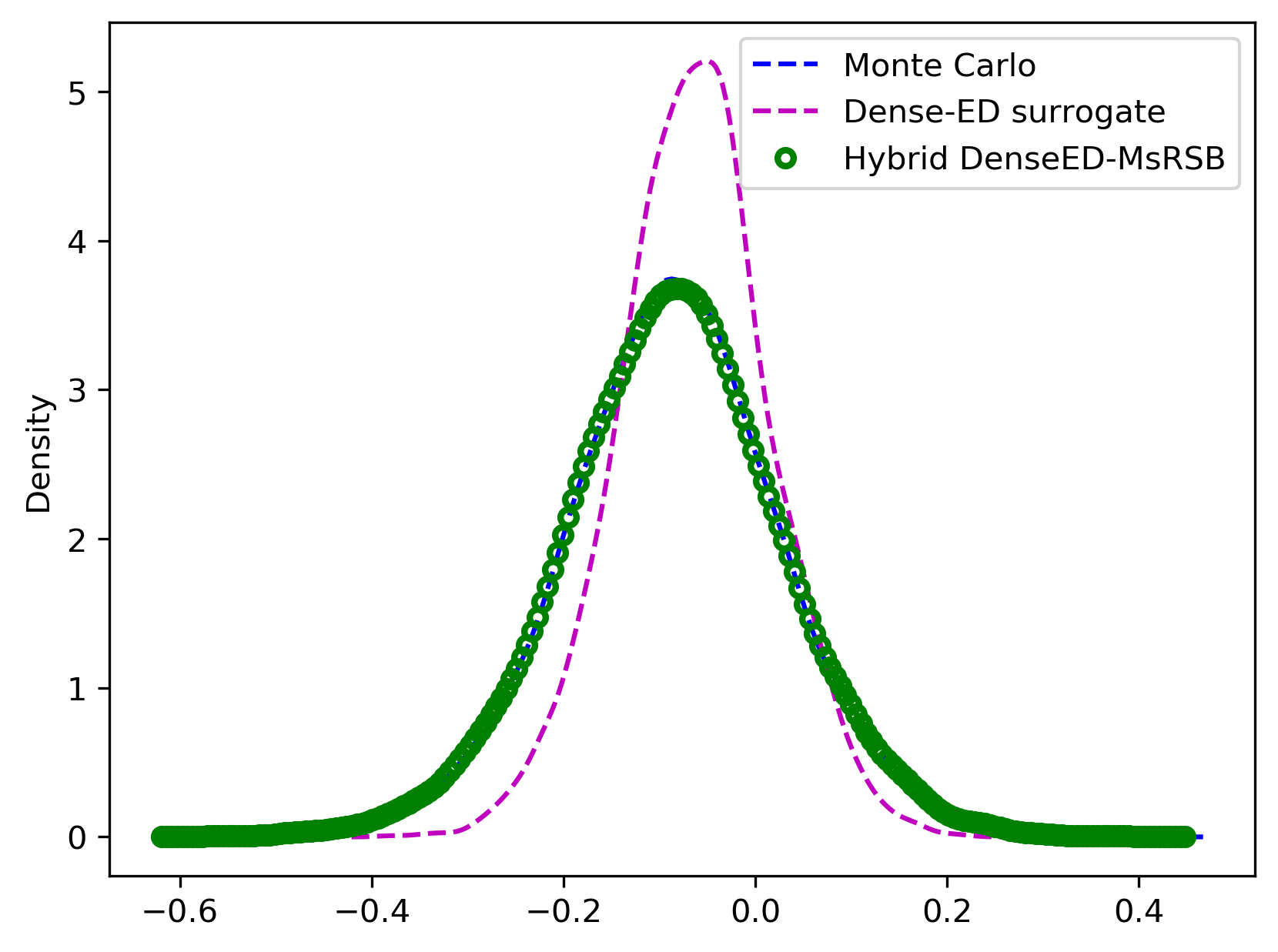}
        \label{fig:first_sub_7}
    }
    \subfigure[KLE$-16384$ ($160$~data)]
    {
        \includegraphics[width=0.3\textwidth]{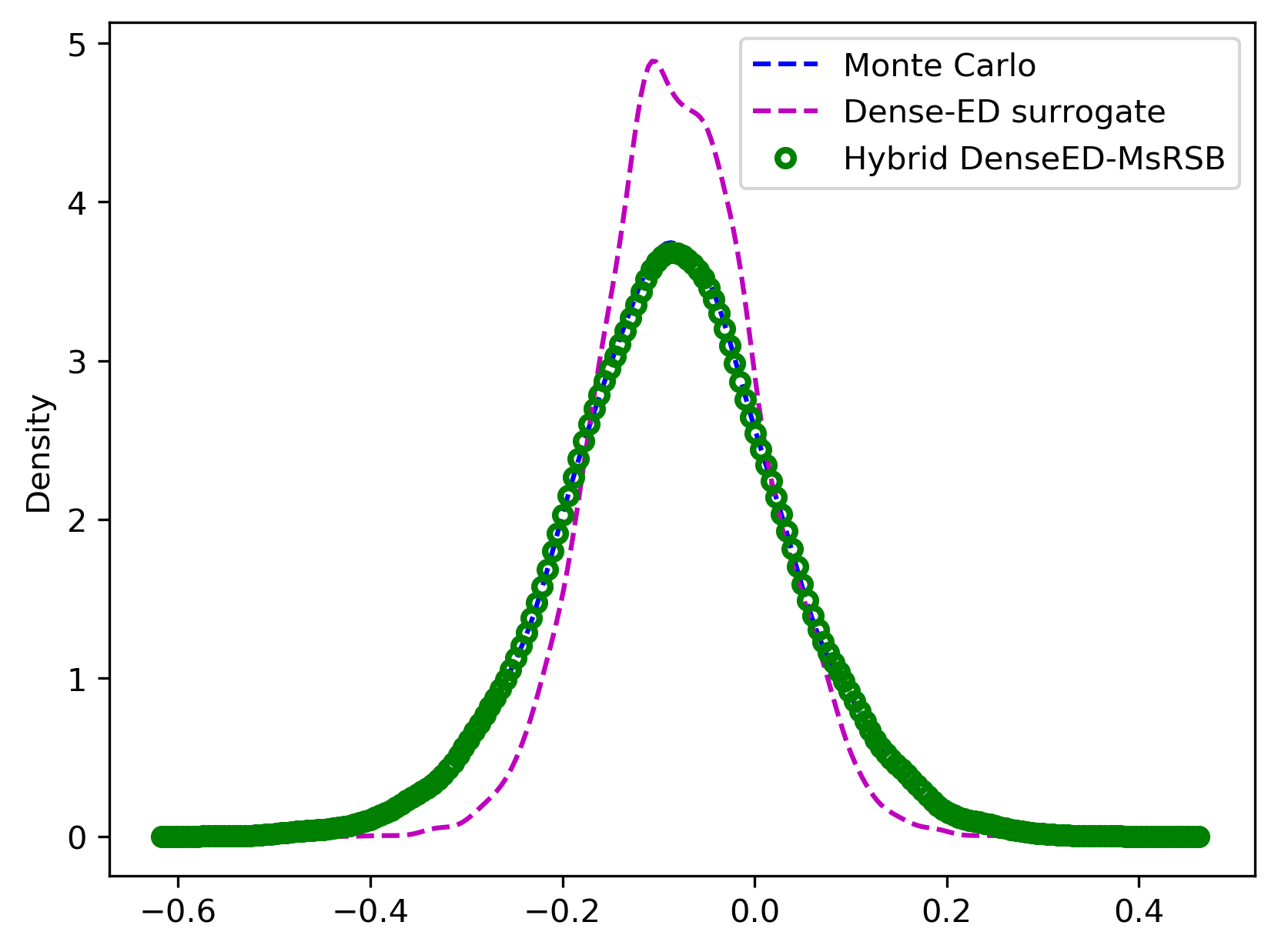}
        \label{fig:second_sub_1}
    }
        \caption{Distribution estimate for the pressure for KLE$-16384$. Location: $(0.6,0.4)$. Here, the Monte Carlo result is shown in the blue dashed line; the hybrid DenseED result is shown in green circles, and the DenseED model result is shown in magenta dashed line.}
    \label{fig:pressure_PDF_16384}
\end{figure}

\begin{figure}[H]
    \centering
        \subfigure[Channelized ($160$~data)]
    {
        \includegraphics[width=0.3\textwidth]{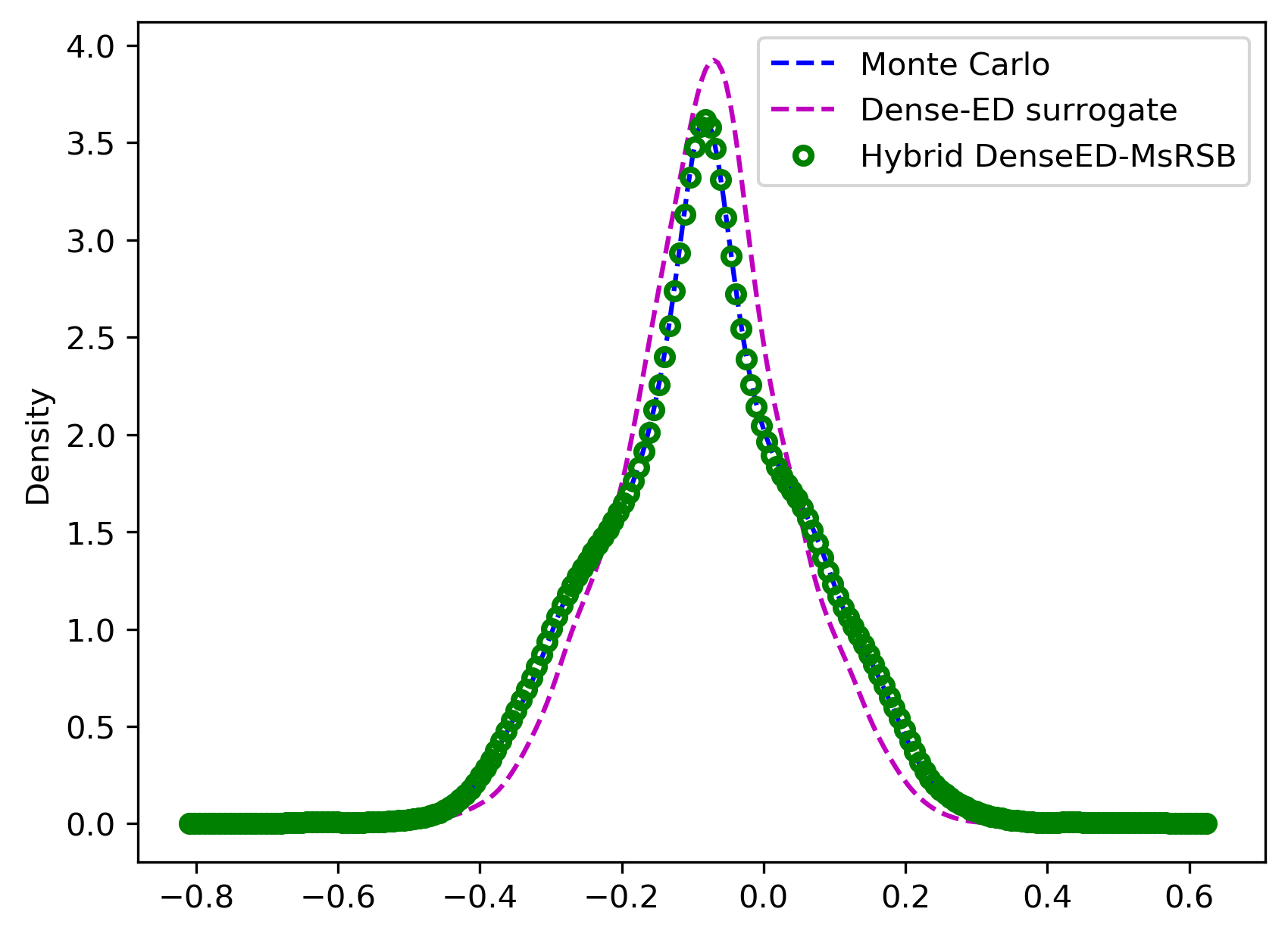}
        \label{fig:third_sub1_CH}
    }
        \caption{Distribution estimate for the pressure for channelized flow. Location: $(0.6,0.4)$. Here, the Monte Carlo result is shown in the blue dashed line; the hybrid DenseED result is shown in green circles, and the DenseED model result is shown in magenta dashed line.}
    \label{fig:pressure_PDF_channel}
\end{figure}

\begin{figure}[H]
    \centering
        \subfigure[KLE$-100$ ($32-$data)]
    {
        \includegraphics[width=0.3\textwidth]{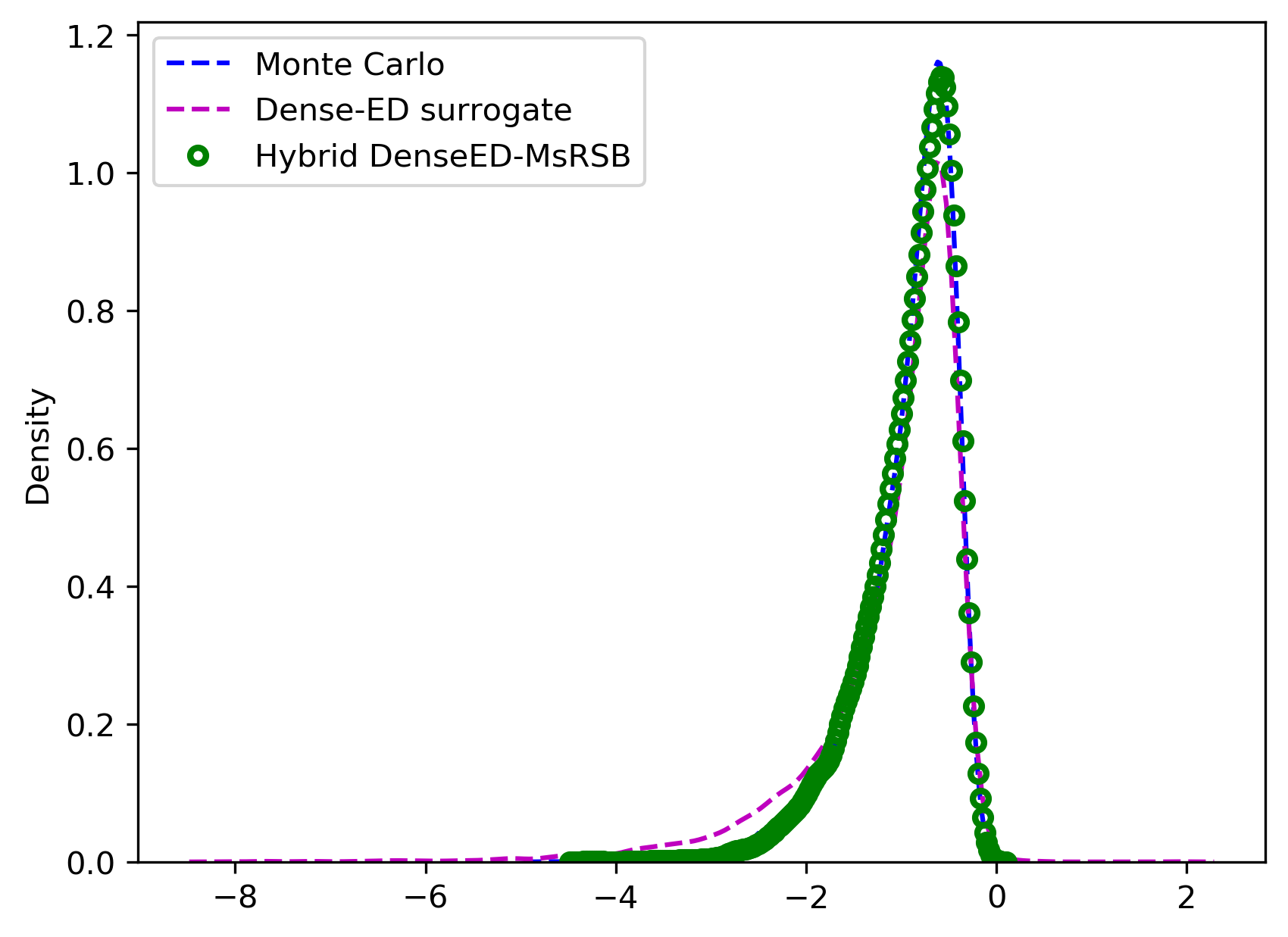}
        \label{fig:first_sub1_2}
    }
    \subfigure[KLE$-100$ ($64-$data)]
    {
        \includegraphics[width=0.3\textwidth]{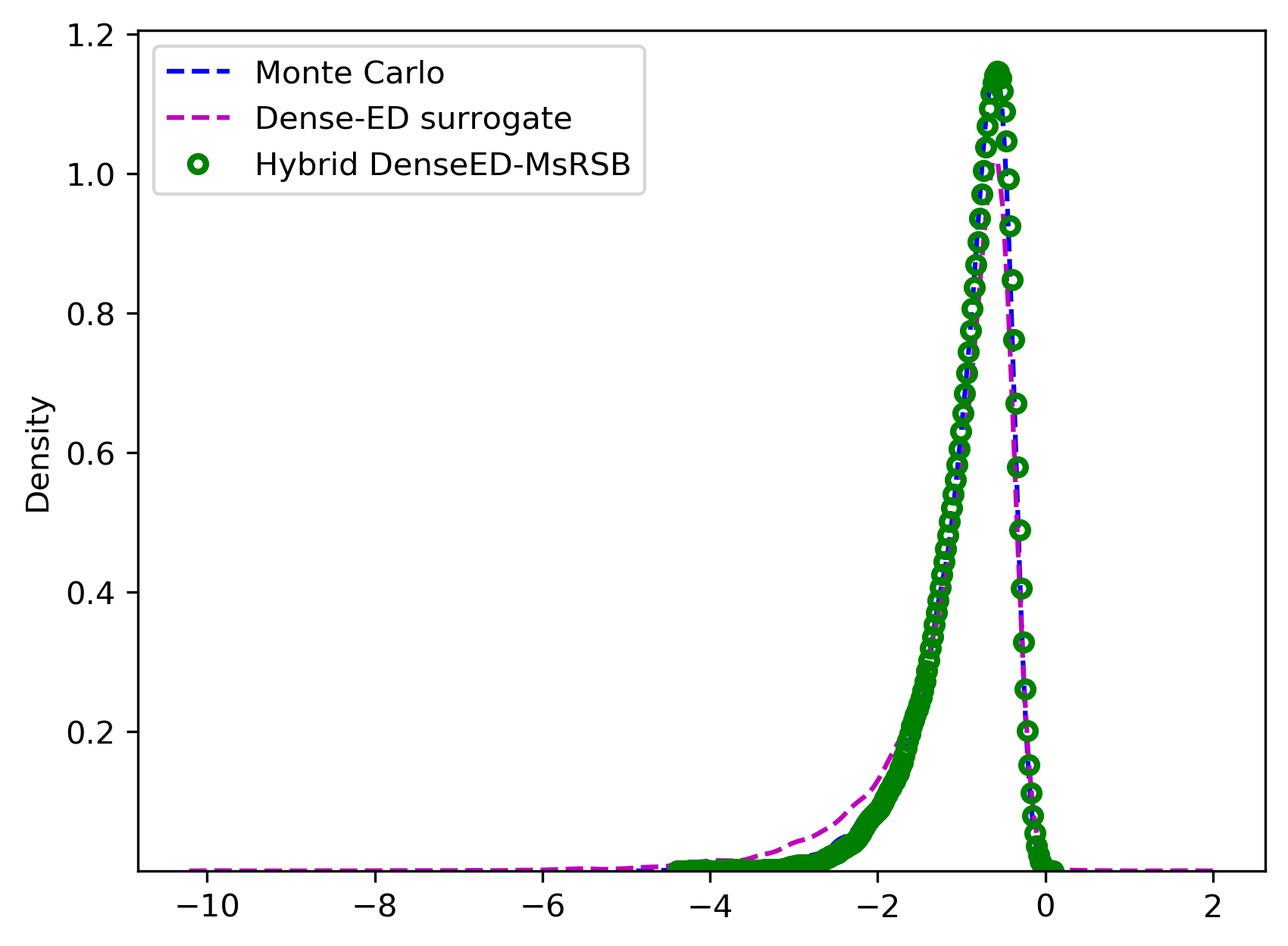}
        \label{fig:first_sub_8}
    }
    \subfigure[KLE$-100$ ($96-$data)]
    {
        \includegraphics[width=0.3\textwidth]{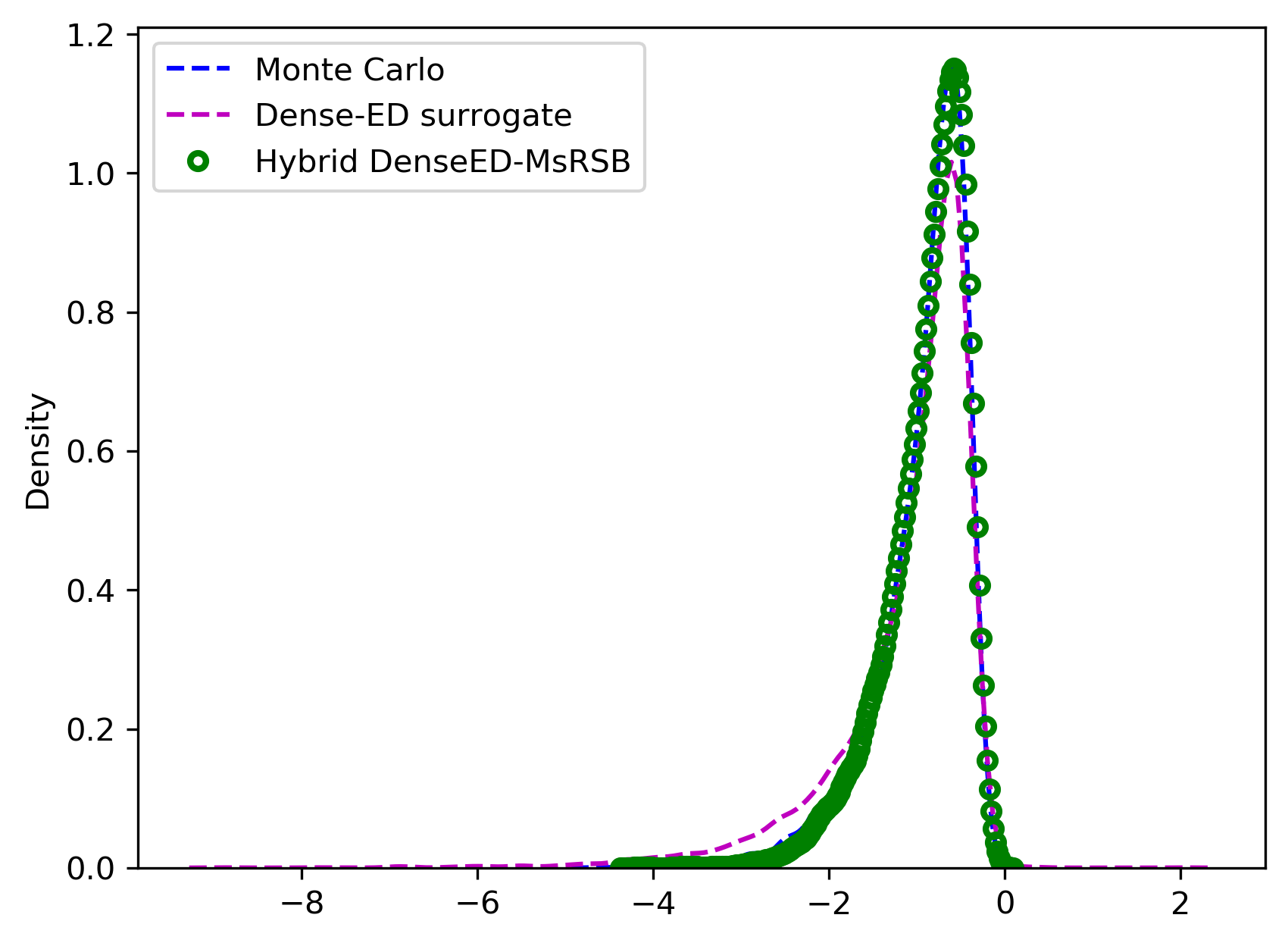}
        \label{fig:first_sub_9}
    }
        \caption{Distribution estimate for the $x-$velocity component (horizontal flux) for KLE$-100$. Location: $(0.6,0.4)$. Here, the Monte Carlo result is shown in the blue dashed line; the hybrid DenseED result is shown in green circles, and the DenseED model result is shown in magenta dashed line.}
    \label{fig:velocity_x_KLE_100}
\end{figure}

\begin{figure}[H]
    \centering
        \subfigure[KLE$-1000$ ($64$~data)]
    {
        \includegraphics[width=0.3\textwidth]{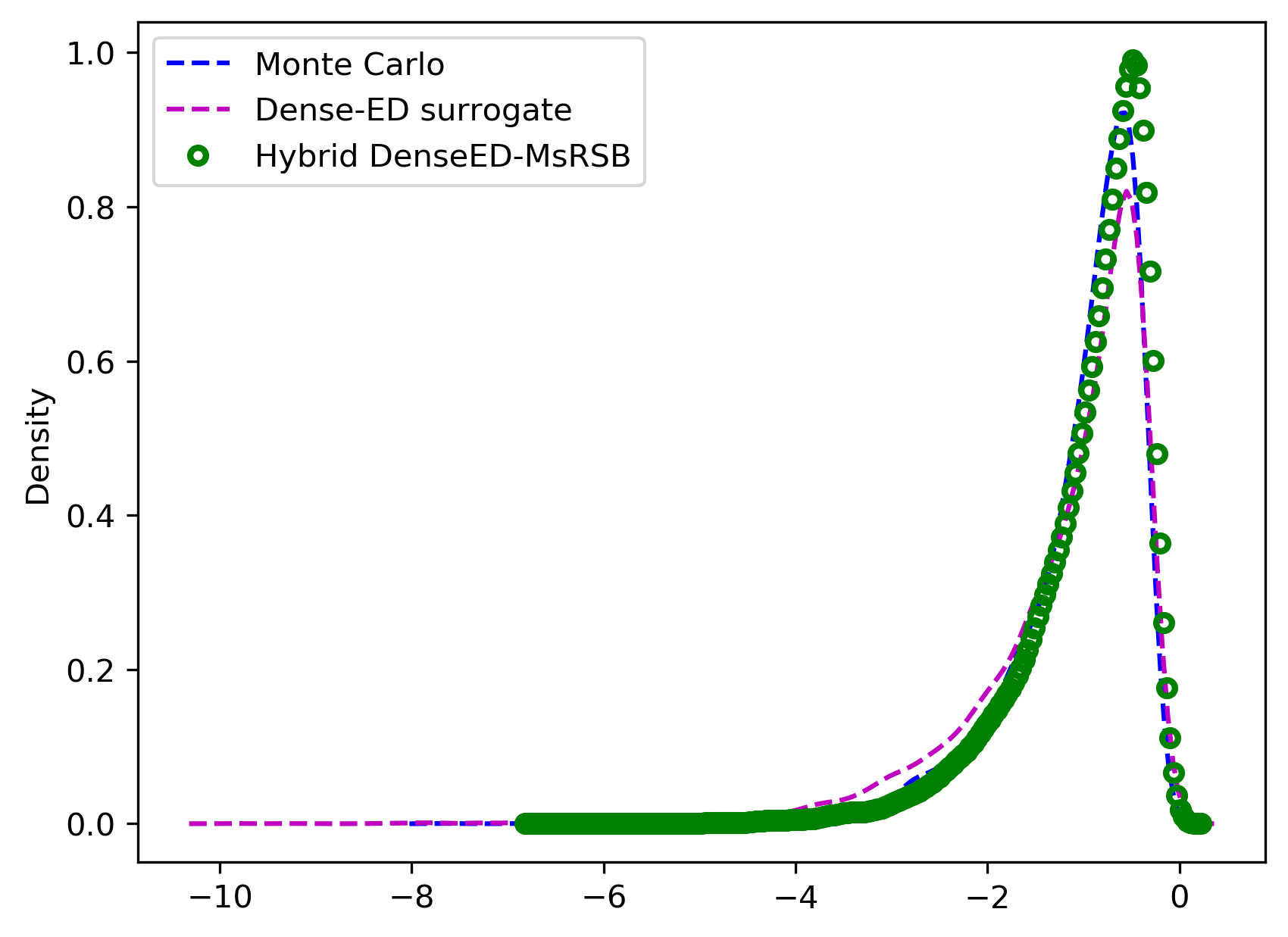}
        \label{fig:first_sub_10}
    }
    \subfigure[KLE$-1000$ ($96$~data)]
    {
        \includegraphics[width=0.3\textwidth]{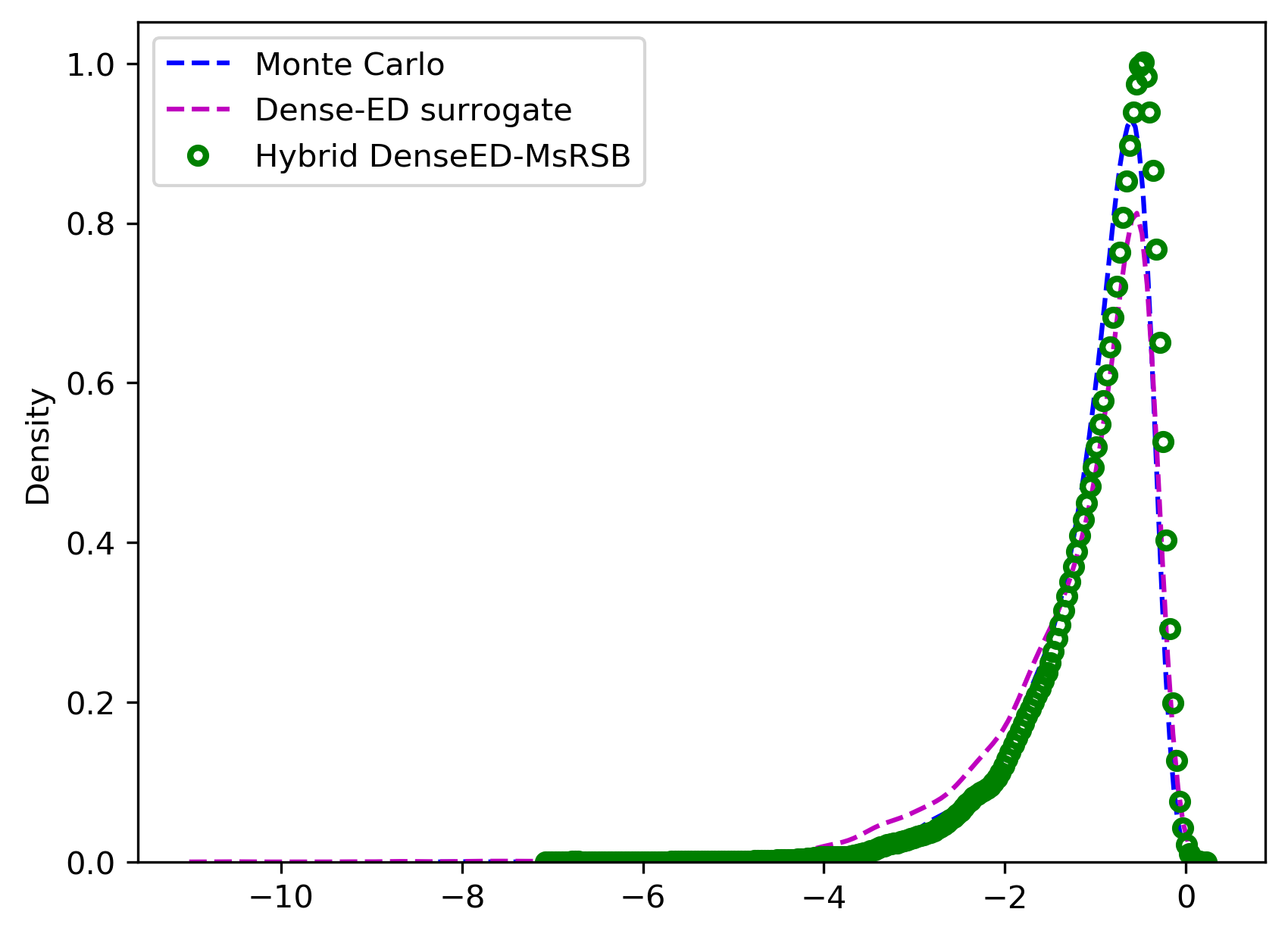}
        \label{fig:first_sub_11}
    }
    \subfigure[KLE$-1000$ ($128$~data)]
    {
        \includegraphics[width=0.3\textwidth]{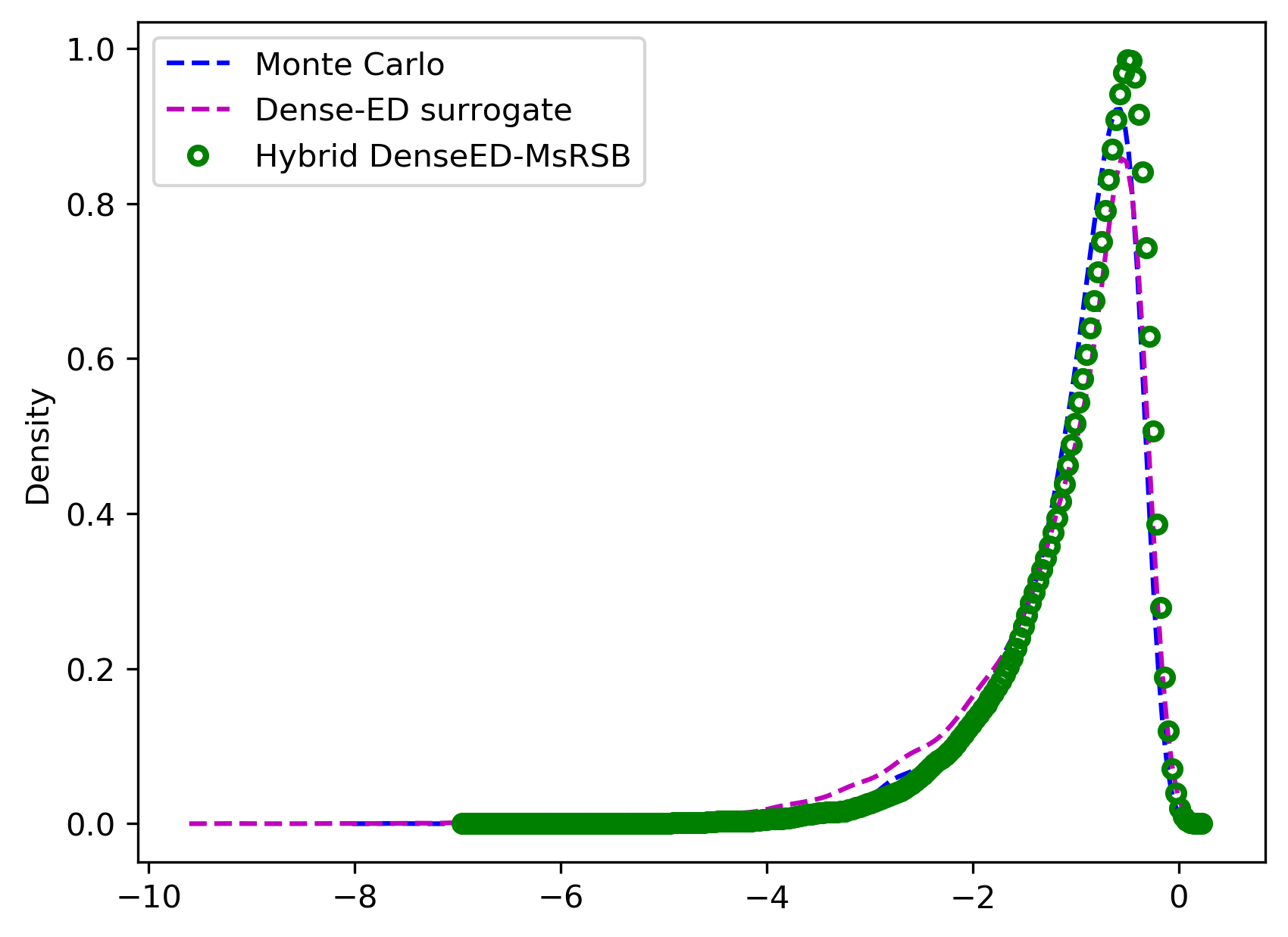}
        \label{fig:first_sub_12}
    }
        \caption{Distribution estimate for the $x-$velocity component (horizontal flux) for KLE$-1000$. Location: $(0.6,0.4)$. Here, the Monte Carlo result is shown in the blue dashed line; the hybrid DenseED result is shown in green circles, and the DenseED model result is shown in magenta dashed line.}
    \label{fig:velocity_x_KLE_1000}
\end{figure}

\begin{figure}[H]
    \centering
        \subfigure[KLE$-16384$ ($96$~ data)]
    {
        \includegraphics[width=0.3\textwidth]{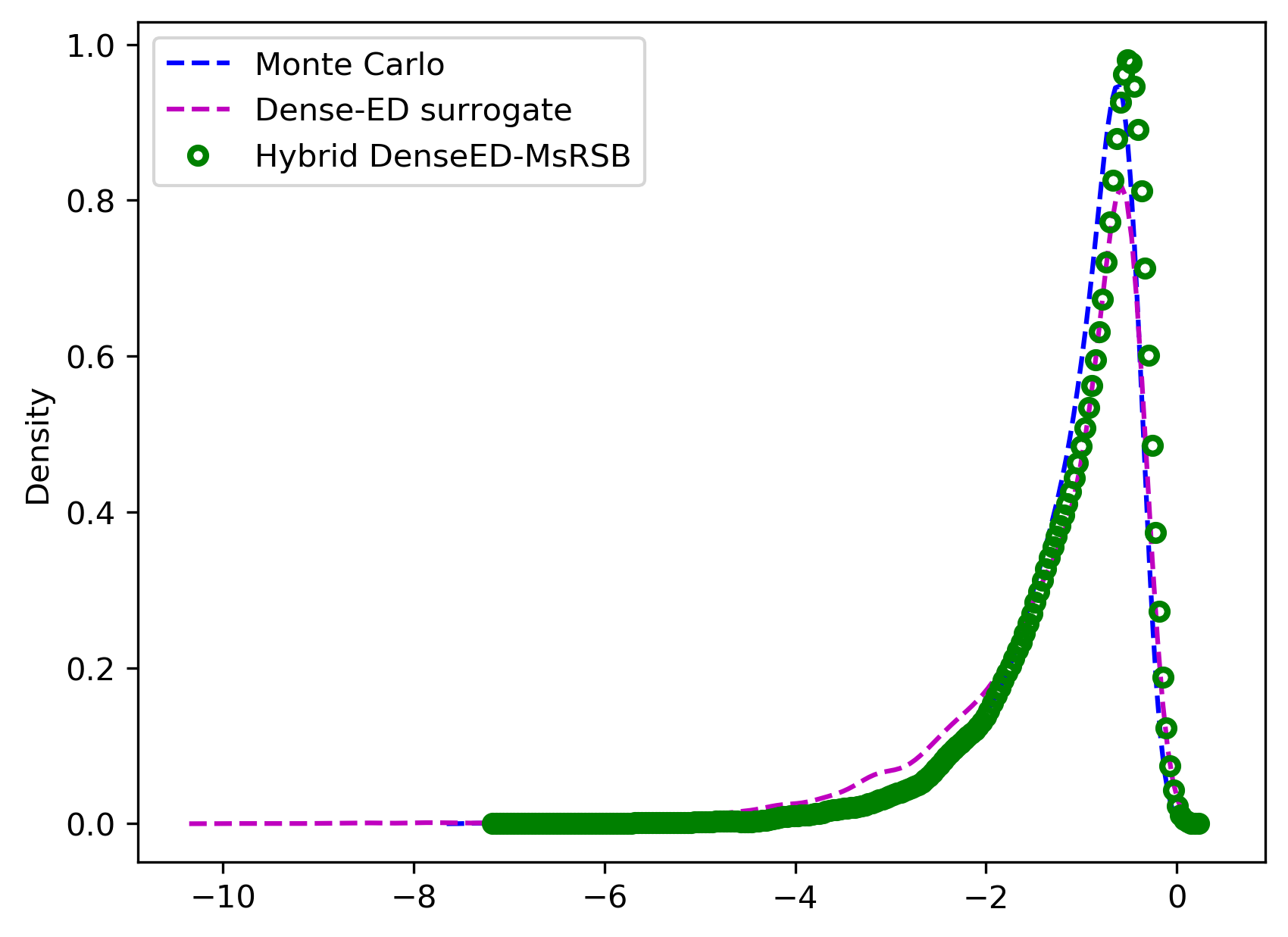}
        \label{fig:first_sub_13}
    }
    \subfigure[KLE$-16384$ ($128$~data)]
    {
        \includegraphics[width=0.3\textwidth]{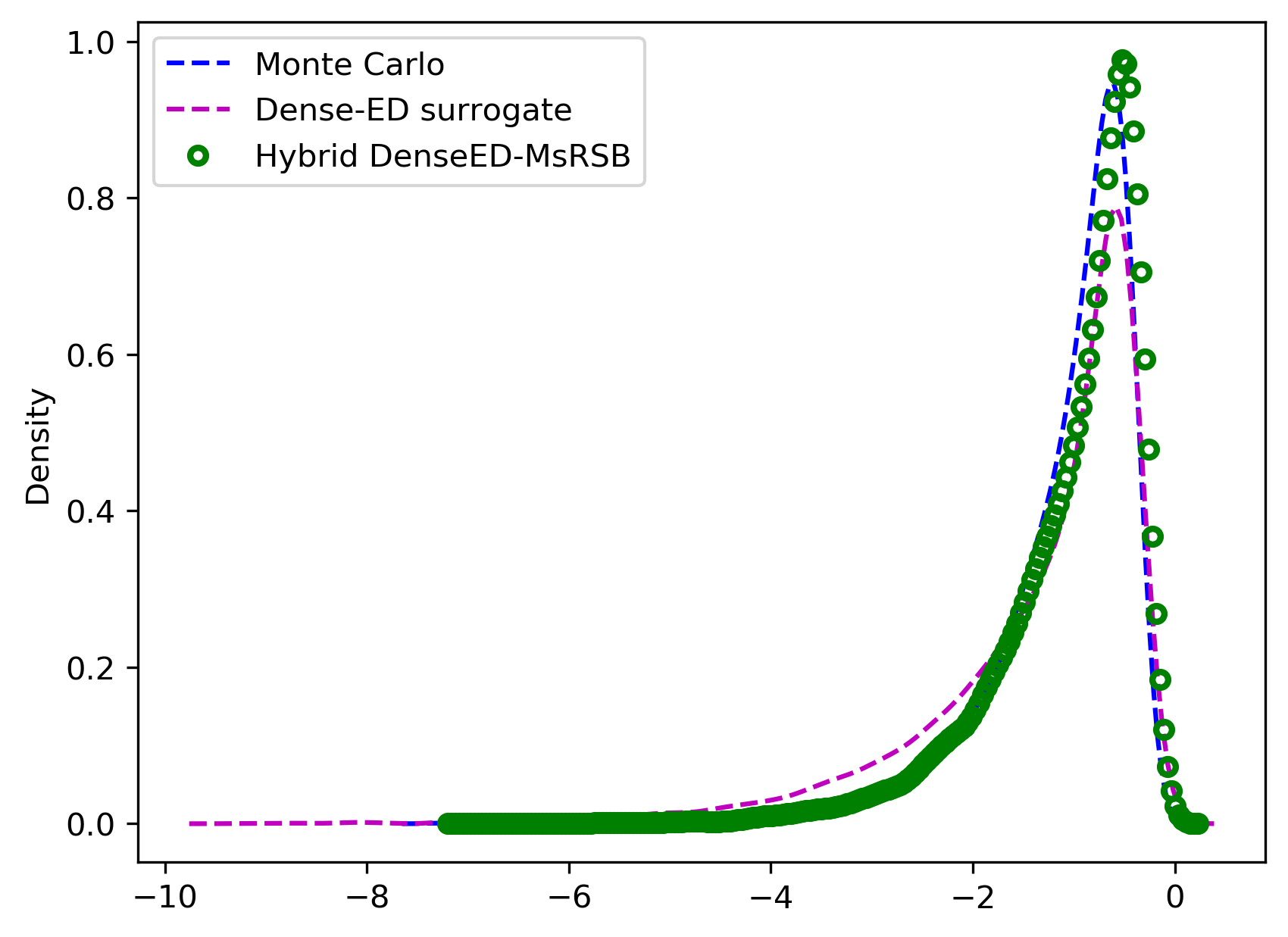}
        \label{fig:first_sub_14}
    }
    \subfigure[KLE$-16384$ ($160$~data)]
    {
        \includegraphics[width=0.3\textwidth]{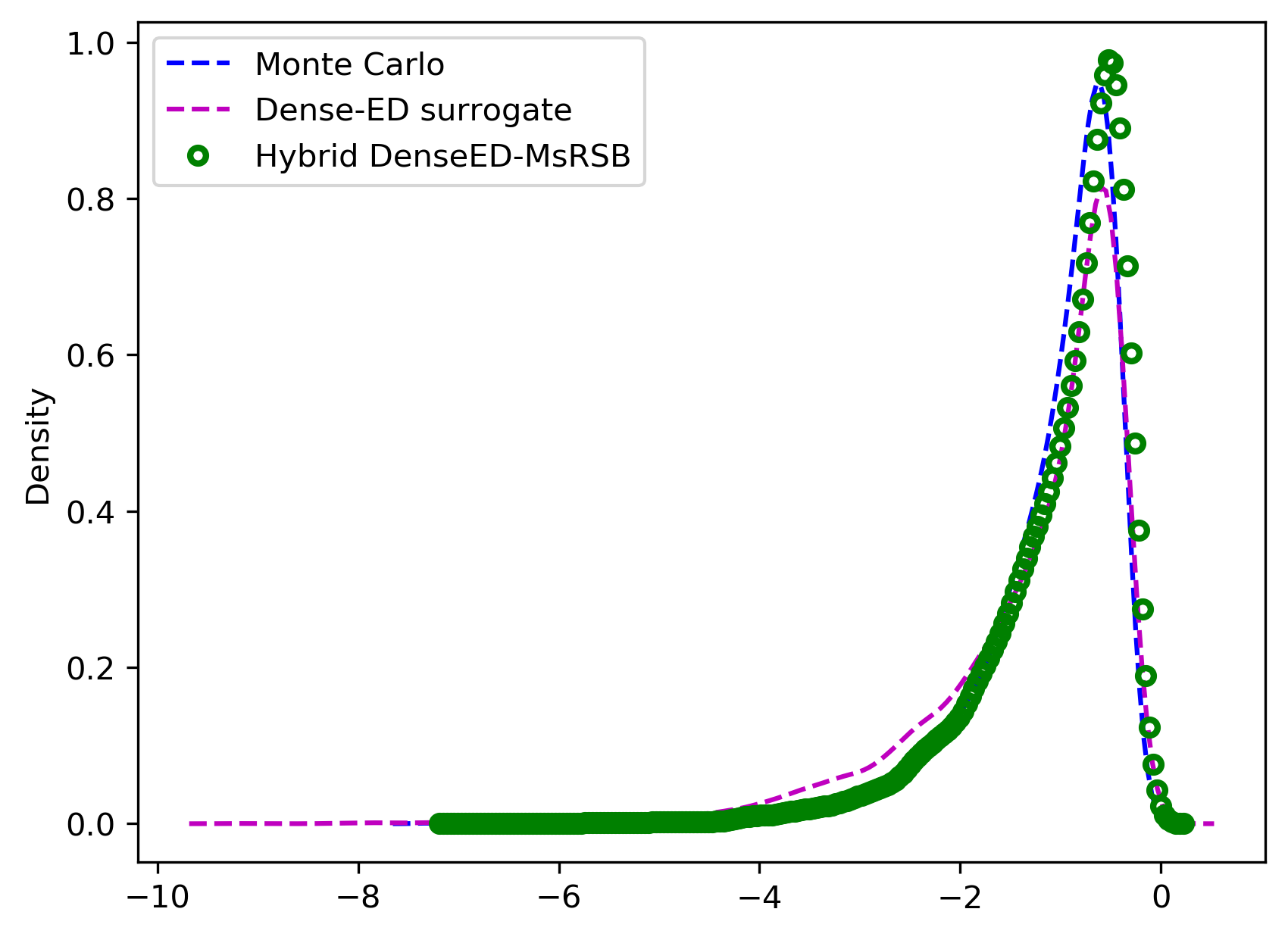}
        \label{fig:second_sub_2}
    }
        \caption{Distribution estimate for the $x-$velocity component (horizontal flux) for KLE$-16384$. Location: $(0.6,0.4)$. Here, the Monte Carlo result is shown in the blue dashed line; the hybrid DenseED result is shown in green circles, and the DenseED model result is shown in magenta dashed line.}
    \label{fig:velocity_x_KLE_16384}
\end{figure}

\begin{figure}[H]
    \centering
      {
        \includegraphics[width=0.3\textwidth]{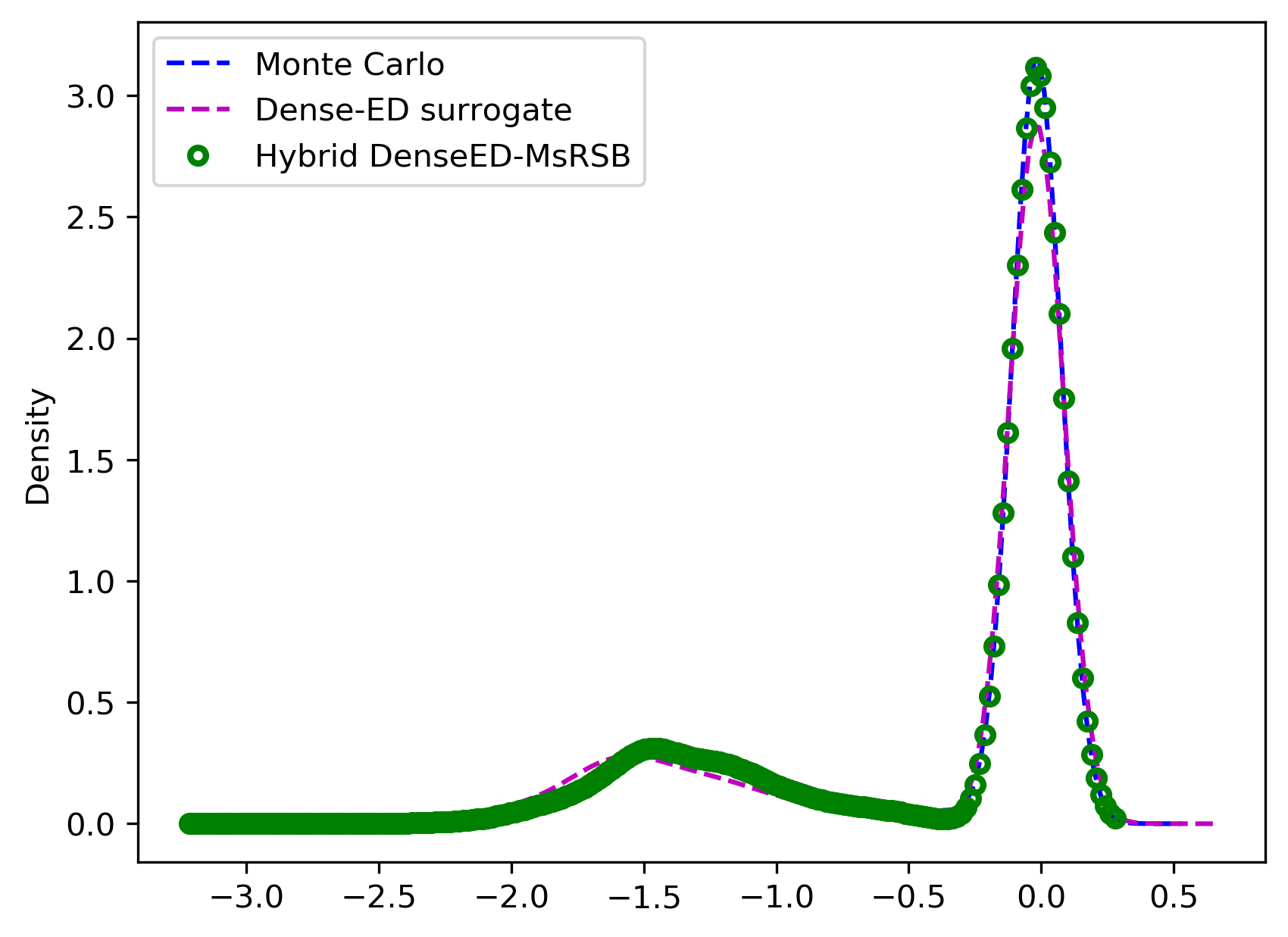}
        \label{fig:third_sub1}
    } 
        \caption{Distribution estimate for the $x-$velocity component (horizontal flux) for channelized flow. Location: $(0.6,0.4)$. Here, the Monte Carlo result is shown in the blue dashed line; the hybrid DenseED result is shown in green circles, and the DenseED model result is shown in magenta dashed line.}
    \label{fig:velocity_x_channel}
\end{figure}

\begin{figure}[H]
    \centering
    \subfigure[KLE$-100$ ($32-$data)]
    {
        \includegraphics[width=0.3\textwidth]{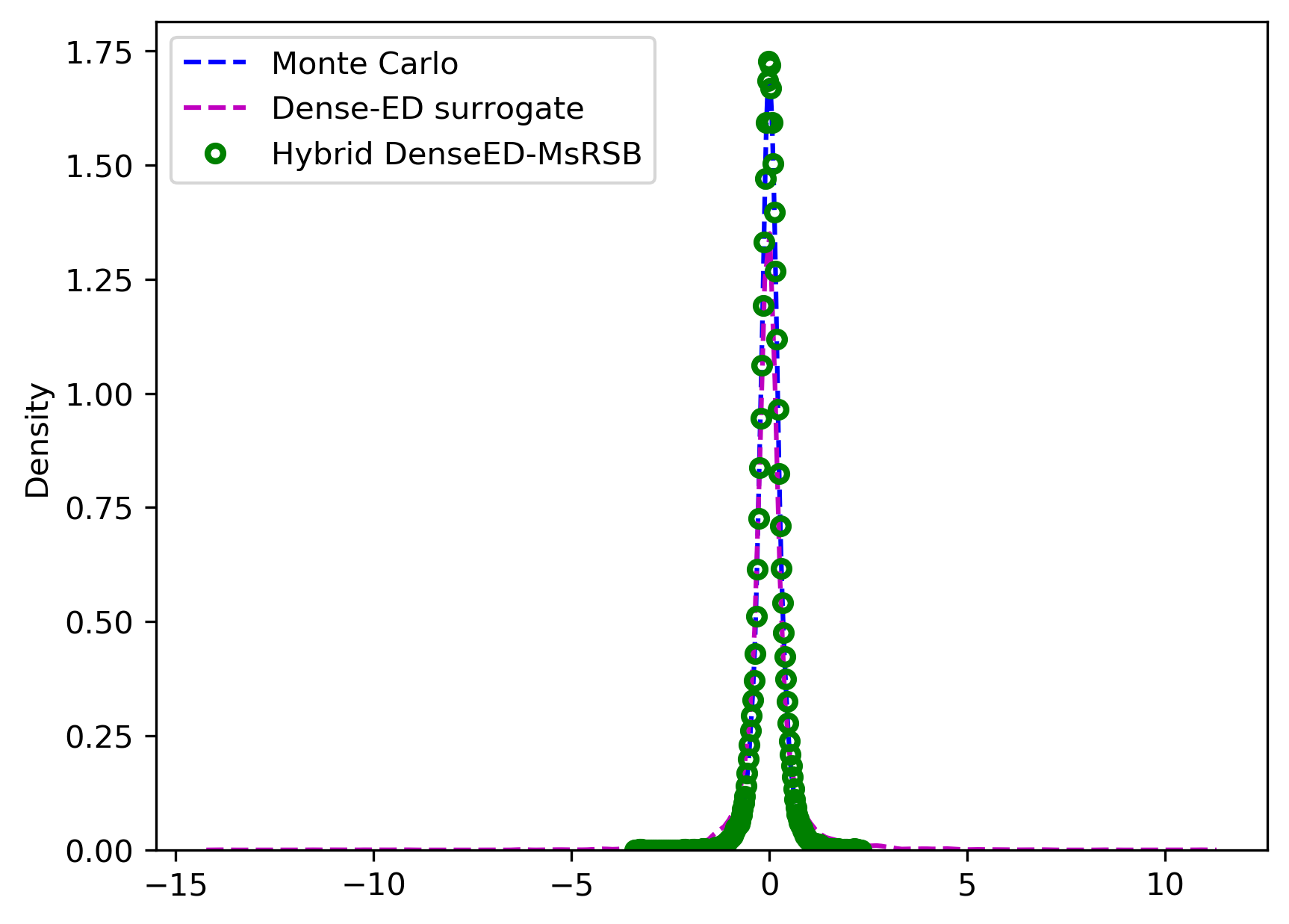}
        \label{fig:first_sub1_3}
    }
    \subfigure[KLE$-100$ ($64-$data)]
    {
        \includegraphics[width=0.3\textwidth]{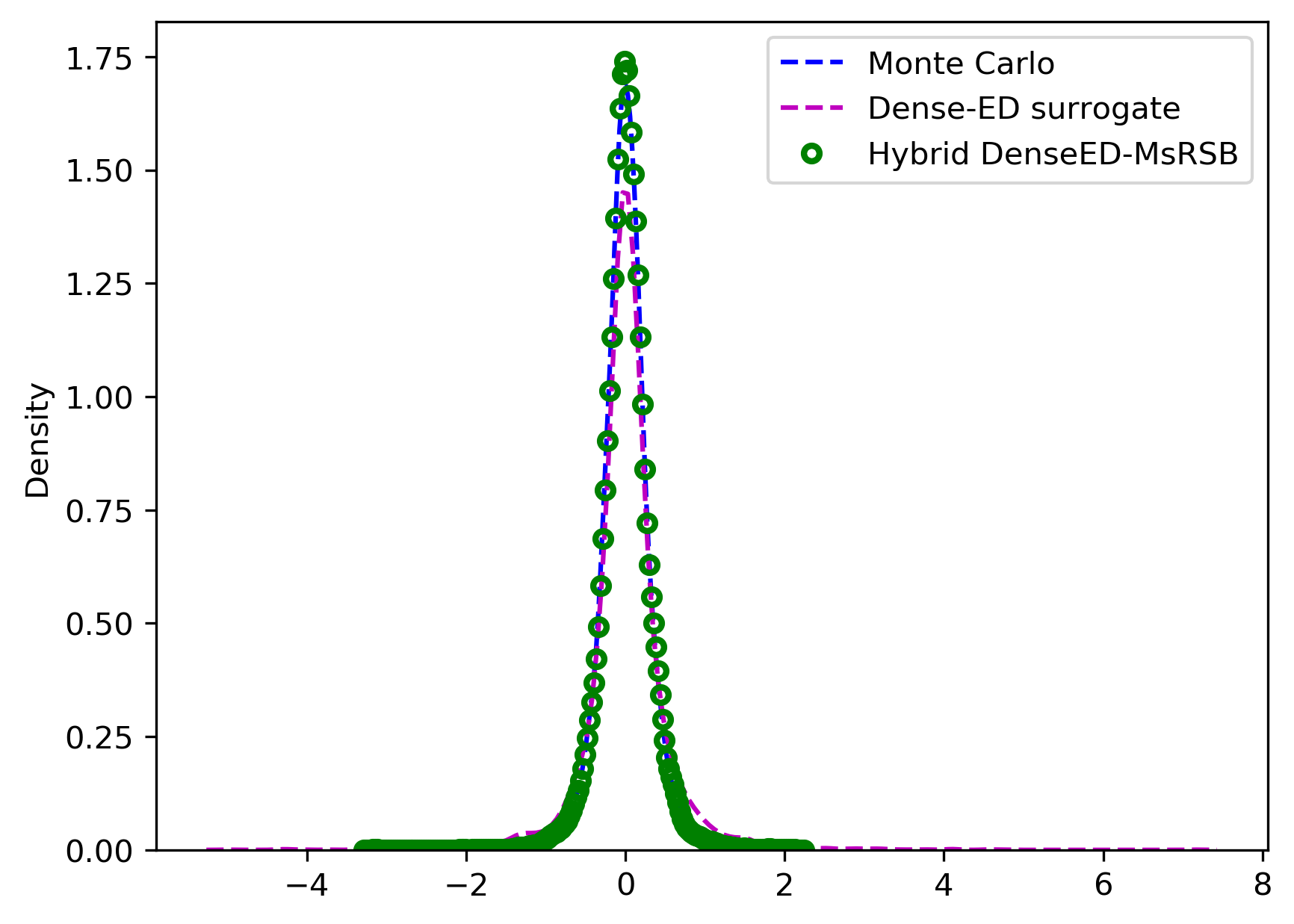}
        \label{fig:first_sub_15}
    }
    \subfigure[KLE$-100$ ($96-$data)]
    {
        \includegraphics[width=0.3\textwidth]{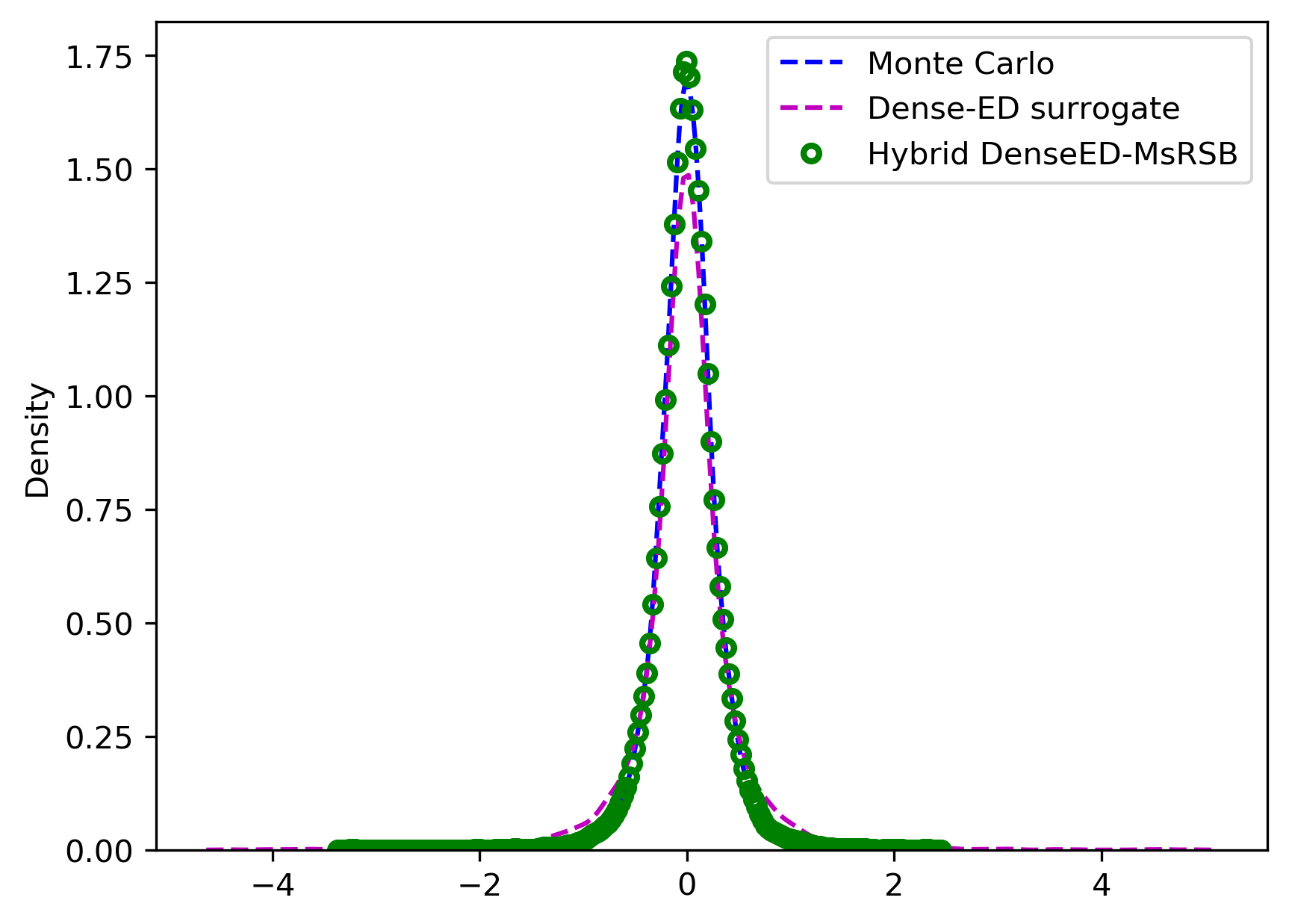}
        \label{fig:first_sub_16}
    }
        \caption{Distribution estimate for the $y-$velocity component (vertical flux) for KLE$-100$. Location: $(0.6,0.4)$. Here, the Monte Carlo result is shown in the blue dashed line; the hybrid DenseED result is shown in green circles, and the DenseED model result is shown in magenta dashed line.}
    \label{fig:velocity_y_100}
\end{figure}

\begin{figure}[H]
    \centering
    \subfigure[KLE$-1000$ ($64$~data)]
    {
        \includegraphics[width=0.3\textwidth]{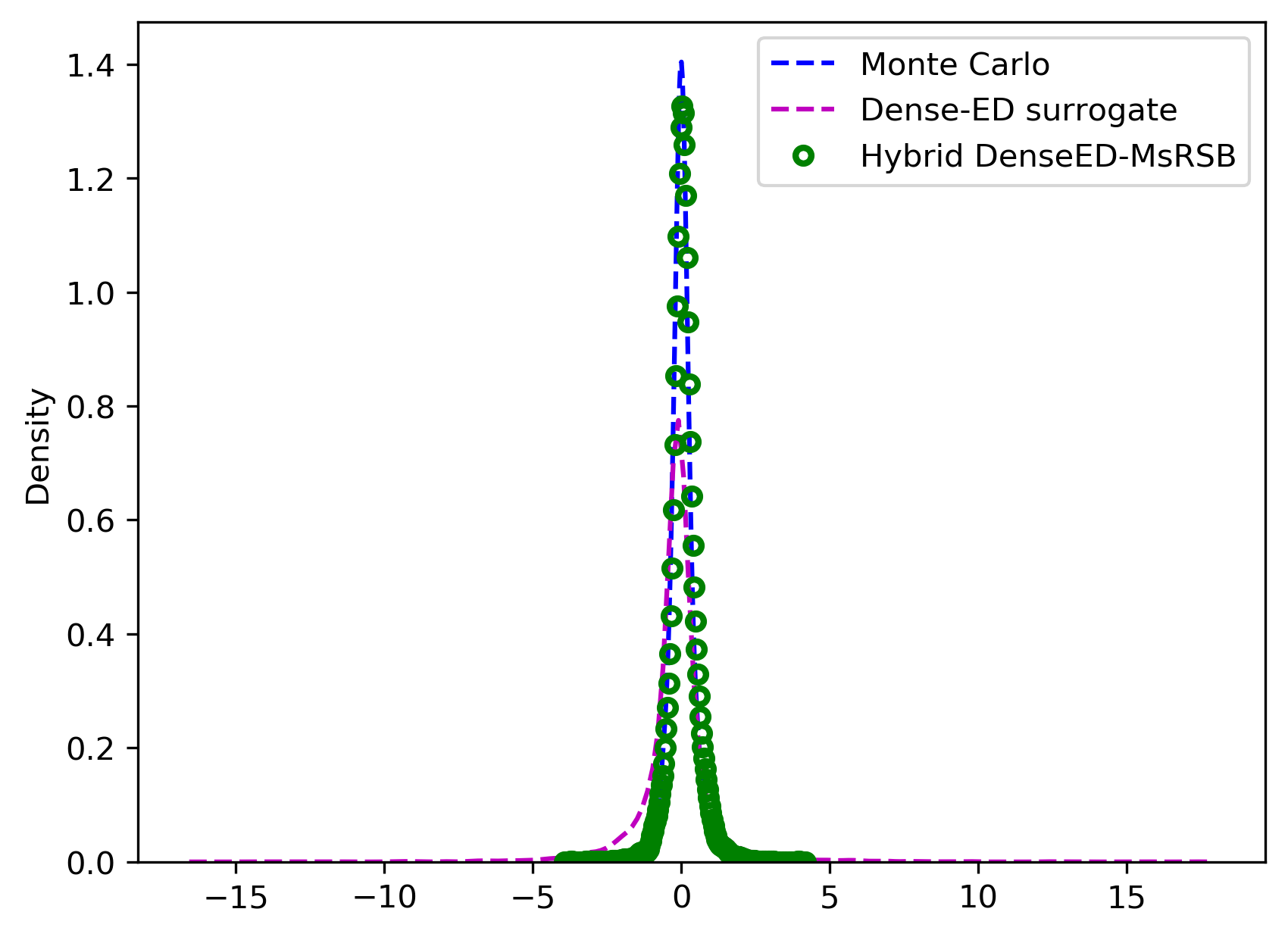}
        \label{fig:first_sub_17}
    }
    \subfigure[KLE$-1000$ ($96$~data)]
    {
        \includegraphics[width=0.3\textwidth]{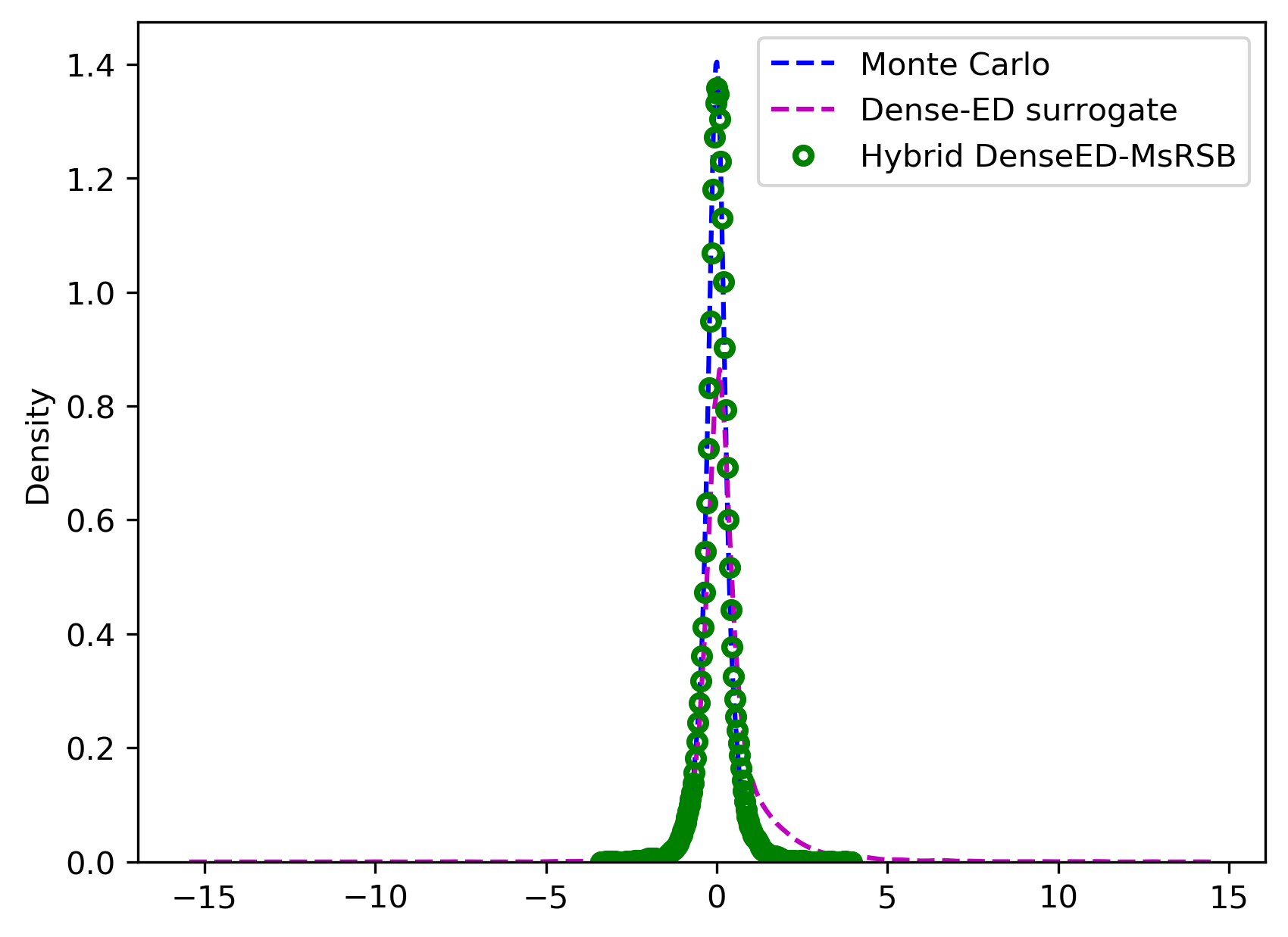}
        \label{fig:first_sub_18}
    }
    \subfigure[KLE$-1000$ ($128$~data)]
    {
        \includegraphics[width=0.3\textwidth]{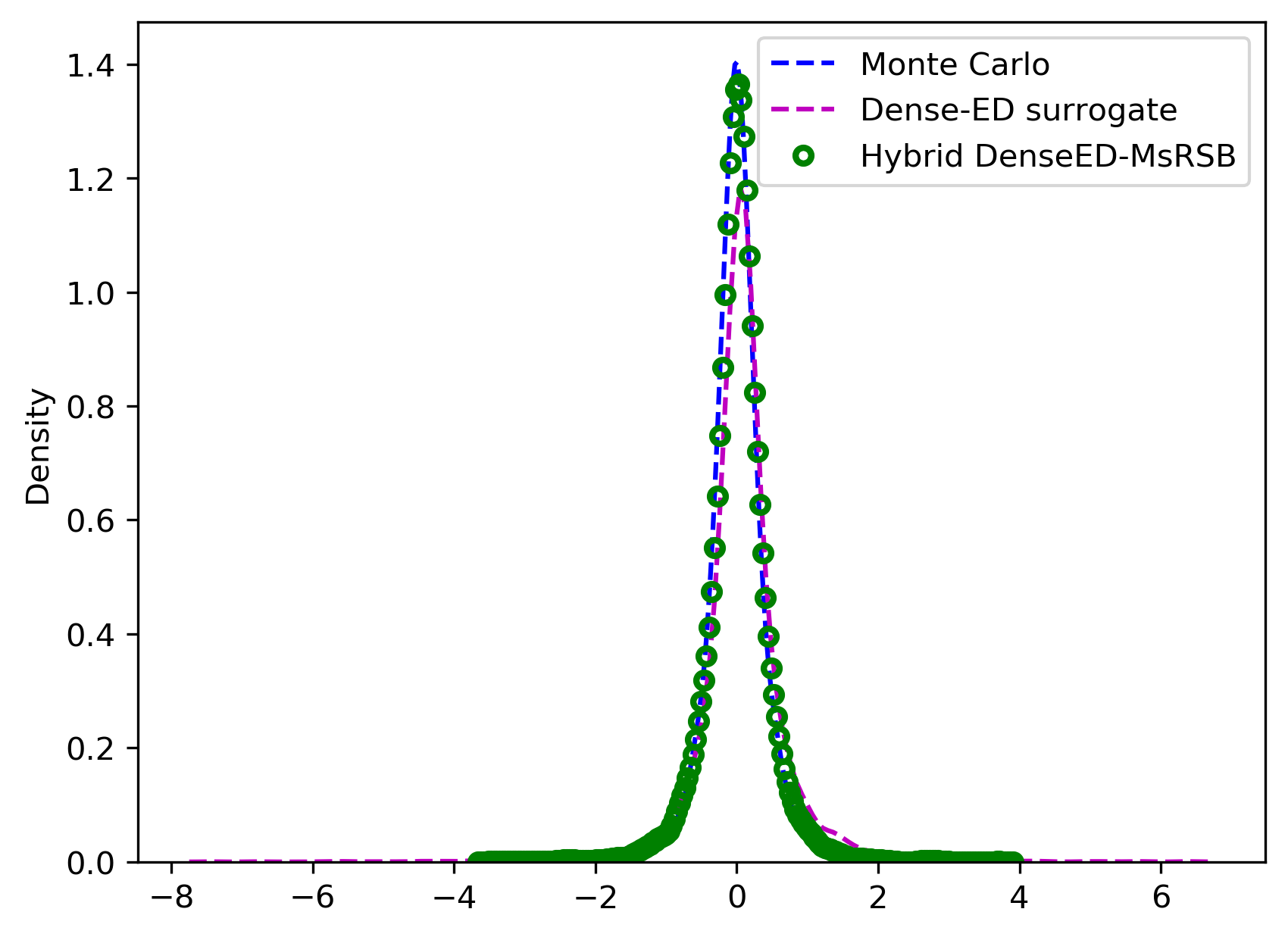}
        \label{fig:first_sub_19}
    }
        \caption{Distribution estimate for the $y-$velocity component (vertical flux) for KLE$-1000$. Location: $(0.6,0.4)$. Here, the Monte Carlo result is shown in the blue dashed line; the hybrid DenseED result is shown in green circles, and the DenseED model result is shown in magenta dashed line.}
    \label{fig:velocity_y_1000}
\end{figure}

\begin{figure}[H]
    \centering
    \subfigure[KLE$-16384$ ($96$~data)]
    {
        \includegraphics[width=0.3\textwidth]{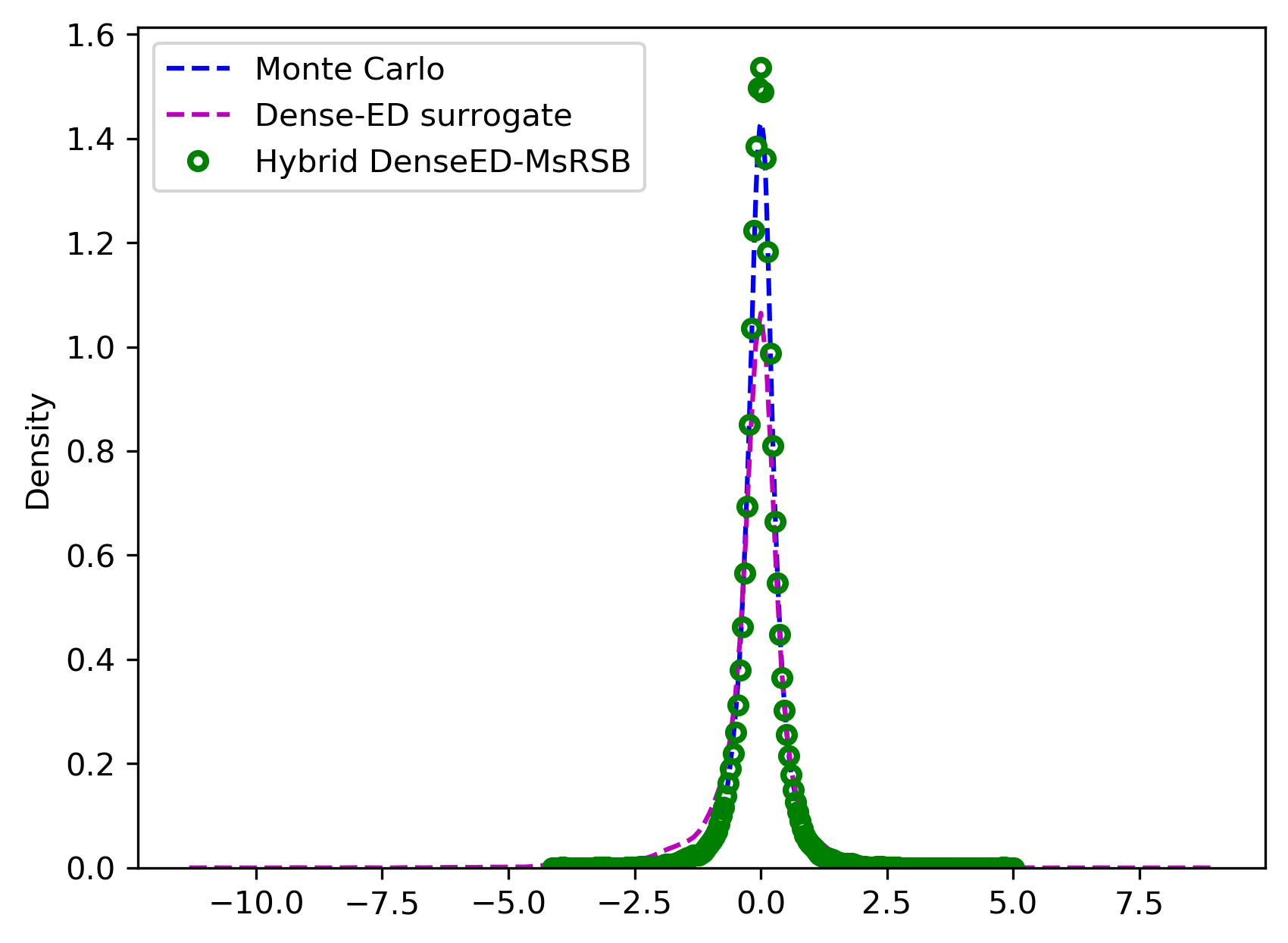}
        \label{fig:first_sub_20}
    }
    \subfigure[KLE$-16384$ ($128$~data)]
    {
        \includegraphics[width=0.3\textwidth]{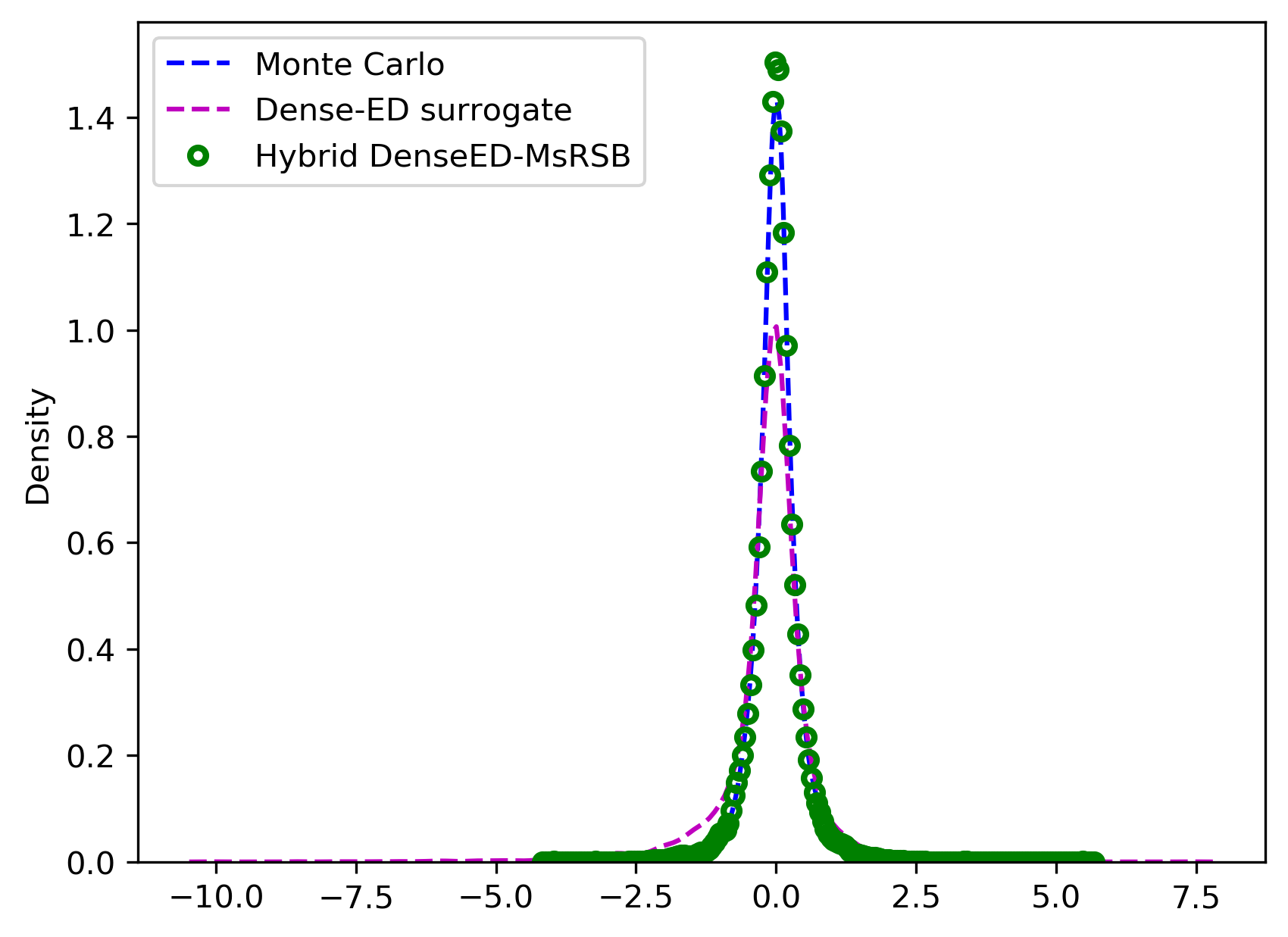}
        \label{fig:first_sub_21}
    }
    \subfigure[KLE$-16384$ ($160$~data)]
    {
        \includegraphics[width=0.3\textwidth]{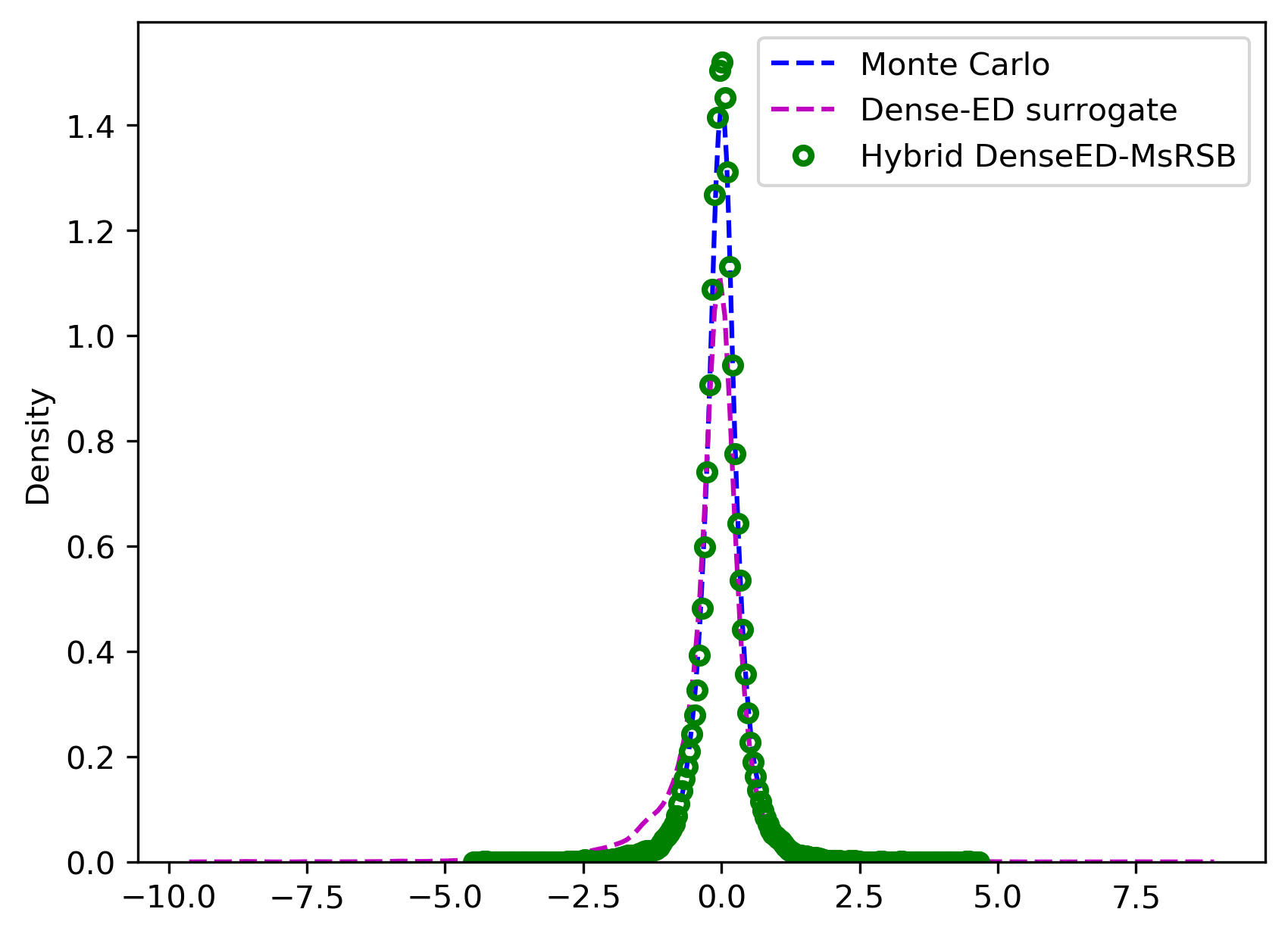}
        \label{fig:second_sub_3}
    }
        \caption{Distribution estimate for the $y-$velocity component (vertical flux) for KLE$-16384$. Location: $(0.6,0.4)$. Here, the Monte Carlo result is shown in the blue dashed line; the hybrid DenseED result is shown in green circles, and the DenseED model result is shown in magenta dashed line.}
    \label{fig:velocity_y_16384}
\end{figure}

\begin{figure}[H]
    \centering
    \subfigure[channelized ($160$~data)]
    {
        \includegraphics[width=0.3\textwidth]{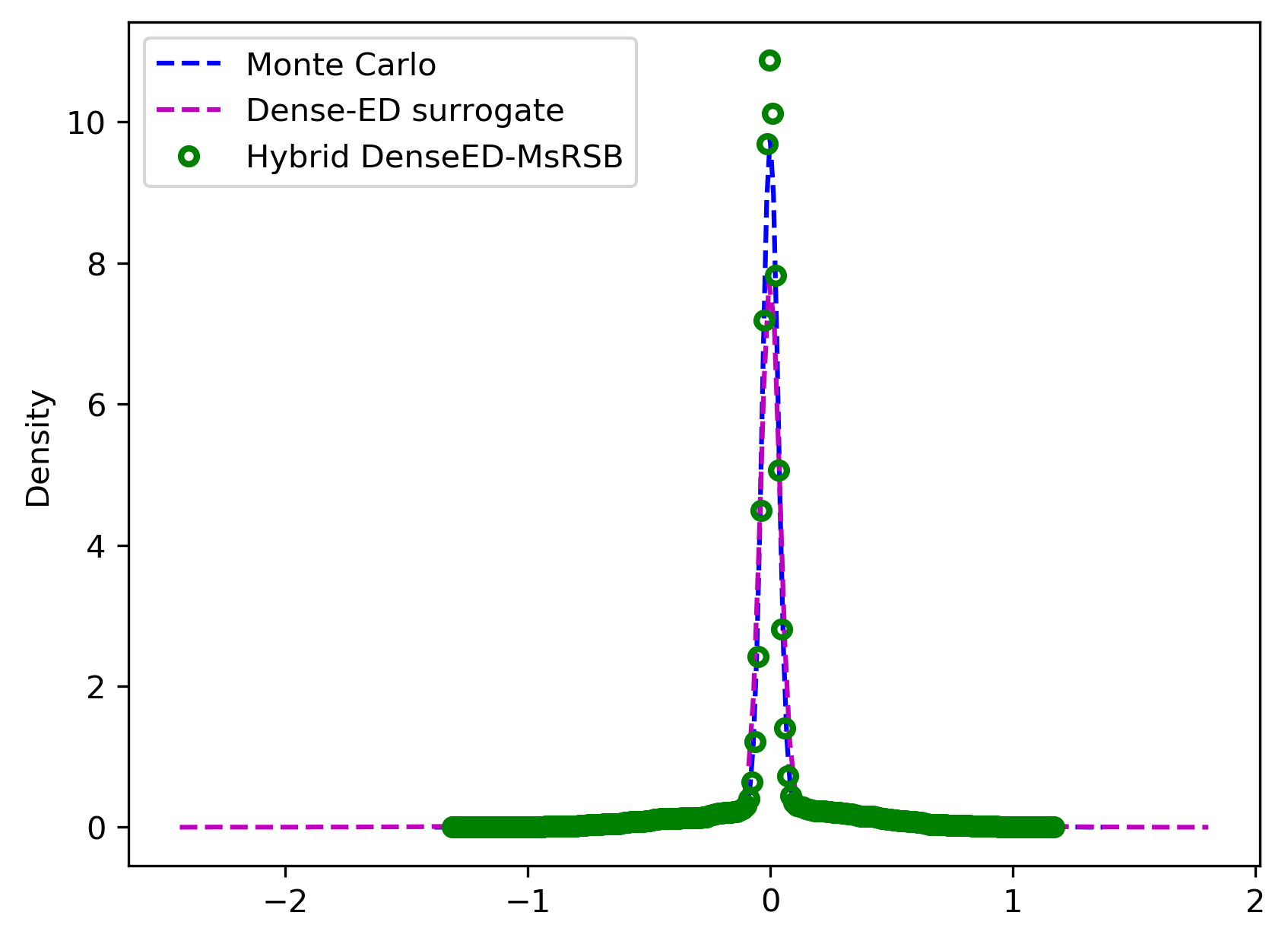}
        \label{fig:third_sub1_velocity}
    }
        \caption{Distribution estimate for the $y-$velocity component (vertical flux) for channelized flow. Location: $(0.6,0.4)$. Here, the Monte Carlo result is shown in the blue dashed line; the hybrid DenseED result is shown in green circles, and the DenseED model result is shown in magenta dashed line.}
    \label{fig:velocity_y_channel}
\end{figure}

\begin{figure}[H]
    \centering
    \includegraphics[scale=0.45]{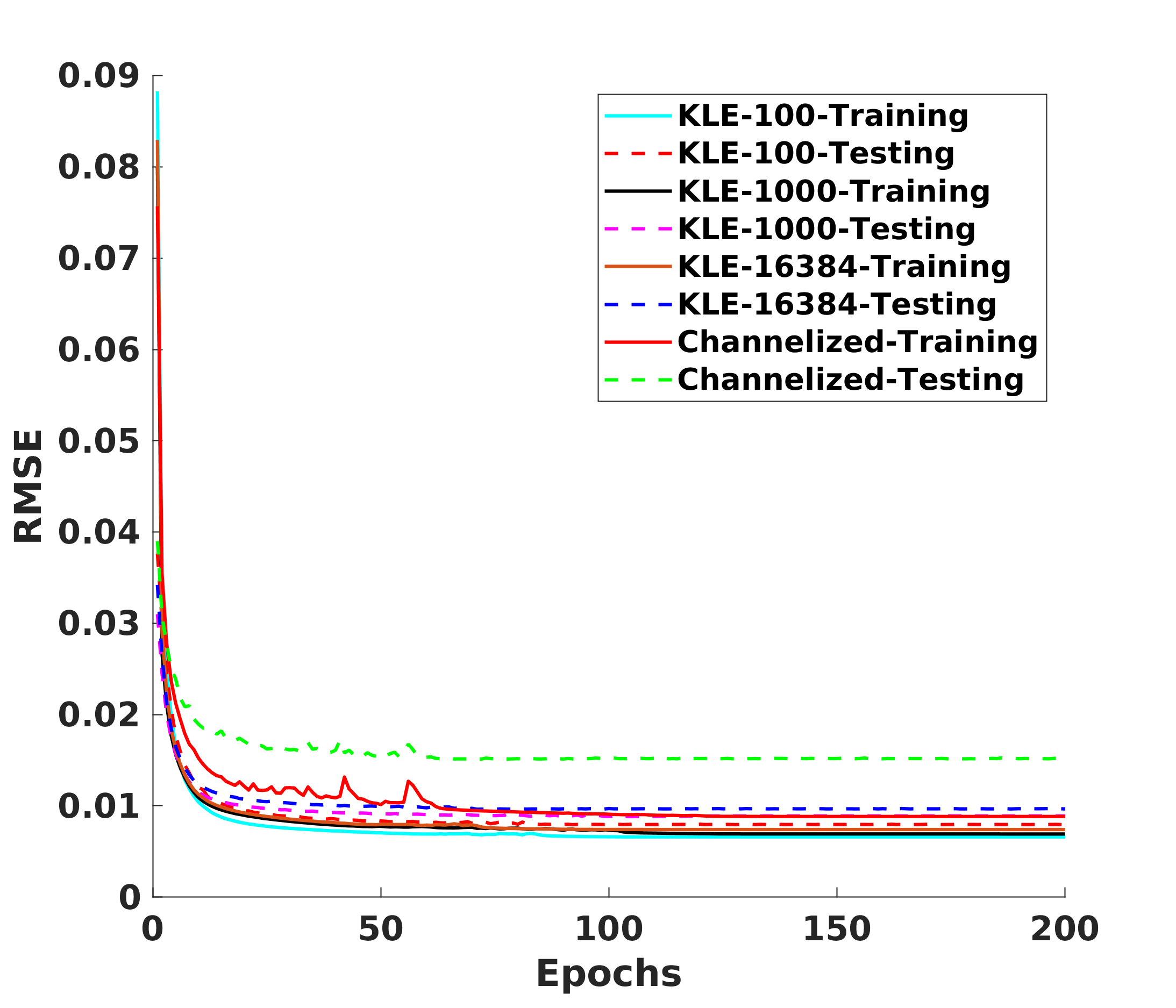}
    \caption{Training and testing RMSE plot for KLE$-100$ ($64-$training data), KLE$-1000$ ($96-$training data), KLE$-16384$ ($128-$training data) and channelized field ($160-$ training data).}
    \label{fig:RMSE_D}
\end{figure}

\begin{figure}[H]
  \centering
  \begin{minipage}[b]{0.48\textwidth}
    \includegraphics[scale=0.38]{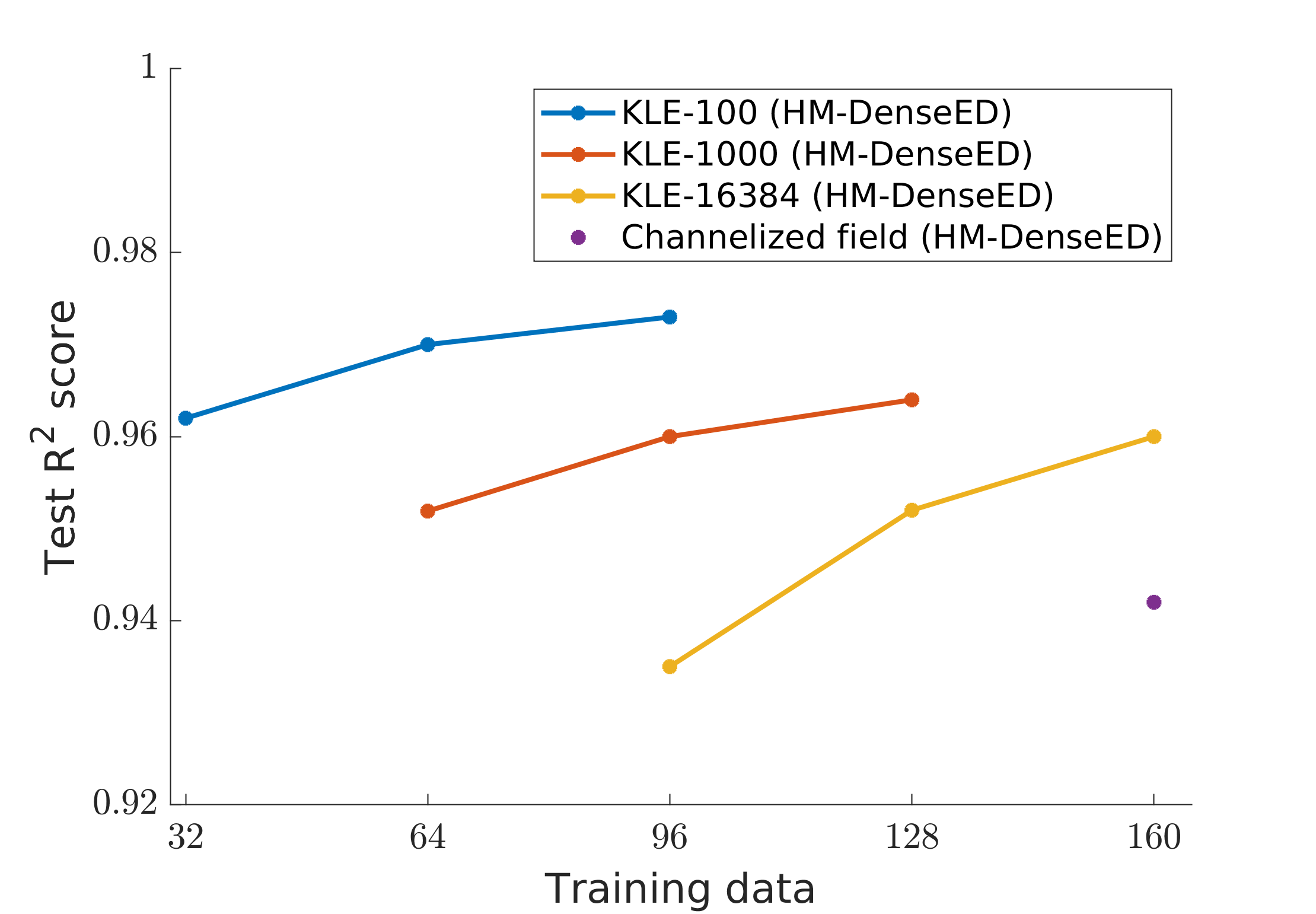}
  \end{minipage}
  \hfill
  \begin{minipage}[b]{0.48\textwidth}
    \includegraphics[scale=0.38]{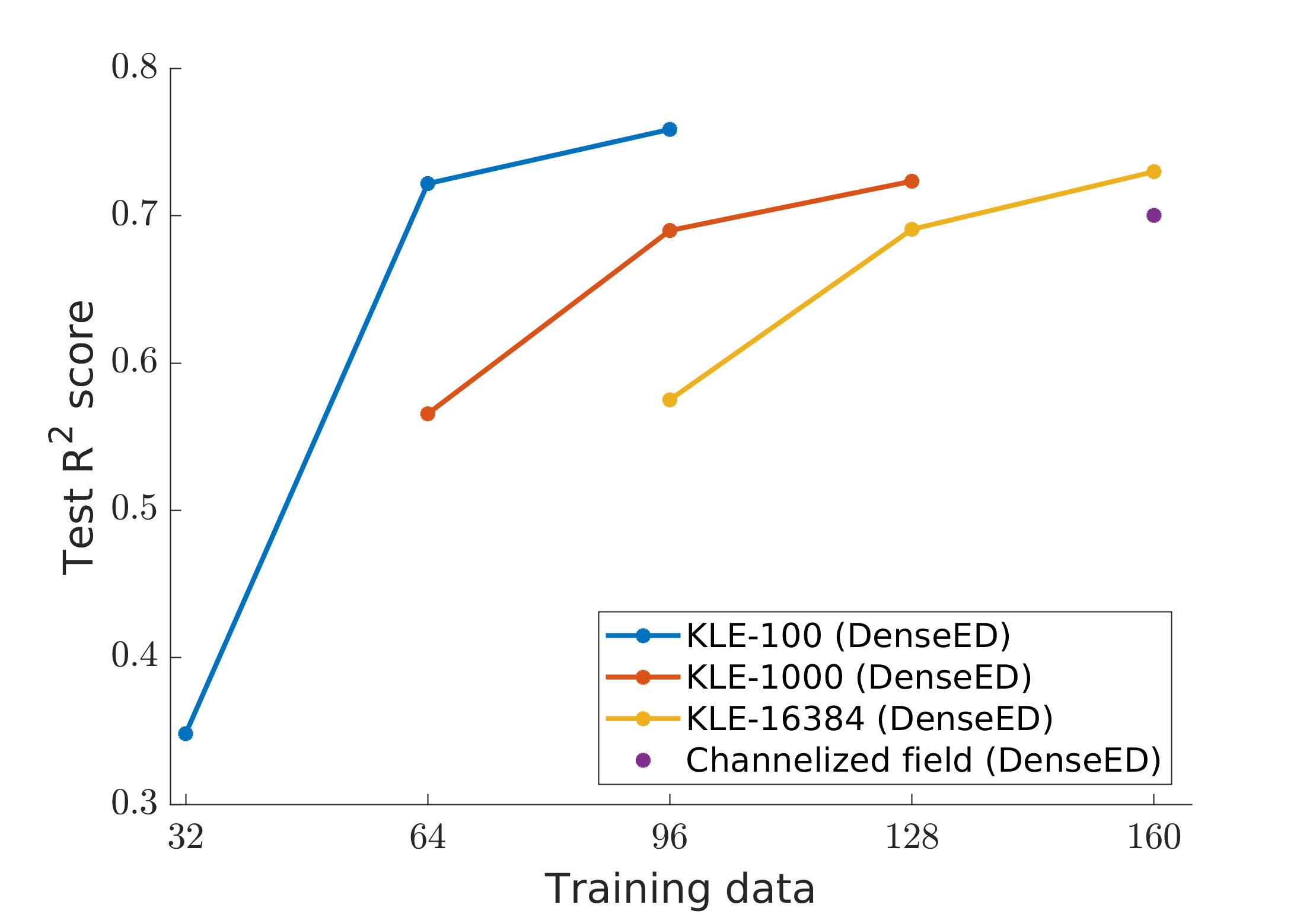}
  \end{minipage}
    \caption{Comparison of test $R^2$ scores (for pressure) for HM-DenseED (left)  and DenseED (right) for KLE$-100$, $-1000$, $-16384$ and channelized permeability field and for various training data.}
    \label{fig:R2_fig}
\end{figure}

To evaluate the performance of the HM-DenseED and DenseED models, we show the estimate of the pressure, $x-$velocity and $y-$velocity components at the location $(0.6, 0.4)$  with $10,000$ input realizations and compare with the Monte Carlo based solution as illustrated in Figs.~\ref{fig:pressure_PDF_100}-\ref{fig:pressure_PDF_channel}  for pressure,  Figs.~\ref{fig:velocity_x_KLE_100}-\ref{fig:velocity_x_channel} for horizontal flux and Figs.~\ref{fig:velocity_y_100}-\ref{fig:velocity_y_channel} for vertical flux. We observe that the density estimate of the hybrid model (shown in green dots) is close to the Monte Carlo result (fine-scale, shown in blue dashed line)  even for the high-dimension GRF input and channelized field when compared to the standard surrogate model shown in magenta dashed line. 
To evaluate the performance of training, we present the RMSE plot and also the test $R^2$ score. From Fig.~\ref{fig:RMSE_D} we observe that the training and the testing for various datasets converged around $50-60$ epochs. The training data is $96$ for KLE$-100$ and KLE$-1000$, and for KLE$-16384$, the training data is $128$. Test $R^2$ scores for the various datasets are illustrated in Fig.~\ref{fig:R2_fig} for both the HM-DenseED and DenseED models.  For the HM-DenseED model, we observe that as we increase the dataset size, the $R^2$ scores tend to $1$. For high-dimensional data (KLE$-16384$), we observe that the  $R^2$ score is $0.93$, which is reasonably good since we are considering limited fine-scale data for training the model. Finally, we can see that the $R^2$ score for the HM-DenseED model is closer to $1$ compared to the standard surrogate model (DenseED model).
\subsection{Bayesian Surrogate Model}
The Bayesian neural network is trained by applying the SVGD algorithm with $20$ samples. Let $\bm{\theta}$ be the uncertain parameters that consist of $20$ DenseED model parameters $\bm{w}$ and noise precision $\beta$. We initialize the $20$ different DenseED networks and  update the parameters using the SVGD algorithm.  Adam optimizer is used to update the parameter $\bm{\theta}$ using the gradient $\bm{\phi}$ as explained in Section~\ref{sec:SVGD} for $100$ epochs. The learning rate for training the network is $2e-3$ and the learning rate for $\beta$ is  $0.01$. The total training time was $8-10$ hours when implementing this algorithm on a single NVIDIA Tesla V$100$ GPU card GPU. We can speedup training by parallelizing the code and using multiple GPUs. We show the results for the Bayesian framework of the HM-DenseED model. Note that, in this section, we present the results for the predictive mean and variance of the pressure. Recall that the model is trained for predicting the pressure given the permeability field.
\begin{figure}[H]
  \centering
  \begin{minipage}[b]{0.48\textwidth}
    \includegraphics[width=\textwidth]{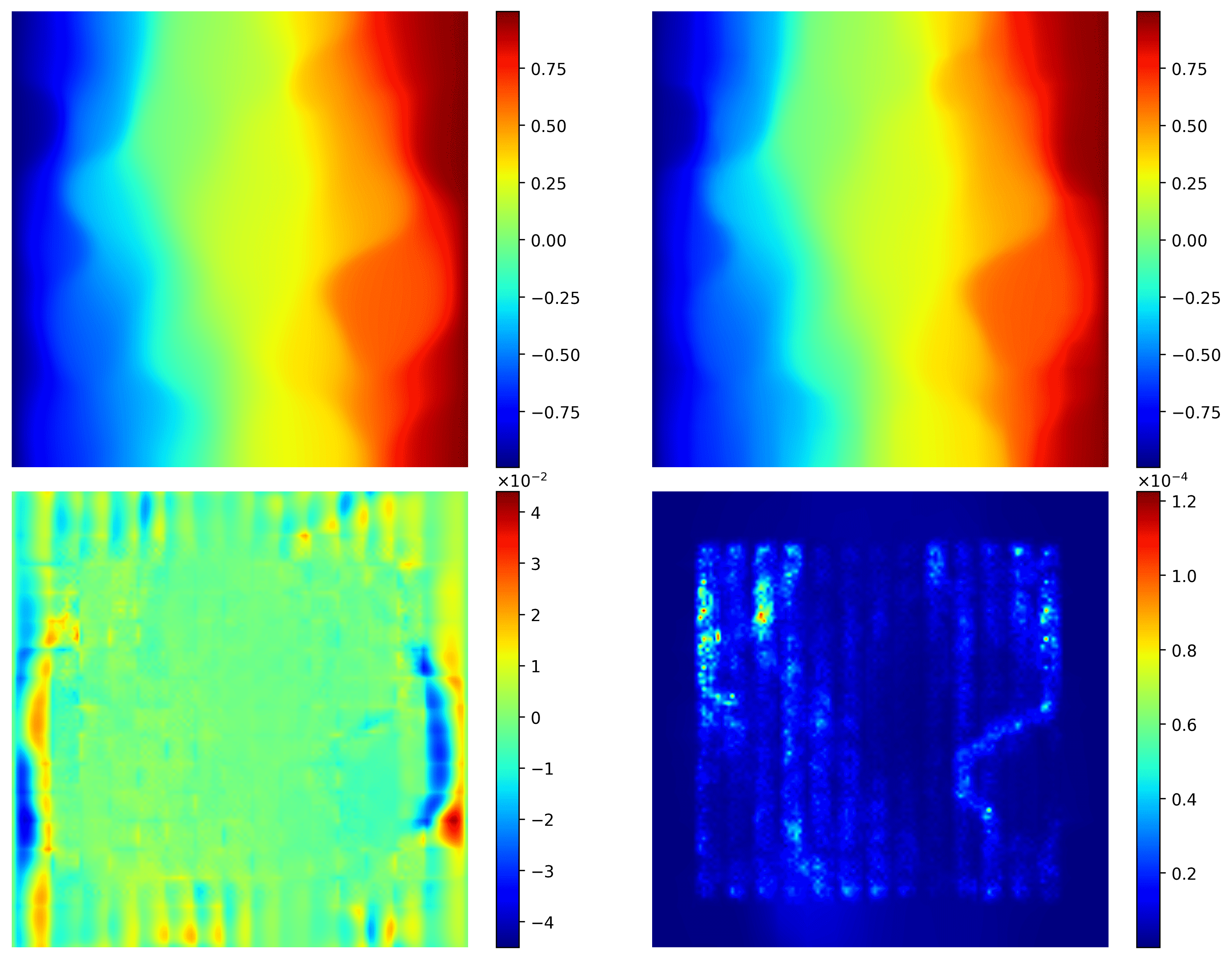}
  \end{minipage}
  \hfill
  \begin{minipage}[b]{0.48\textwidth}
    \includegraphics[width=\textwidth]{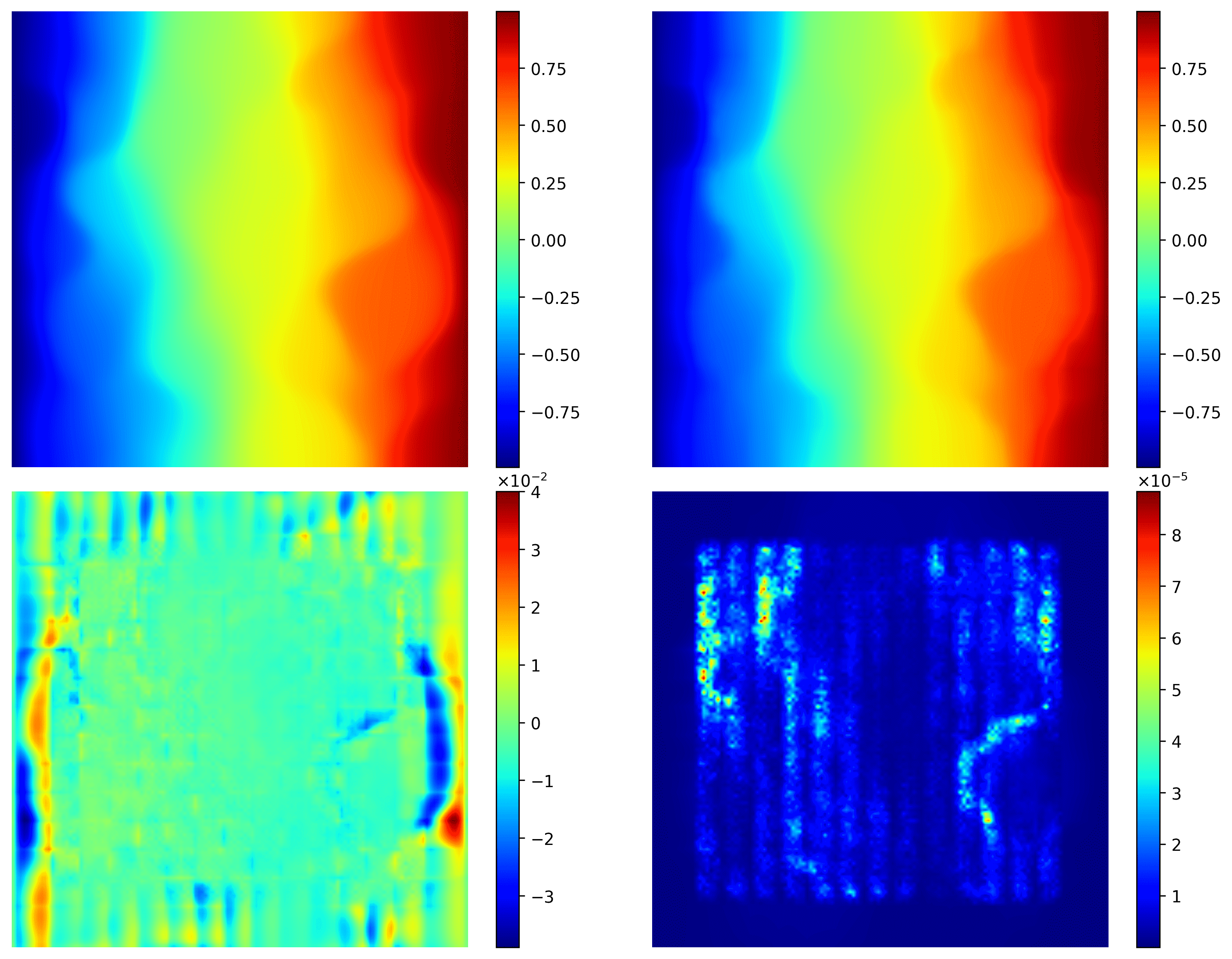}
  \end{minipage}
    \caption{Prediction of KLE$-100$ with $32$ training data (left), and $96$ training data (right); For individual prediction statistics, from left to right (first row): For single test input $\bm{K}^{*}$, test output (ground truth) $t^{*}$, Predictive mean $\mathbb{E}[\widehat{{\bm{P}}_f}^{*}|\bm{K}^{*},\mathcal{\bm{D}}]$, from left to right (second row) Error: Predictive mean and test output (ground truth), and predictive variance $\text{Var}(\widehat{{\bm{P}}_f}^{*}|\bm{K}^{*},\mathcal{\bm{D}})$.}
    \label{fig:BS_11}
\end{figure}
\begin{figure}[H]
  \centering
  \begin{minipage}[b]{0.48\textwidth}
    \includegraphics[width=\textwidth]{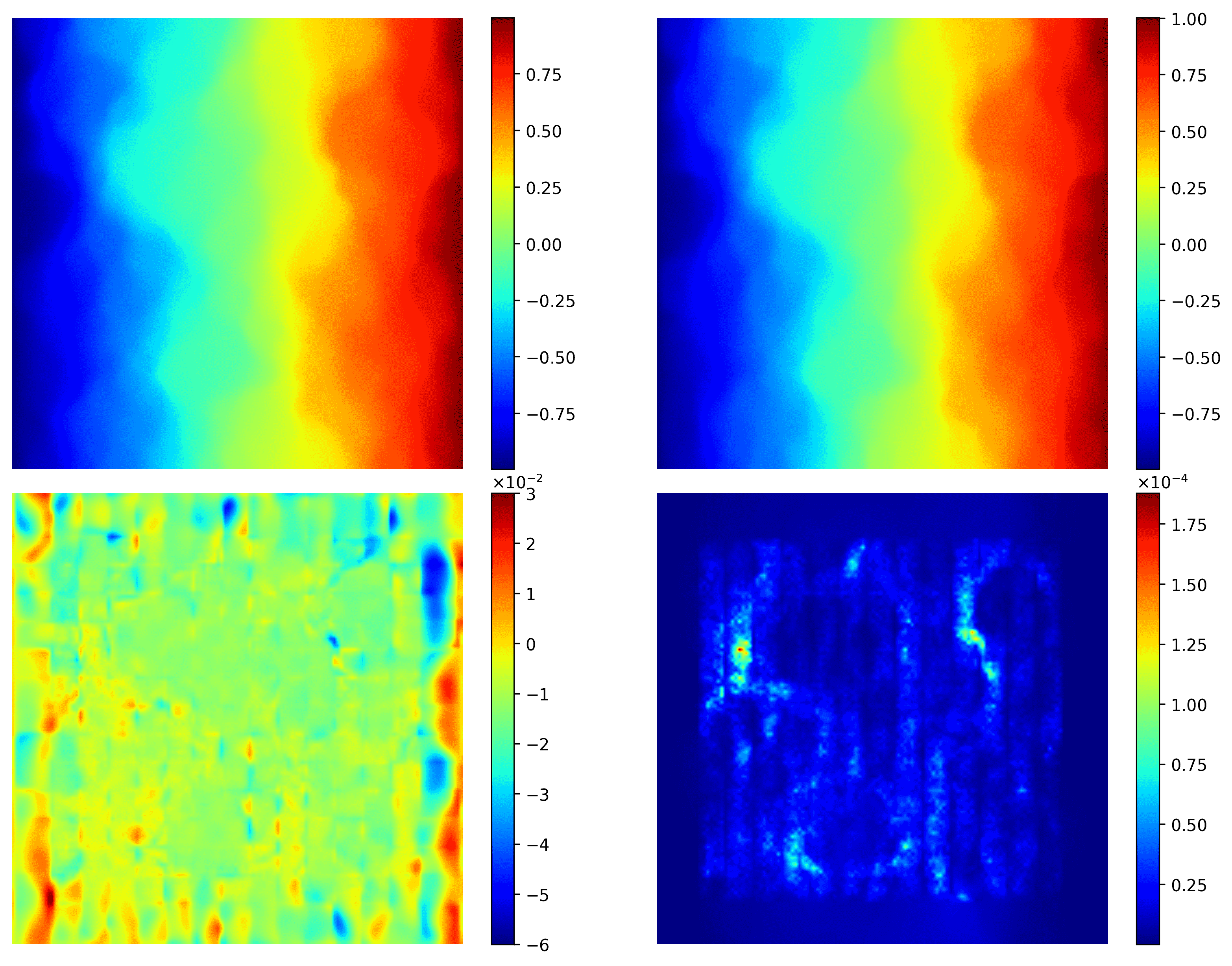}
  \end{minipage}
  \hfill
  \begin{minipage}[b]{0.48\textwidth}
    \includegraphics[width=\textwidth]{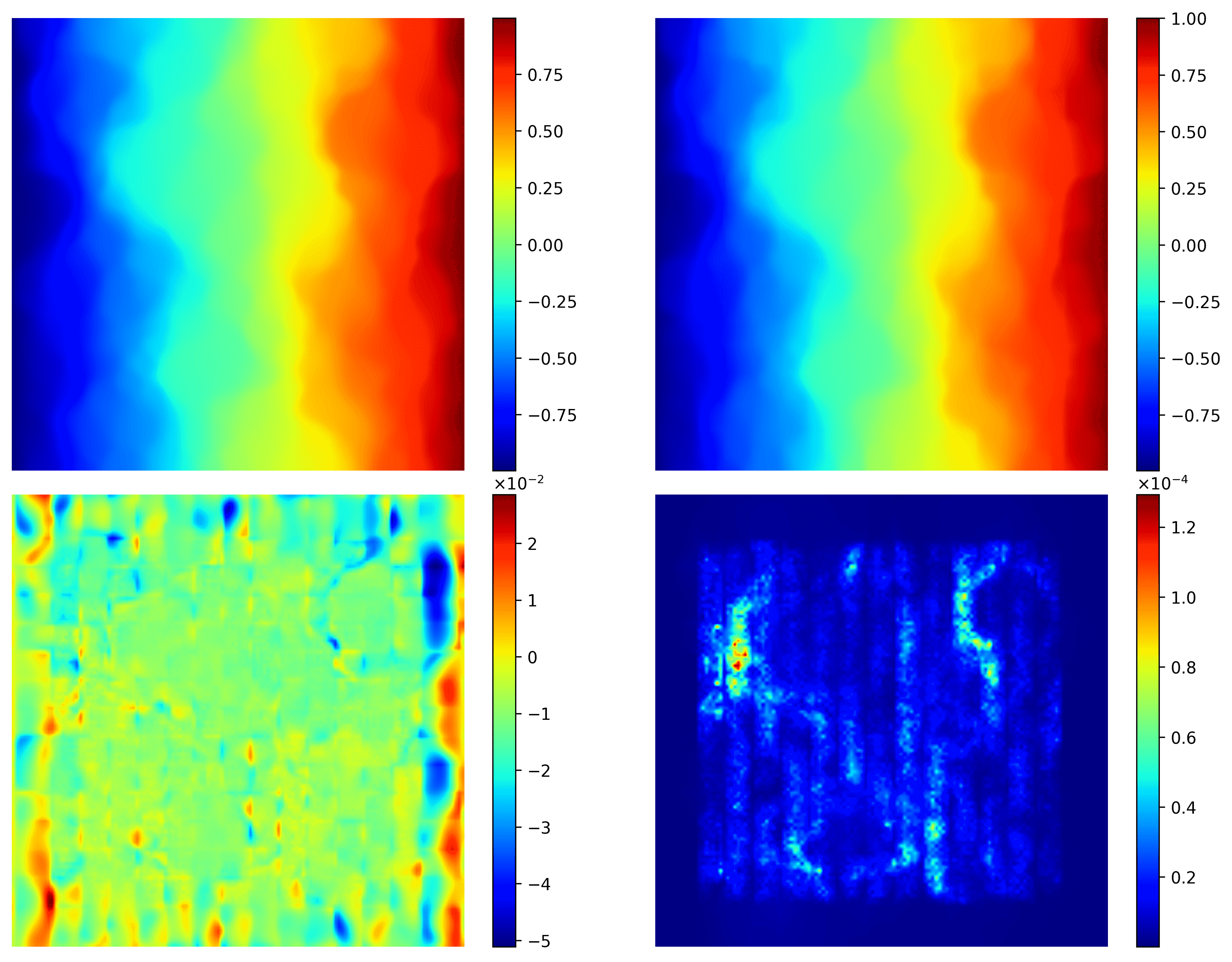}
  \end{minipage}
    \caption{Prediction of KLE$-1000$ with $64$ training data (left), and $128$ training data (right); For individual prediction statistics, from left to right (first row): For single test input $\bm{K}^{*}$, test output (ground truth) $t^{*}$, predictive mean $\mathbb{E}[\widehat{{\bm{P}}_f}^{*}|\bm{K}^{*},\mathcal{\bm{D}}]$, from left to right (second row) Error: Predictive mean and test output (ground truth), and Predictive variance $\text{Var}(\widehat{{\bm{P}}_f}^{*}|\bm{K}^{*},\mathcal{\bm{D}})$.}
    \label{fig:BS_1_64}
\end{figure}
\begin{figure}[H]
  \centering
  \begin{minipage}[b]{0.48\textwidth}
    \includegraphics[width=\textwidth]{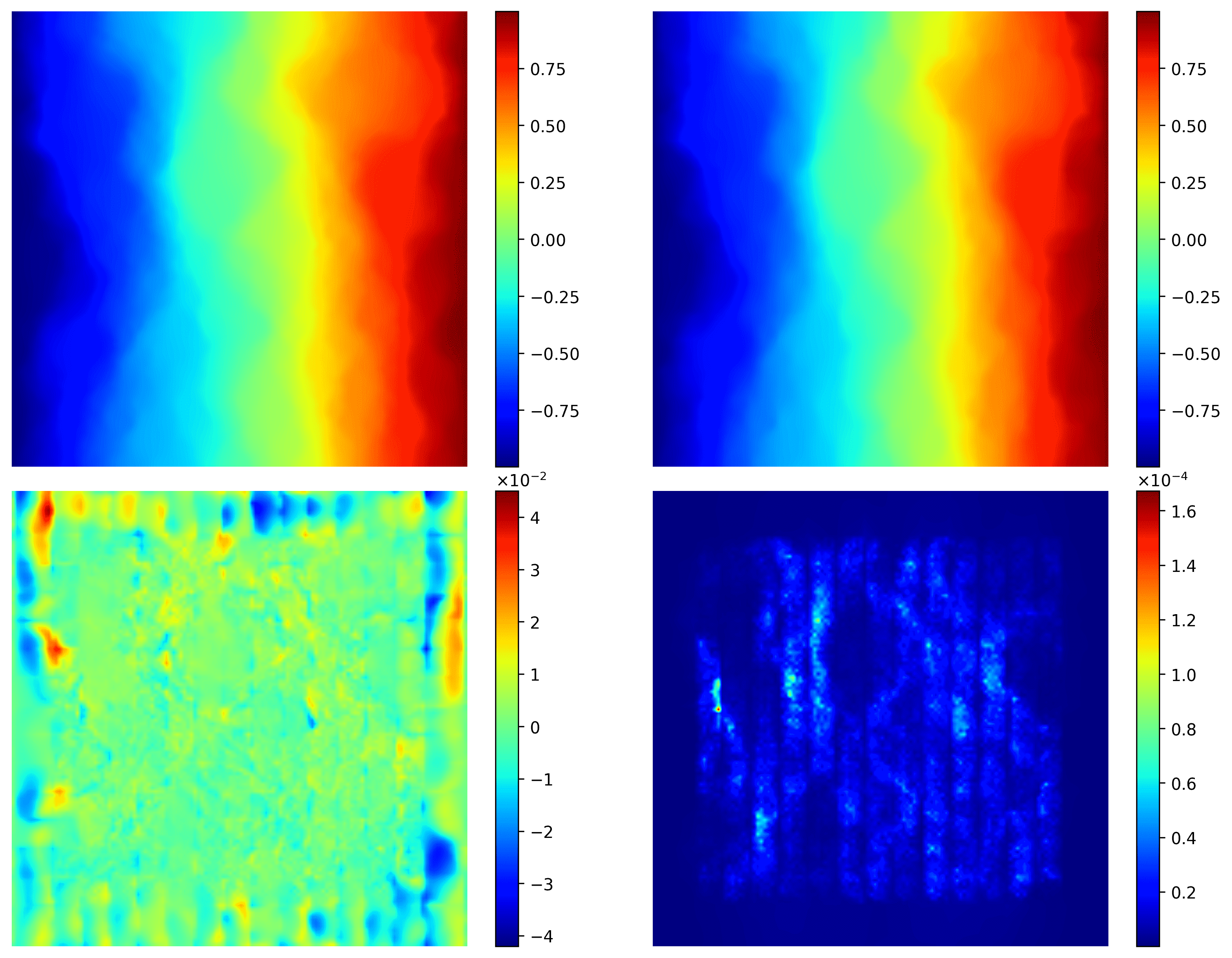}
  \end{minipage}
  \hfill
  \begin{minipage}[b]{0.48\textwidth}
    \includegraphics[width=\textwidth]{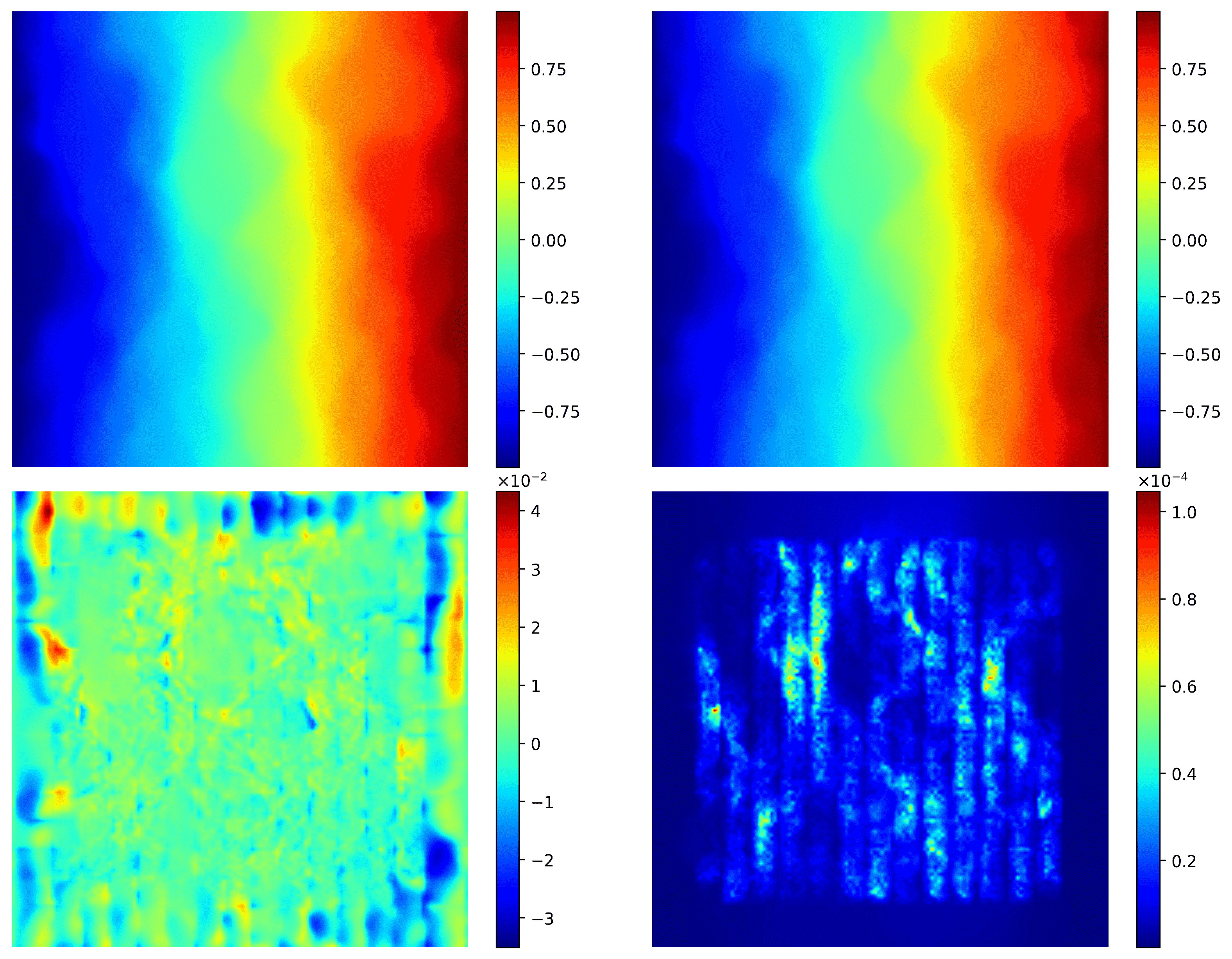}
  \end{minipage}
    \caption{Prediction of KLE$-16384$ with $96$ training data (left),and $160$ training data (right); For individual prediction statistics, from left to right (first row): For single test input $\bm{K}^{*}$, test output (ground truth) $t^{*}$, Predictive mean $\mathbb{E}[\widehat{{\bm{P}}_f}^{*}|\bm{K}^{*},\mathcal{\bm{D}}]$, from left to right (second row) Error: Predictive mean and test output (ground truth), and Predictive variance $\text{Var}(\widehat{{\bm{P}}_f}^{*}|\bm{K}^{*},\mathcal{\bm{D}})$.}
    \label{fig:BS_1}
\end{figure}
\begin{figure}[H]
  \centering
    \includegraphics[width=0.55\textwidth]{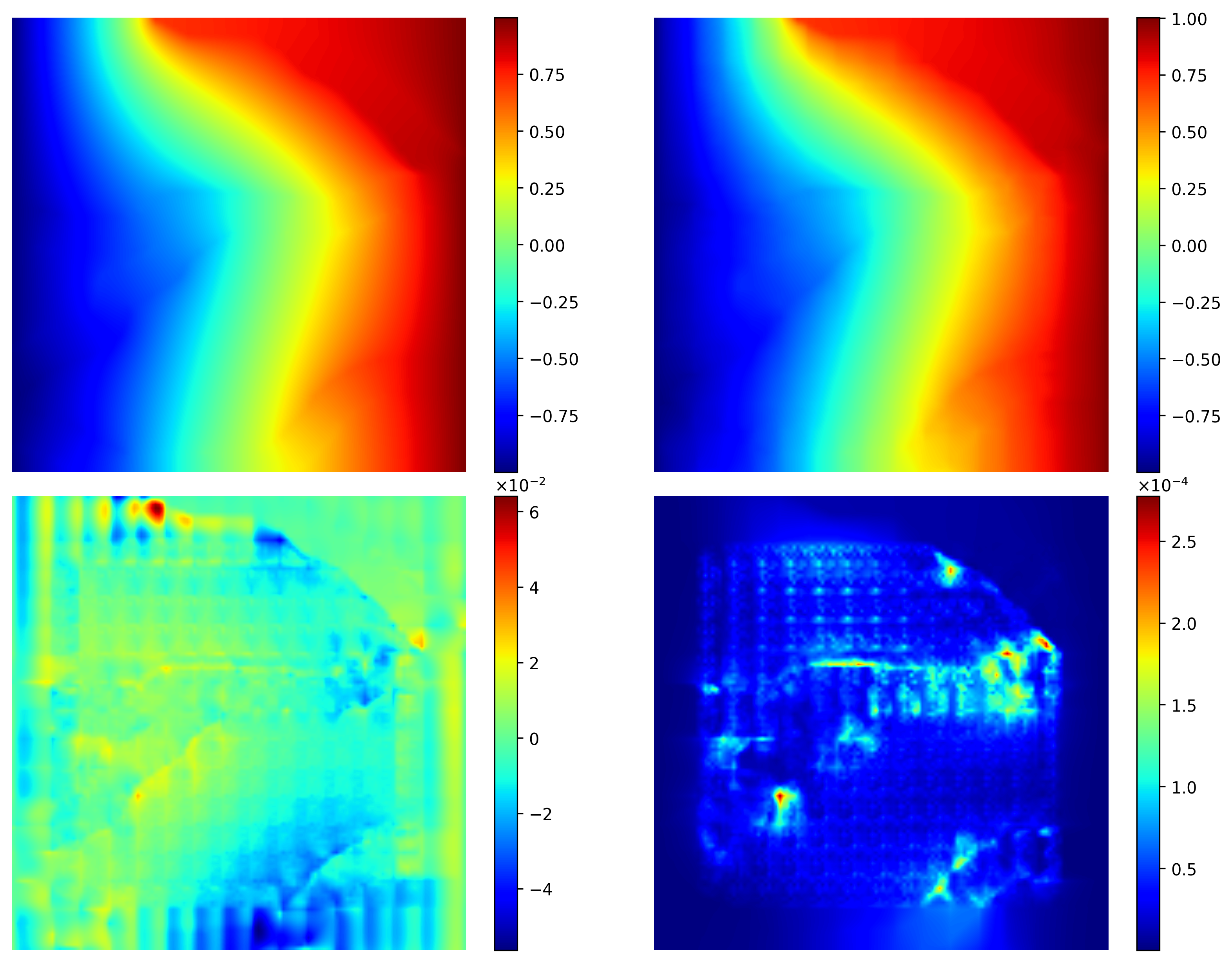}
    \caption{Prediction for channelized field; For individual prediction statistics, from left to right (first row): For single test input $\bm{K}^{*}$, test output (ground truth) $t^{*}$, predictive mean $\mathbb{E}[\widehat{{\bm{P}}_f}^{*}|\bm{K}^{*},\mathcal{\bm{D}}]$, from left to right (second row) Error: Predictive mean and test output (ground truth), and predictive variance $\text{Var}(\widehat{{\bm{P}}_f}^{*}|\bm{K}^{*},\mathcal{\bm{D}})$.}
    \label{fig:BS_2}
\end{figure}
From Figs.~\ref{fig:BS_11}-\ref{fig:BS_1}, we observe that as the training data increases the predictive accuracy increases. For example, in Figs.~\ref{fig:BS_11} or ~\ref{fig:BS_1}, we note that the predictive uncertainty reduces as we increase the training data. \par
\begin{figure}[H]
\begin{center}
    \includegraphics[scale=0.45]{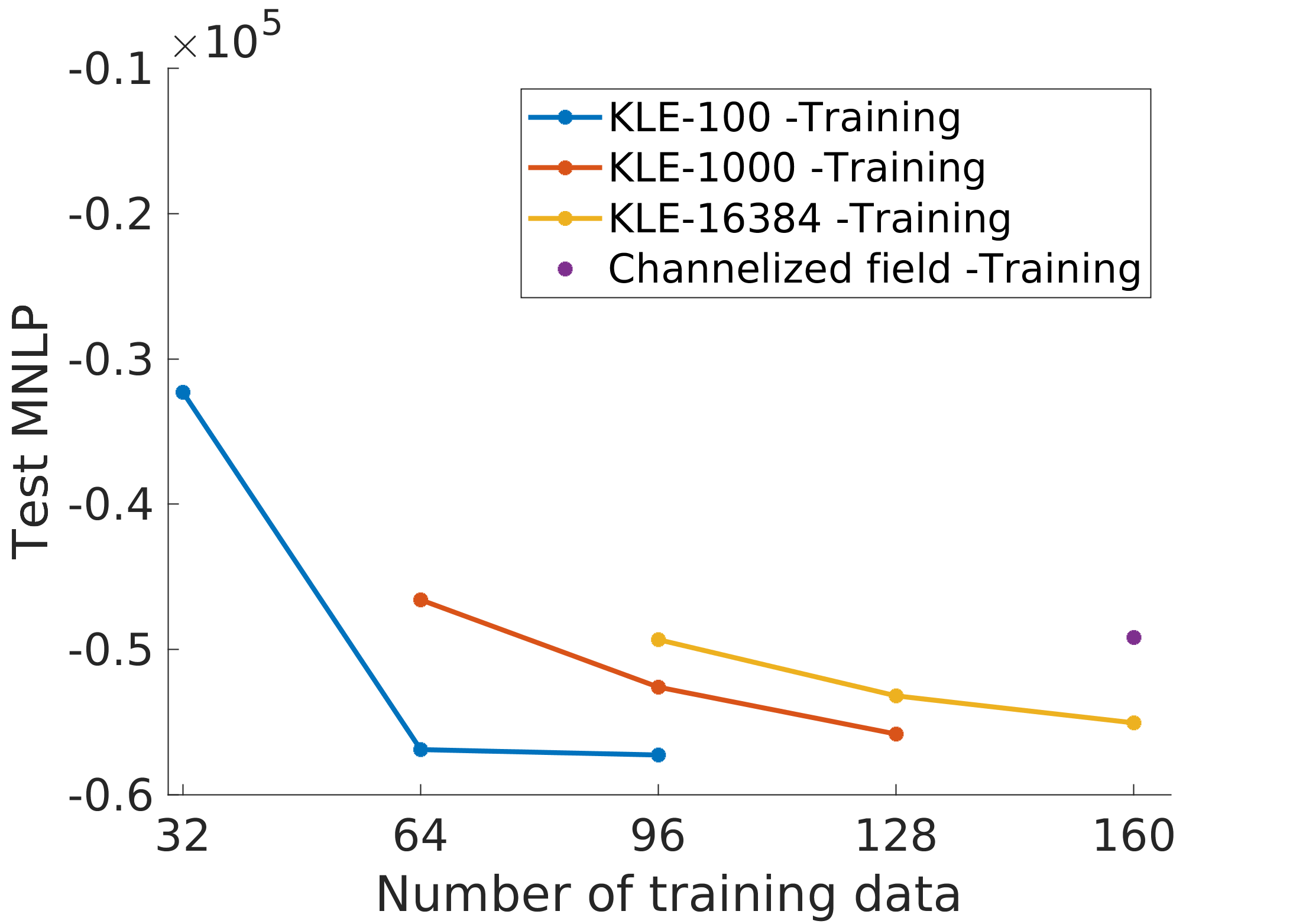}
    \caption{MNLP of test data.}
    \label{fig:MNLP}
  \end{center}
\end{figure}
\begin{figure}[H]
    \centering
    {
        \includegraphics[width=0.62\textwidth]{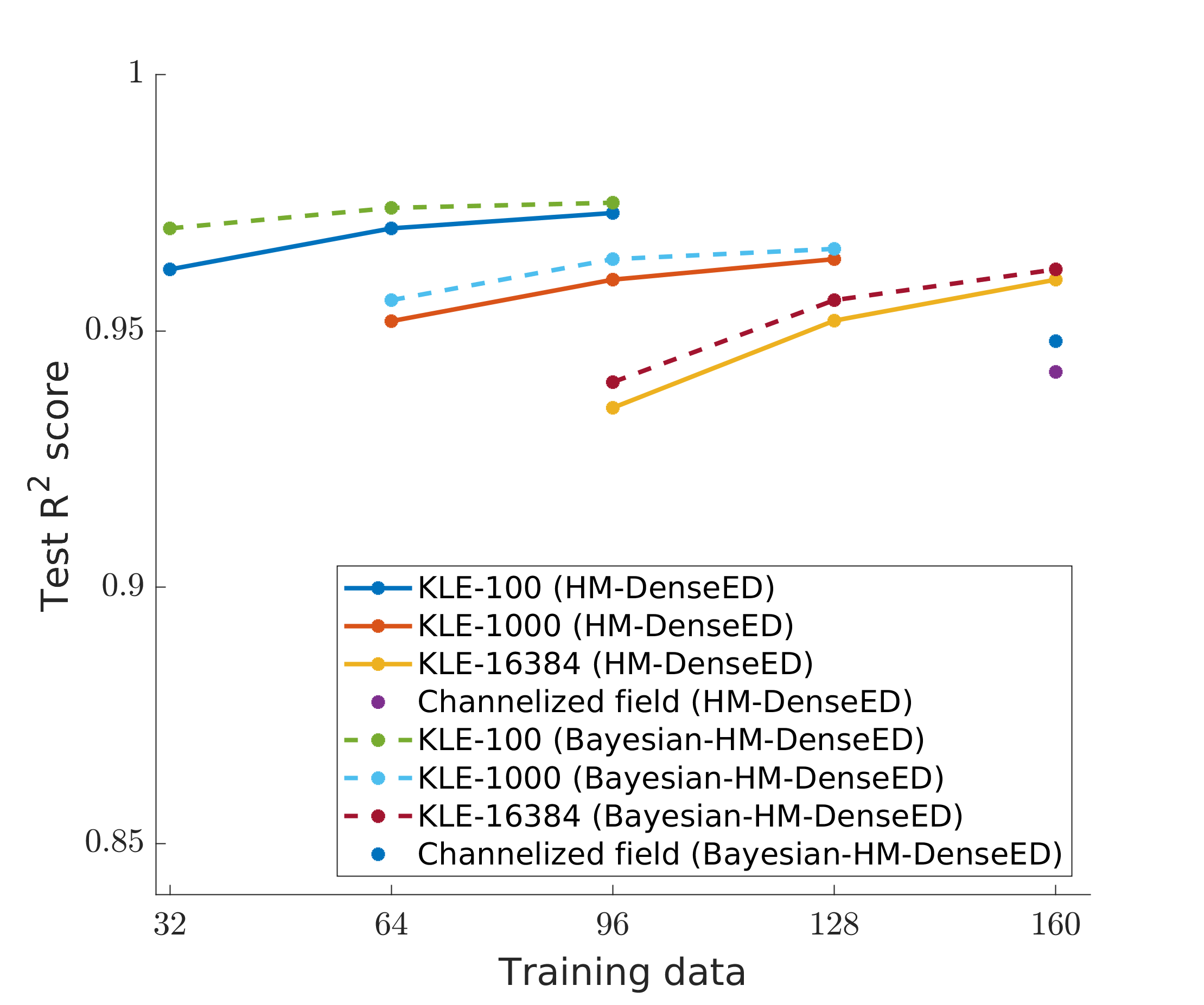}
    }
        \caption{Non-Bayesian and Bayesian test $R^2$ scores for KLE$-100$, $-1000$, $-16384$ and channelized field (Hybrid DenseED model).}
    \label{fig:comp_r2}
\end{figure}
\begin{figure}[H]
    \centering
    \includegraphics[scale=0.35]{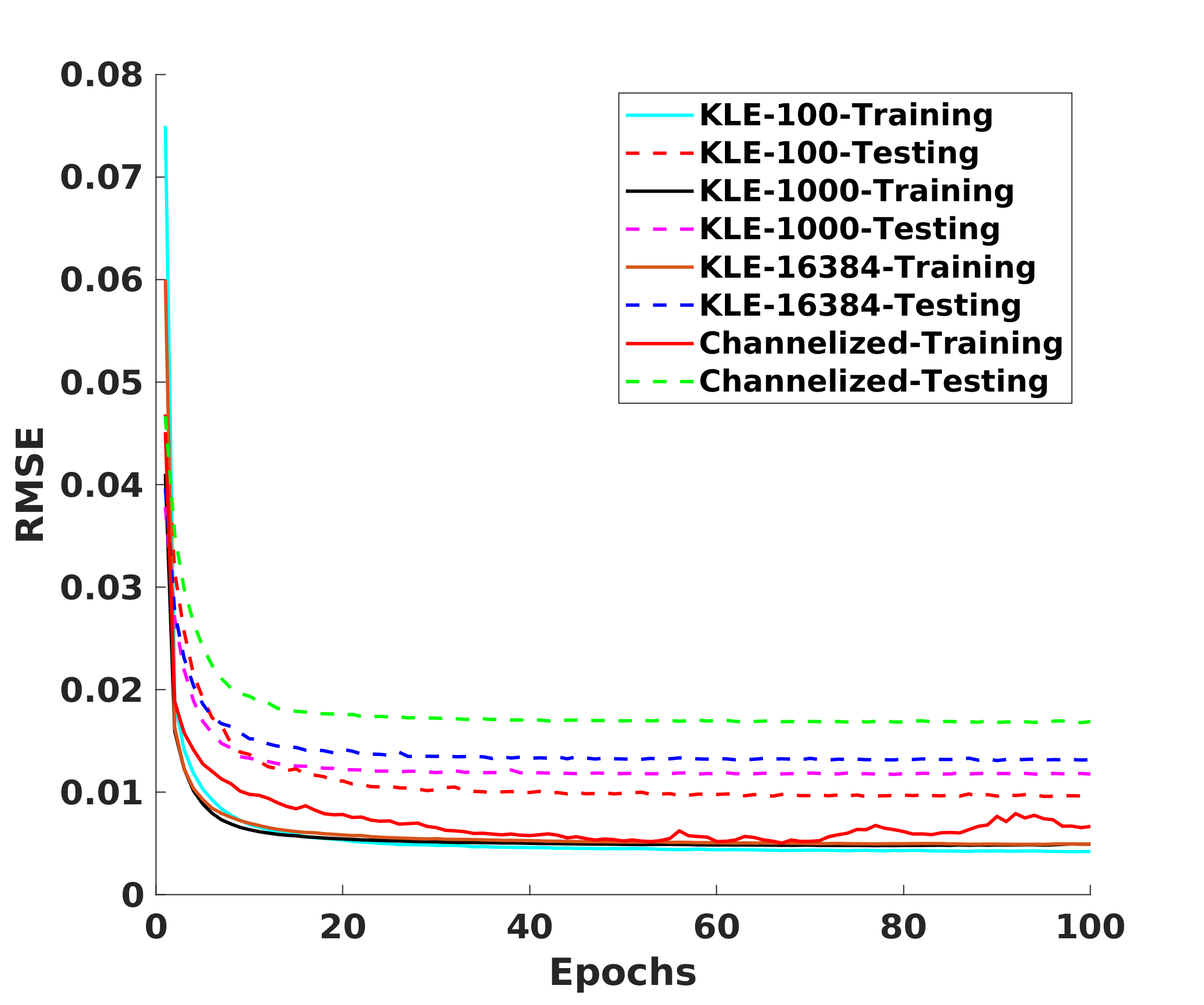}
    \caption{Bayesian HM-DenseED: Training and testing RMSE plot for KLE$-100$ ($64-$training data), KLE$-1000$ ($96-$training data), KLE$-16384$ ($128-$training data) and channelized field ($160-$ training data).}
    \label{fig:RMSE_Bayesian}
\end{figure}
\begin{figure}[H]
  \centering
  \begin{minipage}[b]{0.48\textwidth}
    \includegraphics[width=\textwidth]{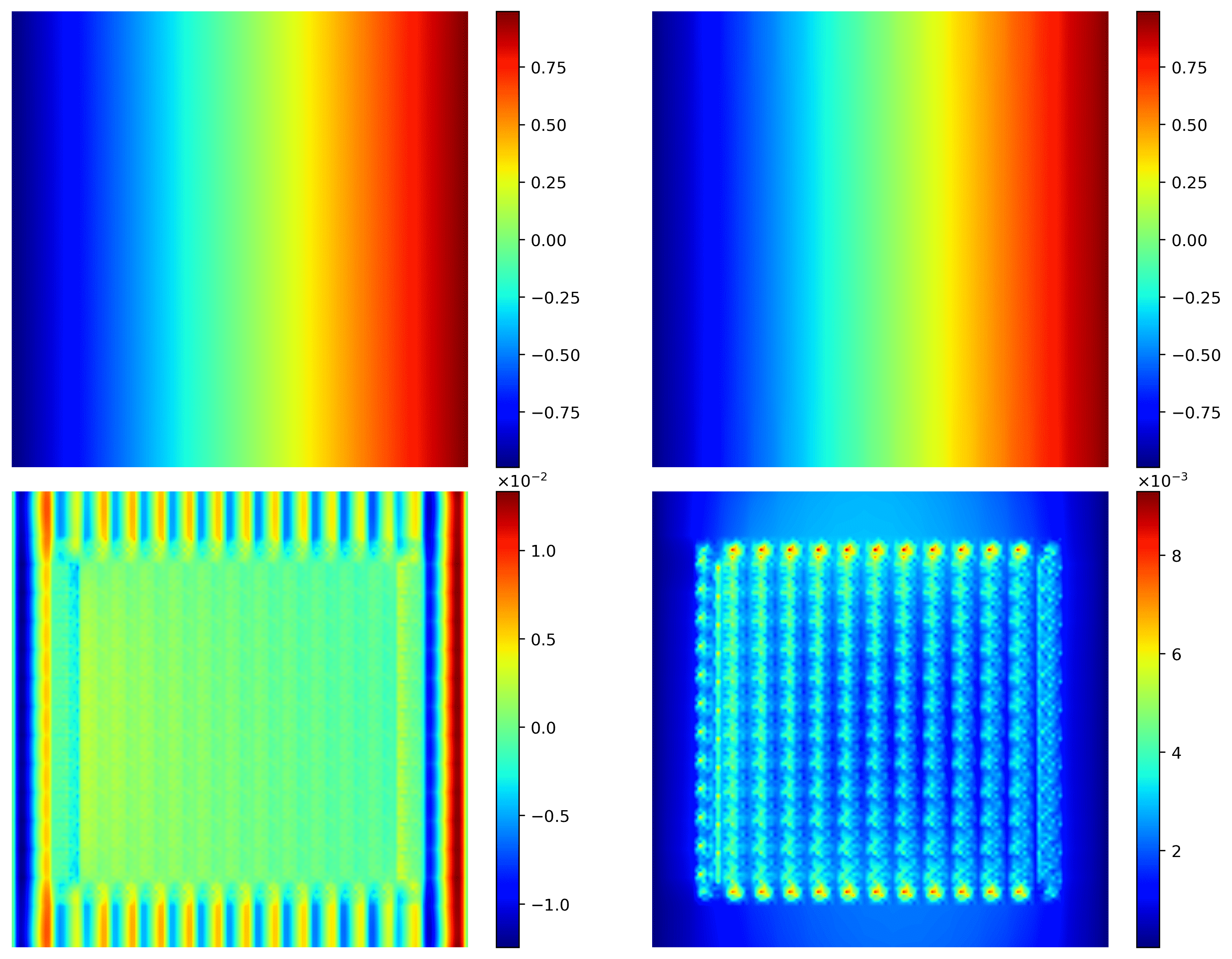}
  \end{minipage}
  \hfill
  \begin{minipage}[b]{0.48\textwidth}
    \includegraphics[width=\textwidth]{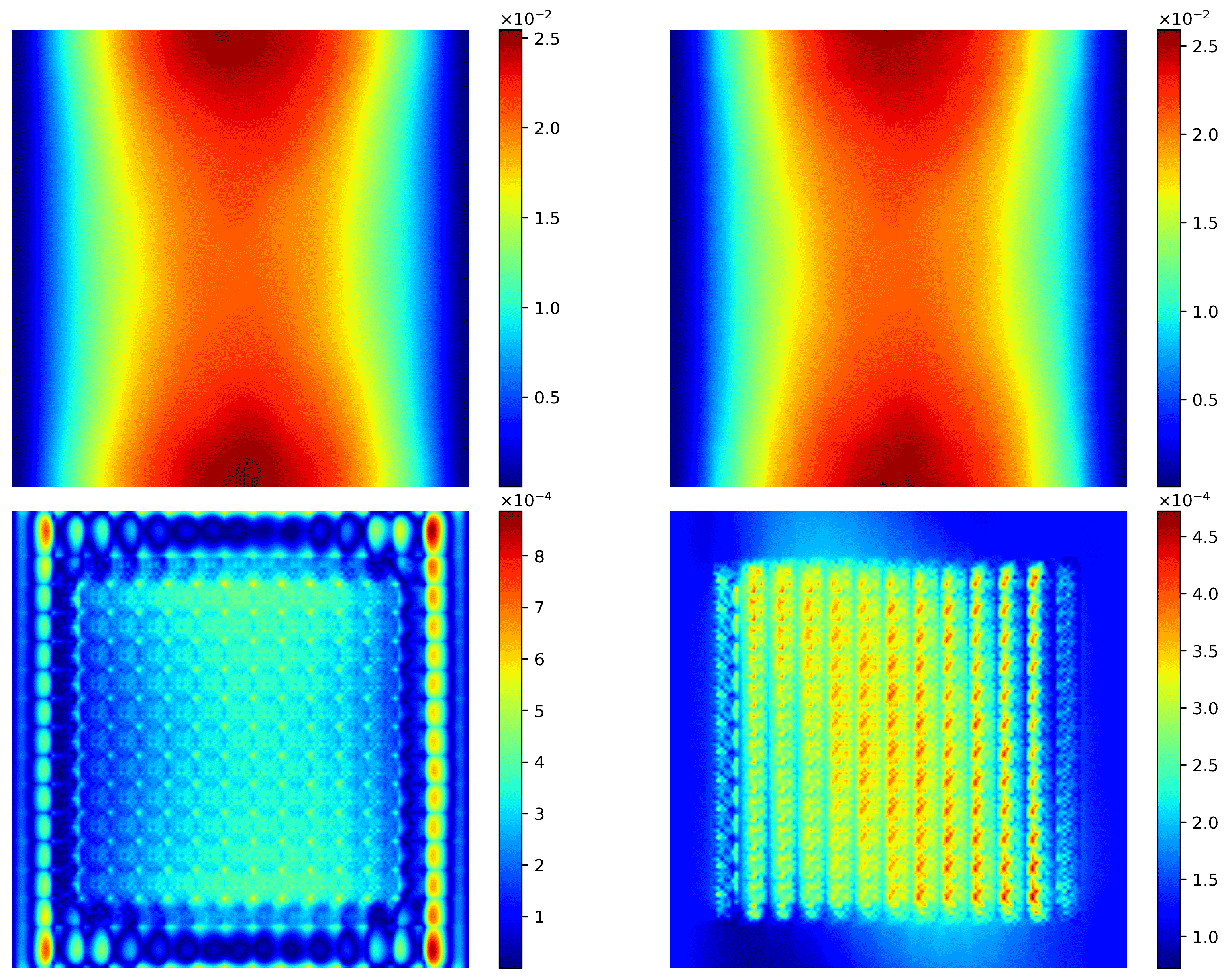}
  \end{minipage}
    \caption{(Left) Uncertainty propagation for KLE$-100$ ($32$ training data). We show the Monte Carlo output mean, predictive output mean $\mathbb{E}_{\bm{\theta}}[\mathbb{E}[\bm{y}|\bm{\theta}]]$, the error of the above two, and two standard deviations of the conditional  predictive mean $Var_{\bm{\theta}}[\mathbb{E}[\bm{y}|\bm{\theta}]]$. (Right) Uncertainty propagation for KLE$-100$: ($64$~training data) we show the Monte Carlo output variance, predictive output variance $\mathbb{E}_{\bm{\theta}}[Var(\bm{y} | \bm{\theta})]$, the error of the above two, and two standard deviations of the conditional  predictive variance $\text{Var}_{\bm{\theta}} (\text{Var}(\bm{y} | \bm{\theta}))$.}
    \label{fig:UP_32}
\end{figure}
\begin{figure}[H]
  \centering
  \begin{minipage}[b]{0.48\textwidth}
    \includegraphics[width=\textwidth]{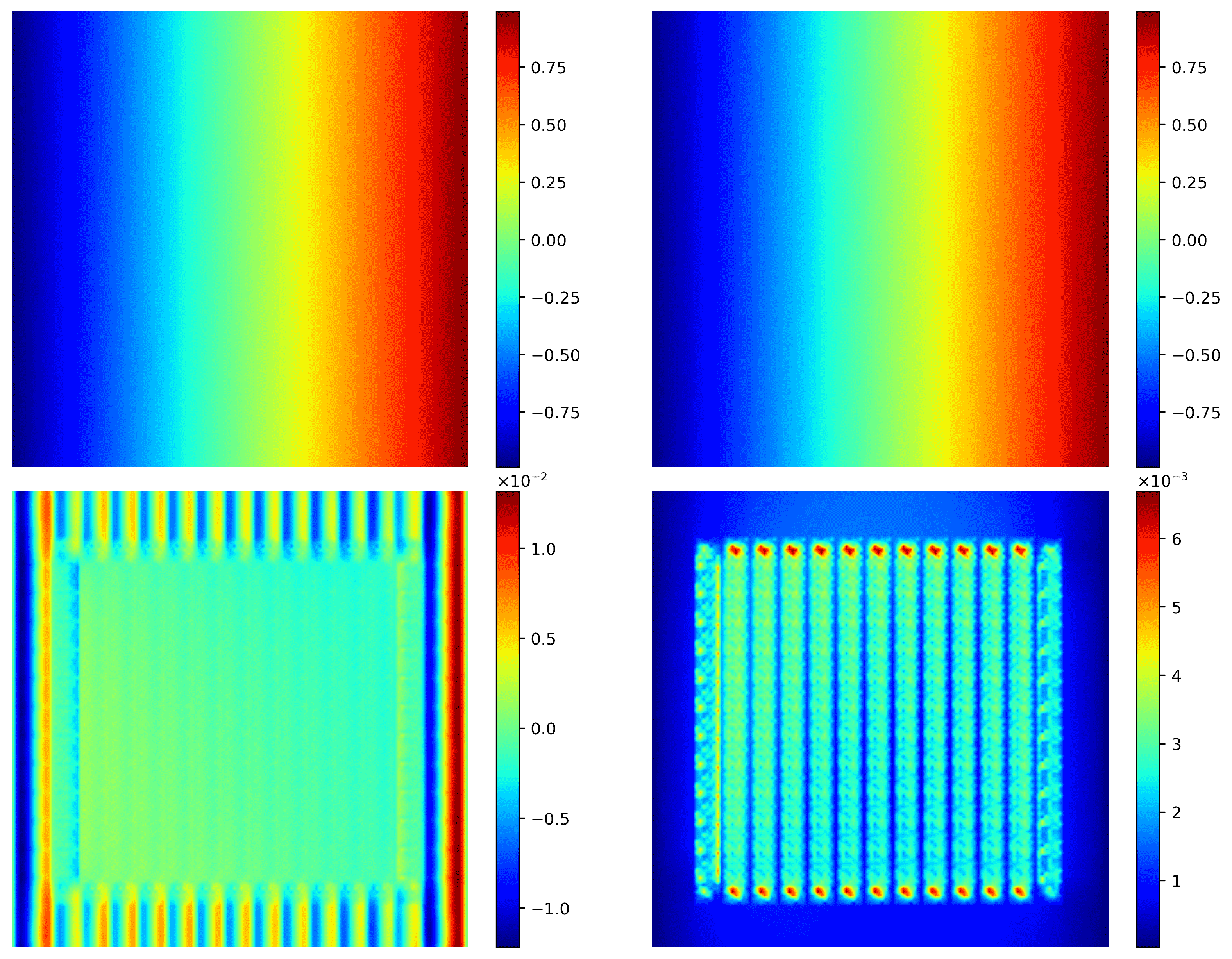}
  \end{minipage}
  \hfill
  \begin{minipage}[b]{0.48\textwidth}
    \includegraphics[width=\textwidth]{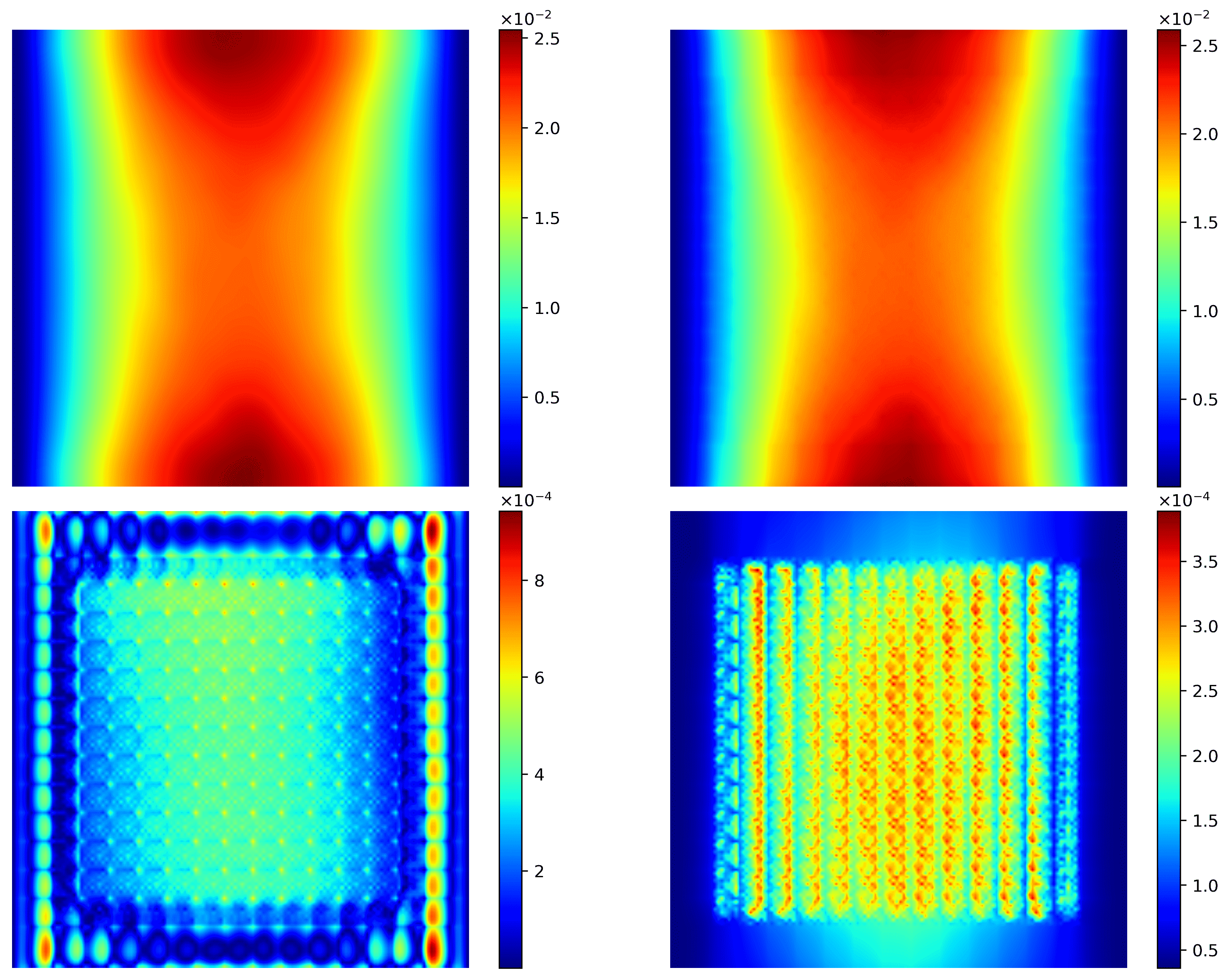}
  \end{minipage}
    \caption{ Uncertainty propagation for KLE$-100$ ($96$~ training data). (Left)  We show the Monte Carlo output mean, predictive output mean $\mathbb{E}_{\bm{\theta}}[\mathbb{E}[\bm{y}|\bm{\theta}]]$, the error of the above two, and two standard deviations of the conditional  predictive mean $Var_{\bm{\theta}}[\mathbb{E}[\bm{y}|\bm{\theta}]]$. (Right)  We show the Monte Carlo output variance, predictive output variance $\mathbb{E}_{\bm{\theta}}[Var(\bm{y} | \bm{\theta})]$, the error of the above two, and two standard deviations of the conditional  predictive variance $\text{Var}_{\bm{\theta}} (\text{Var}(\bm{y} | \bm{\theta}))$.}
    \label{fig:UP_96}
\end{figure}
\begin{figure}[H]
  \centering
  \begin{minipage}[b]{0.48\textwidth}
    \includegraphics[width=\textwidth]{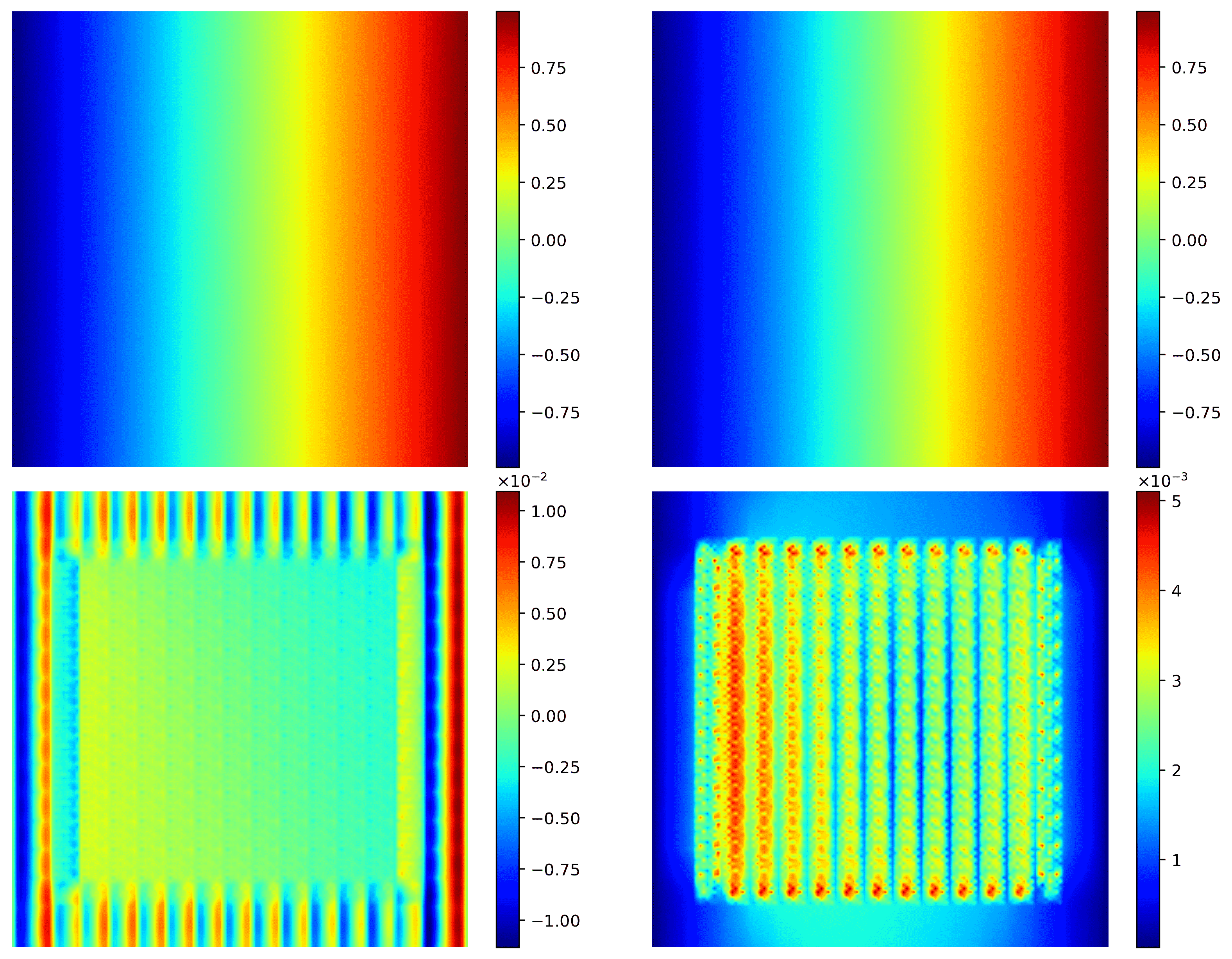}
  \end{minipage}
  \hfill
  \begin{minipage}[b]{0.48\textwidth}
    \includegraphics[width=\textwidth]{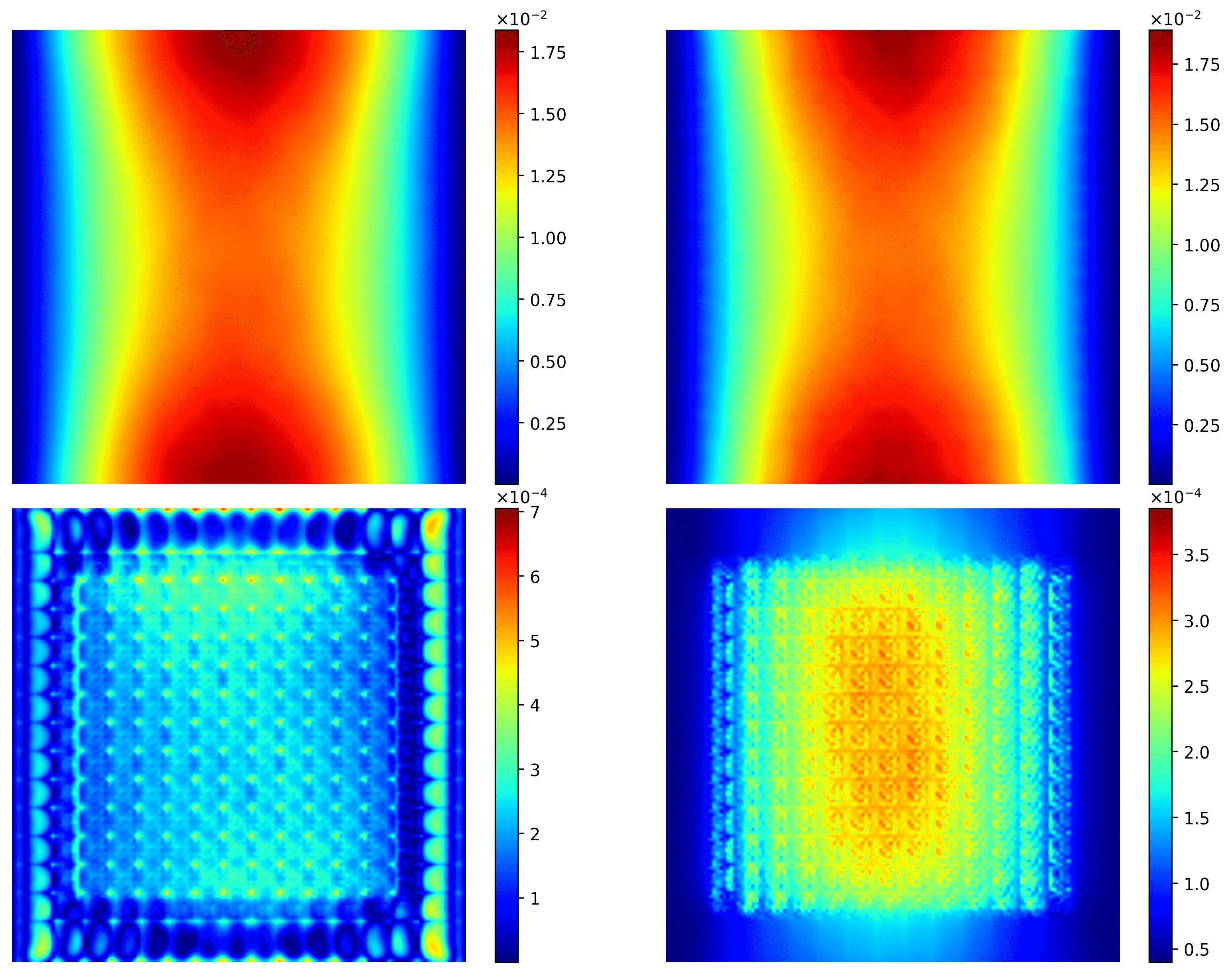}
  \end{minipage}
    \caption{Uncertainty propagation for KLE$-1000$ ($64$~training data). (Left)  We show the Monte Carlo output mean, predictive output mean $\mathbb{E}_{\bm{\theta}}[\mathbb{E}[\bm{y}|\bm{\theta}]]$, the error of the above two, and two standard deviations of the conditional  predictive mean $Var_{\bm{\theta}}[\mathbb{E}[\bm{y}|\bm{\theta}]]$. (Right) We show the Monte Carlo output variance, predictive output variance $\mathbb{E}_{\bm{\theta}}[Var(\bm{y} | \bm{\theta})]$, the error of the above two, and two standard deviations of the conditional  predictive variance $\text{Var}_{\bm{\theta}} (\text{Var}(\bm{y} | \bm{\theta}))$.}
    \label{fig:UP_1000}
\end{figure}
\begin{figure}[H]
  \centering
  \begin{minipage}[b]{0.48\textwidth}
    \includegraphics[width=\textwidth]{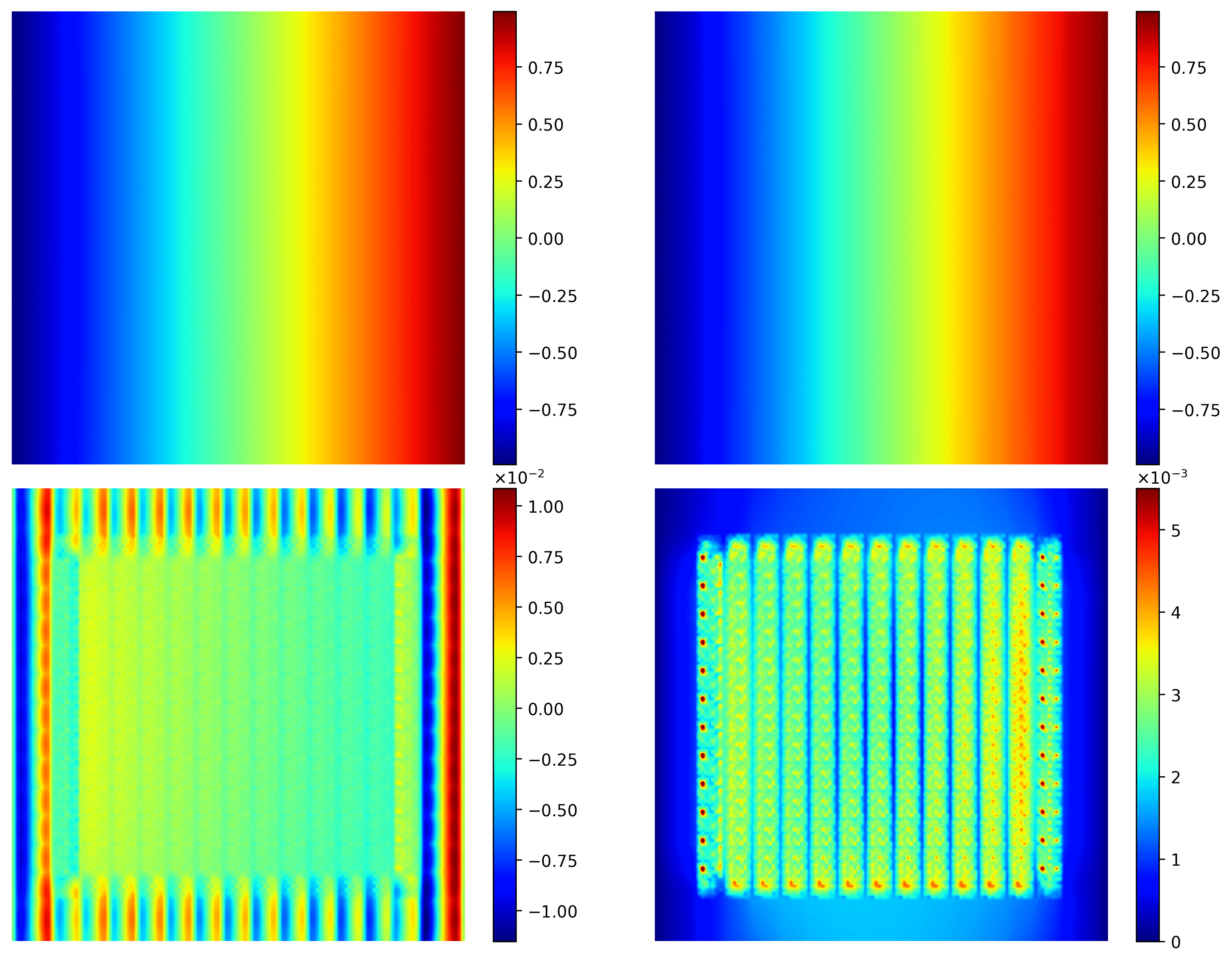}
  \end{minipage}
  \hfill
  \begin{minipage}[b]{0.48\textwidth}
    \includegraphics[width=\textwidth]{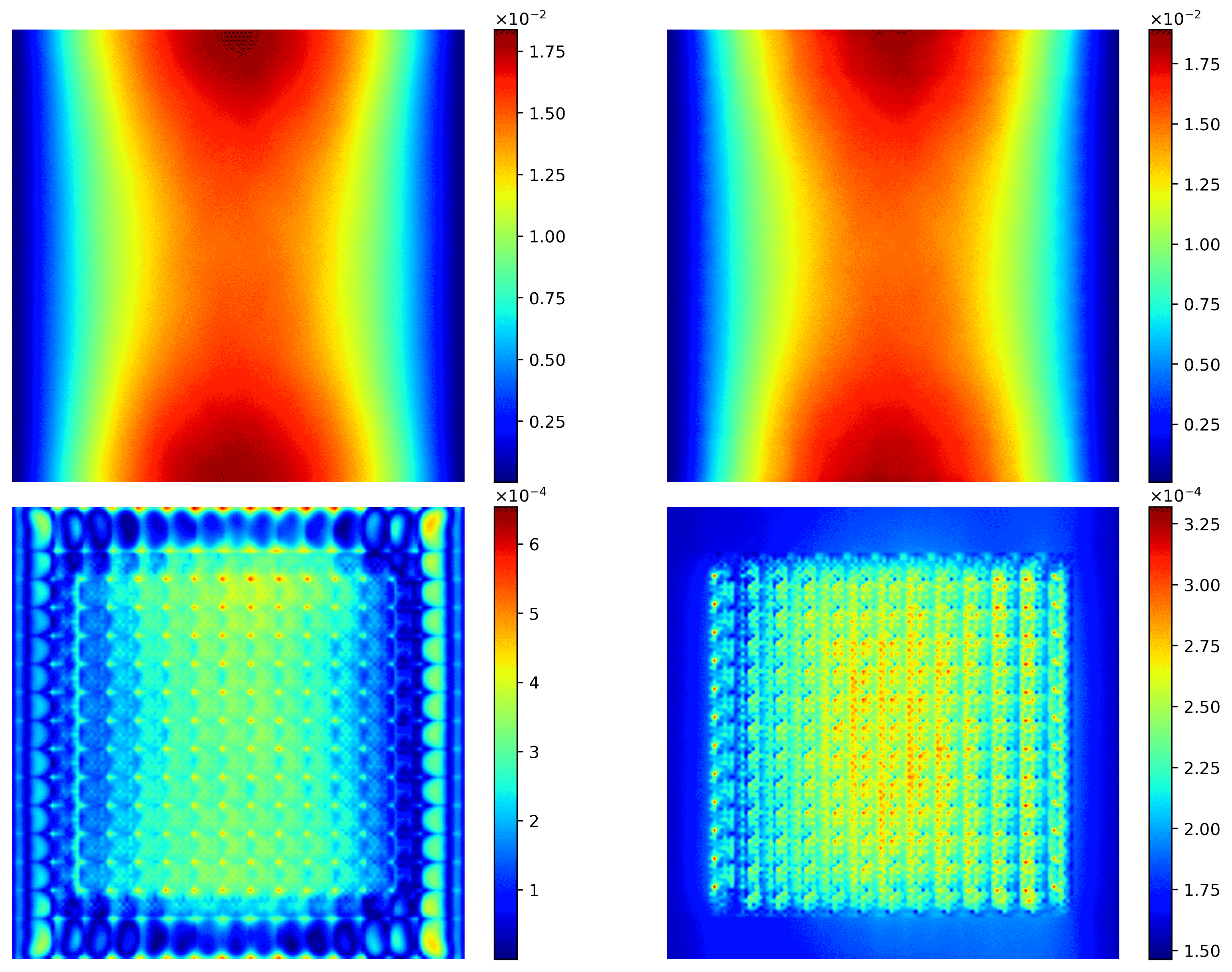}
  \end{minipage}
    \caption{ Uncertainty propagation for KLE$-1000$  ($128$~training data). (Left) We show the Monte Carlo output mean, predictive output mean $\mathbb{E}_{\bm{\theta}}[\mathbb{E}[\bm{y}|\bm{\theta}]]$, the error of the above two, and two standard deviations of the conditional  predictive mean $Var_{\bm{\theta}}[\mathbb{E}[\bm{y}|\bm{\theta}]]$. (Right)  We show the Monte Carlo output variance, predictive output variance $\mathbb{E}_{\bm{\theta}}[Var(\bm{y} | \bm{\theta})]$, the error of the above two, and two standard deviations of the conditional  predictive variance $\text{Var}_{\bm{\theta}} (\text{Var}(\bm{y} | \bm{\theta}))$.}
    \label{fig:UP_1000_128}
\end{figure}
\begin{figure}[H]
  \centering
  \begin{minipage}[b]{0.48\textwidth}
    \includegraphics[width=\textwidth]{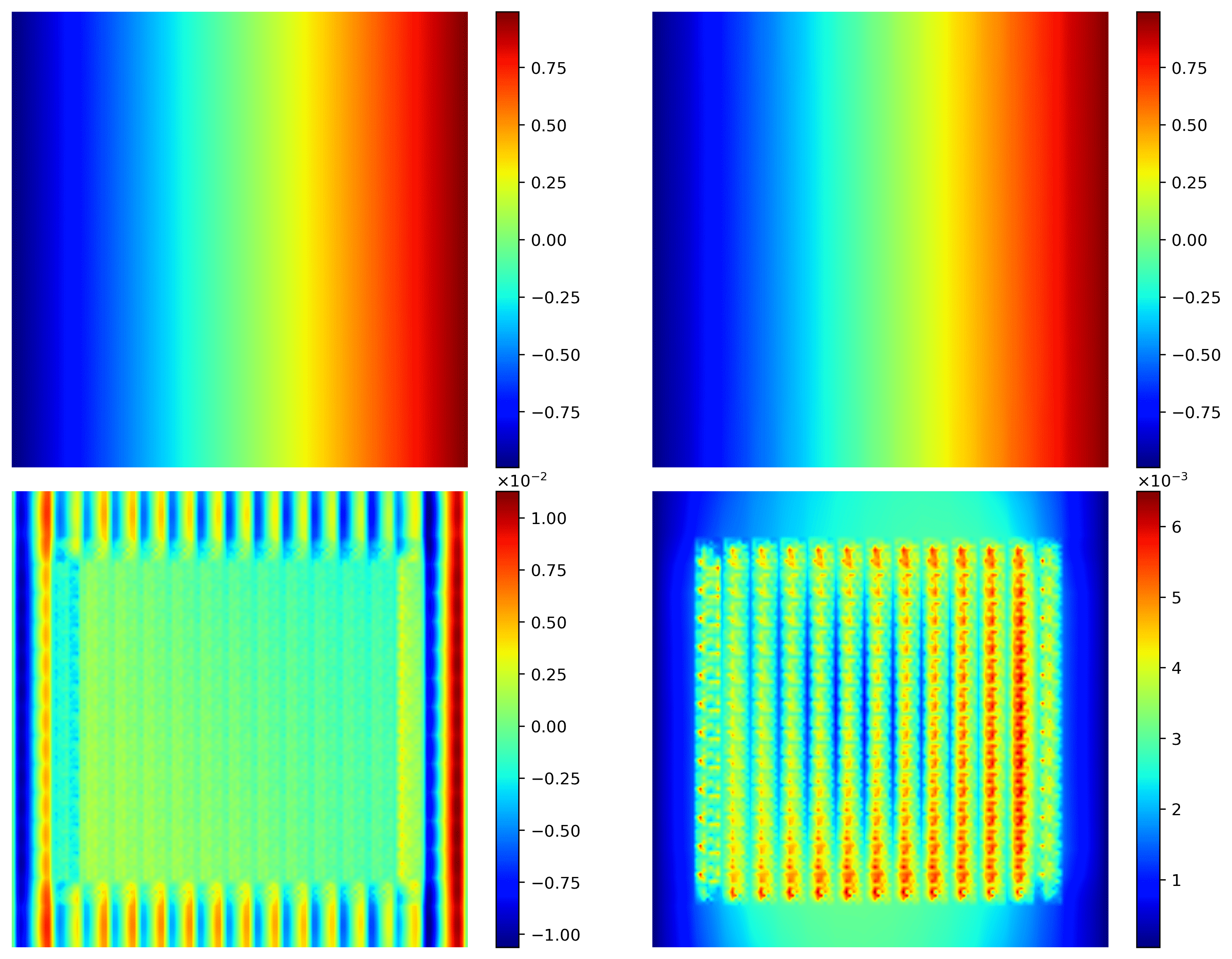}
  \end{minipage}
  \hfill
  \begin{minipage}[b]{0.48\textwidth}
    \includegraphics[width=\textwidth]{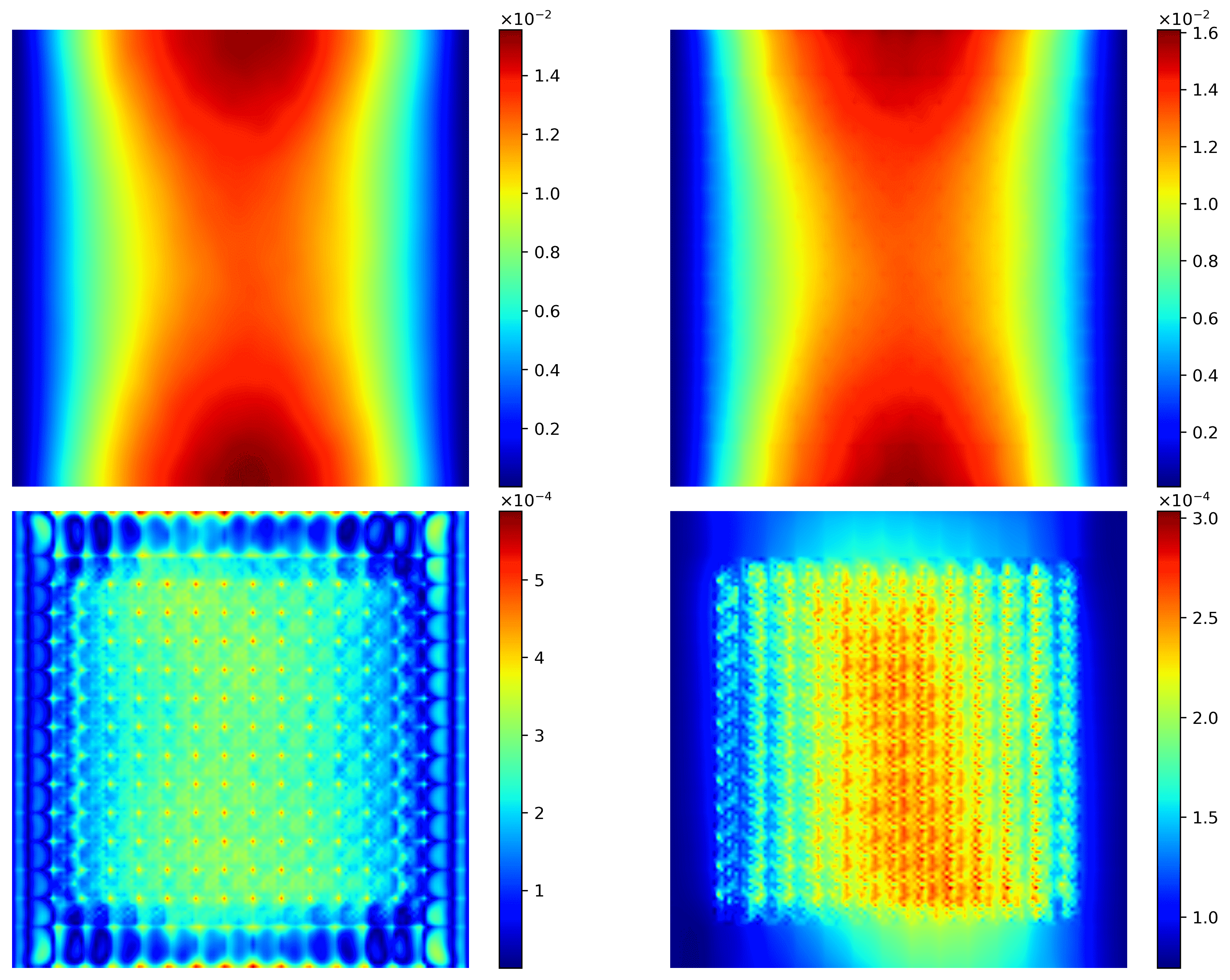}
  \end{minipage}
    \caption{Uncertainty propagation for KLE$-16384$ ($96$~ training data). (Left)  We show the Monte Carlo output mean, predictive output mean $\mathbb{E}_{\bm{\theta}}[\mathbb{E}[\bm{y}|\bm{\theta}]]$, the error of the above two, and two standard deviations of the conditional  predictive mean $Var_{\bm{\theta}}[\mathbb{E}[\bm{y}|\bm{\theta}]]$. (Right) We show the Monte Carlo output variance, predictive output variance $\mathbb{E}_{\bm{\theta}}[Var(\bm{y} | \bm{\theta})]$, the error of the above two, and two standard deviations of the conditional  predictive variance $\text{Var}_{\bm{\theta}} (\text{Var}(\bm{y} | \bm{\theta}))$.}
    \label{fig:M1_UQ3}
\end{figure}
\begin{figure}[htp]
  \centering
  \begin{minipage}[b]{0.48\textwidth}
    \includegraphics[width=\textwidth]{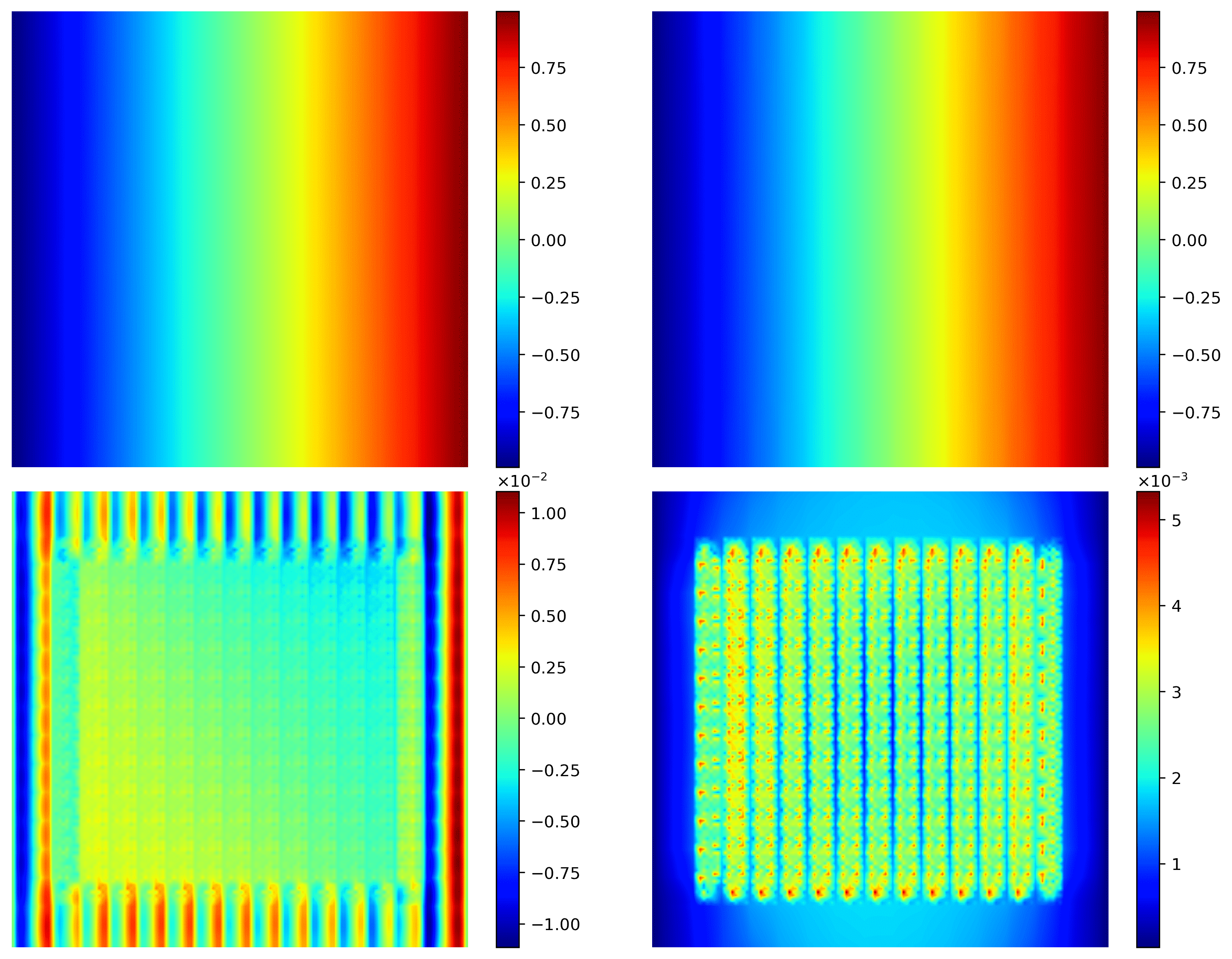}
  \end{minipage}
  \hfill
  \begin{minipage}[b]{0.48\textwidth}
    \includegraphics[width=\textwidth]{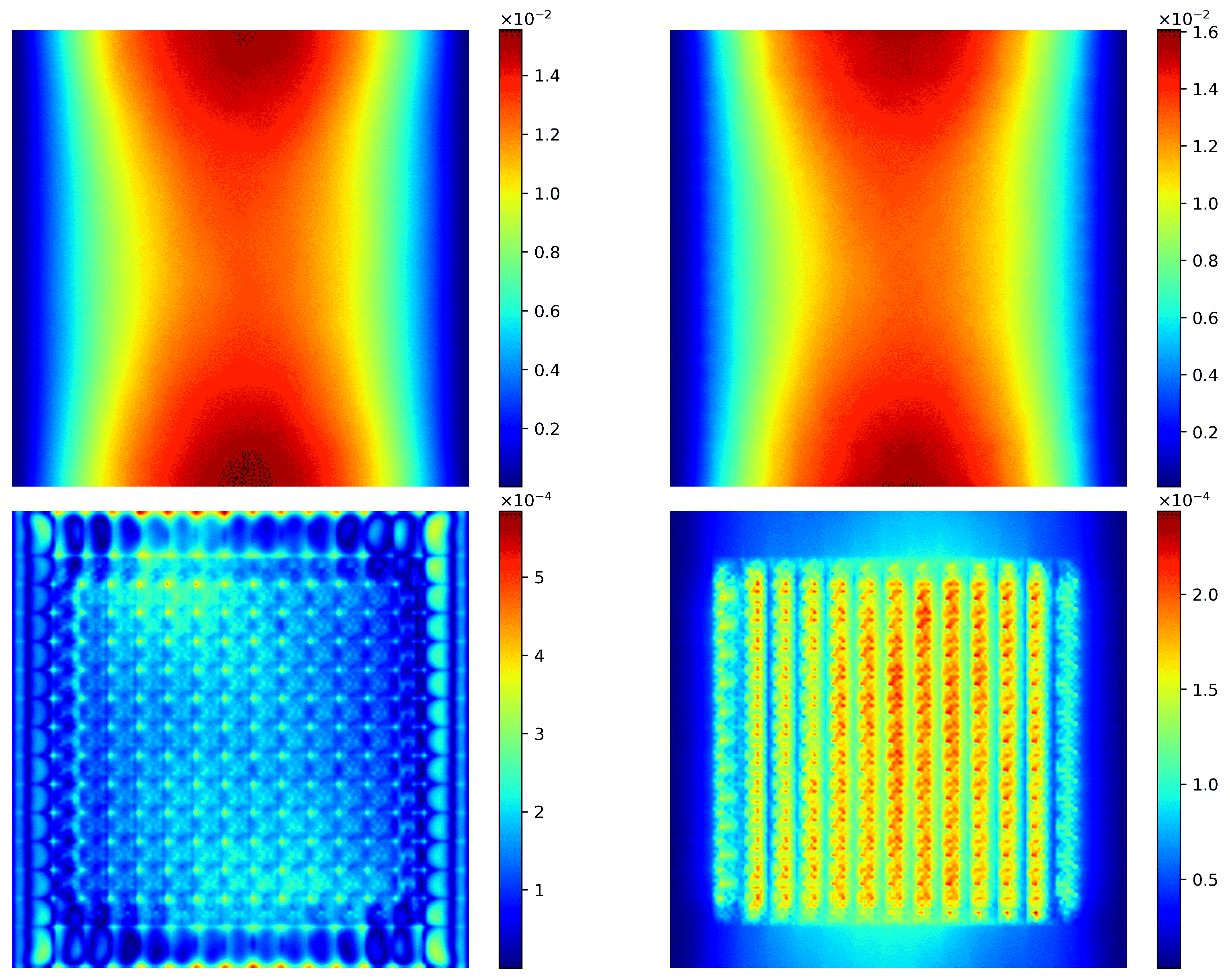}
  \end{minipage}
    \caption{Uncertainty propagation for KLE$-16384$ ($160$ training data).  (Left) We show the Monte Carlo output mean, predictive output mean $\mathbb{E}_{\bm{\theta}}[\mathbb{E}[\bm{y}|\bm{\theta}]]$, the error of the above two, and two standard deviations of the conditional  predictive mean $Var_{\bm{\theta}}[\mathbb{E}[\bm{y}|\bm{\theta}]]$. (Right) We show the Monte Carlo output variance, predictive output variance $\mathbb{E}_{\bm{\theta}}[Var(\bm{y} | \bm{\theta})]$, the error of the above two, and two standard deviations of the conditional  predictive variance $\text{Var}_{\bm{\theta}} (\text{Var}(\bm{y} | \bm{\theta}))$.}
    \label{fig:M1_UQ5}
\end{figure}
\begin{figure}[H]
  \centering
  \begin{minipage}[b]{0.48\textwidth}
    \includegraphics[width=\textwidth]{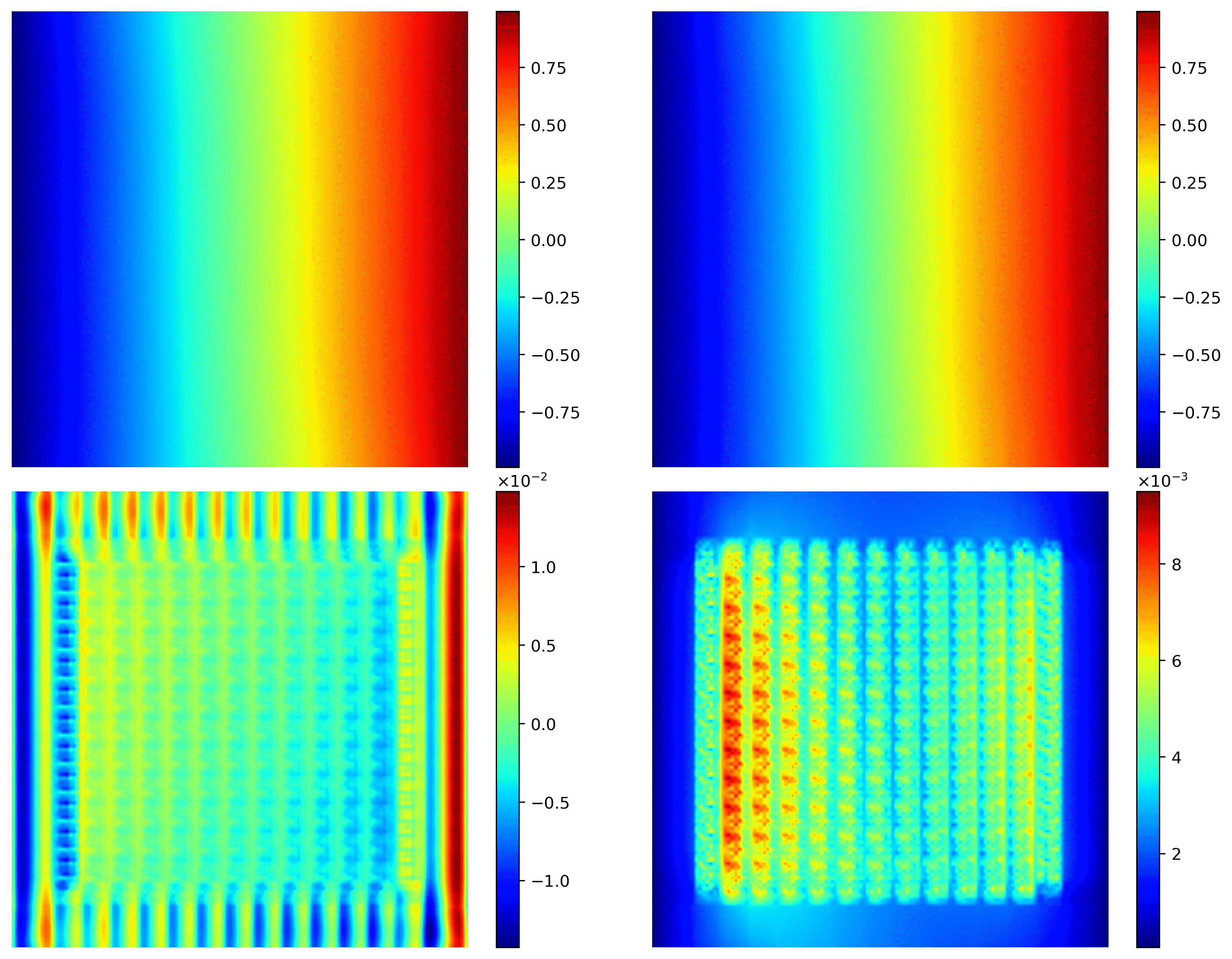}
  \end{minipage}
  \hfill
  \begin{minipage}[b]{0.48\textwidth}
    \includegraphics[width=\textwidth]{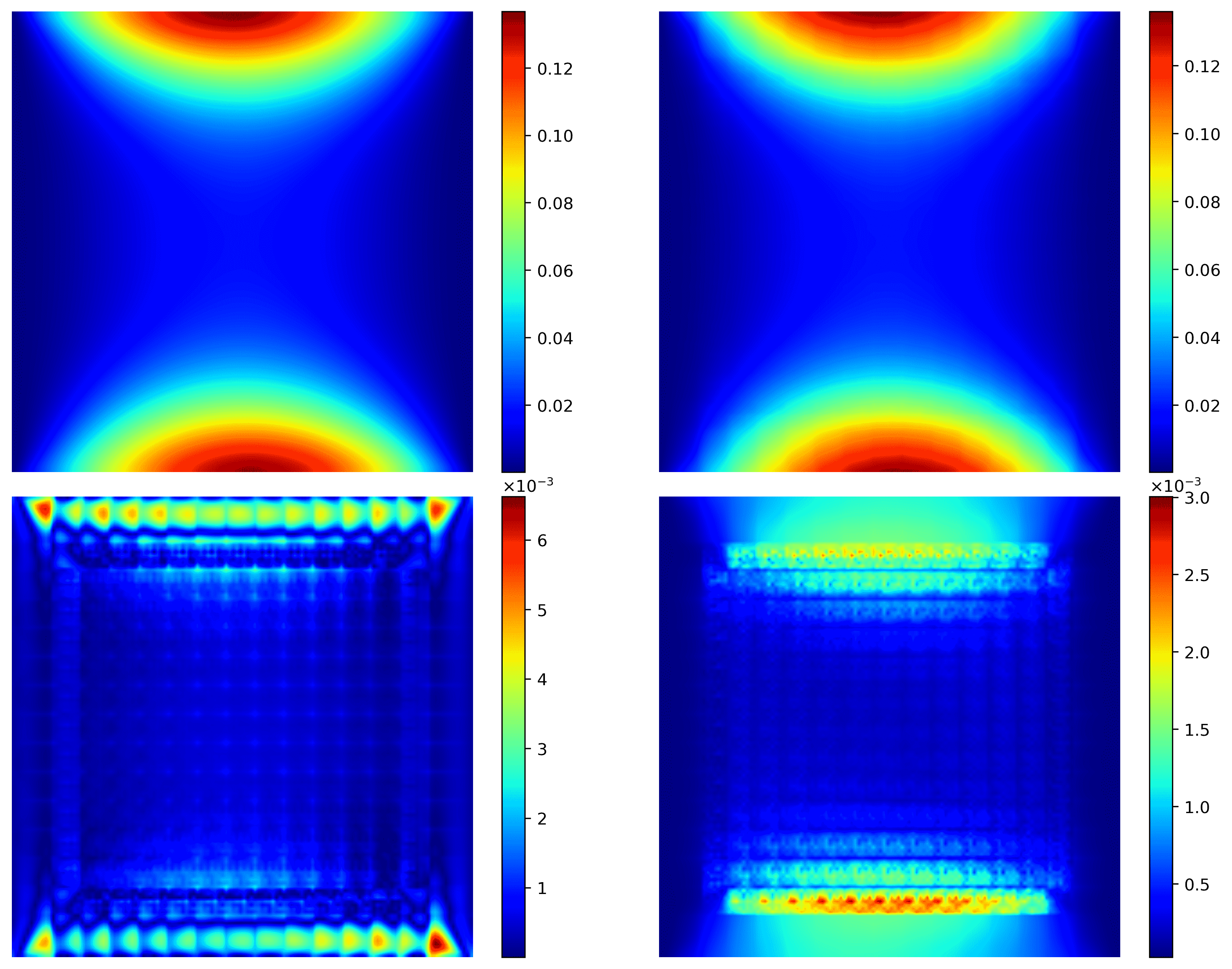}
  \end{minipage}
    \caption{Uncertainty propagation for channelized field  ($160$ training data). (Left)  We show the Monte Carlo output mean, predictive output mean $\mathbb{E}_{\bm{\theta}}[\mathbb{E}[\bm{y}|\bm{\theta}]]$, the error of the above two, and two standard deviations of the conditional  predictive mean $Var_{\bm{\theta}}[\mathbb{E}[\bm{y}|\bm{\theta}]]$. (Right) We show the Monte Carlo output variance, predictive output variance $\mathbb{E}_{\bm{\theta}}[Var(\bm{y} | \bm{\theta})]$, the error of the above two, and two standard deviations of the conditional  predictive variance $\text{Var}_{\bm{\theta}} (\text{Var}(\bm{y} | \bm{\theta}))$.}
    \label{fig:M_UQ}
\end{figure}

In order to assess the quality of the predictive model, we use the mean negative log-probability metric, which evaluates the likelihood of the observed data. Test MNLP for various training data for KLE$-100$, KLE$-1000$ and KLE$-16384$ are shown in Fig.~\ref{fig:MNLP}. From Fig.~\ref{fig:comp_r2}, the test $R^2$ for the Bayesian surrogate is obtained by comparing the predictive output mean with the target pressure. We observe that the test $R^2$ for the  Bayesian surrogate is better when compared to the non-Bayesian surrogate model. The training and testing loss for the Bayesian HM-DenseED is shown in Fig.~\ref{fig:RMSE_Bayesian}. Here, we consider the root mean square between the mean of the predicted pressure from $S = 20$ samples and the fine-scale pressure.  
We perform uncertainty propagation using the trained Bayesian surrogate with $10,000$ input realizations and compare with the Monte Carlo output (fine-scale output), as illustrated in Figs.~\ref{fig:UP_32}-\ref{fig:M_UQ}.  In the variance plot, we observe that the maximum variance with square pattern appears in the interior support regions. This is due to the fact that the Bayesian-HM-DenseED model (where we consider $S$ set of HM-DenseED deterministic models) is trained for the permeability field corresponding to the interior support region as the input, and the output is the basis functions for the interior support region. In the error plots (Figs.~\ref{fig:UP_32}-\ref{fig:M_UQ}),  the error is evaluated by considering the difference between the Monte Carlo output mean and predictive output mean. Also, the error is evaluated between the Monte Carlo output variance and predictive output variance. Since the Bayesian-HM-DenseED model has direct influence over the basis functions corresponding to the interior support regions than the non-interior support region, we observe that the error is high near the non-interior support regions when compared to interior support regions.  This is because the basis functions for the non-interior support region $\bar{\bm{\Phi}}^{non-int}$  was generated without satisfying the partition of unity property~\cite{moyner2016multiscale}. We satisfy this property after predicting the basis functions corresponding to the interior support regions $\tilde{\bm{\Phi}}^{int}$ using the Bayesian-HM-DenseED model and combining it with the basis functions for the non-interior support region $\bar{\bm{\Phi}}^{non-int}$. 
\par
Finally, to further evaluate the performance of the Bayesian model, we show the kernel density estimates for the pressure at the location $(0.64, 0.54)$ in Appendix~\ref{sec:App-D}.

\section{Conclusions}\label{sec:Conclusions}
This work outlines a Bayesian hybrid deep-learning - multiscale approach for uncertainty quantification and propagation in problems governed by stochastic PDEs (SPDEs)  with high-dimensional stochastic input. In this work, we present a novel hybrid framework by combining deep learning and multiscale approaches. The MsRSB method is used to generate the basis functions, which are then combined to obtain a prolongation operator that maps  quantities such as pressure from coarse-scale to fine-scale. In this work, we use a dense convolutional encoder-decoder network (Dense-ED) to learn the basis functions with the loss function defined between the predicted pressure (using the hybrid framework) and the fine-scale pressure. Using Stein variational gradient descent, we extend the hybrid deep-learning-multiscale approach to a Bayesian framework~\cite{zhu2018bayesian}. This allows us to provide uncertainty estimates for our neural network predictions that include the epistemic uncertainty that occurs when training with a small dataset as well as the loss of information while performing the multiscale approach. We show that this hybrid method yields accurate results in high-dimensional settings.   
\par
The model was demonstrated in a two-dimensional, single-phase, steady-state flow through a random permeability field.  To further develop this model, one could implement multiphase flow in porous media as well as fractured reservoirs. Another possible extension is using the combination of data-driven and physics-constrained learning in this framework.  


\section*{Acknowledgements}
The work reported here was supported  from the Defense Advanced Research Projects Agency (DARPA) under the Physics of Artificial Intelligence (PAI) program (contract HR$00111890034$) and  by ARPA-E award \# DE-AR0001204. 
The authors acknowledge computing resources provided by the AFOSR Office
of Scientific Research through the DURIP program  and by the University of Notre Dame’s Center for Research Computing (CRC). 
\appendix
\section{DenseED Configurations}\label{sec:App-B1}
We train the DenseED for different configurations in order to obtain an optimized architecture  for the HM-DenseED model. To be specific, we select the DenseED configuration that gives   better performance with less number of parameters to train. We have trained the following configurations: 
\begin{itemize}
    \item Starting from a simple configuration: $1-1$ where we consider one encoder dense block, and one decoder dense block. 
    \item Increased the number of blocks to $1-1-1$ and $1-1-1-1-1$
    \item Gradually increased the encoder and decoder blocks in the network to $4-8-4$, and $6-12-6$. 
\end{itemize}
We train the above configurations with the same data (also investigated datasets of size $32$, $64$ and $96$), same learning rate, weight decay, batch size and other hyper-parameters. 

\begin{table}[H]
\caption{Test $R^2$ score for different configurations and data for KLE$-100$.}
\begin{tabular}{cccc}
\hline
Data $\longrightarrow$           & \multirow{2}{*}{$\bm{32}$} & \multirow{2}{*}{$\bm{64}$} & \multirow{2}{*}{$\bm{96}$} \\
Configurations \\ \hspace{0.5cm}$\downarrow$&                     &                     &                     \\ \hline
$\bm{1-1}$          & $0.799$              & $0.9317$              & $0.966$               \\
$\bm{1-1-1}$          & $0.859$               & $0.9538$              & $0.968$               \\
$\bm{1-1-1-1-1 }$     & $0.825$               & $0.9198$              & $0.964$              \\
$\bm{4-8-4}$          & $0.962$               & $0.97$                & $0.973$               \\
$\bm{6-12-6}$          & $0.9624$              & $0.972$               & $0.9745$              \\ \hline
\end{tabular}
\label{tab:diff_config}
\end{table}

From Table~\ref{tab:diff_config}, we observe that the $4-8-4$ configuration is optimum as it yields better test $R^2$ compared to the $1-1$, $1-1-1$  and $1-1-1-1-1$ configurations. Also, the $R^2$ for the $4-8-4$ and $6-12-6$ configurations is almost the same.
\section{Comparison of DenseED to a fully-connected network for training the basis functions}\label{sec:App-C}
\begin{figure}[H]
    \centering
    \includegraphics[scale=0.25]{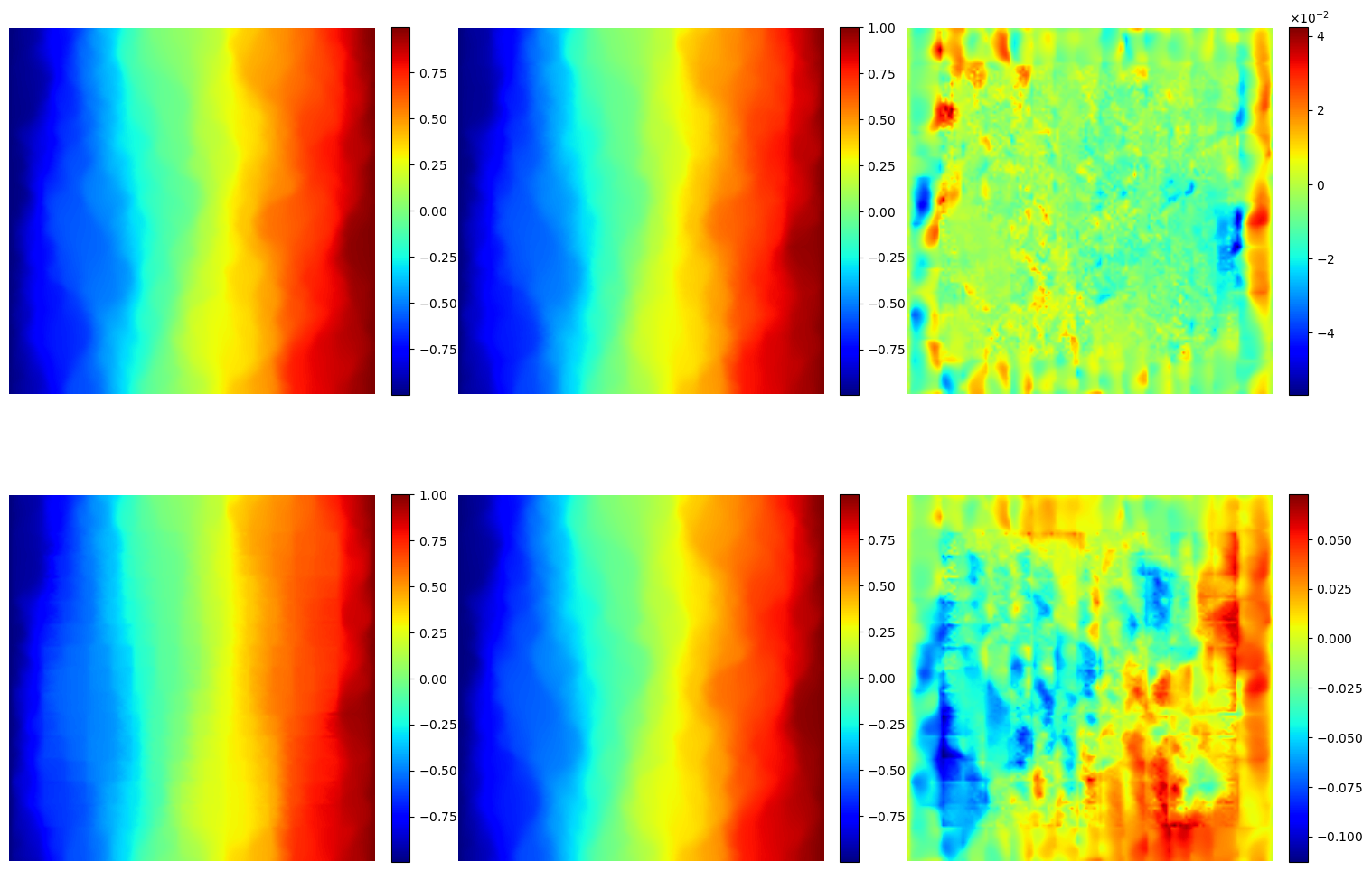}
    \caption{Comparison of DenseED (first row) and fully-connected network (second row) for a test set.}
    \label{fig:C_FC}
\end{figure}

In this appendix, we compare the DenseED and fully-connected network (FCN) for training the basis functions (see Fig.~\ref{fig:C_FC}). For this example, we consider KLE$-16384$ with $160$ fine-scale data for training. The configuration for training both  networks is given in  Table~\ref{tab:Comparison}.
\begin{table}[H]
\caption{Comparison of the DenseED with a fully-connected network for learning the basis functions.}
\centering
\begin{tabular}{ccc}
\hline
                       & \textit{Hybrid DenseED-multiscale} & \textit{Hybrid fully-connected}                                             \\ \hline
\textbf{Configuration} & $4-8-4$                              & $225$ $\rightarrow$ $144$ $\rightarrow 64 \rightarrow 144 \rightarrow 225$ \\ \hline
\textbf{Learning rate} & $1e-5$                               & $1e-4$                                                                 \\ \hline
\textbf{Weight decay}  & $1e-6$                               & $1e-5$                                                                 \\ \hline
\textbf{Optimizer}     & Adam                               & Adam                                                                 \\ \hline
\textbf{Epochs}        & $200$                                & $200$                                                                  \\ \hline
\end{tabular}

\label{tab:Comparison}
\end{table}

\section{Distribution estimate for the pressure (Bayesian surrogate trained)}\label{sec:App-D}
Herein, we show the estimate of the pressure at  location $(0.64, 0.54)$ on the unit square. We obtain the confidence interval (CI) by taking the maximum and minimum density estimate of $20$ samples (DenseED networks). We can see that the density estimate is close to the fine-scale (Monte Carlo) results. Even for high-dimensionality, the results become closer to the fine-scale solution with increased size training datasets.
\begin{figure}[H]
    \centering
    \subfigure[KLE$-100$ ($32-$data)]
    {
        \includegraphics[width=0.3\textwidth]{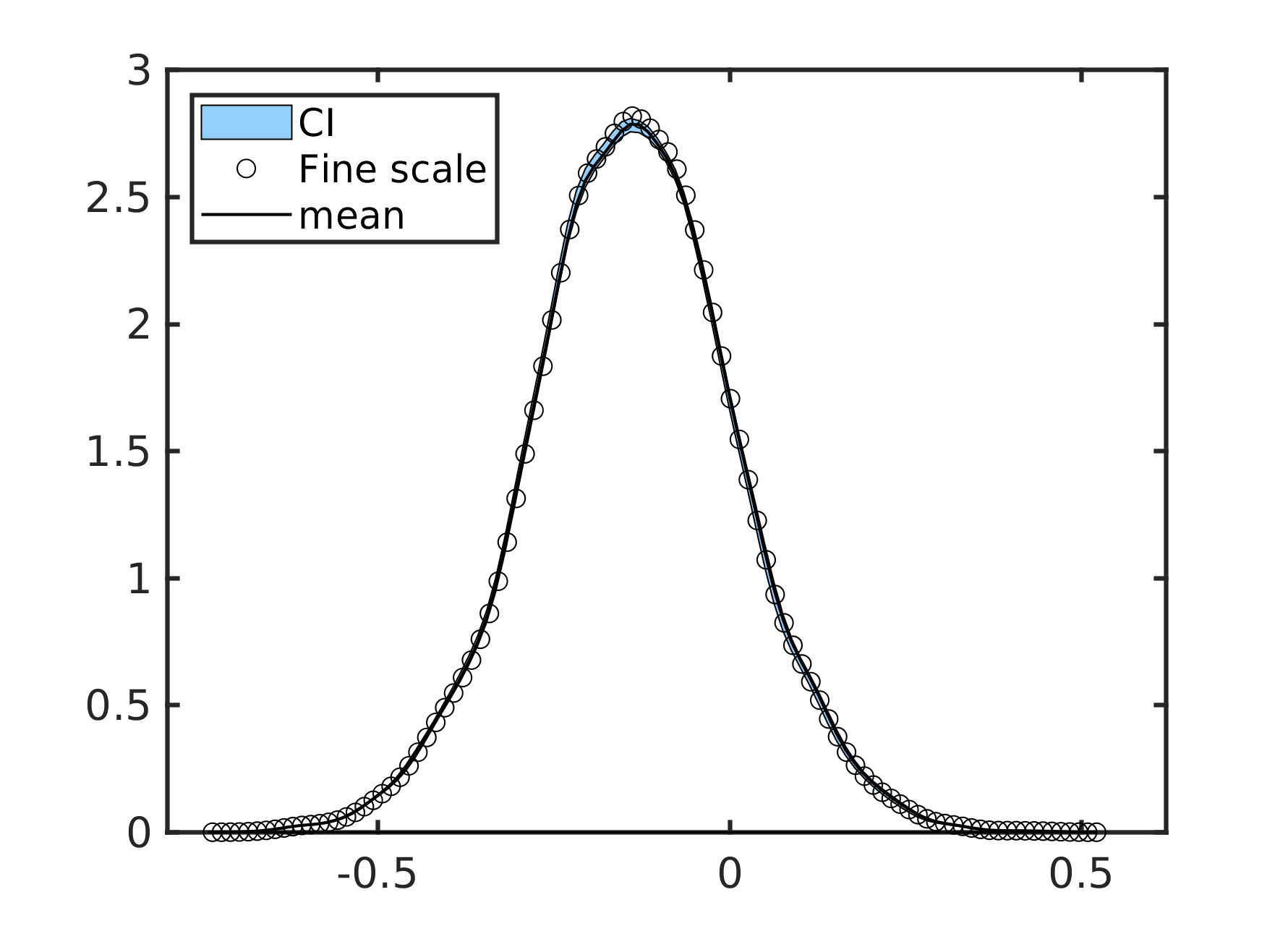}
        \label{fig:first_sub1a}
    }
    \subfigure[KLE$-100$ ($64-$data)]
    {
        \includegraphics[width=0.3\textwidth]{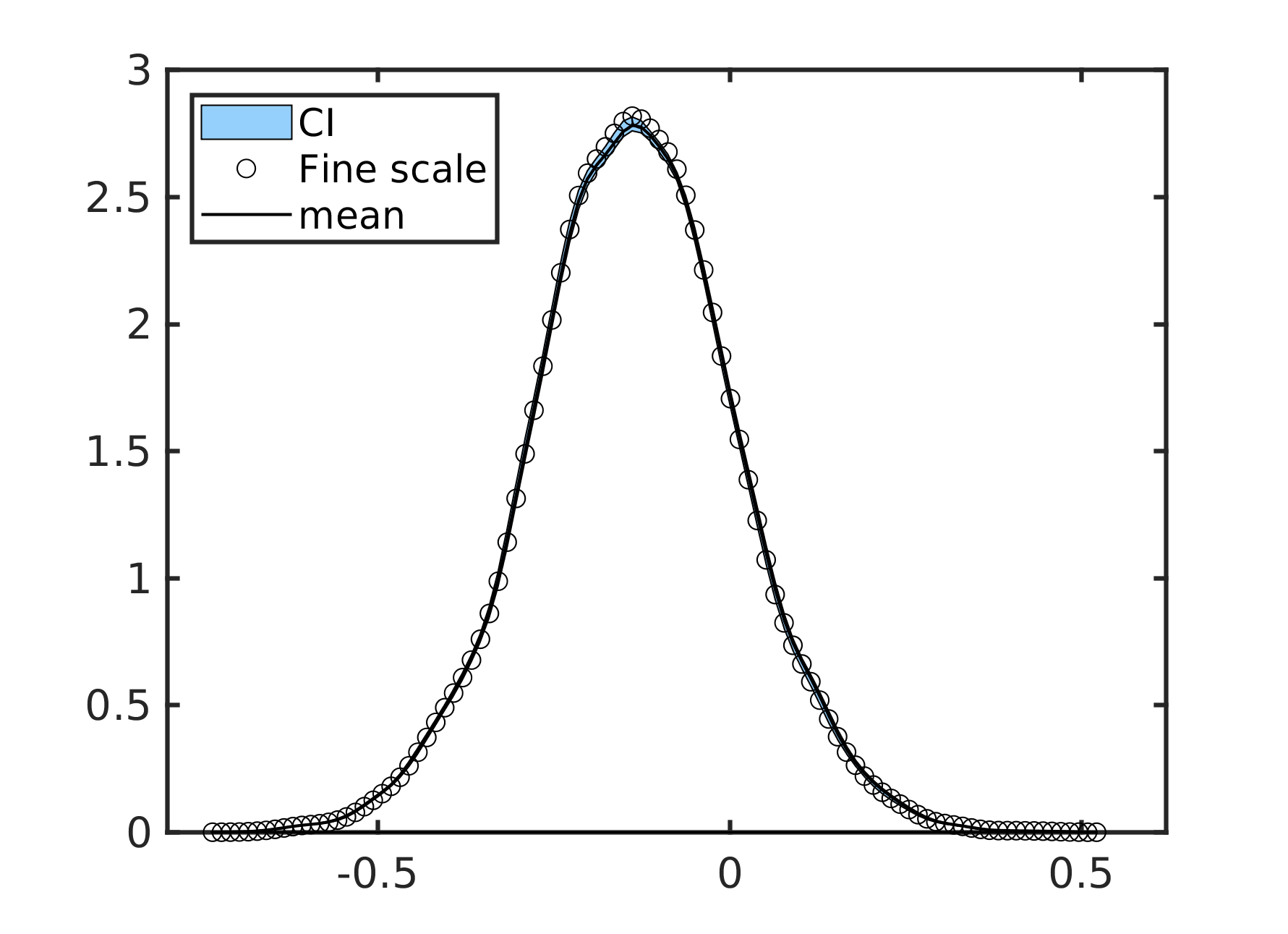}
        \label{fig:first_subb}
    }
    \\
    \subfigure[KLE$-100$ ($96-$data)]
    {
        \includegraphics[width=0.3\textwidth]{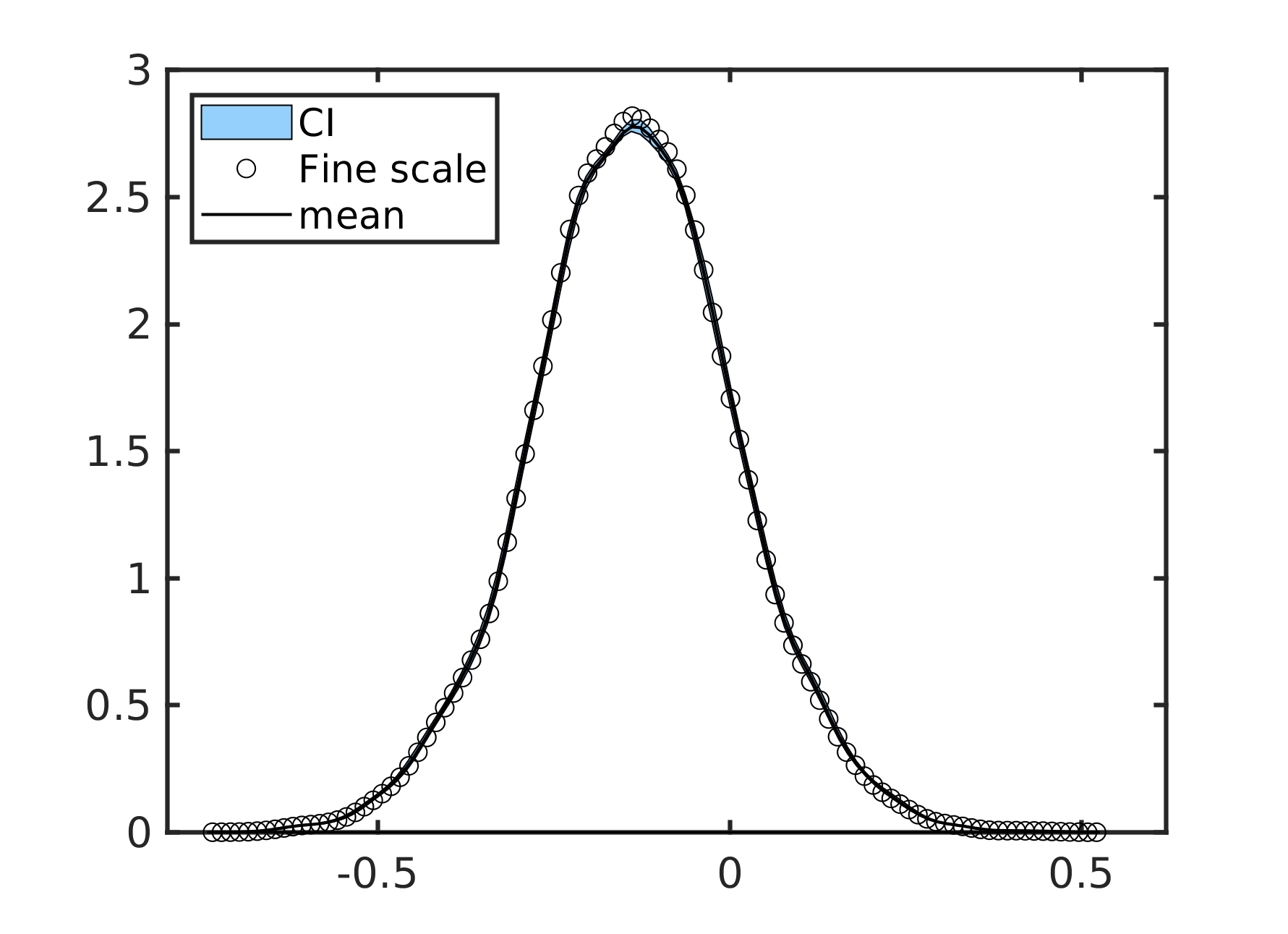}
        \label{fig:first_subc}
    }
    \subfigure[KLE$-1000$ ($64-$data)]
    {
        \includegraphics[width=0.3\textwidth]{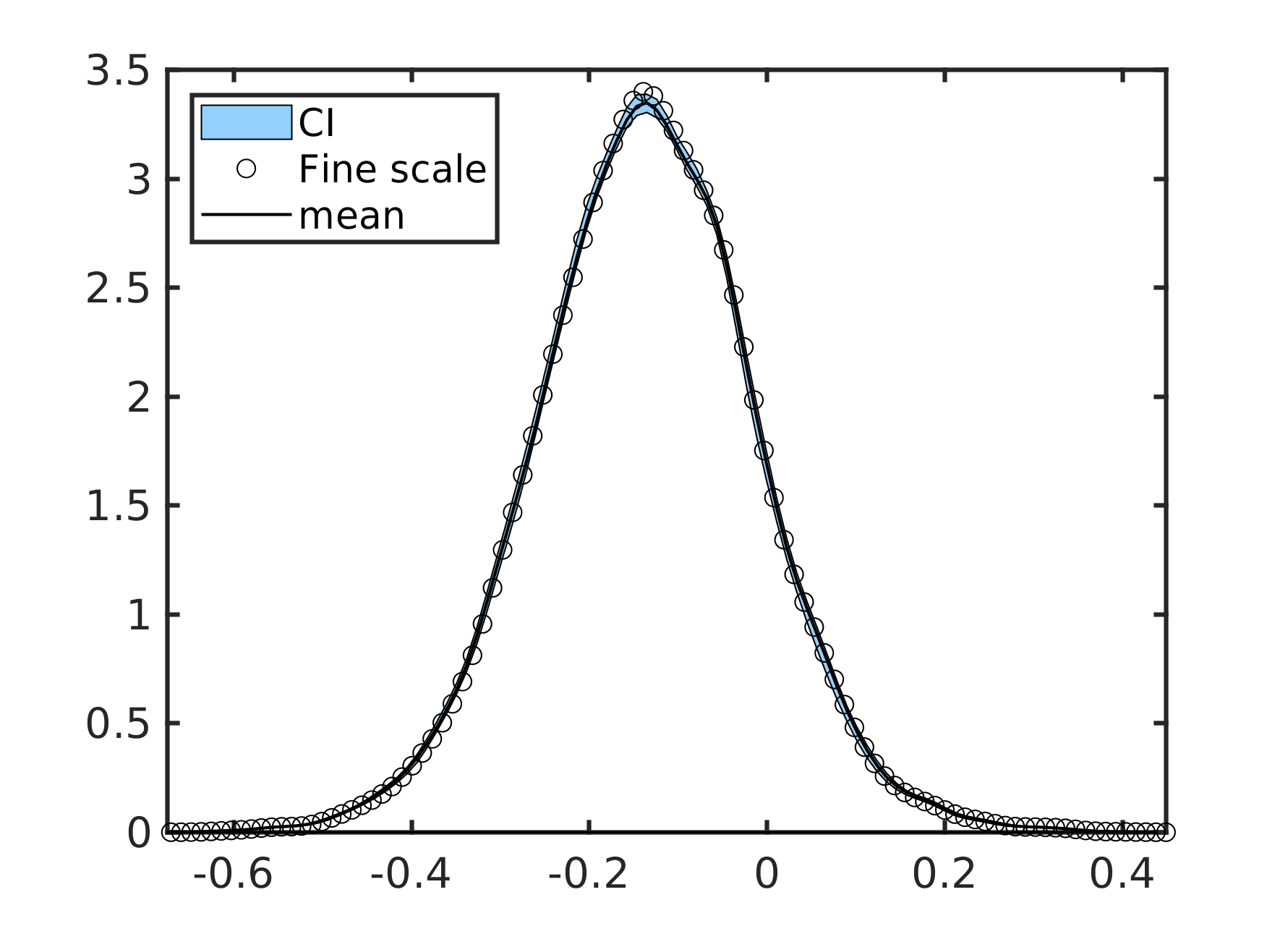}
        \label{fig:first_subd}
    }
    \\
    \subfigure[KLE$-1000$ ($96-$data)]
    {
        \includegraphics[width=0.3\textwidth]{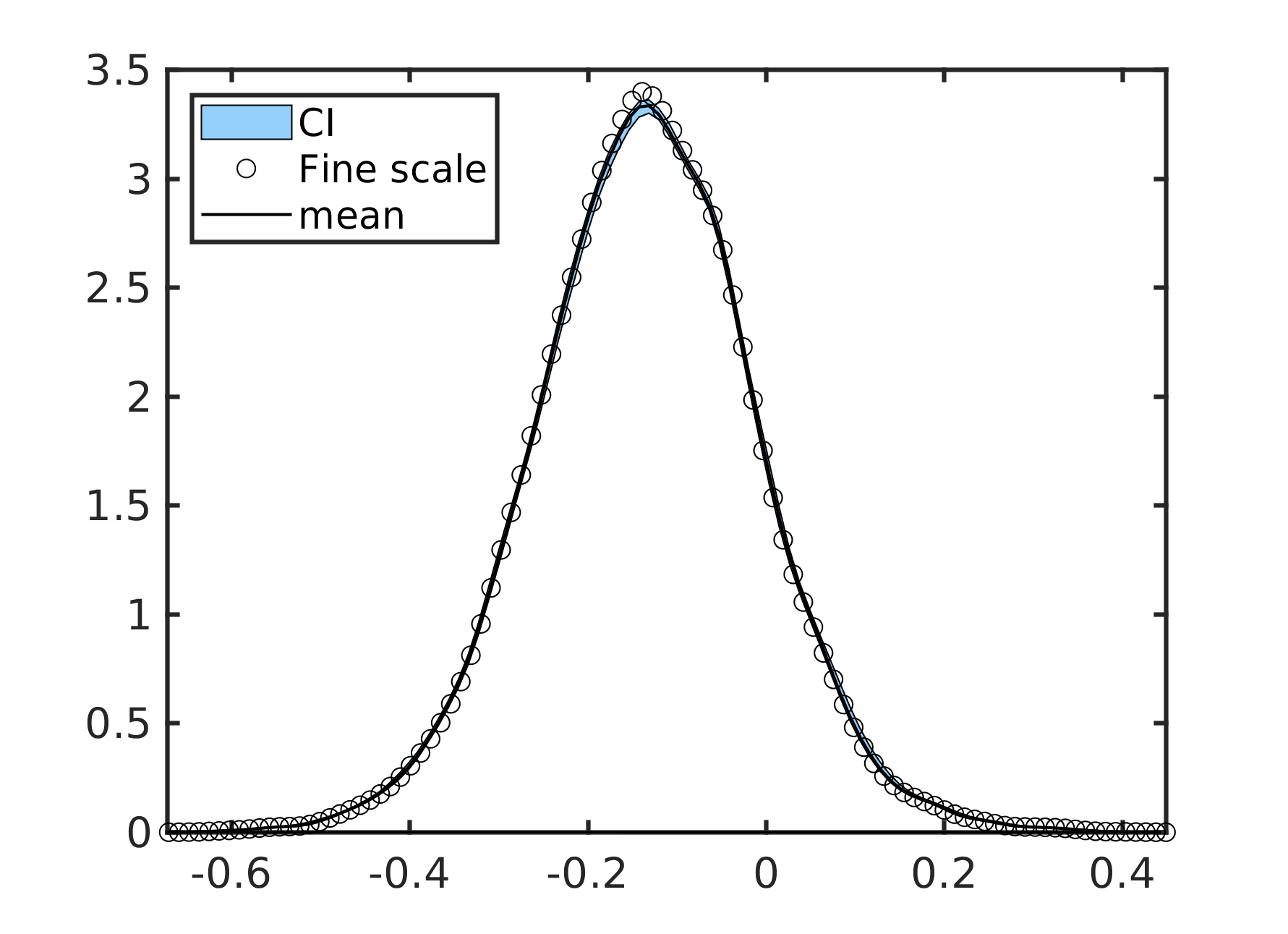}
        \label{fig:first_sube}
    }
    \subfigure[KLE$-1000$ ($128-$data)]
    {
        \includegraphics[width=0.3\textwidth]{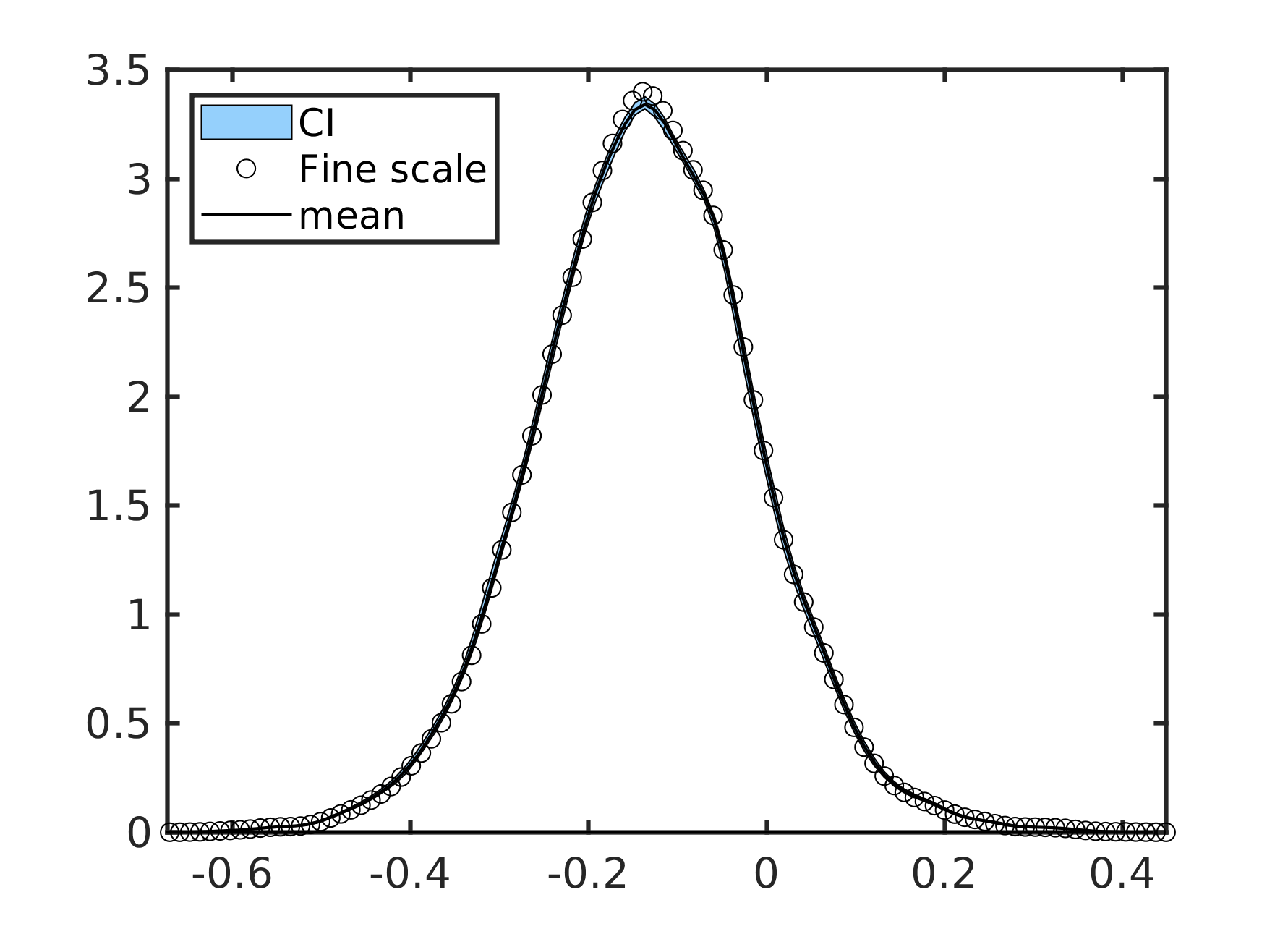}
        \label{fig:first_subf}
    }
    \\
    \subfigure[KLE$-16384$ ($96-$data)]
    {
        \includegraphics[width=0.3\textwidth]{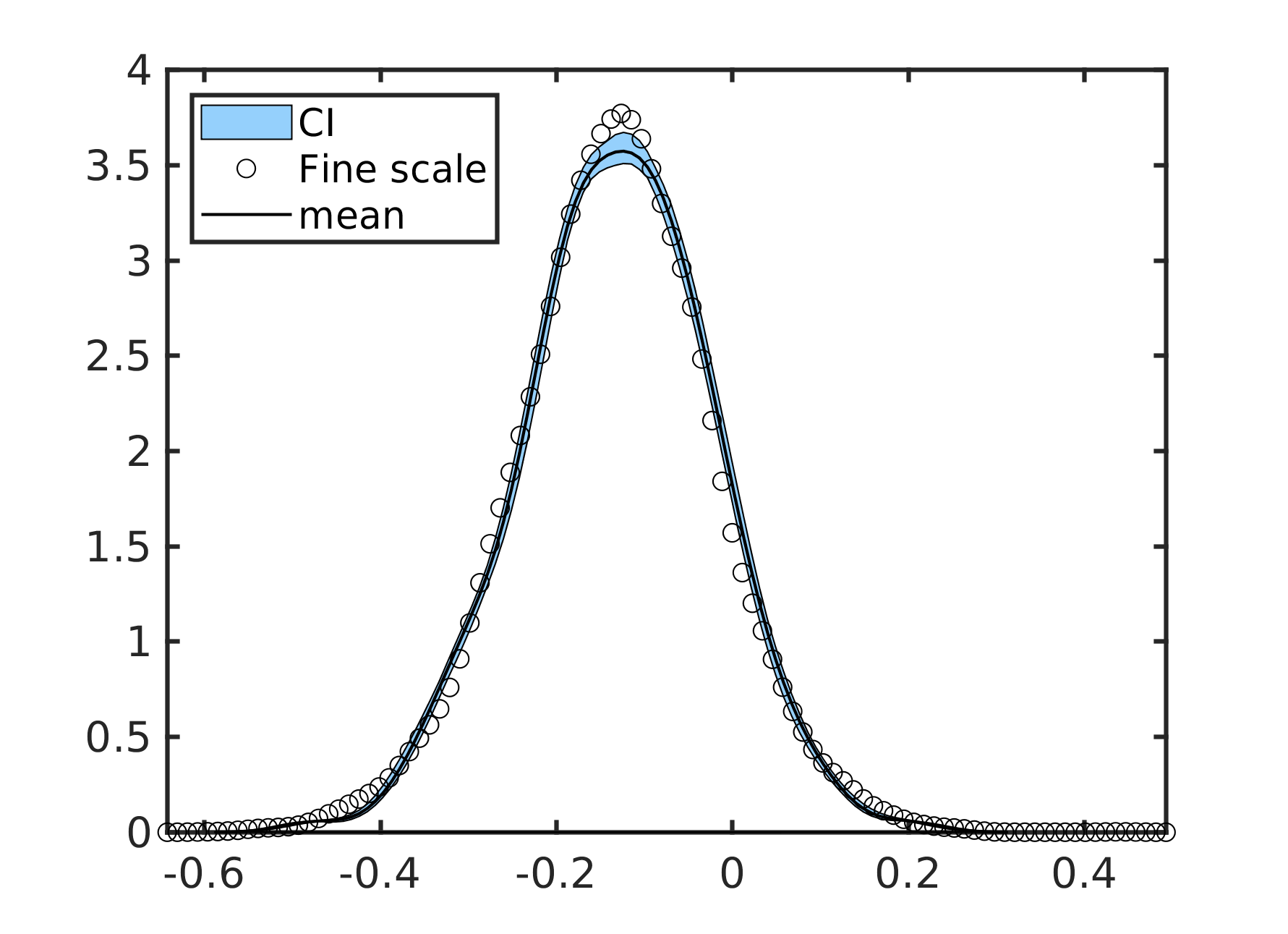}
        \label{fig:first_subg}
    }
    \subfigure[KLE$-16384$ ($128-$data)]
    {
        \includegraphics[width=0.3\textwidth]{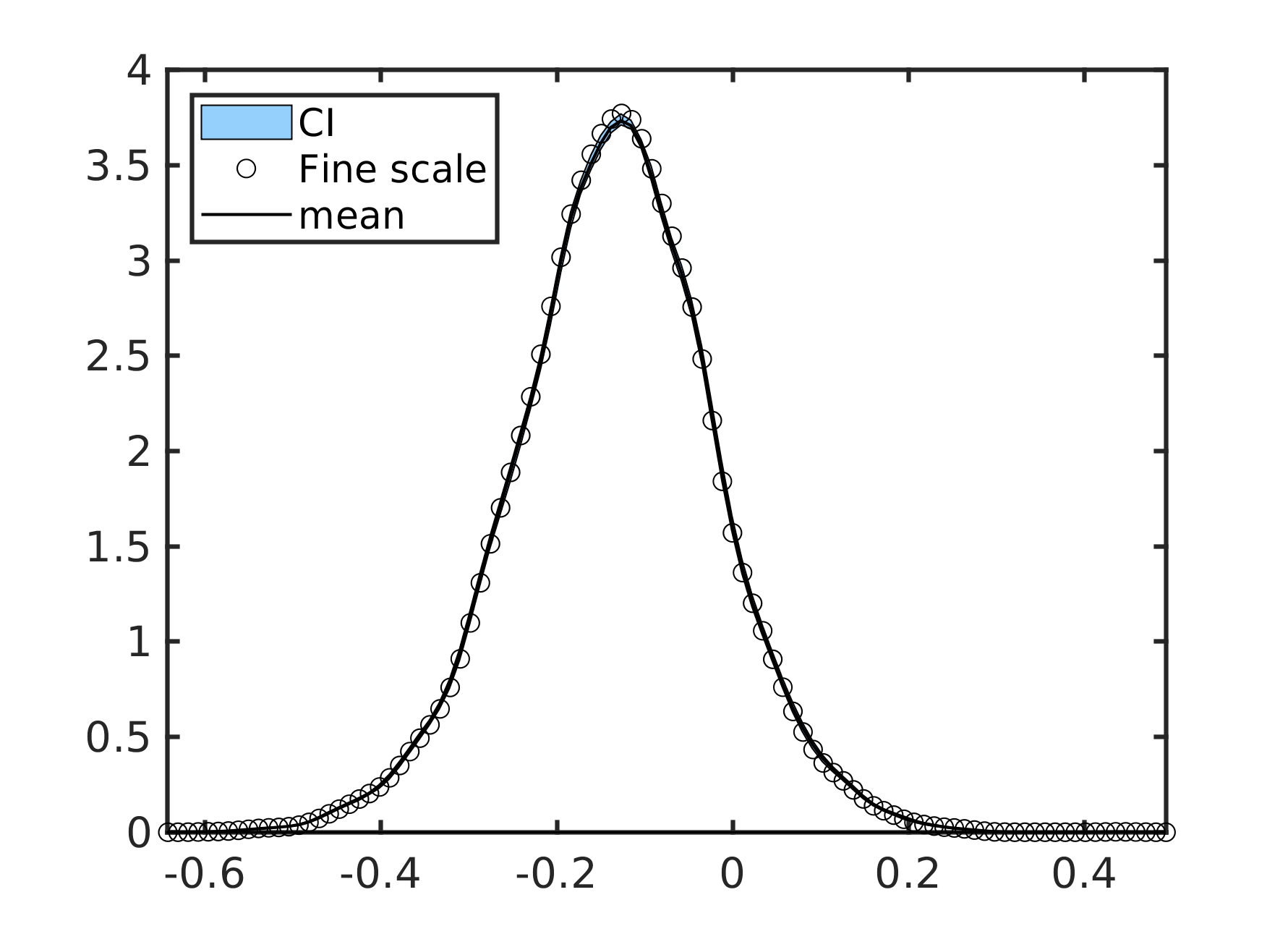}
        \label{fig:first_subh}
    }
    \\
    \subfigure[KLE$-16384$ ($160-$data)]
    {
        \includegraphics[width=0.3\textwidth]{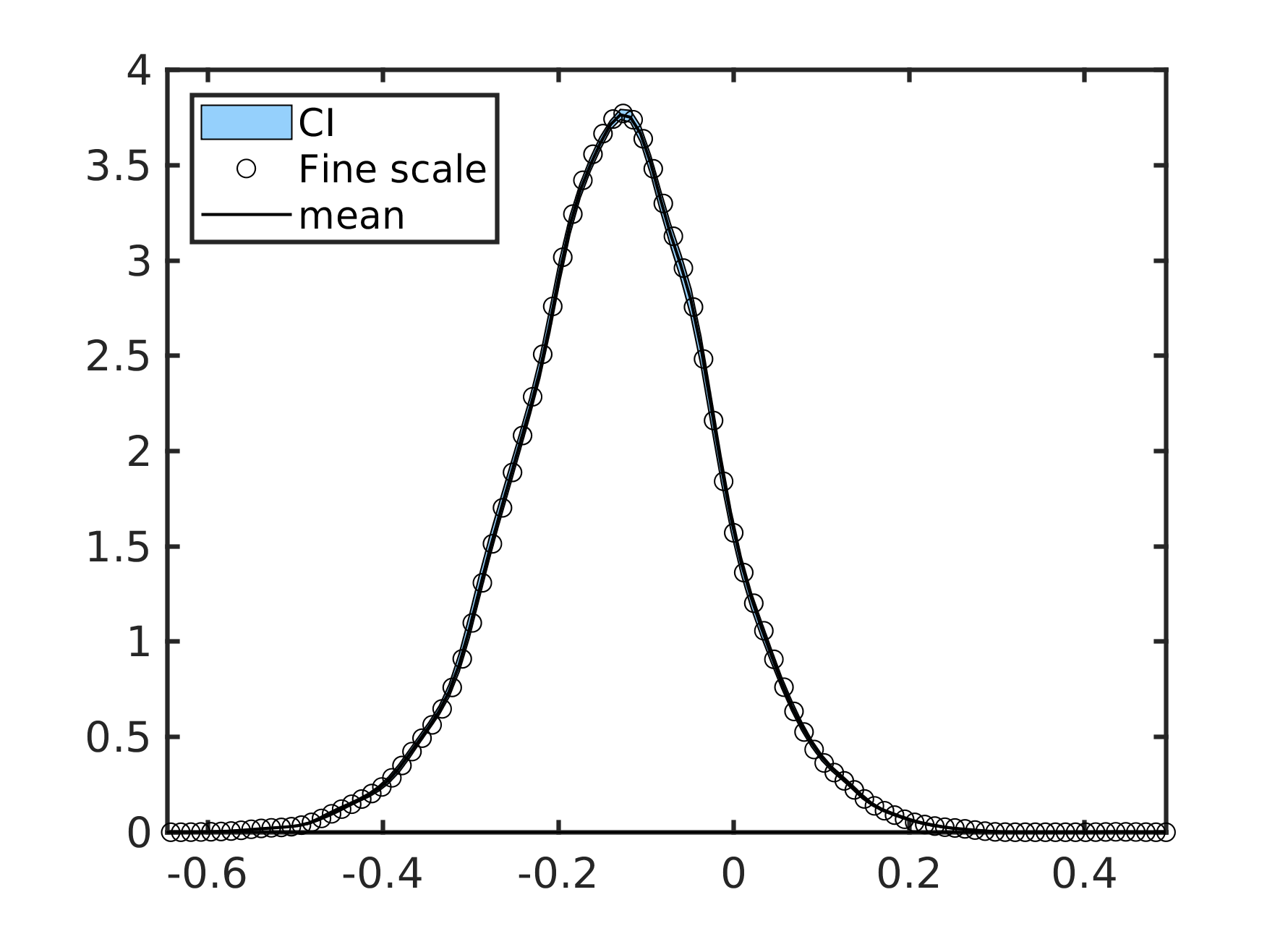}
        \label{fig:second_subi}
    }
    \subfigure[Channelized ($160-$data)]
    {
        \includegraphics[width=0.3\textwidth]{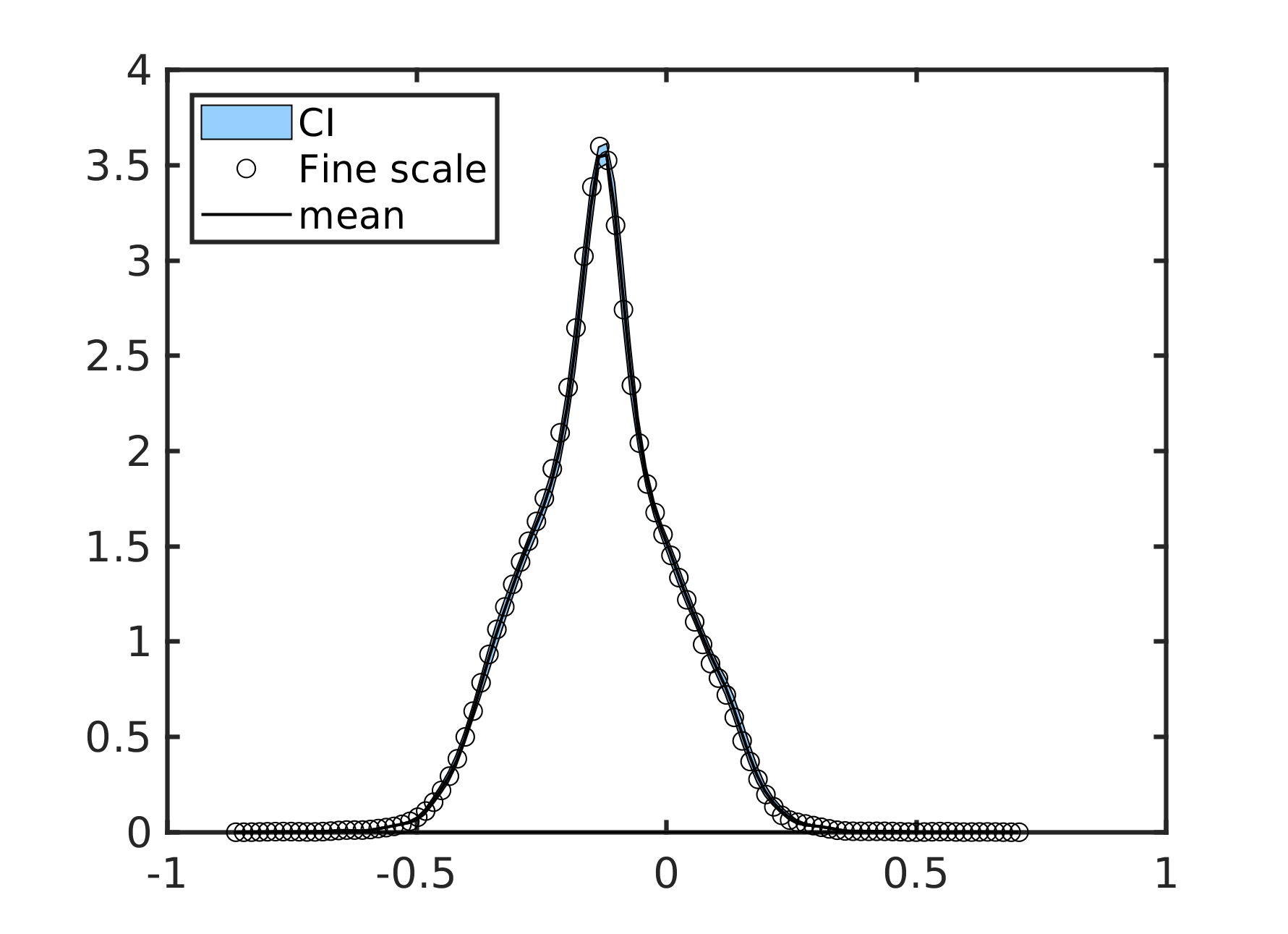}
        \label{fig:third_sub1j}
    }
    \caption{Distribution estimate for the pressure at location $(0.96, 0.54)$.}
    \label{fig:ComparePDFs}
\end{figure}

In order to evaluate the performance of the Bayesian surrogate trained HM-DenseED network, we test the network with $10,000$ input realizations and compare with the Monte Carlo output as illustrated in Fig.~\ref{fig:ComparePDFs}. We observe that the hybrid model is close to the Monte Carlo estimate (fine-scale) as we increase the number of training data. 
\providecommand{\href}[2]{#2}
\providecommand{\arxiv}[1]{\href{http://arxiv.org/abs/#1}{arXiv:#1}}
\providecommand{\url}[1]{\texttt{#1}}
\providecommand{\urlprefix}{URL }

\bibliographystyle{aims}
\medskip
\medskip
\end{document}